\documentclass[11pt,a4paper]{article} 
\usepackage{jheppub}

\usepackage[T1]{fontenc}
\usepackage[normalem]{ulem}
\usepackage{wasysym}
\usepackage{amscd,amsmath,amssymb,amsfonts,xspace,mathrsfs,amsthm}
\usepackage{color}
\usepackage{mathtools}

\usepackage{extarrows}
\usepackage{braket}
\usepackage[dvipsnames]{xcolor}
\usepackage{url}

\usepackage{latexsym}
\usepackage{graphicx}
\usepackage{dsfont}
\usepackage{longtable}
\usepackage{enumitem}
\usepackage{mathdots}
\usepackage{float}
\usepackage[utf8]{inputenc}
\usepackage{multirow}
\usepackage{caption}
\usepackage{subcaption}
\usepackage{tikz-cd}
\usetikzlibrary{bending}
\usepackage{tikz,textcomp}
\usetikzlibrary{decorations.pathreplacing,calc,fadings,fit,shapes,arrows,positioning,chains,matrix}
\usepackage{cleveref}
\crefname{subsection}{subsection}{subsections}
\usepackage{bbold}

\newtheorem{Thm*}{Theorem}

\newtheorem{thm-int}{Theorem}
\theoremstyle{definition}
\newtheorem{Def-s}[Thm]{Definition}

\newenvironment{enumerate*}{\begin{enumerate}}{\end{enumerate}}

\makeatletter
\newtheorem*{rep@theorem}{\rep@title}
\newcommand{\newreptheorem}[2]{%
\newenvironment{rep#1}[1]{%
 \def\rep@title{#2 \ref{##1}}%
 \begin{rep@theorem}}%
 {\end{rep@theorem}}}
\makeatother

\def\bea{\begin{eqnarray}}
\def\eea{\end{eqnarray}}
\def\be{\begin{equation}}
\def\ee{\end{equation}}
\def\ba{\begin{align}}
\def\ea{\end{align}}
\def\bse{\begin{subequations}}
\def\ese{\end{subequations}}

\def\det{\,{\rm det}\, }

\def\tr{\,{\rm tr}\, }

\def\Im{\,{\rm Im}\,}

\DeclareMathOperator{\ch}{ch}

\def\({\left(}
\def\){\right)}
\def\[{\left[}
\def\]{\right]}
\def\<{\left\langle}
\def\>{\right\rangle}

\newcommand{\e}{\mathrm{e}}

\newcommand{\cF}{\mathcal{F}}

\newcommand{\cO}{\mathcal{O}}

\newcommand{\R}{{\mathbb R}}
\newcommand{\Z}{{\mathbb Z}}

\renewcommand{\frak}[1]{\mathfrak{#1}}

\newcommand{\abs}[1]{\left\lvert {#1}\right\rvert}

\newcommand{\N}{\mathbb{N}}
\renewcommand{\Z}{\mathbb{Z}}
\renewcommand{\R}{\mathbb{R}}
\newcommand{\C}{\mathbb{C}}
\newcommand{\Q}{\mathbb{Q}}

\newcommand{\ii}{\mathrm{i}}
\renewcommand{\P}{\mathbb{P}}
\renewcommand{\vec}[1]{\underline{#1}}

\newcommand{\mat}[1]{\left(\begin{matrix}#1 \end{matrix}\right)}

\newcommand{\dd}{\mathrm{d}}
\newcommand{\td}{\mathrm{Td}}

\newcommand{\uz}{{\underline z}}
\newcommand{\ut}{{\underline t}}
\newcommand{\uq}{{\underline q}}
\newcommand{\ude}{{\underline \delta}}
\newcommand{\uth}{{\underline \theta}}
\newcommand{\uX}{{\underline X}}
\newcommand{\uPi}{{\underline \Pi}}
\renewcommand{\cal}{\mathcal}
\newcommand{\Xm}{\cal X} 
\newcommand{\Mcs}{{\cal M}_{cs}(X)}

\numberwithin{equation}{section}

\title{Calabi--Yau Period Geometry and Restricted Moduli in Type II Compactifications}

\author[a]{Janis Dücker,}
\author[a,b]{Albrecht Klemm}
\author[a]{and Julian F.\ Piribauer}
\affiliation[a]{Bethe Center for Theoretical Physics, Universit\"at Bonn, D-53115, Germany}
\affiliation[b]{Hausdorff Center for Mathematics, D-53115, Germany}
\abstract{The period geometry of Calabi--Yau $n$-folds---characterised by their variations of Hodge structure governed by Griffiths transversality, a graded Frobenius algebra, an integral monodromy and an intriguing arithmetic structure---is analysed for applications in string compactifications and to Feynman integrals.
In particular, we consider type IIB flux compactifications on Calabi--Yau three-folds and elliptically fibred four-folds.
After constructing suitable three-parameter three-folds, we examine the relation between symmetries of their moduli spaces and flux configurations.
Although the fixed point loci of these symmetries are projective special Kähler, we show that a simultaneous stabilisation of multiple moduli on the intersection of these loci need not be guaranteed without the existence of symmetries between them.
We furthermore consider F-theory vacua along conifolds and use mirror symmetry to perform a complete analysis of the two-parameter moduli space of an elliptic Calabi--Yau four-fold fibred over $\mathbb{P}^3$.
We use the relation between Calabi--Yau period geometries in various dimensions and, in particular, the fact that the antisymmetric products of one-parameter Calabi--Yau three-fold operators yield four-fold operators to establish pairs of flux vacua on the moduli spaces of the three- and four-fold compactifications.
We give a splitting of the period matrix into a semisimple and nilpotent part by utilising the Frobenius structure.
This helps bringing $\epsilon$-dimensional regulated integration by parts relations between Feynman integrals into $\epsilon$-factorised form and solve them by iterated integrals of the periods. 
}

\emailAdd{jduecker@uni-bonn.de}
\emailAdd{aoklemm@th.physik.uni-bonn.de}
\emailAdd{piribauer@uni-bonn.de}

\begin{document}
\preprint{\begin{minipage}[t]{8cm}\begin{flushright} 
BONN-TH-2025-06 \\
      \end{flushright}\end{minipage}}

\setlength{\parskip}{0.2cm} 

\maketitle

\newpage
\section{Introduction}

\emph{Pure variations of Hodge structures}~\cite{MR2393625} of Calabi--Yau manifolds $X$ determine important terms in the four dimensional effective supergravity actions of type II string theory\footnote{ 
The same is true for the $N=1$ supergravity action of heterotic strings with standard embedding of the tangent bundle of the Calabi--Yau manifold in the gauge bundle.} in terms of the periods 
of $X$ which are determined as the kernel of the flat Gauss--Manin connection or equivalently 
a \emph{homogenous Picard--Fuchs differential ideal}. In $N=2$ supergravity theories obtained from type IIB theories, the
variation of Hodge structure is expressed in periods, which determine the complex structure moduli dependence  of the real  K\"ahler potential and the holomorphic prepotential for the vector multiplets. One of essential problems of string theory compactifications is the stabilisation of these moduli, which correspond to massless fields in the effective actions, at phenomenological valuable vacua in the string landscape.  
In flux compactifications of type IIB theories, one considers non-vanishing NS--NS and R--R three-form background fields, with Dirac--Zwanziger integrality conditions. In this case, the periods determine the holomorphic superpotential that stabilises these vector moduli as well as the axio-dilaton at critical points in the moduli space. While, generically, the supersymmetry is broken
completely, the moduli can sometimes be stabilised at special values to $N=2$ vacua~\cite{Polchinski:1995sm,Michelson:1996pn}. As we explain below, the latter are often 
related to special symmetries in the Calabi--Yau geometry. Loci of $N=2$ vacua always originate from a 
splitting of the middle cohomology over $\mathbb{Q}$ into a rank two subspace of Hodge type $(2,1)+(1,2)$ and a rank $2h^{2,1}$ remainder. Symmetries of finite order in the complex structure moduli space imply the vanishing of elements in the middle homology over quadratic field extensions of $\Q$ at the fixed point locus. For symmetries of appropriate order, the field extension is trivial and a superpotential can be constructed by choosing the background fields to be dual to these vanishing cycles \cite{DeWolfe:2004ns,DeWolfe:2005gy}, which will be further developed in subsection \ref{sec:hodgesplitting}.
For models with only one complex structure modulus, the remainder is necessarily of rank two with Hodge type $(3,0)+(0,3)$. Such a splitting is equivalent to solving the equations governing the attractor phenomenon arising in the context of extremal black hole solutions to supergravity \cite{Moore:1998pn}. For this reason, supersymmetric IIB flux vacua in one-parameter models occur at the  attractor points~\cite{Kachru:2020sio,Bonisch:2022slo}. These are generalisations of points of complex multiplications on an elliptic curve and can be found by studying persistent factorisations of the local zeta function. The latter is briefly reviewed in \cref{sec:app7}.    
It was pointed out in \cite{Taylor:1999ii,Curio:2000sc} that the
combination of R--R fluxes $F$ and NS--NS fluxes $H$ avoids the no-go statements 
for partial supersymmetry breaking to $N=1$ vacua~\cite{Polchinski:1995sm,Michelson:1996pn} at the boundary of the complex structure moduli space, where fields become non-dynamical and Fayet--Iliopoulos terms can be turned on. These non-compact limits exhibit rigid non-abelian gauge symmetry enhancements~\cite{Kachru:1995fv,Katz:1996ht,Klemm:1996kv,Curio:2000sc}, more 
exotic non-critical string spectra~\cite{Ganor:1996mu,Klemm:1996hh}, heterotic string spectra or, trivially in the complete decompactification limit, Kaluza--Klein spectra. Except for the last, they have been all discovered
within $N=2$ heterotic type II duality~\cite{Kachru:1995wm,Klemm:1995tj,Klemm:1996hh}. 
Locally, the mathematical description at the
corresponding singularities becomes a \emph{mixed Hodge structure} due to the additional \emph{monodromy weight filtration}~\cite{MR2393625,MR1416353}. A link to brane and global $N=1$ compactifications was provided  by the local analysis of large $N$ dualisation of branes to fluxes and analysis of the emerging wrapped geometries~\cite{Giddings:2001yu}. In the corresponding $N=1$ compactifications with supersymmetric branes~\cite{Witten:1996bn}, the superpotential is determined by chain integrals fulfilling an \emph{inhomogenous Picard--Fuchs differential ideal}  encoding the \emph{variation of a relative mixed Hodge structure}~\cite{MR2393625,Aganagic:2001nx,Lerche:2002yw,Walcher:2006rs,Grimm:2008dq}. The global tadpole conditions are satisfied in brane compactifications with orientifolds and, more generally, 
in F-theory compactifications, where, on elliptically fibred Calabi--Yau four-folds, one generically gets $N=1$ supergravity vacua with unstabilised moduli as well as an $N=1$ heterotic type II string duality~\cite{Friedman:1997yq}, see~\cite{Weigand:2018rez} for a review. 
In F-theory, the complex structure moduli are stabilised by 4-form fluxes leading to a superpotential depending likewise on the periods~\cite{Gukov:1999ya}, see \cite{Kim:2022jvv} for the moduli dependence in the Sen limit. Symmetry considerations~\cite{Braun:2011zm,Braun:2020jrx} and number theoretic methods are also useful to find the critical loci of this superpotential~\cite{Kachru:2020abh}, 
which generically correspond to Hodge loci~\cite{MR4689371,Grimm:2024fip}.       
Also, these fluxes have to satisfy a quantisation condition \cite{Witten:1996md} and tadpole conditions, see (\ref{eq:eomA3}) and also \cite{Mayr:1996sh}, which can be inferred from M-theory. The question whether the latter allows to stabilise all moduli generically was recently critically discussed in \cite{Bena:2020xrh}. Mirror symmetry~\cite{Lerche:1989uy,Candelas:1990rm}, see \cite{Hori:2003ic} for 
a review,  relates type IIB on $X$ to type IIA theory on the mirror manifold $\hat X$ and complex structure moduli of $X$ to complexified K\"ahler moduli of $\hat X$, which are vector moduli in type IIA theory. Similarly, there is an open mirror map connecting the scalars in the chiral $N=1$ multiplets~\cite{Aganagic:2001nx,Lerche:2002yw,Grimm:2008dq}. The prepotential and superpotential in type IIA theory depend on the dual coordinates and, in (large radius) degenerations, they become counting functions for closed~\cite{Candelas:1990rm} and open~\cite{Aganagic:2001nx,Walcher:2006rs} world-sheet instanton contributions and govern moduli 
stabilisation at the vacuum solutions. A related possibility to stabilise the K\"ahler moduli is the five-brane superpotential \cite{Witten:1996bn}. 

\emph{The variation of pure and mixed  Hodge structures} with their  
Riemann as well as the Griffiths bilinear relations \eqref{Griffiths} on Calabi--Yau $n$-folds lead 
to special (K\"ahler) geometry, which we recall in \cref{sec:CYPeriodgeometry}, in particular in  \cref{sec:periodstructure}. The existence of the unique holomorphic $(n,0)$-form and the induced isomorphism $H^1(X,T_X)\sim H^{n-1,1}(X)$ is essential in the proof that infinitesimal deformations of the complex structures in $H^1(X,T_X)$ are unobstructed~\cite{MR915841,MR1027500}. Together with the Griffith bilinear relations, it implies that the holomorphic part of the special geometry is in all dimensions $n$ captured by a 
graded Frobenius algebra structure ${\cal A}=\oplus_{i=0}^n {\cal A}^{(i)}\sim \oplus_{i=0}^n H^{n-i,i}(X)$ 
\cite{Klemm:1996ts} as explained in subsection~\ref{sec:frobeniusalgebra}. The space ${\cal A}^{(0)}$ is generated by  $\Omega\in H^{n,0}(X)$ and~\cite{MR915841,MR1027500} implies that ${\cal A}^{(1)} \sim H^{n-1,1}(X)$. Elements in 
${\cal A}^{(1)}$ are identified with the $(N,\bar N)=(2,2)$ chiral-chiral  marginal 
operators in the topologically twisted $(2,2)$ supersymmetric string world-sheet theory and ${\cal A}$ with the full chiral-chiral ring~\cite{Lerche:1989uy}. Moreover,
mirror symmetry allows to identify ${\cal A}^{(1)}$ with the complexified K\"ahler deformations of the mirror $\hat X$ and ${\cal A}$ with the chiral-antichiral ring, which in turn is isomorphic to the even cohomology $\oplus_{i=0}^n H^{i,i}(\hat X)$~\cite{Lerche:1989uy}. The Frobenius structure and mirror symmetry imply essential properties of the periods. In particular, they exhibit a maximal unipotent degeneration as well as further typical degenerations, which allow to relate an integral basis for the monodromies via the $\hat \Gamma$-class to the topological data of $\hat X$, as explained in subsection~\ref{sec:GammaClass}.   

In addition to before mentioned arithmetic properties of the periods, the last 
mentioned structures are not only relevant for the moduli stabilisation problem or swampland 
conjectures~\cite{Ooguri:2006in} like, in particular, the landscape distance conjecture~\cite{Ooguri:2006in,Grimm:2018cpv,Joshi:2019nzi,Lee:2019wij}, 
but also determine the degeneration of maximal cut Feynman integrals,  
which have been recently identified with Calabi--Yau periods\footnote{For example, the 
$l$-loop Banana integrals in their critical dimension $d=2$ have been identified as 
periods of the mirror of a complete intersection Calabi--Yau $(l-1)$-fold family 
defined by two degree $(1,\ldots,1)$ constraints in $(\mathbb{P}^1)^{l+1}$~\cite{Bonisch:2020qmm}, whose $l=5$ incarnations feature in section \ref{sec:AntisymProducts}.  The $l+1$ dimensionless 
parameters $m_i^2/p^2$ of the graph are identified with the quantum volumes of the $\mathbb{P}^1$'s.}~\cite{Vanhove:2014wqa,Bourjaily:2018yfy,Bonisch:2020qmm,Bonisch:2021yfw}. 
It is a consequence of Griffiths transversality that in the critical 
dimensions full Feynman integrals, being solutions to an inhomogenous Picard--Fuchs 
differential ideal, like the superpotential, can be written as an iterated integral 
of the corresponding closed periods~\cite{Bonisch:2021yfw}. More significantly, their expansion coefficients in the dimensional regularisation parameter $\epsilon$ are also 
typically nested iterated integrals of periods~\cite{Duhr:2022dxb,Gorges:2023zgv,Klemm:2024wtd,Duhr:2024bzt,Frellesvig:2024rea,Driesse:2024feo} 
and are therefore described by the variation of the mixed Hodge structures associated 
to the integration by parts (IBP) relations of the graph~\cite{Henn:2013pwa}. 
In order to see this, their $\epsilon$-extended Gauss--Manin connection has to be brought 
into $\epsilon$-factorised form \cite{Henn:2013pwa}\cite{Bonisch:2021yfw,Gorges:2023zgv,Klemm:2024wtd,Duhr:2024bzt,Frellesvig:2024rea,Driesse:2024feo}, see section \ref{sec:periodmatrixnfolds}. 
To do this in the Calabi--Yau blocks of the latter, one uses in a crucial way a 
decomposition of the \emph{period matrix} into a semisimple and a unipotent part  \cite{Duhr:2022dxb,Gorges:2023zgv,Klemm:2024wtd,Duhr:2024bzt,Driesse:2024feo}, which is precisely 
possible due to the Frobenius structure of the underlying Calabi--Yau variation 
of Hodge structure. The latter implies a canonical form\footnote{See earlier~\cite{Greene:1993vm} and later work\cite{Bogner:2013kvr}  on one-parameter specialisations.} of the Gauss--Manin connection~\cite{Klemm:1996ts,Mayr:1996sh,CaboBizet:2014ovf}, and using this, a
canonical splitting will be performed explicitly in section~\ref{sec:periodmatrixnfolds}.
Subsection \ref{ss:ellK3} yields a warm-up of the problem for the elliptic curve and K3, \cref{ss:CY3folds,sec:CY4} cover three-folds and four-folds, respectively, while \cref{sec:fivefoldsandgeneral} discusses five-folds and higher dimensions. Some of the necessary calculations for the latter two  subsection are relegated to \cref{app:periodgeometry}. We also comment on the one-parameter specialisations and the situation that the degree-one part ${\cal A}^{(1)}$ of the Frobenius algebra does not generate ${\cal A}$ in subsection~\ref{sec:CY4}, in which case the cohomology is not \emph{purely horizontal}, i.e.\ not generated by derivatives of the holomorphic $(n,0)$-form. The latter situation is very common in the higher dimensional case, see in particular \cite{Duhr:2023eld}, where it is explained using representation theory in higher anti-symmetric powers of Calabi--Yau operators.

In \cref{sec:CYcomp}, we lay out the conditions for flux vacua in type IIB and F-theory. 
\Cref{sec:quotientbuilding} contains a review of hypersurfaces in toric ambient spaces. We construct their complex structure moduli spaces and repeat Griffiths' construction of the rational middle cohomology group.
As mentioned above, one can use group-theoretic properties of the periods to study flux vacua at orbifold points, which are characterised by having a monodromy transformation of finite order. 
This was established and used in \cite{DeWolfe:2004ns} for one-parameter models. In a later article \cite{DeWolfe:2005gy}, these methods were extended to multi-parameter models that exhibit flux vacua along a locus in codimension one of the moduli space. Due to the similarity of rational matrices of equal finite order, statements about the splitting of integral Hodge structure can be made purely from a local basis, without the need to perform an analytical continuation of the integral basis obtained at the point of large complex structure. On the one hand, it will allow us to show in \cref{sec:hodgesplitting} that flux vacua on a family with restricted moduli are also present in the family over the entire moduli space. On the other, this explains the existence of the sets of vacua we encounter in \cref{sec:CYcompSUSY}. We will then define and analyse one of the three three-parameter models in \cref{sec:sextic} with supplementary data in \cref{sec:app6}. The discussion of the other two can be found in \cref{sec:app10,sec:app8}, respectively.  \Cref{sec:FVsym} gives an account of the F-term equations coming from non-invariant moduli, which are satisfied automatically on the symmetric locus, and on vacua in codimension one together with their (in-)compatibility.
If the moduli space possesses an involution symmetry, there are vanishing periods on the fixed point locus. 
Conjecturally, a symmetry reduces the representations of the Galois group action on the $l$-adic middle cohomology $H_l$. 
This happens, for example, in the symmetric three-parameter family discussed in \cref{sec:symexample}. 
For this specific family, one finds three sets of vacua along loci where two of the three moduli are equal.
Given such a supersymmetric flux vacuum, the axio-dilaton is fixed up to an SL(2,$\mathbb{Z}$) transformation of the fluxes. This means that successive or simultaneous stabilisations of moduli presumably require a symmetry between the vacua to guarantee an agreement of the axio-dilaton values. 
In the symmetric model of \cref{sec:symexample}, the symmetries between the moduli ensure that the intersections of the vacuum loci, i.e.\ the symmetric one-parameter locus, is a consistent flux vacuum.
This in contrast to the flux vacua we find in \cref{sec:CYcompSUSY} for the three-parameter model $\mathcal{X}_6^{(3)}$ constructed in \cref{sec:sextic}: while there exist two moduli that are symmetric under a $\Z_2$-action $a_\pm\mapsto -a_\pm$, the moduli space is not symmetric under an exchange of the two. This implies that the axio-dilaton values for flux vacua on the loci $a_+=0$ or $a_-=0$ are not guaranteed to coincide on $a_+=a_-=0$. And indeed, we find that the two vacua are incompatible on generic points of their intersection locus. This shows that a family over a fix point locus of a symmetry is not necessarily a supersymmetric flux vacuum. In \cref{sec:FtheoryVacua}, we consider F-theory vacua. We begin with a thorough analysis of the moduli space of the elliptically fibred Calabi--Yau four-fold $\P_{12,8,1,1,1,1}[24]$ and then describe supersymmetric flux vacua along conifold loci for this model in \cref{sec:X24} and for the model $\P_{18,12,3,1,1,1}[36]$ in \cref{sec:X36}. This section is supplemented by \cref{sec:appD,sec:appE} containing the data of the two models. We build a bridge between the previous two chapters in \cref{sec:AntisymProducts} by considering Calabi--Yau four-fold operators arising as antisymmetric products of three-fold operators to establish pairs of flux vacua on one-parameter manifolds. The minors of the Wronskian in a rational basis of the three-fold form a rational basis of the four-fold operators. This allows us to show that, under the operation of taking the antisymmetric product, attractor points in the Calabi--Yau three-fold moduli space map to supersymmetric flux vacua in the M-theory setup. Conversely, we also analyse the points of supersymmetric vacua of Calabi--Yau four-folds on their corresponding three-folds motives. We find that four-fold vacua generally correspond to a splitting of the three-fold's Hodge structure over a quadratic field extension. Another way of obtaining new supersymmetric compactifications is by considering so-called conifold transitions. At the conifold discriminant, the vanishing $S^3$ can be replaced with an $S^2$, giving rise to a Calabi--Yau family with different Hodge numbers. The three-parameter models introduced in \cref{sec:sextic,sec:app8,sec:app10} have the hypergeometric families $\mathcal{X}_6$, $\mathcal{X}_8$, and $\mathcal{X}_{10}$ as subfamilies and, as we will show in \cref{sec:conifoldtrans}, transition to the hypergeometric models $\mathcal{X}_{3,2,2}$, $\mathcal{X}_{4,2}$ and $\mathcal{X}_{6,2}$, respectively.

\section{Frobenius algebras, period geometries and integer bases}  
\label{sec:CYPeriodgeometry}
Calabi--Yau $n$-folds $(X,\omega,\Omega)$ are $n$ complex dimensional K\"ahler manifolds with K\"ahler $(1,1)$-form $\omega$, which 
have trivial canonical class $K_X=0$ implying that there exists 
a nowhere-vanishing holomorphic $(n,0)$-form $\Omega$ on $X$, see \cite{MR1963559} and \cite{Hubsch:1992nu,MR3965409} for reviews
from the mathematical and physical point of view.  To address vacuum selection problems we consider families $\Xm$ of such Calabi--Yau $n$-folds with fibres $X_{\uz^*}$ over their complex structure moduli space $\Mcs$ parametrised by $\uz=\{z_1,\ldots, z_r\}$. Here, $r={\rm dim}_\mathbb{C}(\Mcs)$  is the number of Beltrami forms   
$r={\rm dim}(H^1(X,T_X))$, which, as can be seen by contraction with $\Omega$, equals the dimension  $h^{n-1,1}(X)$ of the Hodge group $H^{n-1,1}(X)$. 
By the theorem of Tian \cite{MR915841} and Todorov \cite{MR1027500}, these local deformations are unobstructed and yield the actual complex dimension of the smooth strata of $\Mcs$. 

Eventually, we are also interested in the complexified  K\"ahler moduli ${\cal M}_{cK}(X)$ 
parametrised in the large volume limit by the complexified K\"ahler parameters 
\be 
\label{eq:Kaehler}
t^\alpha=\int_{{\cal C}_\alpha} (b^{(2)} +i \omega)\,,\qquad  \alpha=1,\ldots, h^{1,1}(X)\,,
\ee 
where the complex curves ${\cal C}_\alpha$ span the Mori cone of $X$ and 
$b^{(2)}$ is the antisymmetric Neveu--Schwarz B-field, which can be related to a harmonic form in $H^{1,1}(X)$ by the
solution of its equation of motion. Note that ${\rm Im}(t^\alpha)={\rm vol}({\cal C}_\alpha)\ge 0$ and that we introduce the dual  
basis $\omega_\alpha$ of K\"ahler cone with  $\int_{{\cal C}_a} \omega_b=\delta^a_b$ such that $\omega=\sum_{\alpha=1}^{h^{1,1}}  {\Im}( t^\alpha)\, \omega_\alpha$.
As explained in detail below, the complexified and instanton corrected K\"ahler moduli space ${\cal M}_{cK}(\hat X)$  can be analysed using mirror symmetry and its asserted identification of ${\cal M}_{cK}(\hat X)$ with ${\cal M}_{cs}(X)$ by the mirror map $t^a(\uz)$. Since the mirror $\hat X$ is easily constructable torically for our examples of $X$, we can describe it using again the structures on ${\cal M}_{cs}(\hat X)$, which are encoded in the periods over middle dimensional cycles 
of the corresponding Calabi--Yau manifold $X$.  

\subsection{Periods bilinears, Gauss--Manin connection and monodromies} 
\label{sec:periodstructure}
To understand the structure of these complex structure moduli spaces, 
recall that there are two bilinear pairings 
\be
\Sigma:H_n(X,\mathbb{Z})\times 
H_n(X,\mathbb{Z})\rightarrow \mathbb{Z}\,, \qquad \boldsymbol{\Pi}: H_n(X,\mathbb{Z}) \times H^n(X,\mathbb{C}) \rightarrow \mathbb{C}\,. \   
\label{eq:pairings}
\ee
We define the intersection form and the matrix of period integrals as their matrix representations
\be 
\Sigma_{ij}:=\Gamma_i\cap \Gamma_j=:\langle \Gamma_i,\Gamma_j\rangle  \in \mathbb{Z}\,, \qquad \boldsymbol{\Pi}_{ij}(\uz_0):=\int_{\Gamma_i} \gamma^j(\uz_0)\in \mathbb{C}\,,  
\label{eq:SigmaPeriods} 
\ee
respectively\footnote{If we refer to this 
topological integral basis of $H^n(X,\mathbb{Z})$ we indicate no argument $\uz^\star$ is the complex structure of the 
fibre $X_{\uz_0}$. If we refer to elements in the with $\uz$ varying rational basis of the middle cohomology  
given by the Griffiths residuum forms, as explained in \cref{sec:rationalcohomology}, we write 
$\e^j(\uz)$.}.  To state it in this useful form, one needs to provide the topological integral basis of real $n$ dimensional cycles in this middle homology $\Gamma_i\in H_n(X,\mathbb{Z})$ and a dual integral basis in the middle cohomology $\gamma^j\in H^n(X,\mathbb{Z})$ with $\int_{\Gamma_i} \gamma^j=\delta^{ij}$ and $i,j=1,\ldots ,b_n(X)$. The latter induces the dual pairing  ${\hat \Sigma}^{ij}=\langle \gamma^i,\gamma^j\rangle =\int_X \gamma^j\wedge \gamma^j$. The intersection form $\Sigma$ is even if $n$ is even and if $n$ is odd it is odd and can in particular be chosen to be symplectic. The integrality is mathematically important for the structure of $\Mcs$ and physically to fulfil the Dirac--Zwanziger quantisation condition for the flux quanta of 
the flux compactifications. The integral basis will be provided using the monodromy properties of the periods and/or using homological mirror symmetry and the $\hat \Gamma(T_{\hat X})$ class of the mirror. To discuss the first method, which we apply in detail in \cref{sec:X24}, let us first note that the Griffiths residue form in \eqref{eq:OmegaRes}, and generalisations see \cite{MR3965409}, yields a holomorphic $(n,0)$-form $\Omega(\uz)$ in rational cohomology 
over ${\cal M}_{cs}(X)$ parametrised by $\uz$. With respect to the topological basis of  $H_n(X,\mathbb{Z})$,
it can be expanded by the first  column ($\gamma^0=\Omega(\uz_0)$) of the period matrix, the so-called {\sl period vector}
\begin{equation}\label{eq:periodvector}
    \uPi=\left(\int_{\Gamma_1}\Omega(\uz),\,\ldots, \int_{\Gamma_{b_n(X)}} \Omega(\uz)\right)^T.
\end{equation} Let further $\widehat{ \Mcs}$ be 
a resolution of $\Mcs$ so that the proper transform $\hat \Delta_k=0$ of all critical divisors $\Delta_k=0\in \Mcs$ together with the exceptional divisors of the resolution $D_i$ are normal crossing divisors.
By a slight abuse of notation, we call these normal crossing 
divisors $\Delta_k=0\in \widehat{ \Mcs}$ again. The entries of $\uPi$ 
are multivalued functions over $\widehat{ \Mcs}$ fulfilling regular singular systems of differential equations specified below. We fix a base point and denote the monodromies along suitable oriented paths around the $\Delta_k=0$ by $\mathfrak{M}_k$. Up to conjugation in ${\cal O}(\Sigma, \mathbb{Z})$, the integral basis for the cycles  $H_n(X,\mathbb{Z})$ is determined as the basis $\uPi$ of solutions in which the monodromies $\mathfrak{M}_k$ around all singularities of $\uPi(\uz)$ are in 
\be 
{\cal O}(\Sigma, \mathbb{Z})=\{\mathfrak{M}_{b_n \times b_n}\in {\rm GL}(b_n,\mathbb{Z})|\mathfrak{M}^T \Sigma \mathfrak{M}=\Sigma\}\, .
\label{eq:OZZ}  
\ee
In particular, one expects the global monodromy group $\Gamma_X\subset {\cal O}(\Sigma, \mathbb{Z})$ generated by the $\mathfrak{M}_k$ up to van Kampen relations to be irreducible. 
Note that, for $n$ odd, $\Sigma$ can be chosen to be symplectic
and the group in \eqref{eq:OZZ} becomes Sp$(b_n,\mathbb{Z})$, 
while for the even case it will be determined from
an intersection calculation. For example, the intersection is calculated 
for $n=2$ in the Picard Lattice of the mirror and for $n=4$ in the primitive part of  $H^{2,2}(X,\mathbb{Z})$. Determining $\Gamma_X$ is complicated due to the need of 
(numerical) analytic continuation and we will use further structure explained next. 

Restrictions on the Calabi--Yau period geometry are given by the real Griffiths bilinear inequalities
\be
e^{-K(\uz)}=\langle\bar {\underline \Pi}, {\underline \Pi}\rangle=\ii^{n-2} \int_{X} \bar\Omega(\uz) \wedge  \Omega(\uz) = \ii^{n-2} \uPi^\dagger \Sigma \uPi> 0\, ,
\label{Kahler}
\ee
where the real K\"ahler potential $K(\uz)$ determines the Weil--Petersson 
metric on $\widehat{\Mcs}$ as $G_{\bar \imath j}=\bar\partial_{\bar \imath}  \partial_j K$.
Further restrictions are the properties of a holormophic bilinear due to Griffiths transversality 
\be
\int_{X} \Omega \wedge  {\underline \partial}^k_{I^{(k)}} \Omega = \uPi^T \Sigma {\underline \partial}^k_{I^{(k)}} \uPi=  \left\{
\begin{array}{cl} 0\, , & \ {\rm if} \ k < n\, ,\\[2 mm] C_{\uz_{I^{(n)}}}(\uz)\in \mathbb{Q}(\uz)\,, & \ {\rm if} \  k=n \, . \end{array}\right.
\label{Griffiths}
 \ee
Here, $I^{(k)}$ is an index set of $k$ not necessarily distinct indices 
and 
\be 
{\underline \partial}^k_{I^{(k)}} \Omega=\partial_{z_{I^{(k)}_1}}\ldots\partial_{z_{{I^{(k)}_k}}}\Omega \in F^{k}:=\bigoplus_{p=0}^k H^{n-p,p}(X)
\label{eq:Fn}\ee 
are arbitrary combinations of derivatives w.r.t.\ the complex structure moduli 
$z_i$, $i=1,\ldots, r$.  Note that the $\binom{n+r-1}{r-1}$ different $n$ point couplings  $C_{z_{i_1},\ldots, z_{i_n}}$ are sections of ${\cal L}^2 \otimes {\rm Sym}^{r} (T^*{\cal M}_{cs})$.
They  can be obtained from \eqref{Griffiths} using \eqref{eq:GaussManin} and \eqref{eq:PFI} and are rational, if the coefficients 
of the latter are rational. We write ${\cal L}$ for the K\"ahler line bundle, in which $\Omega$ is a section $\Omega\in {\cal L}$ so that under $\Omega\rightarrow \Omega e^{f(\uz)}$ with $f(\uz)$ holomorphic $K(\uz)\rightarrow K(\uz)-f(\uz)-{\bar f}(\bar \uz)$ 
undergoes a K\"ahler transformations leaving $G_{i\bar \jmath}$  invariant. To express the bilinears in terms of the period 
vector, we used the expansion of the holomorphic $(n,0)$-form as $\Omega=\sum_{k=1}^{b_n(X)} \uPi_k \gamma_k$ and their properties follow from the definitions of the bases above. 

The $b_n$ functions in the period vector $\uPi(\uz)$ have only regular singularities, i.e.\ branch cut and 
logarithmic singularities along critical divisors $\Delta_i=0$ in $\widehat{\Mcs}$ and  are determined as linear combinations over $\mathbb{C}$ by the flatness of the Gauss--Manin connection 
\be 
(\partial_{z_i} - A_i(\uz)) \uPi(\uz)=0\,, \qquad i=1,\ldots,r\, , 
\label{eq:GaussManin}
\ee
where $A_i(\uz)\in \mathbb Q[\mathbb{Z}]$ are $r\times r$ matrices of rational functions of the moduli. Equivalently, the periods 
are specified as spanning the kernel of the Picard--Fuchs differential ideal (PFDI)  $\{{\cal L}\}=\{{\cal L}_k^{(d_k)}(\uth,\uz)|k=1,\ldots,l \} $ 
\be
{\cal L}_k^{(d_k)}(\uth,\uz)\uPi(\uz)=0\,, \qquad k=1, \ldots , l:=\# {\rm Gen}({\cal L})\,  
\label{eq:PFI}
\ee
over $\mathbb{C}$. Here, the ${\cal L}^{(d_k)}(\uth,\uz)$ are 
degree $d_k$ reduced differential operators with polynomial coefficients left of the powers of the logarithmic derivatives $\theta_i=z\frac{d}{dz_i}$ and $\{{\cal L}\}$ is the left differential ideal generated by them. We always assume that the PFDI has at least one point of maximal unipotent monodromy (MUM), which we choose to lie at $\uz=0$. That means that this point 
given by $z_i=0$, $i=1,\ldots, r$ all  $r$ monodromies around  $z_i=0$  are maximal unipotent, i.e.\ 
\be 
(\frak{M}_i-{\bf{1}})^{m}\neq 0\,, \ \  m<n+1\,, \ \ {\rm  but}\ (\frak{M}_i-{\bf{1}})^{n+1}=0\,, \ \forall\, i=1,\ldots,r  \,.
\ee
This implies that the formal equations $\{{\cal L(\uth,\uz)}|_{\uz=0}=0\}$ in $\uth$ have an $b_n$-fold degenerate solution implying complete degeneration for the local indices of their solutions and the logarithmic structure leading to the maximal unipotency of the monodromies. We
also note that inserting the operators in $\{{\cal L(\uth,\uz)}\}$ in \eqref{Griffiths} yields a holonomic
system of linear differential equations for the $C_{\uz_{I^{(n)}}}(\uz)$ which allow to solve them 
up to one multiplicative constant fixed by one 
intersection number in  \eqref{eq:classical}.   

We obtain the generators  ${\cal L}^{(d_k)}(\uth,\uz)$ for the examples below either as reduction of the  Gelfand--Kapranov--Zelevinsky system for complete intersections or hypersurfaces in toric ambient spaces as in \cite{MR1316509,MR1319280} or by performing explicitly a period integral defined by a residue expression and determining the operators by ensuring that they generate a complete PFDI with the expected solution structure. 

\subsection{Frobenius Algebra structure}
\label{sec:frobeniusalgebra}
As was pointed out in \cite{Klemm:1996ts}, the variation of Hodge structures encoded both in 
the Gauss--Manin connection \eqref{eq:GaussManin} and the  differential ideal \eqref{eq:PFI} with equation \eqref{Griffiths} determine, for $n=3$, the triple couplings $C_{ijk}(\uz)$,  which provide a graded Frobenius algebra structure that extends not only to higher dimensions $n>3$ but is in all dimensions also manifest in the mirror description. The central 
data is a graded vector space ${\cal A}=\oplus_{i=0}^n {\cal A}^{(i)}$ with maps
\be \label{eq:Frobeniusalgebra} 
C^{(a,b,c)}: {\cal A}^{(a)} \times {\cal A}^{(b)}\times {\cal A }^{(c)} \rightarrow \mathbb{C}\,.  
\ee
For each ${\cal A}^{(k)}$, we introduce bases $e^{(k)}_j$ and denote the components of the triple couplings by $C^{(a,b,c)}_{ijk}$. For example, the three-folds couplings $C_{ijk}$ are given by $C^{(1,1,1)}_{ijk}$. The properties of the graded 
Frobenius algebra are
\begin{itemize}
\item Permutation symmetry: $C_{ijk}^{(a,b,c)}=C^{(\sigma(a,b,c))}_{\sigma(ijk)}$ for all permutations $\sigma$
\item Grading: $C^{(a,b,c)}=0$ unless $a+b+c=n$.
\item Unit: ${\cal A}^{(0)}=\mathbb{C}\cdot 1$, where $1$ is a unit and $C^{(0,a)}_{1ij}=\eta^{(a)}_{ij}$.
\item Associativity $$C^{(a,b,n-a-b)}_{ijp} \eta_{(n-a-b)}^{pq} C^{(a+b,c,n-1-b-c)}_{qkl}=C^{(a,k,n-a-k)}_{ikr}\eta^{rs}_{(n-a-c)} C^{(a+c,b,n-a-b-c)}_{sjl}.$$ 
\end{itemize} 
Then ${\cal A}$ is a Frobenius algebra with fusion product 
\be \label{eq:FrobAlg}
e_i^{(a)}\cdot e^{(b)}_j=C^{(a,b,n-a-b)}_{ijk} \eta_{(n-a-b)}^{kp}e^{(a+b)}_p . 
\ee
For a fixed fibre $X_{\uz_0}$ of a complex family the  $e^{(a)}_j$ can be thought of
as elements in the homology $H^p(X_{\uz_0},\wedge^p TX_{\uz_0})$. Using the contraction with the holomorphic $(n,0)$ form $\Omega_0$ in that fibre,
we get the definition for the three-point functions
\be 
C^{(a,b,n-a-b)}_{ijk}=\int_{X_{\uz_0}} \Omega_0 \wedge  \Omega_0(e^{(a)}_i\wedge e^{(b)}_j\wedge e^{(n-a-b)}_k) =\int_{X_{\uz_0}}\gamma_0^{(a)}\wedge \gamma_0^{(b)}\wedge \gamma_0^{(a+b)} . 
\label{eq:3pointfunctions} 
\ee
Note that contraction $\dashv$ of the vector indices of $e^{(a)}$ 
with $\Omega_0$ induces an isomorphism between $H^p(X_{\uz_0},\wedge^p TX_{\uz_0})$ and $H^{n-p,p}(X_{\uz_0})$ and that we denoted $\gamma_0^{(p)}=\Omega_0 \dashv e^{(p)}\in 
H^{n-p,p}(X_{\uz_0})$. 

The most interesting part of the B-model application arises when we
extend the Frobenius algebra over the complex family ${\cal X}$ and 
define the $C^{(a,b)}_{ijk}(\uz)$ as holomorphic section of suitable bundles over ${\cal M}_{cs}({\cal X})$. The latter should be governed by  the flatness of 
the Gauss--Manin connection \eqref{eq:GaussManin} or equivalently the Picard--Fuchs differential 
ideal \eqref{eq:PFI} as well as Griffiths transversality \eqref{Griffiths}. 
There is also an interesting interplay with the non-holomorphic equation \eqref{Kahler}, but we focus first on the holomorphic equations. Here, one sees in particular the $C^{(a,b)}_{ijk}(\uz)$ as invariants of these holomorphic differential structures. For example, the $C_{z_{i_1}\ldots z_{i_n}}(\uz)$ $=:C_{i_1\ldots i_n}(\uz)=\int_{X_\uz}\Omega\wedge\Omega\bigl(
\wedge_{k=1}^n e^{(1)}_{i_k}\bigr)$  in \eqref{Griffiths} are 
rational functions that can be calculated from \eqref{Griffiths} and 
\eqref{eq:PFI}. The Frobenius structure  implies that they 
can be decomposed in all possible ways compatible with the weights into the three-point functions $C^{(a,b)}_{ijk}(\uz)$ as, for example, in \eqref{eq:fourpointfactorization} and it is a very interesting 
question what is the transcendentality of the latter. This is relevant for the  the string applications but also for the recently discovered connection  between multi parameter maximal cut Feynman integrals and Calabi--Yau period and chain integrals as explained~\cite{Bonisch:2021yfw}. 

The decisive step is to extend the $e^{(k)}_a$ globally over ${\cal M}_{cs}({\cal X})$ 
to sections $e^{(k)}_a(\uz)$ of the bundles of holomorphic sheaves ${\cal F}^k$ defined in \eqref{eq:Fn}.
In a given fibre $X_{\uz^*}$ with fixed Hodge structure, we can always pick in \eqref{eq:SigmaPeriods} a graded topological 
basis\footnote{We drop the reference index $^*$ 
to the fibre in the following.}  $\Gamma^{(n)}_a$ and $\gamma_a^{(p)}$, $ a=1,\ldots, h^{n-p,p}$  so that  $\langle \gamma^{(p)}_a,\gamma^{(q)}_b \rangle =0 $ if $p+q\ge n$. For example, the unique $e^{(0)}(\uz)=\Omega(\uz)$ is the one
given by the Griffiths residue form as an element of rational cohomology e.g.\ for hypersurfaces in \eqref{eq:OmegaRes}. 
By the Tian--Todorov theorem, the elements $e^{(1)}_i(\uz)$, $i=1,\ldots,h^{n-1,1}$ are simply related to derivatives 
of $\Omega(\uz)$ w.r.t.\ to the $h^{n-1,1}$ moduli spanning ${\cal A}^{(1)}$.    
In the examples in \cite{Klemm:1996ts}, ${\cal A}^{(1)}$ further generates the Frobenius algebra, 
which is true for smooth hypersurfaces in projective spaces~\cite{Greene:1993vm,Mayr:1996sh,Klemm:2007in,Klemm:1996ts,CaboBizet:2014ovf} and further complete intersection examples \cite{CaboBizet:2014ovf}. As already 
stated in \cite{Klemm:1996ts}, this is not an essential feature of the Frobenius structure for Calabi--Yau n-folds, but, as proven there, it is equivalent to the additional axiom 
\begin{itemize}
\item Non-degeneracy: If $C^{(1,a)}=0$ $\forall$ vectors in ${\cal A}^{(1)}$ and ${\cal A}^{(n-1-a)}$ then  $v^{(a)} \in {\cal A}^{(a)}$ is zero.    
\end{itemize} 
Indeed, in \cite{Honma:2013hma}, it was found that even for one-parameter 
four-folds this is not necessarily the case. The indication was that the Picard--Fuchs operator obtained in \cite{Honma:2013hma} for a gauged linear $\sigma$-model representing a four-fold  
was of order six, which suggests 
that  instead of a middle cohomology of Hodge type $(1,1,1,1,1)$, the geometry associated to 
it should have a middle cohomology of Hodge type $(1,1,2,1,1)$. A careful analysis of the geometry 
performed in \cite{Gerhardus:2016iot} shows that this is indeed the case. In the case of non-degeneracy we speak of \emph{purely horizontal cohomology}. Irrespective of the non-degeneracy axiom   
we expand
\be 
e^{(k)}_a(\uz)  = \gamma^{(k)}_a + \sum_{p>0} \hat \Pi^{(p)b}_a(\uz)  \gamma_b ^{(p)} \in F^k.
\ee
Note that for $e^{(0)}(\uz)=\Omega(\uz)$, the $\hat \Pi(\uz)^{(p)b}(\uz)$ constitute the period vector \eqref{eq:periodvector}
in a gauge fixed form ($\Pi_{0}=1$) and inherit a grading from the Hodge decomposition of the fibre $X_{{\uz}_0}$.  

To see this, notice that, thanks to \eqref{eq:Fn} and consideration
of Hodge type, one may view $e^{(a)}(\uz)$ as section of the holomorphic sheaves ${\cal F}^k$ defined as extension of the $F^k$ on the total space of the complex family ${\cal X}$ making
the $C^{(a,b)}_{ijk}(\uz)$ sections of suitable holomorphic bundles 
on the complex structure moduli space  ${\cal M}_{cs}(X)$ of the family
${\cal X}$.

\subsection{Mirror Symmetry and \texorpdfstring{$\hat \Gamma$}\ -class}\label{sec:GammaClass}
Locally, e.g.\ near $\uz=\uz_*$, we introduce
coordinates $\delta_i=z_i-z_{i*}$. We can then   
define from a given $(n,0)$-form $\Omega(\ude)$ in the fibre $X_{\uz_*}$ the $(n-1,1)$-forms $\chi_{i}=(\partial_{\delta_i}-\partial_{\delta_i} K)\Omega(\ude)$, $i=1,\ldots,r$. In a suitable completion of the forms $\{\Omega,\chi_{i},i=1,\ldots,r\}=\{\chi_{I},I=0,\ldots,r\}$ to a 
basis of $H^n(X)$, we find $n$-cyles $A^I\in H_n(X,\mathbb{Z})$ dual to the $\chi_I$, so that, by the local Torelli theorem, $X^I=\int_{A^I} \Omega$ are  homogeneous 
local coordinates on $\Mcs$ near $\uz_*$\footnote{Note that the $\chi_i$, $A^I$, $B_I$, $t^i$ and ${\cal F}_i$ all 
depend on the choice of the local point $\uz_*$. To avoid
cluttering the notation we suppress the $*$ label on these quantities.}. The homogeneity is due to the freedom of making a K\"ahler gauge transformation $\Omega\rightarrow e^{-f(\ude)}\Omega$ with $f(\ude)$
holomorphic, without changing the K\"ahler metric, i.e.\ $\Omega$ is a section of the K\"ahler line bundle ${\cal L}^{-1}$ over $\Mcs$. 
We  can choose $e^{f(\ude)}\Omega$ to be a period 
which is regular at $\ude=\underline 0$. This defines a choice of inhomogeneous coordinates on $\Mcs$  given by 
\be 
    t^i(\ude) = \frac{X^i(\ude)}{X^0(\ude)}\,,\quad i=1,\ldots ,h_{n-1,1}\, .
\label{eq:inhomogeneous}
\ee
Note that  \eqref{eq:inhomogeneous}  defines a locally invertable map between 
the $\ut$ and the  $\uz$ ($\ude$)  coordinates.

In order to simplify the choice of the integral basis for the periods, we use the fact that our families have a least one point of maximal unipotent monodromy, which we choose to be at $\uz=\uz_*={\underline 0}$. As reviewed in \cref{sec:quotientbuilding} for hypersurfaces or complete intersection in toric ambient spaces, the Batyrev mirror construction via reflexive polyhedra yields also a parametrisation of $\Mcs$ given by the Batyrev coordinates
\eqref{eq:batyrevcoordinates} in which that is the case.
The maximal degeneration of the local indices and maximal
unipotency enforces a particular logarithmic degeneration 
of the periods, which  simplifies the identification of 
the integral basis. In particular, $X^0$ can be chosen to 
be the unique holomorphic period and normalised to start with $X^0=1+{\cal O}(\uz)$. Furthermore, there are $r$ single logarithmic periods $X^i=\frac{1}{2 \pi i} X^0 \log(z_i)+\text{holomorphic}$. Mirror symmetry identifies the point of maximal unipotent monodromy in $\Mcs$ of $X$ with a large radius point in ${\cal M}_{cK}(\hat X)$ of a mirror $\hat X$ and  
\be 
t^i(\uz)=\frac{X^i(\uz)}{X^0(\uz)}=\frac{\log(z_i)}{2 \pi i}+\text{holomorphic}\,,\ \ i=1,\ldots,h_{n-1,1}(X)=h_{1,1}(\hat X)=r
\label{eq:mirrormap}
\ee
with the  complexified Kähler parameters \eqref{eq:Kaehler} of this mirror manifold $\hat X$. According to \eqref{eq:Kaehler},
the imaginary part of $t^{i=\alpha}$ describes the area ${\rm Area}({\cal C}_\alpha)=\int_{{\cal C}_\alpha} \omega$ of that curve ${\cal C}_\alpha$ in the Mori cone. Maximal unipotency implies that $\uPi$ has a grading in the leading logarithms of the Frobenius basis of $\uz$ with powers $p$ running from $p=0,\ldots,n$, which reflects the degeneration of the horizontal mixed Hodge structure at that point, which in turn can be identified with the vertical Hodge structure of the mirror~\cite{MR1416353}. Let us recall the argument to construct
the integral basis in the three- and four-fold case used first in~\cite{CaboBizet:2014ovf}.

As Deligne pointed out in \cite{MR1416353}, using the monodromy weight filtration at the MUM point, 
the degenerate periods over $n$-cycles $\Gamma^{(p)}_r$, $r=1,\ldots, h_{n-p,p}$ in $H^{\text{hor}}_{n-p,p}(X,\mathbb{Z})$ are identified with the periods of degree $p$ in the logarithms and are paired with the ones of degree $n-p$ in the logarithms over cycles $\hat \Gamma^{(n-p)*}_i$ in $H^{\text{hor}}_{p,p-n}(X,\mathbb{Z})$ by the pairing of the cycles with the intersection pairing $\Sigma$ in \eqref{eq:pairings}. Moreover, homological 
mirror symmetry identifies these periods with the central charges of A-branes, which are  
mathematically coherent sheaves ${\mathcal E}^{(p)}_r$ and ${\mathcal E}^{(n-p)}_r$ in the K-theory group $K(\hat X)$, whose maximal support is on even degree cycles of real dimension $2k$ and $2(n-k)$ in $H^{\text{prim}}_{p,p}(\hat X,\mathbb{Z})$ and $H^{\text{prim}}_{n-p,n-p}(\hat X,\mathbb{Z})$, respectively. The map  \cite{MR2683208}
\be 
\begin{array}{rl}
\mu: K(\hat X)&\rightarrow H^{*}(\hat X,\mathbb{Z})\,,\\[2 mm]
     {\mathcal E}\ \ \ & \mapsto {\rm ch}({\mathcal E}) \hat \Gamma(T\hat X)
\end{array}
\label{eq:mu} 
\ee
allows to define the mirror intersection $\hat \Sigma_{i,j}$ of the coherent sheaves ${\mathcal E}_i$ as their Hirzebruch--Riemann--Roch pairing
\be 
\hat \Sigma_{ij}=\chi({\mathcal E}_i,{\mathcal E}_j)=\int_{\hat X} \mu({\mathcal E}^\vee_i) \wedge  \mu({\mathcal E}_j) \in \mathbb{Z} \,,
\label{eq:dualintersection} 
\ee
where we suppressed the grading $^{(k)}$, which can be obtained in the mirror dual intersection \eqref{eq:dualintersection} as in \eqref{eq:sigma} by choosing the  ${\mathcal E}^{(k\le n)}_r$ appropriately, as explained below. 
Here, the $\hat \Gamma(T\hat X)$ class is the multiplicative class obtained from $\Gamma\left(1- \frac{x}{2 \pi i}\right)$ in terms of the Chern classes of $c_i=c_i(T\hat X)$. 
Specifically for Calabi--Yau manifolds, where $c_1(T\hat X)=0$, one gets \cite{CaboBizet:2014ovf}
\be 
\hat \Gamma(T\hat X)=1+\frac{c_2}{24}+\frac{i c_3 \zeta(3)}{8 \pi^3} + \frac{7 c_2^2-4 c_4}{5760}+\frac{\pi^2 c_2 c_3\zeta(3)+ 6(c_2 c_3-c_5)\zeta(5)}{192 \pi^5}+{\cal O}(6)\,.
\label{eq:hatGamma}  
\ee
With $\hat{\Gamma}(T\hat{X}^\vee)\,\hat{\Gamma}(T\hat{X})=\td(\hat{X})$, we can also write the mirror intersection as
\begin{equation}
\label{eq:Asideperiods}
    \hat\Sigma_{ij}= \int_{\hat{X}} \td(\hat{X})\wedge \ch(\mathcal{E}_i)\wedge\ch(\mathcal{E}_j^\vee)\,.
\end{equation}
From the identification of the central charges given on the B-side by the periods 
and on the A-side by the map \eqref{eq:mu}  
\be 
\Pi^{(p)}_r=\int_{\Gamma_r^{(p)}} \Omega = Z({\mathcal E}^{(p)}_r)=\int_{\hat X} e^{\sum_i t^i\cdot J_i}\wedge \hat \Gamma(T\hat X) \wedge {\rm ch}( {\mathcal E}^{(p)\vee}_r)=\int_{\hat X} e^{\sum_i t^i\cdot J_i}\wedge \mu( {\mathcal E}^{(p)\vee}_r)\,.
\label{eq:ident}
\ee
Note that, using the identifications \eqref{eq:mirrormap}, $X^0=\int_{\Gamma^{(0)}_1}\Omega $, $X^r=\int_{\Gamma^{(1)}_r}\Omega$ we can identify the coefficient of the $p'\le p$ logarithmic 
Frobenius solution on the left hand side by comparing the $t^i$ powers in \eqref{eq:ident} in terms of classical intersection numbers on $\hat X$ appearing on the right hand side 
of~\eqref{eq:ident}. 

If $\hat X$ is a hypersurfaces or complete intersections in toric geometry obtained by the Batyrev~\cite{Batyrev:1993oya}  or Batyrev--Borisov \cite{MR1463173} mirror construction one can identify $\tilde {\mathcal E}^{(p)}_r$ in terms  of toric sheaves restricted  
to $\hat X$ with complex maximal dimension of support  $p$ and calculate $\mu(\tilde{\mathcal E}^{(p)}_r)$ explicitly~\cite{MR2683208,Gerhardus:2016iot}. In general, the corresponding intersection form does not obey the $^{(k)}$ grading, exemplified in \eqref{eq:sigma},  that naturally corresponds to the grading of the $U(1)_{L/R}$ charges of the topological metric  between the chiral-chiral and chiral-antichiral ring operators of the twisted  $N=(2,2)$ non-linear $\sigma$-model. It is important to notice that the corresponding Frobenius structure with its $U(1)_{L/R}$ charge grading read out here at the MUM point  from the logarithmic grading is rooted structurally in the $\sigma$-model. Since the infinitesimal complex and quantum K\"ahler moduli deformations are $(2,2)$ marginal deformations, which are unobstructed, this structure extends over the corresponding moduli spaces. 

Therefore, it is natural for Calabi--Yau manifolds to choose ${\mathcal E}^{(p\le 1)}_r= \tilde{\mathcal E}^{(p\le 1)}_r$ as in~\cite{MR2683208}, where in particular ${\mathcal E}^{(1)}_r$ $r=1,\ldots, h^{n-1,1}(X)$ corresponds to the marginal deformations. For the K-theory objects corresponding to the  higher logarithmic periods, we choose a basis so that the intersection form obeys the grading. This can be achieved by a lower triangular transformation 
\be 
{\mathcal E}^{(k)}_r=\sum_{l\le k} c_l \tilde {\mathcal E}^{(l)}_r
\ee
with respect to the grading. We shall give an example immediately: for the highest logarithmic period we identify 
\be
\label{eq:shift}
{\mathcal E}^{(n)}_1={\mathcal O}_{\hat X}-\frac{1}{2}(1+(-1)^n) {\mathcal O}_{\hat X_{\rm pt}}
\ee
for even and odd dimension $n$ as in~\cite{CaboBizet:2014ovf}. Here, ${\mathcal O}_{\hat X \rm pt}$ is the skyscraper sheaf corresponding to the type IIA brane 
$D_0$ supported at a point and ${\mathcal O}_{\hat X}$ is the structure sheaf corresponding to the $D_{2n}$ brane supported on all of $\hat X$. Since $\int_{\hat X} \mu({\mathcal O}^\vee_{\hat X}) \wedge  \mu({\mathcal O}_{\hat X})=(1+(-1)^n)$, $\int_{\hat X} \mu({\mathcal O}^\vee_{\hat X}) \wedge  \mu({\mathcal O}_{{\hat X}_{pt}})=1$, the shift \eqref{eq:shift} is necessary for $\Sigma$ to be block-anti-diagonal and compatible with the grading in both cases. In \cref{sec:CY4}, we will give a basis choice for the periods of Calabi--Yau four-folds that respects this grading, cf.\ \eqref{eq:sigma}.
\paragraph{}
In four or more dimensions, the absence of a prepotential prompts us to use \eqref{eq:ident} to obtain a rational basis for the periods. One starts with a basis where the $\mathrm{D}(2n)$- and $\mathrm{D}(2n-2)$-branes correspond to the structure sheaves of $\hat X$ and on divisors $D_i$. The lower dimensional branes can be described by structure sheaves on intersections of these divisors. For the structure sheaf on $\hat{X}$, we have $\ch(\cO_{\hat{X}})=1$. To compute the Chern characters of the structure sheaf on an intersection of divisors $S=\bigcap_{i\in \mathcal{I}}D_i$, we use the long exact sequence 
\begin{align}
\begin{split}
\label{eq:Koszul}
    &0\longrightarrow \cO_{\hat{X}}(-\sum_{i\in\mathcal{I}}D_i) \longrightarrow\bigoplus_{j\in\mathcal{I}}\cO_{\hat{X}}(-\sum_{i\in\mathcal{I}\backslash j}D_i)\longrightarrow\ldots\\
    &\hspace{3cm}  \ldots\longrightarrow \bigoplus_{i\in\mathcal{I}}\cO_{\hat{X}}(-D_i) \longrightarrow \cO_{\hat{X}}\longrightarrow \cO_{S}\longrightarrow 0
\end{split}
\end{align}
and the fact that the alternating sum over the elements' Chern characters vanishes. For example, for the $\mathrm{D}(2n-2)$-branes, we choose $S=D_i$ and obtain
\begin{equation}
    \ch(\cO_{D_i}) = \ch(\cO_{\hat{X}})-\ch(\cO_{\hat{X}}(-D_i))=1-e^{-J_i},
\end{equation}
while for the $\mathrm{D}(2n-4)$-branes with $n>3$, we use $S=D_1\cap D_2$ and obtain
\begin{align}
\begin{split}
    \ch(\cO_{D_i\cdot D_j}) &= \ch(\cO_{\hat{X}})-(\ch(\cO_{\hat{X}}(-D_i))+\ch(\cO_{\hat{X}}(-D_j))) + \ch(\cO_{\hat{X}}(-D_i-D_j))\\
    &=1-e^{-J_i}-e^{-J_j}+e^{-J_i-J_j}.
\end{split}
\end{align}
The polynomial contributions in the mirror coordinates $\vec{t}$ to the brane charges can then be computed via \eqref{eq:ident}. The D2-branes are wrapped on curves $\mathcal{C}^i$ dual to the divisors $D_i$ and for the D0-brane one uses $[\text{pt}]$. It follows that
\begin{align*}
    \Pi_{\mathcal{C}^i}^\text{asy} = (-1)^{n+1}t^i,\\
    \Pi_{\text{pt}}^\text{asy} = 1\,.
\end{align*}

\section{Canonical form of the period matrix}
\label{sec:periodmatrixnfolds}
In this section, we make explicit the relation between the real Riemann bilinear inequality \eqref{Kahler},
the holomorphic  bilinear called Griffiths transversality \eqref{Griffiths}, the Frobenius algebra 
structure \eqref{eq:Frobeniusalgebra} and an integer choice of the basis of the periods induced 
from the $\hat \Gamma$-class \eqref{eq:hatGamma} for Calabi--Yau $n$-folds. 
The focus is to determine a basis for the middle cohomology in which the period matrix  
takes a special unipotent form and the Gauss--Manin connection simplifies drastically. This 
is of general interest. It is
important to study the Dwork deformation method in the arithmetic geometry of Calabi--Yau  
$n$-folds with higher dimensional moduli spaces \cite{Candelas:2024vzf}.  

It also  plays an important role in the applications of Calabi--Yau periods to Feynman integrals. The integration by parts relations between the master integrals $I(\uz,\epsilon)$ 
can always be recast in a first order form~\cite{Henn:2013pwa}, representing a flat 
connection on a finite-dimensional graph cohomology~\cite{Laporta:2000dsw}, 
which determine in principle the master integrals in a Laurent expansion in $\epsilon$. 
The corresponding Gauss--Manin connection depends on the dimensional regularisation parameter $\epsilon$
\be 
\label{eq:IBPGM}
\nabla I(\uz,\epsilon):= ( d_\uz - {\bf \underline{B}}(\uz,\epsilon))I(\uz,\epsilon)=0\,.
\ee
To proceed, it is essential, but not proven to be possible in general, to bring it by 
an $(\uz,\epsilon)$-dependent transformation $J(\uz,\epsilon)={\bf R}(\uz,\epsilon)I(\uz,\epsilon)$ in an $\epsilon$-factorised form~\cite{Henn:2013pwa} (see also \cite{Bonisch:2021yfw,Gorges:2023zgv,Klemm:2024wtd,Duhr:2024bzt,Frellesvig:2024rea,Driesse:2024feo}) 
\be
\label{eq:epsilonfactorised}
\nabla_{\epsilon} J(\uz,\epsilon) := ( d_\uz - \epsilon {\bf \underline{A}}(\uz)) J(\uz,\epsilon) =0 \,,
\ee 
where the main task is finding a rotation ${\bf R}(\uz,\epsilon)$ (or a sequence thereof) such that the transformed connection $\tilde{\bf \underline{B}}$ satisfies in the final step 
\be 
  \left({\bf R}(\uz,\epsilon){\bf \underline{B}}(\underline{z},\epsilon)+d_\uz {\bf R}(\uz,\epsilon)\right) {\bf R}^{-1}(\uz,\epsilon)\eqqcolon \tilde{\bf \underline{B}}(\uz,\epsilon )=\epsilon {\bf \underline{A}}(\uz)\, .
\ee
In leading order in $\epsilon$ w.r.t.\ the critical dimension, the Gauss--Manin system \eqref{eq:IBPGM}, written here symbolically for arbitrary number of moduli, can coincide with the Gauss--Manin system for the variation of \emph{the pure Hodge structure} or \emph{the mixed Hodge structure} as can be either seen by the appearance of blocks with Calabi--Yau Gauss--Manin connection matrices  
in suitably written IBP relations~\cite{Bonisch:2021yfw,Klemm:2024wtd,Gorges:2023zgv,Driesse:2024feo} or by 
relating the Feynman integrals directly to Calabi--Yau geometries as e.g.\ in the Symanzik representations \cite{Bonisch:2021yfw} or more 
systematically in the Baikov representation, see e.g.~\cite{Frellesvig:2023bbf},  
where maximal cut and non-maximal cut integrals can be more easily distinguished.     

Suppose that an $\epsilon$-factorised form has been achieved~\eqref{eq:IBPGM} and we have boundary conditions $J(\uz_0,0)=J_0$, then we can write the solution vector $J(\uz,\epsilon) $
\be 
\label{eq:wavefunction}
J(\uz,\epsilon)=\mathbb{P} \exp\left[\epsilon\int_{\uz_0}^{\uz} {\bf \underline{A}} (\uz) d z \right] J_0   
\ee 
in terms of iterated integrals in the form 
\be 
\label{eq:iterated}
J(\uz,\epsilon)=\left[
{\bf 1}+ \epsilon \int_{\uz_0}^\uz {\bf \underline{A}}(\uz')  dz' + \epsilon^2    
\int_{\uz_0}^\uz \int_{\uz_0}^{\uz'} {\bf \underline{A}}(\uz')dz' {\bf \underline{A}} (\uz'') dz''  +\ldots \right]J_0\, .
\ee

The quite involved details of the steps to go from the IBP relations, as provided by computer programs, like \texttt{Kira}~\cite{Klappert:2020nbg} and  \texttt{FireFly}~\cite{Klappert:2020aqs}  in its raw form, over \eqref{eq:IBPGM} to \eqref{eq:epsilonfactorised} depend very crucially on the choice for the basis the master integrals in the initial stage. 
The subsequent procedure is not completely algorithmic, but 
a step-by-step strategy has been outlined in~\cite{Gorges:2023zgv}, in the case when a block of the pure variation of a 
Calabi--Yau Hodge structure and the corresponding Gauss--Manin connection appears.

In analogy to the situation for polylogarithms in one-loop integrals or in the variations of mixed Hodge structures~\cite{MR2393625},
iterated integrals involving Calabi--Yau periods appear in producing the $\epsilon$-form
and raise the transcendentality in  \eqref{eq:iterated} in a systematic way. As described in~\cite{Gorges:2023zgv}, oftentimes, a splitting of the period matrix\footnote{In our convention of the period matrix for the one-parameter case it is up to a rational basis change equal to the Wronskian, see for example \eqref{eq:3foldperiodmatrix}.} into a \emph{semisimple and unipotent part} serves as one of the essential steps to obtain the $\epsilon$-factorised form of~\eqref{eq:epsilonfactorised}. The inverse of the semisimple part is used in the top sectors as one of the rotation matrices ${\bf R}_*(\uz,\epsilon)$. This method has been used to obtain an analytic expression for the post-Minkowskian (PM) perturbative approximation to black hole scattering in the 5PM, one self-force sector \cite{Klemm:2024wtd,Driesse:2024feo}. In general, the IBP relation will not 
lead to a pure but rather to a mixed variation of Calabi--Yau Hodge structure with correspondingly larger Calabi--Yau blocks, as e.g.\ for the Calabi--Yau that appears for example in the two self-force sector~\cite{Frellesvig:2023bbf}. A general theory of what 
kind of integration kernels and transcendentality will occur in this case will be discussed in \cite{Duhr:inprep}. However, also in this case it is advantageous to split the  
period matrix involving no relative cohomology into a semisimple and nilpotent part\footnote{We thank the authors of~\cite{Duhr:inprep} for sharing insights, which motivates the discussion below.}    .

We therefore discuss the latter problem in some generality. We introduce the notation and general idea, we review the splitting for elliptic curves, K3 surfaces and Calabi--Yau three-folds. We then extend the list to four-folds and give instructions to derive such a splitting in higher dimensions, exemplified in \cref{sec:appCY6} for six-folds.

\subsection{Elliptic curves and K3 surfaces}
\label{ss:ellK3}
Let us introduce the concept with the family of elliptic curves ${\cal E}_z$ and families of polarised 
K3 surfaces ${\cal K}_{\uz}$  as examples. In the 
former case, we introduce the symplectic basis $\alpha_0,\beta^0$ lattice basis  
in $H^1({\cal E}_z, \mathbb{Z})$ with dual basis $(A,B)$ of $H_1({\cal E}_z,\mathbb{Z})$ with $A\cap B=-B\cap A=1$. The latter choice is called {\sl a marking}. 
We normalise the period vector to ${\underline \Pi}^T=(1,t(z))$ with $\underline{\tilde \Pi}^T= (\int_A \Omega(z), \int_B\Omega(z))=(X^0(z),F_0(z))=X^0(z)(1,t(z))$. 
Here, $X^0,F_0$ is a Frobenius basis, e.g.\ 
at a point of maximal unipotent monodromy at $z=0$, $t=F_0/X^0=\frac{\log(z)}{2 \pi i}+ \sigma(z)$, with $X_0(z)$ and $\sigma(z)$ holomorphic, and define 
\be
\Omega_0=\alpha_0+t \beta^0, \Omega^0=\beta^0\, .
\label{basisperiods} \ee
The period matrix $\boldsymbol{\Pi}=\Pi_{\text{ss}} \Pi_{\text{up}}$ factorises into a semi simple part 
\be 
\Pi_\text{ss}=\begin{pmatrix} X^0& 0\\ \partial_z X^0& \partial_z F_0-\frac{\partial_z X^0}{X^0} F_0
\end{pmatrix}
\ee 
and the unipotent part 
\be
{\Pi_\text{up}}=\begin{pmatrix} 1&t\\0 & 1\end{pmatrix},
\qquad 
\partial_t \begin{pmatrix} \Omega^0\\ \Omega_0\end{pmatrix}=
\begin{pmatrix} 0&1\\0&0 \end{pmatrix} \begin{pmatrix} \Omega^0\\ \Omega_0\end{pmatrix}\ \ee
so that the Gauss--Manin connection in the inhomogenous $t$ variable is in a canonical  form.
The first Riemann bilinear relation \eqref{Kahler} implies $2 {\rm Im}(t)>0$, i.e.\ the 
normalised period $t(z)$ lives in the period domain, which is the upper half space
${\cal H}=\{t\in \mathbb{C}| {\rm Im}(t)>0\}$, a symmetric domain analytically equivalent to the unit disc, which in turn is equivalent to 
$\cal H\simeq {\rm Sl}(2,\mathbb{R})/U(1)$. The local and the global Torelli theorem 
identify the latter with the complex structure moduli space of the marked family ${\cal E}_z$ 
and the redundancy of the choice of marking, which is up to a $\Gamma={\rm Sl}(2,\mathbb{C})$ transformations, acting as a fractional  linear transformation on $t$,   
is finally removed by identifying the fundamental region of moduli space with  ${\cal F}=\Gamma\backslash{\cal H}$.       
The $k=0$ piece  of the second bilinear relation \eqref{Griffiths} is trivial and the $k=1$ piece  becomes the {\sl Legendre relation}, which takes for the normalised periods in the $t$ coordinate the form 
\be 
\vec{\Pi}^T\Sigma \partial_t \vec{\Pi}= \frac{\partial z}{\partial t}\frac{X^0 \partial_z F_0-F_0 \partial_z X^{0}}{(X^0)^2} =\frac{\partial z}{\partial t}\frac{\alpha(z)}{(X^0)^2}=1 \, .
\ee
Let $\Lambda$ be the even unimodular lattice of signature $(3,19)$   
\be
\Lambda=U\oplus U\oplus U\oplus E_8(-1)\oplus E_8(-1)\, , 
\ee
where $E_8(-1)$ is given by the negative of the Cartan matrix 
of the Lie algebra $E_8$ and $U={\tiny{\begin{pmatrix}0&1\\1&0\end{pmatrix}}}$. One 
further defines for each $\lambda\in\Lambda$ with $(\lambda)^2=2 d$ the orthogonal complement of $\lambda$ in $\Lambda$  $\Lambda_d=\lambda^\perp\simeq E_8(-1)\oplus E_8(-1)\oplus U\oplus U\oplus \mathbb{Z}(-2 d)$. Analogous to the elliptic curve case above, a marked K3 family is defined by an isomorphism of  
$H^2({\cal K}_\uz,\mathbb{Z})$ into $\Lambda$. The period domain \cite{MR3586372}
is constrained by \eqref{Kahler} and the non-trivial $k=0$ piece in \eqref{Griffiths} 
and defined to be 
\be 
D=\left\{x \in \mathbb{P}(\Lambda_\mathbb{C})|\langle x,\bar x\rangle>0 \ {\rm and} \    \langle x,x\rangle=0\right\}\, .
\ee
For any lattice $\Lambda$, we define $\Lambda_\mathbb{C}\coloneqq\Lambda\otimes_{\mathbb{z}}\mathbb{C}$ and $x$ 
will be identified with the period  vector. The projectivisation is to remove the 
redundancy in the choice of $\Omega$ in the K\"ahler 
line bundle ${\cal L}$ and corresponds to the normalisation of the periods. Using this 
freedom and the $k=0$ and $k=1$ part of \eqref{Griffiths}, we see that the  normalised period vector has to be of the form
\begin{align}
    \Omega_0 &= \alpha_0 +t_a \gamma^a -\frac{1}{2} \eta_{ab} t^a t^b\beta^0,\\
    \chi_b &=\partial_{t_b} \Omega_0=\gamma^b - \eta_{ba}t^a \beta^0,\\
    \Omega^0 &= -\beta^0,
\end{align}
where $\eta_{ab}$ is the intersection form of the Picard lattice of the mirror K3 \cite{Dolgachev:1996xw}. As a consequence, the Gauss--Manin connection becomes in the projectivised coordinates  
\begin{equation}
    \partial_a \begin{pmatrix}
        \Omega_0 \\ \chi_b\\ \Omega^0
    \end{pmatrix}
    = \begin{pmatrix}
        0 & \delta_{a}^k & 0  \\
        0 & 0 & \eta_{ab}  \\
        0 & 0 & 0 &\\
    \end{pmatrix}
    \begin{pmatrix}
        \Omega_0 \\ \chi_k\\ \Omega^0
    \end{pmatrix}.
\end{equation}
The above basis directly translates to the unipotent part of the 
period matrix,
\begin{align}
       \Pi_\text{up}=\begin{pmatrix}
        1 & t^a & -\frac{1}{2}\eta_{ab}t^a t^b  \\
        0 & \delta^a_b & -\eta_{ab}t^b  \\
        0 & 0 & -1 \\
    \end{pmatrix} 
\end{align}
and the semisimple part reads
\begin{align}
\scalebox{0.93}{$
\Pi_\text{ss}=\begin{pmatrix}
        X^0 & 0 & 0  \\
        \partial_{z_i}X^0 & \frac{1}{X^0}W^{0j}_{k}& 0 \\
        \eta^{ij}\partial_{z_i}\partial_{z_j}X^0 & \frac{1}{X^0}\eta^{ij}W^{0k}_{ij} & \frac{\eta^{ij}\eta_{kl}W^{0k}_{i}W^{0l}_{j}}{(X^0)^3} \\
    \end{pmatrix}, 
    $}
\end{align}
where $\eta^{ab}$ is the inverse of $\eta_{ab}$ and for brevity we write for the minors of the Wronskian $W^{0i}_j\equiv X^0\partial_{z_j}X^i-X^i\partial_{z_j}X^0$ and $W^{0i}_{jk}\equiv X^0\partial_{z_j}\partial_{z_k}X^i-X^i\partial_{z_j}\partial_{z_k}X^0$.

The $k=2$ relation of Griffiths transversality \eqref{Griffiths} becomes 
\be
{\underline \Pi}^T \Sigma \partial_{t_a} \partial_{t_b} {\underline \Pi}=\frac{1}{(X^0)^2} 
\frac{\partial z_i}{\partial_{t_a}} \frac{\partial z_j}{\partial_{t_b}} {\tilde {\underline \Pi}}^T\Sigma \partial_{z_i}\partial_{z_j} {\tilde {\underline \Pi}}= \frac{1}{(X^0)^2} 
\frac{\partial z_i}{\partial_{t_a}} \frac{\partial z_j}{\partial_{t_b}} \alpha_{ij}=\eta_{ab}\,.
\ee
This expresses the triviality of the quantum cohomology of K3 due to the negative virtual dimensions of the moduli space of holomorphic maps into K3 surfaces and reflects the fact  
that the period domain is a symmetric space \cite{MR3586372}. The Kuga--Sato construction 
described also in \cite{MR3586372} relates this period domain to the one of a rank one Hodge 
structure, i.e.\ the one of a higher genus curve. This implies modular expressions for the 
K3 periods, which are worked out using the Borcherds lift in  \cite{MR1386841}. Some relevance of the aspects of the K3 structure to Feynman integrals
was pointed out in \cite{Bonisch:2021yfw}.

\subsection{Calabi--Yau three-folds}
\label{ss:CY3folds}
To study the implications of \eqref{Kahler} and \eqref{Griffiths}, which are known as {\sl special geometry} for $n=3$, we write\footnote{Note that the labels on 
the $B$-cycles and $\beta$-forms run in reverse order $r,\ldots,0$, 
to make $\Sigma$ anti-diagonal with non-vanishing entries 
$\Sigma_{b_n-k,k+1}=1$, $k=0,\ldots,r$ and $\Sigma_{b_n-k,k+1}=-1$ for $k=r+1,\ldots,b_n-1$.}     $\{\underline \Gamma\}=\{A^I,B_{r-J}$, $I,J=0,\ldots, r\}$ and 
$\{\underline \gamma\}=\{\alpha^I,\beta_{r-J}$, $I,J=0,\ldots, r\}$ with  $A^I\cap B_J = - B_J\cap A^I =\delta_J^I
=\int_{A^J} \beta^I=\int_{B_I} \alpha_J$ and other intersections- and integral pairings are zero, leading to
\begin{equation}\label{eq:sigmaCY3}
    \Sigma=\begin{pmatrix}
        0 & 0 & 0 & 1\\
        0 & 0 & 1 & 0\\
        0 & -1 & 0 & 0\\
        -1 & 0 & 0 & 0\\
    \end{pmatrix}.
\end{equation}
The period vector is denoted  
$\uPi(\uz)=(X^I,F_{r-J})^T=(\int_{A^I}\Omega(\uz), \int_{B_{r-J}}\Omega(\uz))^T$, $I,J=0,\ldots,r$ and 
\eqref{Kahler} specialises with
\begin{equation}
    \Omega=X^I\alpha_I + F_I \beta^I
\end{equation}
to $e^{-K(\uz)}=\ii\left(\overline{X}^I F_I - X^I \overline{F}_I\right)$. Moreover \eqref{Griffiths}  implies 
$F_I=\partial_{X_I} F(\uX)$, where $F(\uX)=2 X^I F_I$ is the holomorphic prepotential of degree two in $\uX$. We can hence write the period vector as
\be 
\label{eq:leadingorder}
    \vec{\Pi} = X^0\begin{pmatrix}1\\t^i\\\partial_{t^i}\mathcal{F}(\ut)\\2\mathcal{F}(\ut)-t^j\partial_{t^j}\mathcal{F}(\ut)\end{pmatrix}\ 
\ee    
in inhomogenous coordinates in terms of the inhomogeneous prepotential ${\cal F}(\ut):=F(\uX)/(X^0)^2 $, which 
fulfils 
\be 
-\partial_{t_i}\partial_{t_j}\partial_{t_k} {\cal F}(\ut)=C_{ijk}(\ut)=\frac{1}{(X^0(\uz(\ut))^2} 
\sum_{a,c,b} C_{z_az_bz_c}(\uz(\ut)) \frac{\partial z_a}{\partial t^i} \frac{\partial z_b}{\partial t^j} \frac{\partial z_c}{\partial t^k}\ .
\label{eq:triplecoupling}
\ee
The $C_{z_az_bz_c}\in \mathbb{Q}(\uz)$ 
are the rational functions determined by  \eqref{Griffiths} and \eqref{eq:PFI} up to a multiplicative constant. The transformation \eqref{eq:triplecoupling} reflects the general fact that $C_{\uz_I^{(n)}}(\uz)$ is a section of ${\cal L}^{-2} \otimes {\rm Sym}^n((T_{\Mcs}^*)^{\otimes n})$ over $\Mcs$.
For $n=3$, we fix then all coefficients in the unique highest 
logarithmic period $\Pi^{(3)}_1$ by \eqref{eq:ident} purely in terms 
of intersection numbers on $\hat X$ and use special geometry as encoded  
in \cref{eq:inhomogeneous,eq:leadingorder,eq:triplecoupling} and integrality of the monodromy at the MUM point to conclude that ${\cal F}$ has to be cubic in the $t^i$ and has the form
	\begin{equation}\label{eq:prepotential}
	    \mathcal{F}(\ut) = -\frac{1}{3!}\kappa_{ijk}t^{i}t^{j}t^{k}+\frac{1}{2!}A_{ij}t^it^j+\frac{[c_2(T_{\hat X})]\cdot D_i}{24}t^i+\frac{\chi\,\zeta(3)}{2(2\pi\ii)^3}-\mathcal{F}_{\text{inst}}(q_1,\ldots,q_{h_{2,1}})\,.
	\end{equation}
where the $\kappa_{ijk}$ are identified with the classical triple intersection numbers
\begin{equation}
	    \kappa_{ijk}=\int_{\hat{X}}J_{i}\wedge J_j \wedge J_{k}=D_{i}\cdot D_{j} \cdot D_{k}\,.
\label{eq:classical} 
\end{equation}
Here, the $J_i$ form a basis of $H^{1,1}(\hat X,\Z)$ and are Poincaré dual to divisor classes 
$[D_i]$ in $H_4(\hat X,\mathbb{Z})$. According to \eqref{eq:ident}, the lower order terms are given by the Euler number $\chi(\hat X)$, the evaluation
of the second Chern class on a basis of $H_4(\hat X,\mathbb{Z})$ 
and\footnote{Modulo Sp$(b_3,\mathbb{Z})$ only the half integral part of the $A_{ij}$ are fixed and determined by demanding integral monodromies around  $z_i=0$ given the $C^{(0)}_{ijk}$ and $[c_2(\hat X)]\cdot D_i$ and using \eqref{eq:leadingorder}. For an
explicit calculation see~\cite{Huang:2006hq}.}   
\be 
\chi=\int_{\hat X} c_3(T_{\hat X})\,,\qquad [c_2(\hat X)]\cdot D_i=\int_{\hat X} c_2(T_{\hat X})\wedge J_i\,,\qquad A_{ij}=\frac{1}{2} \int_{\hat X} i_* c_1(D_i)\wedge J_j\, .
\label{eq:subleading} 
\ee
As explained in \cref{sec:quotientbuilding} and \cite{MR3965409}, these topological intersection numbers \eqref{eq:classical} and \eqref{eq:subleading} are in particular easily calculable, if $\hat X$ is embedded as hypersurface or complete intersection in a toric ambient space. 

The last term $\mathcal{F}_\text{inst}$ collects the exponentially suppressed instanton contributions in the $q$-coordinates $q^i=e^{2\pi\ii\,t^i}$ and has the following form
	\begin{equation}
	    \mathcal{F}_\text{inst}(\uq) = \frac{1}{(2\pi\ii)^3}\sum_{\vec{\beta} >0}n_0^{\beta_1,\ldots,\beta_{h_{2,1}}}\,\mathrm{Li}_3\left(\prod_i e^{2\pi\ii\,\beta_i t_i}\right).
	\end{equation}
	The integers $n_0^{\beta_1,\ldots ,\beta_{h_{2,1}}}\in \mathbb{Z}$ are the genus zero or rational curve counting 
	BPS numbers. Where ${\underline \beta} \in H_2(\hat X,\mathbb{Z})$ denotes a class in the integral second homology of $\hat X$.
Mirror symmetry identifies not only the number of solutions with 
leading power $p$ to $h^{p,p}_{\text{vert}}(\hat X)$, but the Fukaya category of Lagrangian $n$-cycles on $X$ 
with the derived category of coherent sheaves with $2p$-dimensional support on $\hat X$ and 
${\cal O}(\Sigma,\mathbb{Z})$ with auto-equivalences of the derived category. In particular 
now $\uz$ have a power series expansion in $q_i=e^{2\pi\ii t_i}$ and are hence 
invariant under $t_i\rightarrow t_i+1$.

We can complete $\Omega$ to a basis of $H^3(X,\Z)$ to give the Gauss--Manin connection in \eqref{eq:GaussManin} explicitly. We perform a Kähler gauge transformation on $\Omega$ and express the $(3,0)$-form as
\begin{equation}
    \Omega_0 = \alpha_0 + t^i \alpha_i - \partial_i \cF \beta^i - (2\cF - t^i\partial_i \cF)\beta^0,\label{eq:Omega0CY3}
\end{equation}
where $\partial_i\equiv \partial_{t^i}$. The remaining generators are chosen as
\begin{align}
    \chi_i &= \partial_i \Omega_0 = \alpha_i - \partial_i \partial_j \cF \beta^j - (\partial_i \cF - t^j\partial_i \partial_j \cF)\beta^0,\\
    \chi^i &= -\beta^i + t^i \beta^0,\\
    \Omega^0 &= \beta^0.
\end{align}
With \eqref{eq:triplecoupling}, it follows immediately that the Gauss--Manin connection acts on this basis as
\begin{equation}
    \partial_i \begin{pmatrix}
        \Omega_0 \\ \chi_j\\ \chi^j\\ \Omega^0
    \end{pmatrix}
    = \begin{pmatrix}
        0 & \delta_{i}^k & 0 & 0 \\
        0 & 0 & C_{ijk} & 0 \\
        0 & 0 & 0 & \delta_i^j\\
        0 & 0 & 0 & 0
    \end{pmatrix}
    \begin{pmatrix}
        \Omega_0 \\ \chi_k\\ \chi^k\\ \Omega^0
    \end{pmatrix}.
\end{equation}
To give a closed form for the semisimple part of the period matrix, we introduce a new object $\mathfrak{C}^{ijk}$, which is defined as an inverse of $C_{ijk}$ in the following sense:
\begin{align}
   \mathfrak{C}^{ijk}C_{ijl}=\delta^k_l\,.
\end{align}
In the unipotent part of the period matrix such inverse couplings appear only for $n>4$. Below \eqref{eq:inverseCWronskian} we comment on the existence of such inverse couplings in the $n$-fold case. Following \cite{Candelas:2024vzf} we define the period matrix as\footnote{The minus signs are in accordance with the first row of the unipotent part of period matrix. Further note that in \cite{Candelas:2024vzf} only the inverse classical couplings are used. We explain below why this makes no difference for the purpose of calculating the local zeta functions as in \cite{Candelas:2024vzf}.} 
\begin{align}
\label{eq:3foldperiodmatrix}
\boldsymbol{\Pi}=\begin{pmatrix}
        X^0 & X^l & -F_l & -F_0 \\
        \partial_{z_i}X^0 & \partial_{z_i}X^l & -\partial_{z_i}F_l & -\partial_{z_i}F_0 \\
        \mathfrak{C}^{ijk}\partial_{z_i}\partial_{z_j}X^0 &  \mathfrak{C}^{ijk}\partial_{z_i}\partial_{z_j}X^l & - \mathfrak{C}^{ijk}\partial_{z_i}\partial_{z_j}F_l & -\mathfrak{C}^{ijk}\partial_{z_i}\partial_{z_j}F_0 \\
    \mathfrak{C}^{ijk}\partial_{z_i}\partial_{z_j}\partial_{z_k}X^0 &  \mathfrak{C}^{ijk}\partial_{z_i}\partial_{z_j}\partial_{z_k}X^l & - \mathfrak{C}^{ijk}\partial_{z_i}\partial_{z_j}\partial_{z_k}F_l & - \mathfrak{C}^{ijk}\partial_{z_i}\partial_{z_j}\partial_{z_k}F_0 \\
    \end{pmatrix}.
\end{align}
It then follows, that
\begin{align}
\Pi_\text{ss}=\begin{pmatrix}
        X^0 & 0 & 0 & 0 \\
        \partial_{z_i}X^0 & \frac{1}{X^0}W^{0l}_i& 0 & 0 \\
        \mathfrak{C}^{ijk}\partial_{z_i}\partial_{z_j}X^0 & \frac{1}{X^0}\mathfrak{C}^{ijk}W^{0l}_{ij} & -\frac{1}{X^0}\frac{\partial z_k}{\partial t^m} & 0 \\
     \mathfrak{C}^{ijk}\partial_{z_i}\partial_{z_j}\partial_{z_k}X^0 & \frac{1}{X^0}\mathfrak{C}^{ijk}W^{0l}_{ijk} & -\frac{1}{X^0}\mathfrak{C}^{ijk} E_{m,ijk} & -\frac{h^{2,1}}{X^0} \\
    \end{pmatrix},
\end{align}
with
\begin{align}
\begin{split}
     E_{m,ijk}&=\frac{\partial_{t^m}X^0}{X^0} C_{ijk}-(\partial_{t^m}\vec{\Pi}^T)\Sigma\partial_{z^i}\partial_{z^j}\partial_{z^k}\vec{\Pi}\\
     &=\frac{\partial_{t^m}X^0}{X^0} C_{ijk} -\partial_{t^m}C_{ijk} + 2 \frac{\partial z^l}{\partial t^m}\partial_{z^{(l}}C_{ijk)},
\end{split}
\end{align}
where we used the identity \eqref{eq:delYuk}.
 
The Picard--Fuchs ideals for families of Calabi--Yau $n$-folds with one complex structure deformation are generated by so-called Calabi--Yau operators which are analysed in detail in \cite{van2018calabi}. As a consequence of Griffiths transversality such Calabi--Yau operators are in particular \textit{essentially selfadjoint}. Writing an operator $\mathcal{L}^{(n+1)}(z)$ in the form
\begin{align}
   \mathcal{L}^{(n+1)}(z)=\sum_{i=0}^{n+1}a_i(z) \partial^i_z \quad \text{with} \quad a_{n+1}=1
\end{align}
this is the statement 
\begin{align}
\label{eq:selfadjointness}
   \mathcal{L}^{(n+1)}\alpha=(-1)^{n+1}\alpha \mathcal{L}^{(n+1)\vee}
\end{align} 
for some non-zero function $\alpha(z)$, where the dual operator $\mathcal{L}^{(n+1)\vee}$ is defined as 
\begin{align}
   \mathcal{L}^{(n+1)\vee}(z)=\sum_{i=0}^{n+1}(-\partial_z)^i a_i(z)\,.
\end{align}
If such an $\alpha$ exists, it fulfils the differential equation \cite{van2018calabi} 
\begin{align}
\label{eq:yukawaode}
    \alpha^\prime=-\frac{2}{n+1}a_{n}\alpha
\end{align}
and $\alpha(z)$ is, up to normalisation, the Yukawa coupling $C_{z\dots z}(z)$. Equivalently to self-adjointness, the coefficients of the operator fulfil certain relations. In general, we have from comparing the coefficients of the $k$-th derivative in \eqref{eq:selfadjointness} \cite{Almkvist:2004kj,bogner}
\begin{align}
\sum_{j=k}^{n+1} \binom{j}{k} \left\{ \frac{\alpha^{(j-k)}}{\alpha} a_j + (-1)^{n+j} a_j^{(j-k)} \right\} = 0\,,
\end{align}
which is a constraint in the $a_i$ only, upon using \eqref{eq:yukawaode} to write
\begin{align}
    \frac{\alpha^{(j)}(z)}{\alpha(z)}=j!\sum_{p\in \mathcal{P}(j)}\prod_{i=1}^{n}\frac{\left(-\frac{2}{n+1}a_n^{(i-1)}\right)^{\#_i(p)}}{i!^{\#_i(p)}\#_i(p)!}\,,
\end{align}
where $p\in\mathcal{P}(j)$ are the partitions of $j$ and $\#_i(p)$ is the number of times $i$ appears in $p$. For $n=1$, self-adjointness is automatically fulfilled, while for $n=2$, it imposes one independent relation:
\begin{align}
     0=4 a_2^3+9a_2^{\prime\prime} + 18 a_2 a_2^\prime  + 54 a_0 -18 a_1 a_2 - 27a_1^\prime\,.
\end{align}
For $n=3$, there is also a single constraint given by
\begin{align}
\label{eq:1111Q}
    0=a_3^3+4a_3^{\prime\prime}+6a_3a_3^{\prime}+8a_1-4a_2a_3-8a_2^{\prime}\,.
\end{align}
These conditions can alternatively be obtained by solving the system of equations given by Griffiths transversality \eqref{Griffiths} together with their derivatives and using the differential operator to express derivatives of the periods of the order of the differential operator in terms of lower order derivatives. As shown in \cite{Almkvist:2004kj}, a consequence of this relation is that the minors of the Wronskian of solutions of the Picard--Fuchs operator fulfil a differential equation of order five, which will be used in \cref{sec:AntisymProducts} to reduce F-theory vacua on Calabi--Yau four-folds to type IIB vacua on the three-fold whose antisymmetric product in the above sense gives rise to the four-fold operator.

\subsection{Calabi--Yau four-folds}
\label{sec:CY4}
As mentioned in \cref{sec:GammaClass}, from dimension four on, we need to specify a basis for the middle homology obtained from intersections of divisors.
For the periods given by \eqref{eq:ident}, we use the following ordering of sheaves ($r=h_{3,1}(X)$)
\begin{align}\begin{split}\label{eq:basis}
    ([\text{pt}],\mathcal{C}_1,\ldots ,\mathcal{C}_r,\cO_{D_r\cdot D_r}, \cO_{D_{r-1}\cdot D_r},&\cO_{D_{r-1}\cdot D_{r-1}},\ldots \\
    &\ldots ,\cO_{D_1\cdot D_r},\ldots,\cO_{D_1\cdot D_1},\cO_{D_r},\ldots,\cO_{D_1},\cO_{\hat{X}})\,.\end{split}
\end{align}
Note that the basis for the sheaves with support in codimension two is generally redundant. Eventually, we will therefore need to project onto an independent basis (cf.\ \eqref{eq:nondegint}).
We define the matrix
\begin{equation}\label{eq:Lambdabc}
    \Lambda=\begin{pmatrix}
    1 & 0 & D & C & -1\\
    0 & \mathbb{1}_{r\times r} & B & A & 0 \\
    0 & 0 & \mathbb{1}_{\frac{r(r+1)}{2}\times \frac{r(r+1)}{2}} & 0 & 0 \\
    0 & 0 & 0 & \mathbb{1}_{r\times r} & 0\\
    0 & 0 & 0 & 0 & 1
    \end{pmatrix}
\end{equation}
with
\begin{align}
    A_{ij}&=\begin{cases}
    0, & i+j >r+1\,,\\
    \frac{1}{2}\chi(\cO_{D_i},\cO_{D_i}) & i+j=r+1\,,\\
    \chi(\cO_{D_{i}},\cO_{D_{r+1-j}}) & \text{else}.
    \end{cases}\\
    B_{i\alpha}&= \chi(\cO_{D_{i}},\cO_{S_\alpha})\,,\\
    C_{i} &= - \chi(\cO_{{\hat{X}}},\cO_{D_{r+1-i}})\,,\\
    D_{\alpha} &= -\chi(\cO_{{\hat{X}}},\cO_{S_\alpha})\,,
\end{align}
where $S_\alpha=D_l\cdot D_k$ for $l$ and $k$ given by the ordering in \eqref{eq:basis}. For example, $S_1=D_r\cdot D_r$ and $S_{h_{2,2}-1} = D_{1}\cdot D_2$. One can show that this matrix brings the intersection $\hat\Sigma$ in the basis \eqref{eq:basis} into block-anti-diagonal form
\begin{equation}
    \Lambda^T\hat\Sigma\Lambda = \begin{pmatrix}
     &        &        &       & 1 \\
     &        &        & \eta^{(1,3)} & \\
     &        & C_{\alpha\beta} &       & \\
     & \eta^{(3,1)}&        &       & \\
   1 &        &        &       &
  \end{pmatrix}
\end{equation}
with $\eta^{(3,1)}_{ij}=\eta^{(1,3)}_{ij}=-\delta^i_{h^{3,1}-j+1}$.
As mentioned above, the set of sheaves is generally redundant, which implies that the the block $C_{\alpha\beta}$ is degenerate and we must project onto an independent basis of $H^{\text{vert}}_{2,2}(\hat{X},\Z)$ with a matrix $P$. In our cases, we fix these bases by omitting certain combinations of $D_i\cdot D_j$. The projection $P$ is then simply an identity matrix where the columns corresponding to sheaves $\mathcal{O}_{D_i\cdot D_j}$ are removed. We denote the restricted form by
\begin{equation}\label{eq:nondegint}
    \Sigma^{-1} = (\Lambda P)^T\hat\Sigma \Lambda P
\end{equation}
In order to have a period basis that is subject to the intersection form $\Sigma$, we perform the change of basis $\Pi= (\Lambda P)^T \Pi^\text{asy}$. Then, $\int_{X} \Omega \wedge \vec{\partial}\Omega = \Pi^T\Sigma \vec{\partial}\Pi$. 
The matrix $\Lambda$ is furthermore guaranteed to have integral coefficients. This is because it consists of intersection numbers which are integral and the identity
\begin{equation}
    \chi(\mathcal{O}_{D_j},\mathcal{O}_{D_j}) = 2 \chi(\mathcal{O}_{{\hat{X}}},\mathcal{O}_{D_j})\in 2\Z\,,
\end{equation}
which follows directly from the formulae above.
\par
We therefore have derived an integral basis for the periods with symmetric intersection form
\begin{equation}\label{eq:sigma}
    \Sigma= \begin{pmatrix}
     &        &        &       & 1 \\
     &        &        & \eta^{(1,3)} & \\
     &        & \eta^{(2,2)} &       & \\
     & \eta^{(3,1)}&        &       & \\
   1 &        &        &       &
  \end{pmatrix}.
\end{equation}
We denote this period vector as $\vec{\Pi}^T=(X^I,H_l,F_{h_{3,1}-I})= (\int_{A^I}\Omega,\int_{G_l}\Omega,\int_{B_{h_{3,1}-I}} \Omega)$, where $I=0,\ldots,h_{3,1}$, $l=1,\ldots, h^{\text{hor}}_{2,2}$ and $A^I,G_l$ and $B_I$ are cycles dual to the Hodge cohomology groups $H^{4,0}(X,\Z)\oplus H^{3,1}(X,\Z)$, $H^{2,2}_{\text{hor}}(X,\Z)$ and  $H^{1,3}(X,\Z)\oplus H^{0,4}(X,\Z)$ in the complex stucture of the MUM point.
We read from \eqref{eq:ident} and \eqref{eq:Lambdabc} that the highest logarithmic period $F_0$ has the form
\begin{equation}\label{eq:F0}
F_0=X^0\left(\frac{1}{4!}\kappa_{ijkl}t^i t^j t^k t^l+\frac{1}{2}c_{ij}t^i t^j +c_i t^i+c_0 +\mathcal{O}(q)\right)
\end{equation}
with
\be
c_0=\frac{\zeta(4)}{4(2\pi i)^4}\int_{\hat{X}}(7c_2^2-4c_4)-1\,,\ \ c_i=\frac{\zeta(3)}{(2\pi i)^3}\int_{\hat{X}}c_3 \wedge J_i\,, \ \  c_{ij}=-\frac{\zeta (2)}{(2\pi i)^2}\int_{\hat{X}}c_2 \wedge J_i\wedge J_j\,.
\ee
While the shift \eqref{eq:shift} makes the metric block-anti-diagonal, for $\tilde {\mathcal E}^{(4)}={\mathcal O}_{\hat X}$, the expression for $c_0$ becomes $\tilde c_0=\frac{\zeta(4)}{4(2\pi i)^4}\int_{\hat X}(7c_2^2-4c_4)=\int_{\hat X}(7c_2^2-4c_4)/5760$ and yields a period over the $S^4$ 
that vanishes at the closest conifold with a square root cut and an $\Z/2 \Z$ monodromy 
according to the Lefschetz monodromy formula as pointed out in~\cite{CaboBizet:2014ovf}. 
The classical intersection numbers are denoted by $\kappa_{ijkl}$, the Chern classes and the $\eta^{(p,q)}$ 
are calculated using  classical intersection theory on the mirror $\hat X$ of $X$.

An alternative way to obtain the integral basis  is to fix the  subleading terms in \eqref{eq:prepotential} and \eqref{eq:F0} and the rest of the periods by conjugating all monodromy matrices into ${\mathcal O}(\Sigma,\mathbb{Z})$ up to an ${\mathcal O}(\Sigma,\mathbb{Z})$ choice.  Due to the need of extensive analytic continuation this methods is extremely cumbersome in the multi moduli case.

Similar to the case of three-folds, one may choose a basis for $H^{4}(X,\Z)$ such that the Gauss--Manin connection has entries only on the secondary diagonal. In \cref{sec:appCY4}, we show that, in the basis given by
\begin{align}
    \Omega_0 &= \alpha_0 + t^i \alpha_i + H^{\alpha} \gamma_{\alpha} + F_i\beta^i + F_0 \beta^0,\label{eq:Omega0}\\
    \chi_i &= \partial_i \Omega_0\,,\label{eq:chii}\\
    h_{\alpha} &= \gamma_{\alpha}+ \partial_i H_{\alpha} \beta^i - (H_{\alpha}-t^i \partial_i H_{\alpha})\beta^0,\label{eq:halpha}\\
    \chi^i &= \beta^i + t^i\beta^0,\label{eq:chihi}\\
    \Omega^0 &= \beta^0\label{eq:Omegah0},
\end{align}
where $H_\alpha=\eta^{(2,2)}_{\alpha\beta}H^\beta$, the connection takes the form
\begin{equation}\label{eq:GMCY4}
    \partial_i \begin{pmatrix}\Omega_0\\ \chi_j \\ h_\alpha \\ \chi^j \\ \Omega^0
    \end{pmatrix} = \begin{pmatrix}
    0 & \delta_i^k & 0 & 0 & 0 \\
    0 & 0 & C_{ij}^\beta & 0 & 0\\
    0 & 0 & 0 & C_{ik}^{\ \ \gamma} \eta^{(2,2)}_{\alpha \gamma}& 0\\
    0 & 0 & 0 & 0 & \delta_{i}^j\\
    0 & 0 & 0 & 0 & 0
    \end{pmatrix}
    \begin{pmatrix}\Omega_0\\ \chi_k \\ h_\beta \\ \chi^k \\ \Omega^0
    \end{pmatrix},
\end{equation}
where we introduced the triple couplings $\big(C^{(112)}\big)_{ij}^{\ \ \alpha} \equiv C_{ij}^{\ \ \alpha}=\partial_i\partial_j H^\alpha$ that is related to the four-point coupling via
\begin{equation}
\label{eq:fourpointfactorization}
    C_{ijkl}=C_{ij}^{\ \ \alpha}\eta^{(2,2)}_{\alpha\beta}C_{kl}^{\ \ \beta}.
\end{equation}
We give again the semisimple part of the connection:
\begin{equation}
\Pi_\text{ss}=\begin{pmatrix}
        X^0 & 0 & 0 & 0 & 0 \\
        \partial_{z_i}X^0 & \frac{1}{X^0}W^{0m}_i& 0 & 0 & 0 \\
       \mathfrak{C}^{ij}_{\ \ \alpha}\partial_{z_i}\partial_{z_j}X^0 & \frac{\mathfrak{C}^{ij}_{\ \ \alpha}}{X^0}W^{0m}_{ij} & \frac{\mathfrak{C}^{ij}_{\ \ \alpha}}{X^0}\left\{W^{ 0\beta}_{ij}-\frac{W^{0m}_{ij}}{(X^0)^2}\frac{\partial z_n}{\partial t^m}W^{ 0\beta}_n\right\} & 0 & 0\\
     \mathfrak{C}^{ijkl}\partial_{z_i}\partial_{z_j}\partial_{z_k}X^0 & \frac{\mathfrak{C}^{ijkl}}{X^0}W^{0m}_{ijk} & \frac{\mathfrak{C}^{ijkl}}{X^0}\left\{W^{ 0\beta}_{ijk}-\frac{W^{0m}_{ijk}}{(X^0)^2}\frac{\partial z_n}{\partial t^m}W^{ 0\beta}_n\right\} &  -\frac{1}{X^0}\frac{\partial z_l}{\partial t^n} & 0 \\
 \mathfrak{C}^{ijkl}\partial_{z_i}\partial_{z_j}\partial_{z_k}\partial_{z_l}X^0  & \frac{\mathfrak{C}^{ijkl}}{X^0}W^{0m}_{ijkl} & \frac{\mathfrak{C}^{ijkl}}{X^0}\left\{W^{ 0\beta}_{ijkl}-\frac{W^{0m}_{ijkl}}{(X^0)^2}\frac{\partial z_n}{\partial t^m}W^{ 0\beta}_n\right\} & -\frac{\mathfrak{C}^{ijkl}}{X^0} E_{n,ijkl} & \frac{h^{3,1}}{X^0}\\ 
    \end{pmatrix}
\end{equation}
with
\begin{align}
\begin{split}
     E_{n,ijkl}&=\frac{\partial_{t^n}X^0}{X^0} C_{ijkl}-(\partial_{t^n}\vec{\Pi}^T)\Sigma\partial_{z^i}\partial_{z^j}\partial_{z^k}\partial_{z^l}\vec{\Pi}\\
     &=\frac{\partial_{t^n}X^0}{X^0} C_{ijkl} -\partial_{t^n}C_{ijkl} + \frac{5}{2} \frac{\partial z^m}{\partial t^n}\partial_{z^{(m}}C_{ijkl)},
\end{split}
\end{align}
and inverse couplings
\begin{align}
    &\mathfrak{C}^{ij}_{\ \ \alpha} C_{ij}^{\ \ \beta}=\delta^\beta_\alpha\,,\\
    &\mathfrak{C}^{ijkl} C_{ijkm}=\delta^l_m\,,
\end{align}
and the same shorthand notation for minors of the Wronskian as used above for the K3 and three-fold cases. For higher dimensional Calabi--Yau families, the expressions for the semisimple part become significantly more complex and we will not give them explicitly. They can be obtained by taking the inverse of the unipotent part and multiplying it by the period matrix.

As mentioned earlier the structure underlying these results is the Frobenius algebra and it was assumed that it is generated by the elements of $\mathcal{A}^{(1)}$. However even for one-parameter models it was found in \cite{Honma:2013hma} that this need not be true. One such example is the model $X_{1,4}\subset \text{Gr}(2,5)$, which was further analysed in \cite{Gerhardus:2016iot}. The middle cohomology is of Hodge type $(1,1,2,1,1)$. The corresponding sixth-order Picard--Fuchs operator is not essentially self-adjoint. Self-adjointness for a $(1,1,1,1,1)$ operator 
\begin{align}
   \mathcal{L}^{(5)}=\sum_{i=0}^{5}a_i(z) \partial^i_z \quad \text{with} \quad a_5=1
\end{align}
is captured by the following constraints:
\begin{align}
\begin{split}
    &0=8a_4^3 - 30a_3a_4 + 60a_4 a_4^{\prime} + 50a_2 + 50a_4^{\prime\prime} - 75 a_3^{\prime},\\
    &0=24a_4^4 -90 a_3 a_4^2 + 132 a_4^2 a_4^{\prime} + 150 a_2 a_4 + 60 a_3 a_4^{\prime}  + 30 a_4 a_4^{\prime\prime} \\
      &\hspace{5cm}-165 a_4 a_3^{\prime}-120 a_4^{\prime 2} + 150 a_3^{\prime\prime} - 100 a_4^{\prime\prime\prime}-100a_2^{\prime}\,. 
\end{split}
\end{align}
For the sixth-order operator of $X_{1,4}$ we find that the coefficients fulfil
\begin{align}
\label{eq:11211Q}
   0=20a_5^3-70a_4a_5+210a_5a_5^{\prime}+98a_3+245a_5^{\prime\prime}-245 a_4^{\prime}\,.
\end{align}
The constraint \eqref{eq:11211Q} together with Griffiths transversality implies that the four-point coupling $C_{zzzz}=\underline{\Pi}^\text{T}\Sigma \partial^4_z\underline{\Pi}$ satisfies the first order differential equation
\begin{align}
    C_{zzzz}^\prime=-\frac{2}{7}a_5C_{zzzz}
\end{align}
that when integrated again gives a rational function expression for $C_{zzzz}$. Further, we note that the factorisation property \eqref{eq:fourpointfactorization} continues to hold as well. Thus, also for this case, the above results carry over when accounting for the extra element in $H^{2,2}(X)$. For the other operators of Hodge type $(1,1,2,1,1)$ discussed in \cite{Honma:2013hma,Gerhardus:2016iot}, we note that, while not all operators fulfil the constraint \eqref{eq:11211Q}, the four-point coupling universally satisfies a third-order differential equation following solely from Griffiths transversality and the order of the operator:
\begin{align}
     C_{zzzz}^{\prime\prime\prime}+a_5C_{zzzz}^{\prime\prime}+\left(\frac{4}{35}a_4+\frac{2}{7}a_5^2+\frac{5}{7}a_5^\prime\right)C_{zzzz}^{\prime}+\left(-\frac{4}{35}a_3+\frac{4}{35}a_4a_5+\frac{2}{7}a_4^\prime\right)C_{zzzz}=0\,.
\end{align}
 
\subsection{From five-folds to the general case}
\label{sec:fivefoldsandgeneral}
The construction of a rational basis of periods for general Calabi--Yau $n$-folds using the $\hat \Gamma$-class formalism is a straightforward generalisation to what was written above for four-folds: one adds sheaves corresponding to intersections of divisors $D_{i_1}D_{i_2}\cdots D_{i_k}$ with $k\leq n-2$ and, using the long exact sequence \eqref{eq:Koszul}, one computes their Chern characters. Then, applying \eqref{eq:ident} and projecting onto a set of independent generators yields the leading order of a rational basis of periods. To obtain the sub-leading terms of the periods, one must match the leading order with that of a local solution of the Picard--Fuchs differential ideal at the MUM point. We will give more details on the local structure and the period matrix in \cref{sec:localstructure}. By taking rational linear combinations of the basis elements, one may achieve a block-anti-diagonal form of the intersection matrix. 

In principle, it is also straightforward to choose a basis for the middle cohomology such that the Gauss--Manin connection is upper-diagonal. The only new ingredients necessary to write down this form for $n>4$ is the appearance of multiple types of three-point functions and inverses with respect to tensor indices of these. In \cref{sec:GMgeneral}, we exemplify this for the case $n=5$ and give an outlook for the generalisation to higher dimensional manifolds. The case $n=6$ is discussed in \cref{sec:appCY6}.

\subsubsection{Special local normal form, Frobenius bases and period matrices}
\label{sec:localstructure}
To adapt the results for three- and four-folds to higher dimensional manifolds, it is useful to recognise that the above is simply a generalisation of the work in \cite{Bogner:2013kvr} for one-parameter Calabi--Yau manifolds with non-degenerate Frobenius structure using the results of \cite{Klemm:1996ts}. Bogner shows that Calabi--Yau operators can always be brought into {\sl special local normal form} given by
\begin{equation}\label{eq:speciallocalnormalform}
    \mathcal{N}\left(\mathcal{L}^{(d)}\right)(t)=\partial_t^2 \frac{1}{C^{(1,1)}}\partial_t \frac{1}{C^{(1,2)}}\partial_t\ldots \partial_t \frac{1}{C^{(1,d-3)}}\partial_t^2,
\end{equation}
where $t$ is the mirror map \eqref{eq:mirrormap} and we abbreviated the triple couplings as $C^{(1,k)} = C^{(1,k,d-k-2)}_{111}(t)$. Due to the symmetry of the couplings, we have $C^{(1,k)}=C^{(1,d-k-2)}$. It may be insightful to verify that this form annihilates naturally all components of $\Omega_0$ given in \cref{eq:Omega0CY3,eq:Omegah0} for three- and four-folds. Note that, in the one-parameter case, the Gauss--Manin connection can be represented by a matrix that has these couplings in this ordering on the secondary diagonal. Therefore, the multi-parameter connection should be in block form of similar type, where the blocks consist of components of the respective triple coupling. Inferring the couplings from the period structure and the expected one-parameter limit becomes increasingly involved, as one can see from the results for $\smash{n=6}$ in \cref{sec:appCY6}. The case $n=5$ is the lowest dimension where we are required to introduce inverse couplings to formulate such a canonical basis for the connection. The next subsection aims at guiding the generalisation to higher dimensional Calabi--Yau families by combining lower-dimensional results with \eqref{eq:speciallocalnormalform}.
\par
Another object that is of interest in the generalisation to higher dimensions is the period matrix \eqref{eq:SigmaPeriods}. One usually starts by constructing a Frobenius basis by computing solutions to the Picard--Fuchs ideal at the MUM point, cf.\ \cref{sec:periodstructure}. We expect that the leading order terms in $\log z_i$ are given by
\begin{equation}
\label{eq:generalfrobeniusbasis}
    \left\{1\right\}\cup \left\{\log z_i\right\}_{i} \cup \left\{\eta_{\alpha A}C_{(1,\ldots ,1,n-k)}^{i_1,\ldots,i_k,A}\log z_{i_1}\ldots \log z_{i_k}\right\}_{2\leq k<n; A}\cup \left\{C_{(1,\ldots ,1)}^{i_1,\ldots ,i_n}\log z_{i_1}\ldots \log z_{i_n}\right\}.
\end{equation}
One then uses the $\hat{\Gamma}$-class to find the leading order behaviour of a basis that has integral monodromies globally, as explained in \cref{sec:GammaClass}. Identifying the mirror map $t_i$ with $\log z_i/2\pi\ii$ at the MUM point, one finds the period vector \eqref{eq:periodvector} consisting of linear combinations of the Frobenius basis.
\par
We note that the couplings $C^{(1,\ldots ,n-k)}_{i_1,\ldots,i_k,A}$ can be computed by choosing a basis for the homology group $H_{n-k,k}$ in terms of $k$-fold intersections of divisors, as used, for example, in \cref{sec:CY4}. This gives a mapping from the index $A$ to a set of indices $\left\{j_1,\ldots ,j_k\right\}$ of weight one. Alternatively, it is straightforward to use \eqref{eq:FrobAlg} to decompose these functions into triple couplings iteratively
\begin{equation}
    C^{(1,\ldots ,n-k)}_{i_1,\ldots,i_k,A} = e^{(1)}_{i_1}\cdot \ldots \cdot e^{(1)}_{i_k}\cdot e^{(n-k)}_A = C^{(1,\ldots ,n-k+1)}_{i_1,\ldots,i_{k-1},\rho}\eta^{\rho A} C^{(1,n-k,k-1)}_{i_k,\alpha,A}.
\end{equation}
The end result is a product of the triple functions appearing in the connection. This relation can also be used to show that the the $n$-point coupling decomposes into the product of all couplings in the connection. While the statement is trivial for three-folds, the decomposition for four- and five-folds are given in \eqref{eq:fourpointfactorization} and \eqref{eq:fivefoldcoupling}, respectively.
\par
Having obtained the period vector, one typically uses linear combinations of derivatives of $\Omega_0$ or equivalently of the period vector to give an expression for the period matrix. In the simplest case of a one-parameter model, it is sufficient to consider the first $n-1$ derivatives. For models with multiple moduli, not all derivatives are independent and one needs to restrict the possible combinations with inverse couplings. In \cite{Candelas:2024vzf}, a basis for the cohomology was used that is obtained by contracting the multi-derivatives with inverses of the classical intersection numbers. Here, we propose a basis that includes higher order corrections by utilising the full couplings instead of their constant terms. 
We define implicitly inverse couplings $\mathfrak{C}$ by the relation
\begin{equation}\label{eq:inverseCWronskian}
    \mathfrak{C}_{(1,\ldots, 1 ,n-k)}^{i_1,\ldots,i_k,\alpha}\cdot C^{(1,\ldots, 1 ,n-k)}_{i_1,\ldots,i_k,\beta} = \delta_{\alpha}^\beta\,.
\end{equation}
Assuming that the middle cohomology is purely horizontal, i.e.\ generated by derivatives of the holomorphic $(n,0)$-form, $h^{n-k,k}$ must be less or equal to the number of degrees of freedom of a symmetric rank $k$ tensor in $h^{n-1,1}$ dimensions. This implies that the number of equations cannot exceed the number of components of the inverse couplings, ensuring their existence for a generic fibre. In general, they are not defined uniquely. A basis of sections is then given by
\begin{equation}
\label{eq:generalperiodmatrix}
    \left\{\Omega\right\}\cup \left\{\partial_i\Omega\right\}_i\cup \left\{\eta_{\alpha A}\mathfrak{C}_{(1,\ldots, 1 ,n-k)}^{i_1,\ldots,i_k,\alpha}\partial_{i_1}\ldots \partial_{i_k}\Omega\right\}_{2\leq k < n; A}\cup \left\{\mathfrak{C}_{(1,\ldots, 1)}^{i_1,\ldots,i_n}\partial_{i_1}\ldots \partial_{i_n}\Omega\right\}.
\end{equation}
We note that this basis may be used also for the computation of the so-called local zeta function using the deformation method as done for multi-parameter three-folds in \cite{Candelas:2024vzf}. However, the basis used there differs insofar that the higher order contributions of the couplings are dropped and only the classical part is used. For the purpose of calculating the local zeta function this does not affect the result as we explain in \cref{sec:app7}.

\subsubsection{Canonical form for Gauss--Manin connection}
\label{sec:GMgeneral}
We start with a rational basis of periods $\Pi=(1,t^i,H^\alpha, K_A,F_i,F_0)^{T}$ with antisymmetric intersection pairing  
\begin{equation}
    \Sigma = \begin{pmatrix}
     &        &        &       &  & 1 \\
     &        &        &  &\eta^{(1,4)} & \\
     &        &        & \eta^{(2,3)} & & \\
     &        & \eta^{(3,2)} &       & \\
     & \eta^{(4,1)}&        &       & \\
   -1 &        &        &       &
  \end{pmatrix}
\end{equation}
with $\eta^{(4,1)}_{ij}=-\eta^{(1,4)}_{ij}=-\delta^i_{h^{4,1}-j+1}$ and $\eta\equiv \eta^{(2,3)}=-\eta^{(3,2)T}$. The Frobenius algebra tells us that there are two types of three-point functions, $\big(C^{(113)}\big)^{\ \ \alpha}_{ij}$ and $C^{(122)}_{iAB}$, and we expect the factorisation 
\begin{equation}\label{eq:fivefoldcoupling}
    C_{ijklm}=\big(C^{(113)}\big)^{\ \ \alpha}_{ij}\eta^{\ A}_\alpha C^{(122)}_{mAB} \eta^{\ B}_\beta \big(C^{(113)}\big)^{\ \ \beta}_{kl}.
\end{equation}
Using the Griffiths transversality relations confirms this and identifies the three-point functions in terms of the periods as 
\begin{align}
    &\big(C^{(113)}\big)^{\ \ \alpha}_{ij}=\partial_i\partial_j H^\alpha,\\
    &C^{(122)}_{iAB}\eta^{\ B}_\alpha\equiv C^{(122)}_{iA\alpha}=\partial_i(\partial_j\partial_k K_A \mathfrak{C}^{jk}_{\ \ \alpha})\,.
\end{align}
Similarly to \eqref{eq:inverseCWronskian}, the auxiliary quantity $\mathfrak{C}^{ij}_{\ \ \alpha}$ is defined as an inverse of $\left(C^{(113)}\right)^{\ \ \alpha}_{ij}$:
\begin{align}\label{eq:inverseCfivefold}
   \mathfrak{C}_{\ \ \alpha}^{ij}\partial_i \partial_j H^\beta=\delta^\beta_\alpha\,.
\end{align}
As before, a sufficient condition for the existence of $\mathfrak{C}$ is the non-degeneracy property of the Frobenius algebra, meaning that the middle cohomology is generated by the horizontal elements.
The Gauss--Manin connection can be brought into the form 
\begin{equation}\label{eq:GMCY5}
    \partial_i \begin{pmatrix}\Omega_0\\ \chi_j \\ h_\alpha \\ k^A \\ \chi^j \\ \Omega^0
    \end{pmatrix} = \begin{pmatrix}
    0 & \delta_i^k & 0 & 0 & 0 & 0\\
    0 & 0 & \big(C^{(113)}\big)^{\ \ \beta}_{ij} &0 & 0 & 0\\
    0 & 0 & 0 & C_{iB\alpha}^{(122)} &  0 & 0\\
    0 & 0 & 0 & 0 & \big(C^{(113)}\big)^{\ \ \gamma}_{ik} \eta^{\ A}_\gamma & 0\\
    0 & 0 & 0 & 0 & 0 & \delta_{i}^j \\
    0 & 0 & 0 & 0 & 0 & 0 \\
    \end{pmatrix}
\begin{pmatrix}\Omega_0\\ \chi_k \\ h_\beta \\ k^B \\ \chi^k \\ \Omega^0
    \end{pmatrix}
\end{equation}
where 
\begin{align}
    &\Omega_0=\alpha_0+t^i\alpha_i+H^\alpha \gamma_\alpha +K_A \delta^A+ F_i \beta^i+F_0\beta^0,\\
    &\chi_i=\partial_i \Omega_0\,,\\
    \begin{split}
    &h_\alpha=\gamma_\alpha + \partial_i\partial_j K_A \mathfrak{C}^{ij}_{\ \ \alpha}\delta^A +\left(\eta^{\ A}_\alpha\partial_k K_A -\partial_k H^\beta \eta^{\ A}_\beta \mathfrak{C}^{ij}_{\ \ \alpha}\partial_i \partial_j K_A\right)\beta^k\\
    &\; +\left(\eta^{\ A}_\alpha K_A-H^\beta \eta^{\ A}_\beta \mathfrak{C}^{ij}_{\ \ \alpha}\partial_i\partial_j K_A +t^k\left\{-\eta^{\ A}_\alpha\partial_k K_A+\partial_k H^\beta \eta^{\ A}_\beta\mathfrak{C}^{ij}_{\ \ \alpha}\partial_i\partial_j K_A\right\}\right)\beta^0,\end{split}\\
    &k^A=\delta^A-\partial_k H^\alpha \eta^{\ A}_\alpha\beta^k +\left(-H^\alpha \eta^{\ A}_\alpha +t^k\partial_k H^\alpha \eta^{\ A}_\alpha\right)\beta^0,\\
    &\chi^i=-\beta^i+t^i\beta^0,\\
    &\Omega^0=\beta^0.
\end{align}

For general dimension $n$, the connection matrix can be written as
\begin{align}
\label{eq:generalgaussmanin}
\scalebox{0.8}{$
\partial_{i}
\begin{pmatrix}
e^{(0)} \\ 
e^{(1)}_{a_1} \\ 
e^{(2)}_{a_2} \\ 
\vdots \\ 
e^{(n-2)}_{a_{n-2}} \\ 
e^{(n-1)}_{a_{n-1}} \\ 
e^{(n)}
\end{pmatrix}
=
\begin{pmatrix}
0 & \delta_{i}^{b_1} & 0 & 0 & \cdots & 0 & 0 \\
0 & 0 & \left(C^{(1,1,n-2)}\right)_{i a_1}^{\ \ \ b_2} & 0 & \cdots & 0 & 0 \\
0 & 0 & 0 & \left(C^{(1,2,n-3)}\right)_{i a_2}^{\ \ \ b_3} & \cdots & 0 & 0 \\
\vdots & \vdots & \vdots & \ddots & \ddots & \vdots & \vdots \\
0 & 0 & 0 & 0 & \cdots & \left(C^{(1,n-2,1)}\right)_{i a_{n-2}}^{\ \ \ b_{n-1}} & 0 \\
0 & 0 & 0 & 0 & \cdots & 0 & \delta_{i, \alpha_{n-1}} \\
0 & 0 & 0 & 0 & \cdots & 0 & 0
\end{pmatrix}
\begin{pmatrix}
e^{(0)} \\ 
e^{(1)}_{b_1} \\ 
e^{(2)}_{b_2} \\ 
\vdots \\ 
e^{(n-2)}_{b_{n-2}} \\ 
e^{(n-1)}_{b_{n-1}} \\ 
e^{(n)}
\end{pmatrix},
$}
\end{align}
where, as before, $e^{(0)}=\Omega_0$ and, for example, the first few three-point couplings are expressed in terms of the periods as 
\begin{align}
\begin{split}\label{eq:generalcouplings}
       &\big(C^{(1,1,n-2)}\big)_{ij}^{\ \ a}=\partial_i\partial_j \hat{\Pi}^{(2)a},\\
        &\big(C^{(1,2,n-3)}\big)_{ia}^{\ \ b}=\partial_i\left\{\big(\mathfrak{C}_{(1,1,n-2)}\big)^{ij}_{\ \ a}\partial_i\partial_j \hat{\Pi}^{(3)b}\right\},\\
        &\big(C^{(1,3,n-4)}\big)_{ia}^{\ \ b}=\partial_i\left\{\big(\mathfrak{C}_{(1,2,n-3)}\big)^{jc}_{\ \ a}\partial_j\left\{\big(\mathfrak{C}_{(1,1,n-2)}\big)^{kl}_{\ \ c}\partial_k\partial_l \hat{\Pi}^{(4)b}\right\}\right\}\\
\end{split}
\end{align}
with the inverse three-point coupling defined by the generalisation of \eqref{eq:inverseCfivefold}
\begin{align}
    \big(\mathfrak{C}_{(1,p,n-p-1)}\big)^{ia}_{\ \ b}\big(C^{(1,p,n-p-1)}\big)_{ia}^{\ \ c}=\delta^{c}_{b}\,.
\end{align}
The form \eqref{eq:generalgaussmanin} determines the basis recursively as 
\begin{align}\label{eq:cohomgen}
    e_{a_{k+1}}^{(k+1)}=\big(\mathfrak{C}_{(1,k,n-k-1)}\big)^{ib_k}_{\ \ a_{k+1}}\partial_i e^{(k)}_{b_k}\,,\quad k=1,\dots, n-3\,.
\end{align}
Griffiths transversality can then be used to simplify the expressions for the basis elements, in particular by reducing the number of inverse three-point functions appearing. Note that, in the four-fold case for example, naive application of the recursion formula yields expressions containing $\mathfrak{C}_{(1,1,2)}$, limiting applicability to those cases where this inverse exists, which is true in particular when the non-degeneracy axiom holds. Owing to Griffiths transversality, the basis elements simplify to \cref{eq:Omega0,eq:chii,eq:halpha,eq:chihi,eq:Omegah0} and are applicable also for the cases where the middle cohomology is not purely horizontal.  Starting from $n=5$ however, the inverse couplings appear also after employing all available constraints with the resulting expressions given for five-folds above and for six-folds in \cref{sec:appCY6}. In general, Griffiths transversality avoids only the appearance of  $\mathfrak{C}_{(1,\lfloor  \frac{n-1}{2}\rfloor, \lfloor \frac{n}{2}\rfloor)}$.

We note that, for families with purely horizontal middle cohomology, we can express the inverse couplings as the inverses of rank-two tensors. These are given by
\begin{align}
    \tilde{\mathfrak{C}}_{(p,n-p)}&:\ \mathcal{A}^{(p)}\times \mathcal{A}^{(n-p)}\rightarrow \C,\\
    \tilde{\mathfrak{C}}_{(p,n-p)}^{\alpha A}&=\left( \left(C^{(1,p-1,n-p)}_{\alpha_1,\alpha^{*},A}\right)_{\alpha A}\right)^{-1},
\end{align}
where we used horizontality of the cohomology to express the index $\alpha$ in terms of one index in $\mathcal{A}^{(1)}$ and one in $\mathcal{A}^{(p-1)}$. Replacing the inverse couplings by these tensors, it is then sufficient to restrict the summations in \eqref{eq:generalcouplings} to basis elements of the respective vector space $\mathcal{A}^{(k)}$. For example, the summation over $i,j=1,\ldots ,h^{n-1,1}$ may be replaced by a sum over tuples $(i,j)$ whose associated sheaves $\mathcal{O}_{D_i\cdot D_j}$ correspond to independent elements in the K-theory group.
In this case, we write for \eqref{eq:generalcouplings}
\begin{align}
\begin{split}
        &\big(C^{(1,2,n-3)}\big)_{ia}^{\ \ b}=\partial_i\left\{\big(\tilde{\mathfrak{C}}_{(2,n-2)}\big)_a^{\ \alpha}\partial_{\alpha_1}\partial_{\alpha_2} \hat{\Pi}^{(3)b}\right\},\\
        &\big(C^{(1,3,n-4)}\big)_{ia}^{\ \ b}=\partial_i\left\{\big(\tilde{\mathfrak{C}}_{(3,n-3)}\big)_{a}^{\ \alpha}\partial_{\alpha_1}\left\{\big(\tilde{\mathfrak{C}}_{(2,n-2)}\big)_{\alpha^*}^{\ \beta}\partial_{\beta_1}\partial_{\beta_2} \hat{\Pi}^{(4)b}\right\}\right\}
\end{split}
\end{align}
and analogous expressions for the other three-point couplings.
This description simplifies the computation of the inverses but also the iterative construction of the cohomology basis in \eqref{eq:cohomgen}.

\section{Effective type II theories from Calabi--Yau compactifications}\label{sec:CYcomp}
This section introduces the effective $N=2$ and $N=1$ action that is 
relevant for flux compactifications in type II and F-theory. We discuss the breaking of supersymmetry in such configurations and revisit the derivation of vacuum conditions in both theories. Inserting fluxes requires the inclusion of brane sources, which restricts the possible flux configurations via tadpole conditions, as we will review in the last subsection. 

\subsection{Type IIB flux compactifications}
\label{sec:Type IIB flux compactifications}
The setting that will concern us is that of flux compactifications to four dimensions in type IIB string theory. We will briefly recall the ten-dimensional type IIB effective supergravity action and discuss general features of flux compactifications such as the resulting superpotential and its expression in terms of periods.

The bosonic action for the effective type IIB supergravity can be written as the sum~ \cite{Becker:2006dvp}
	\begin{equation}\label{eq:actionIIB}
		\mathcal{S}=S_\text{NS} + S_\text{R} + S_\text{CS}\,,
	\end{equation}
	where the terms are the contributions from the NS-NS sector, the R-R sector and the Chern--Simons term, respectively:
	\begin{align}
		S_\text{NS} &= \frac{1}{2\kappa^2} \int \dd x^{10}\sqrt{-g}e^{-2\phi}\left(R+4\partial_\mu \phi\,\partial^\mu \phi -
		\frac{1}{2} \abs{H_3}^2\right),\\
			S_R &= -\frac{1}{4\kappa^2} \int \dd x^{10} \sqrt{-g}\left(\vert F_1 \vert^2 + \vert{\tilde{F}_3}\vert^2 + \frac{1}{2}\vert\tilde{F}_5\vert^2
			\right),\\
		S_\text{CS} &= -\frac{1}{4\kappa^2} \int C_4 \wedge H_3 \wedge F_3\,.
	\end{align}
	Here, $\phi$ is the dilaton, $\kappa^2=8\pi G_{10}=\frac{1}{4\pi}(2\pi \sqrt{\alpha^\prime})^8g_{\text{s}}^2$ is the ten-dimensional gravitational constant and $\tilde{F}_5$ is constrained to be self-dual. The field strengths of the real R--R $p$-form fields $C_p$ and the NS--NS $B$-field are defined as
	$F_{p+1}=\dd C_p$, $H_3=\dd B_2$ and further we set
	\begin{align}
		\tilde{F}_3 &= F_3 -C_0 H_3\,,\\
		\tilde{F}_5 &= F_5 - \frac{1}{2} C_2 \wedge H_3 + \frac{1}{2} B_2 \wedge F_3\,,
	\end{align}
	where $C_0$ denotes the axion.
	\par
	By giving nontrivial background values to the field strengths $F_3$ and $H_3$, some directions of the moduli space may be lifted. In order for this background to be compatible with the ten-dimensional supergravity equations of motion, one uses a warped product ansatz of a maximally symmetric four-dimensional spacetime and an internal Calabi--Yau space. A non-vanishing value for $G_3$ presupposes the insertion of branes, which give rise to a term $S_\text{loc}$ in the action \eqref{eq:actionIIB}. We will comment on the resulting tadpole cancellation condition in \cref{sec:tadpoles}.
	\par
	The main effect of turning on fluxes from the perspective of the effective four-dimensional theory is to introduce a coupling between the hypermultiplet and the vector multiplet sector \cite{Polchinski:1995sm}. Thus the effective theory is described by a gauged $\mathcal{N}=2$ supergravity theory. More precisely, it is given by gauging a specific subgroup of the isometry group in the hypermultiplet sector (the corresponding gauge fields coming from the $h_{21}$ vector multiplets and the graviphoton). As desired, the gauging results in a nontrivial scalar potential. The latter may be computed directly by dimensional reduction of the ten-dimensional action or by considering the change of the superpotential across domain walls in four dimensions given by wrapping D$5$- or NS$5$-branes around $3$-cycles in the Calabi--Yau whose duals correspond to the change of R- and NS-flux, respectively. The result is a scalar potential that takes the form familiar from $\mathcal{N}=1$ supergravity \cite{Gukov:1999ya,Taylor:1999ii}:
	\begin{equation}
	\label{eq:SUGRAscalarpotential}
		V=e^{K_{\text{tot}}}\left(\sum_{A,\overline{B}} g^{A\overline{B}} D_A W D_{\overline{B}} \overline{W} - 3 \abs{W}^2\right),
	\end{equation}
	with the superpotential 
	\begin{equation}\label{superpotential}
		W = \int_X G_3\wedge \Omega\,,
	\end{equation}
	where $G_3\coloneqq F_3-\tau H_3$ with the axio-dilaton field $\tau=C_0 +\ii e^{-\phi}$. The sums in \eqref{eq:SUGRAscalarpotential} run over all scalars $z_A$. The covariant derivative is given by $D_A=\partial_A + \partial_A K_{\text{tot}}$, where the total Kähler potential receives contributions from both vector and hypermultiplet moduli and $	g_{A\overline{B}}$ is the corresponding Kähler metric:
	\begin{equation}\label{eq:metric}
		g_{A\overline{B}}\coloneqq \frac{\partial }{\partial z_A} \frac{\partial }{\partial \overline{z}_{\overline{B}}}
		K_\text{tot}\,.
	\end{equation} 
	\par We note that, while the coupling of the hypermultiplets and the vector multiplets may be expressed in terms of $\mathcal{N}=1$ language, the theory is still $\mathcal{N}=2$ supersymmetric on the level of the action. However, an important question is that of the possibility of partially breaking $\mathcal{N}=2$ supersymmetry down to $\mathcal{N}=1$. Let us first consider the conditions for unbroken supersymmetry in the $\mathcal{N}=1$ formalism. These are the vanishing of the F-terms in $V$:
	\begin{align}
	    e^{K_{\text{tot}}/2}D_AW=0\,.
	\end{align}
	This ensures minimisation of the scalar potential and the cosmological constant is 
	\begin{align}
	\label{eq:cosmologicalconstant}
	    \Lambda = -3e^{K_{\text{tot}}/2} \abs{W}^2=3m_{3/2}^2\,,
	\end{align}
	where $m_{3/2}$ is the mass parameter of the gravitino. But $\Lambda=3m^2_{3/2}$ is precisely the condition for a spin-$3/2$ particle to be physically massless (having only two polarisations) in an AdS background \cite{Deser:1977uq} and hence supersymmetry is indeed unbroken. For a Minkowski background the conditions for a supersymmetric vacuum can also be seen more directly from the requirement of vanishing vacuum expectation values for the supersymmetry variations of the chiral multiplet superpartners $\chi^A$ of the $z_A$ and the gravitino $\psi_\mu$\footnote{For an AdS background $\braket{\delta P_{L}\psi_\mu}$ would contain also a contribution from the torsion-free spin connection.}:
	\begin{align}
        &\braket{\delta \chi^A}=-\frac{1}{\sqrt{2}}e^{K_{\text{tot}/2}}g^{A\overline{B}}D_{\overline{B}}\overline{W}=0\,,\\
        &\braket{\delta P_{L}\psi_\mu}=\frac{1}{2}e^{K_{\text{tot}/2}}W\gamma_\mu=0\,.
    \end{align}
    Since we will only consider Minkowski vacua in the explicit examples, we will restrict our discussion from now on to this case. The eigenvalues of the $\mathcal{N}=2$ gravitino mass matrix differ in general\footnote{One may determine which hypermultiplet isometries are gauged by demanding that the lighter gravitino mass eigenvalue be proportional to the above superpotential \cite{Curio:2000sc} (see also \cite{Jockers:2024ocq} for a recent discussion of the gauging of the quaternionic isometries in this context)}. 
    In \cite{Curio:2000sc}, it was shown that for a supersymmetric Minkowski background, i.e.\ $e^{K_{\text{tot}}/2}W=0$, not only does the lighter gravitino become massless but in fact both eigenvalues of the gravitino mass matrix vanish. Thus, solving the above $\mathcal{N}=1$ supersymmetry conditions corresponds to $\mathcal{N}=2$ vacua and partial supersymmetry breaking from $\mathcal{N}=2$ to $\mathcal{N}=1$ is not possible in this framework.
    
    \par One way of achieving $\mathcal{N}=1$ vacua is to consider orientifolds. We will discuss briefly O$3$/O$7$-orientifolds \cite{Acharya:2002ag}, which are constructed by gauging the discrete symmetry
    \begin{align}
    \mathcal{O}=(-1)^{F_L}\Omega_p\sigma^{\ast},
    \end{align}
    where $F_L$ is the left-moving spacetime fermion number, $\Omega_p$ is the worldsheet parity operator and $\sigma:X\longrightarrow X$ is a holomorphic involution such that
    \begin{align}
    \label{eq:orientifoldOmega}
        \sigma^\ast \omega=\omega\,,\quad \sigma^\ast\Omega=-\Omega\,.
    \end{align}
    The inclusion of the factor $(-1)^{F_L}$ in $\mathcal{O}$ is necessary in order for one linear combination of the gravitini to survive the projection. Orientifolds whose isometry satisfies \eqref{eq:orientifoldOmega} feature O$3$- or O$7$-planes, which fill the non-compact spacetime and in the case of O$7$-planes additionally wrap divisors in the Calabi--Yau. There is another possible choice for the gauged symmetry:
    \begin{align}
    \mathcal{O}=\Omega_p\sigma^{\ast},\quad   \sigma^\ast \omega=\omega\,,\quad \sigma^\ast\Omega=\Omega\,,
    \end{align}
    which gives rise to O$5$/O$9$-orientifolds, however we will restrict ourselves to the O$3$/O$7$ case.
    \par The action of $\Omega_p$ on the ten-dimensional bosonic fields is
    \begin{align}
    \label{eq:wsparity}
    &\Omega_p: (g,\phi,C_2) \mapsto (g,\phi,C_2)\,,\\
    &\Omega_p: (C_0,B_2,C_4) \mapsto (-C_0,-B_2,-C_4)\,.
    \end{align}
    Under $(-1)^{F_L}$, the NS-NS fields $g,\phi,B_2$ are even while the R-R fields $C_0,C_2,C_4$ are odd. As a result, the surviving components of the fields satisfy
    \begin{align}
        \label{eq:involution}
    &\sigma^\ast: (g,\phi,C_0,C_4) \mapsto (g,\phi,C_0,C_4)\,,\\
    &\sigma^\ast: (B_2,C_2) \mapsto (-B_2,-C_2)\,.
    \end{align}
    The cohomology splits into even and odd parts under the involution $\sigma$:
    \begin{align}
        H^{p,q}(X)=H_{+}^{p,q}(X)\oplus H_{-}^{p,q}(X)\,.
    \end{align}
    We will consider only isometries such that $h^{2,1}=h^{2,1}_{-}$ and $h^{1,1}=h^{1,1}_{+}$. Dimensional reduction of the truncated theory then yields the following moduli fields: $h^{2,1}$ complex structure moduli $z_i$, the axio-dilaton $\tau=C_0 +\ii e^{-\phi}$, $h^{1,1}$ real Kähler moduli $v_\alpha$ and $h^{1,1}$ real scalars $c_\alpha$ from the reduction of $C_4$.
    
    \par The orientifold projection breaks half of the supersymmetry. Since only $\mathcal{N}=1$ supersymmetry is present, the structure of the scalar field space of the surviving hypermultiplet scalars, now residing in chiral multiplets, is no longer quaternionic but only Kähler. However, this Kähler structure becomes only manifest when choosing appropriate holomorphic coordinates $T_\alpha$ combining $v_\alpha$ and $c_\alpha$. At tree-level the $T_\alpha$ are given by \cite{Grimm:2004uq}
    \begin{align}
    T_\alpha=-c_\alpha+\frac{\ii}{2}C^{(0)}_{\alpha\beta\gamma}v^\beta v^\gamma,
    \end{align}
    where $C^{(0)}_{\alpha\beta\gamma}$ are the classical intersection numbers on $X$.
	The total Kähler potential after orientifolding is obtained from $K_\text{cs}$ in \eqref{Kahler} by including contributions by both the axio-dilaton and the complexified Kähler moduli:
	\begin{equation}\label{eq:Ktot}
		K_\text{tot}\coloneqq K_\text{cs}+K_{\tau} + K_{\text{Ks}}\,.
	\end{equation}
	Ignoring any perturbative corrections in $\alpha'$ and $g_s$, the first is given by
	\begin{equation}\label{Ktau}
		K_\tau = -\log\left(-\ii (\tau -\overline{\tau})\right)
	\end{equation}
	and the Kähler potential for the complexified Kähler moduli is, to lowest order in $\alpha^\prime$ and up to worldsheet instantons,
	\begin{align}\label{eq:Kaehlerpot}
	    K_{\text{Ks}}=-2\log\left(C^{(0)}_{\alpha\beta\gamma}v^\alpha v^\beta v^\gamma\right),
	\end{align}
	where it is understood that the $v^\alpha$ are functions of the $T_\alpha$ and $c_\alpha$.
	One can show \cite{Giddings:2001yu,Grimm:2009ef} that supergravity obtained from string theory is subject to the so-called ``no-scale property'' \cite{Cremmer:1983bf}
	\begin{equation}
	    g^{\alpha\bar{\beta}}\partial_{\alpha} K_{\text{Ks}}\partial_{\bar{\beta}}K_{\text{Ks}} = 3\,.
	\end{equation}
	This relation simplifies the scalar potential in \eqref{eq:SUGRAscalarpotential} to 
	\begin{equation}
	    \label{eq:V_noscale}
	    V = e^{K_{\text{tot}}}\sum_{a,\overline{b}} g^{a\overline{b}} D_a W D_{\overline{b}} \overline{W}\,,
	\end{equation}
	where $a$ and $b$ run over the complex structure moduli and the axio-dilaton.
	As an example, we give the derivation for cases with a single Kähler modulus $T$. Here, the Kähler potential of \eqref{eq:Kaehlerpot} simplifies to
	\begin{align}\label{KKs}
		K_\text{Ks} &= -2\log\left(C_{111}^{(0)}v_1^3\right)=-6\log\left(v_1\right) +\text{const.}\\
		&=-3\log\left(-\ii (T-\bar{T})\right) + \text{const.}\,,
	\end{align}
	where the constant is real.
	This implies that the term in the scalar potential $V$ involving the Kähler modulus satisfies
	\begin{equation} 
	\begin{split}
		g^{T \overline{T}}D_T W D_{\overline{T}} \overline{W} &= \left(\partial_T \partial_{\overline{T}}
		K_{\text{tot}} \right)^{-1} \left(\partial_T K_{\text{tot}}\right) W
		\left(\partial_{\overline{T}}K_{\text{tot}}\right) \overline{W}\\
		&= 3\left(T - \overline{T}\right)^2 \frac{W}{T -\overline{T}} \frac{\overline{W}}{T -\overline{T}} =3 \abs{W}^2,
	\end{split}
	\end{equation}
	cancelling the appearance of $-3\abs{W}^2$.
	\par The no-scale structure is broken by perturbative corrections in $g_s$ and $\alpha^\prime$. As we will see below in the supersymmetric case, this does not affect the existence or non-existence of vacua at a given point in the moduli space. The superpotential does not receive perturbative corrections but may receive non-perturbative corrections by Euclidean D-branes and gaugino condensation \cite{Gorlich:2004qm}. In proposals for the construction of realistic dS-vacua such as the KKLT scenario \cite{Kachru:2003aw} or the Large Volume Scenario \cite{Balasubramanian:2005zx}, these effects play a central role in the stabilisation of Kähler moduli. We will not consider Kähler moduli stabilisation and hence ignore these effects in the following. For an extensive discussion and estimates of the different types of perturbative and non-perturbative corrections we refer to \cite{McAllister:2024lnt}.

\subsection{Vacuum criteria in type IIB}\label{sec:vacuum_criteria}
    We are interested in the flux configurations that preserve supersymmetry. The vacuum condition $V=0$ can be translated into a splitting of the integral middle cohomology of the Calabi--Yau manifold. We also comment on the necessary gauge independence of the vacuum configuration.
    \par
    One begins by considering the F-term equation for a Kähler modulus $T$ that parametrises a direction in which the Kähler potential is not flat, yielding
    \begin{equation}
    \label{eq:Wequal0}
        0=D_T W = (\partial_T K_{Ks})W
    \end{equation}
    and therefore $W=0$. Here, we used that the superpotential $W$ is independent of the Kähler moduli. While the vacuum for the scalar potential in the no-scale limit \eqref{eq:V_noscale} does not require $D_T W=0$, 
    higher order corrections in $g_s$ and $\alpha^\prime$ would break the no-scale structure of $W$ and thus the minimisation of $V$ \cite{Blumenhagen:2013fgp}.
    Now, the F-term for the axio-dilaton gives
    \begin{align}
        \begin{split}\label{eq:DtauW}
        0&= D_\tau W = \partial_\tau W + (\partial_\tau K_\tau)W \\
        &=-\int_X H_3 \wedge \Omega\,,
        \end{split}
    \end{align}
    where the derivative of $K_\tau$ containing perturbative corrections in $g_s$ does not contribute due to $W=0$. \Cref{eq:DtauW} together with $W=0$ also implies that 
    \begin{equation}
        0=\int_X F_3\wedge \Omega\,.
    \end{equation}
    In terms of cohomology classes, this means that the real forms $F_3$ and $H_3$ span a rank two lattice $\Lambda$ in $H^{3}(X,\Z)\cap \left(H^{2,1}(X)\oplus H^{1,2}(X)\right)$, which yields the splitting of the rational middle cohomology 
    \begin{equation}\label{eq:hodgesplitting}
        H^{3}(X,\Q) = \Lambda_\Q \oplus \Lambda_\Q^\perp
    \end{equation}
    with $\Lambda_\Q=\Lambda \otimes \Q$.
	It remains to satisfy the F-term equations for the complex structure moduli
    \begin{equation}
        0=D_{z_i}W=\int_X G_3\wedge  D_i\Omega\,,\quad i\in\{1,\ldots ,h^{2,1}\}\,.
    \end{equation}
	Since $\{D_i\Omega\,|\,i\in\{1,\ldots ,h^{2,1}\}$ is a basis for the class $H^{2,1}(X)$, this implies that the axio-dilaton must be chosen such that $G_3=F_3-\tau H_3$ is orthogonal to $H^{1,2}(X)$ and thus contained in $H^{2,1}(X)$.
    In practise, one scans the complex structure moduli space for loci, where two $\Q$-linear combinations of periods vanish. Let us assume that this happens at a locus $z_i=0$. Still leaving $\tau$ unfixed, the corresponding superpotential $W$ vanishes identically on the locus $z_i=0$, which also implies that the F-terms along the locus are satisfied. The only left-over relation consists of the F-term for $z_i$, i.e.\ $D_{z_i}W=0$, which forces the axio-dilaton to be a function of the moduli $z_j$, $j\neq i$, parametrising the vacuum 
    \begin{equation}
        \tau(\vec{z}) = \frac{\int_X F_3\wedge D_{i}\Omega}{\int_X H_3\wedge D_{i}\Omega}\,.
    \end{equation}
	With the integral symplectic basis introduced in \cref{sec:CYPeriodgeometry}, we may express the fluxes as
	\begin{align}
		F_3&=f^I\alpha_I + f_{I+h^{2,1}+1}\beta^I,\\
		H_3&=h^I\alpha_I + h_{I+h^{2,1}+1}\beta^I.
	\end{align}
	Analogous to \cref{Kahler}, we may express the superpotential in \cref{superpotential} as
	\begin{equation}\label{eq:generalWvector}
		W = (f-\tau h)^T \Sigma\, \vec{\Pi}
	\end{equation}
	with the flux vectors
	\begin{align}
		f^T&\coloneqq \left(f^0,f^1,\ldots ,f^{h^{2,1}},f_{2h^{2,1}+1},\ldots, f_{h^{2,1}+1}\right)^T,\\
		h^T&\coloneqq \left(h^0,h^1,\ldots ,h^{h^{2,1}},h_{2h^{2,1}+1},\ldots, h_{h^{2,1}+1}\right)^T.
	\end{align}
    \paragraph{\textbf{Gauge independence.}}
    The conditions for (supersymmetric) vacua must be independent of the gauge transformations
    \begin{equation}\label{eq:Kaehlergauge}
    \begin{split}
        \Omega(\{z_i\})&\rightarrow e^{f(\{z_i\})}\Omega(\{z_i\})\\
        K_{\text{cs}}(\{z_i\},\{\bar{z}_j\})&\rightarrow K_{\text{cs}}(\{z_i\},\{\bar{z}_j\})-f(\{z_i\})-\bar{f}(\{\bar{z}_j\})
    \end{split}
    \end{equation}
    for any holomorphic function $f$.
    The scalar potential $V$ in \eqref{eq:V_noscale} is naturally invariant. Only when the metric $g_{i\bar{j}}$ is regular, we may express the condition of $V=0$ as 
    \begin{equation}
    e^{K_{\text{tot}}/2}D_i\,W=e^{K_{\text{tot}}/2}\,W=0\,,\quad i\in\{z_1,\ldots ,z_{h^{2,1}}\}\,,
    \end{equation}
    which, for regular Kähler potential $K_{\text{tot}}$, reduces to the equations derived above.
    
\subsection{F-theory flux vacua}
In this section, we briefly review some aspects of F-theory relevant to later discussions. F-theory is a framework to study non-perturbative vacua in type IIB string theory. The theory may be defined as M-theory on an elliptically fibred four-fold $Y\rightarrow B$, leading to $\mathcal{N}=2$ supersymmetry in three dimensions. In the limit of vanishing volume of the torus fibre, the theory is dual to type IIB string theory on the base $B$. The duality identifies the axio-dilaton with the complex structure modulus of the torus fibre giving rise to a holomorphically varying axio-dilaton profile and a geometric realisation of the $\text{SL}(2,\mathbb{Z})$ symmetry of type IIB string theory. The base of the fibration is not Calabi--Yau but in the so-called Sen limit \cite{Sen:1997gv}, corresponding to weakly coupled type IIB theory, the base is the double cover of a Calabi--Yau space.

The four-dimensional effective action is again that of $\mathcal{N}=1$ supergravity. The superpotential is given by the Gukov--Vafa--Witten superpotential 
\begin{equation}
    W = \int_Y G_4\wedge \Omega\,,
\end{equation}
where $\Omega$ is the holomorphic $(4,0)$ form on $Y$ and $G_4=\text{d}C_3$ is the four-form flux of M-theory. The Kähler potential of the effective action consists of two terms:
\begin{equation}
		K_{\text{tot}}\coloneqq K_\text{cs}+ K_{\text{Ks}}\,,
\end{equation}
where 
\begin{equation}
    K_\text{cs}= -\ln\left(\int_Y \Omega \wedge \bar{\Omega}\right), \quad  K_\text{Ks}=-3\ln\left(\frac{1}{4!}\int_Y J\wedge J \wedge J \wedge J\right).
\end{equation}
The presence of the term $\int_Y G_4 \wedge \ast G_4$ in the eleven-dimensional M-theory action generates a scalar potential for the moduli fields, given in terms of the two functions $W$ and $K_{\text{tot}}$ by the same form as above:
\begin{equation}
	\label{eq:SUGRAscalarpotential3d}
		V=e^{K_{\text{tot}}}\left(\sum_{i,\overline{\jmath}} g^{i\overline{\jmath}} D_i W D_{\overline{\jmath}} \overline{W} - 3 \abs{W}^2\right).
	\end{equation}

The flux vacua condition in F-theory take on a similar form as those for type IIB theory discussed above.
Supersymmetric vacua require \cite{Becker:1996gj}
\begin{equation}
\label{eq:FtermsFtheory}
    D_{z_i} W =W=0,\quad i\in\{0,\ldots ,h^{3,1}\}
\end{equation}
as well as primitivity of the flux 
\begin{equation}
    G_4\wedge J=0\,.
\end{equation}
Similarly to the derivations for type IIB vacua, the conditions \eqref{eq:FtermsFtheory} imply $G_4 \in H^{2,2}(Y,\C)$, so that in total $G_4\in H^{2,2}_\text{prim}(Y,\mathbb{C})$.

In M- and F-theory the flux is quantised according to \cite{Witten:1996md}
\begin{equation}
    G_4+\frac{c_2(Y)}{2}\in H^4(Y,\mathbb{Z})\,.
\end{equation}
Locally the type IIB flux $G_3$ and the flux $G_4$ are related via
\begin{equation}
    G_4 = \frac{1}{\tau-\overline{\tau}}\left(\overline{G}_3\wedge \dd z - G_3\wedge \dd \overline{z}\right).
\end{equation}
The F-term equations for $G_4$ in components read
\begin{align}
    0 = G_4^{3,1} &= \frac{1}{\tau-\overline{\tau}}\left((\overline{G}_3)^{2,1}\wedge \dd z - (G_3)^{3,0}\wedge \dd \overline{z}\right),\\
    0 = G_4^{0,4} &= -\frac{1}{\tau-\overline{\tau}}\, G_3^{0,3}\wedge \dd \overline{z}
\end{align}
and imply the condition $G_3\in H^{2,1}(X,\C)$ we derived earlier. 

\subsection{Tadpole cancellation}\label{sec:tadpoles}
    When compactifying type IIB theory with the action as in \eqref{eq:actionIIB} on a warped background, the ten-dimensional Einstein equations yield a no-go theorem which states, among other things, that $G_3$ must vanish \cite{Becker:2006dvp,Giddings:2001yu}. This can be circumvented by introducing brane sources into the action, which for a D$p$-brane wrapping a $(p-3)$-cycle $\Sigma$ of the Calabi--Yau manifold corresponds to the additional term 
    \begin{equation}
        S_\text{loc} = -T_p\int_{\R^4\times \Sigma}\dd\xi^{p+1}\,\sqrt{-g} + \mu_p \int_{\R^4\times \Sigma} C_{p+1}\,,
    \end{equation}
    where, for positive tension objects, the Einstein frame tension $T_p$ and the D$p$-brane charge $\mu_p$ are related by
    \begin{equation}
        T_p=\abs{\mu_p}e^{(p-3)\phi/3}.
    \end{equation}
    The equation of motion for $C_4$ of the modified action is then given by
	\begin{equation}\label{eq:dF5}
		\dd \tilde{F}_5 = H_3 \wedge F_3 + 2\kappa^2 T_3 \rho_3\,,
	\end{equation}
    where $\rho_3$ contains obtains contributions from D$3$-branes, O$3$-planes and D$7$-branes, which couple to $C_4$ via the worldvolume flux $F_2$ on the D$7$-brane in the term $\int_\Sigma C_4\wedge F_2\wedge F_2$. Integrating \eqref{eq:dF5} over the Calabi--Yau
	three-fold $X$ yields the tadpole cancellation condition
	\begin{equation}
		\int_X H_3 \wedge F_3 = f^T\Sigma  h= -2 \kappa^2 T_3 Q_3
	\end{equation}
	with $Q_3=\int_X\rho_3$.
    For our purposes, we just demand that the l.h.s.\ is non-vanishing which means that $f$ and $h$ are linearly independent.
    \par The tadpole cancellation in F-theory is described via its duality to M-theory. The eleven-dimenional supergravity action \cite{Becker:2006dvp}
    \begin{equation}
        \mathcal{S}=\frac{1}{2\kappa_{11}^2}\int\dd x^{11}\sqrt{-g}\left(R-\frac{1}{2}\abs{G_4}^2\right) -\frac{1}{12\kappa_{11}^2}\int A_3\wedge G_4\wedge G_4
    \end{equation}
    receives quantum corrections of the form
    \begin{equation}
        \delta S = -T_{\text{M}2}\int A_3\wedge X_8\,,
    \end{equation}
    with the eight-form 
    \begin{equation}
        X_8=\frac{1}{(2\pi)^4}\left(\frac{1}{192}\tr R^4 -\frac{1}{768}\left(\tr R^2\right)^2\right).
    \end{equation}
    Including source terms for the M2 and M5 branes, the equations of motion for the field $A_3$
    \begin{equation}\label{eq:eomA3}
        \dd \star G_4=-\frac{1}{2}G_4\wedge G_4-2\kappa_{11}^2T_{\text{M}2}\left(X_8+\sum_i Q^i_{\text{M2}}\delta^{(8)}_i +\sum_i Q^i_{\text{M5}}\delta^{(5)}_i \wedge A_3\right)\,.
    \end{equation}
    Similar to the derivation for type IIB above, the tadpole condition will follow upon integrating \eqref{eq:eomA3} over the internal Calabi--Yau four-fold. For this, one expresses $X_8$ in terms of the Pontryagin classes, which allow us to write
    \begin{equation}
        X_8=\frac{1}{192}\left(c_1^4-4c_1^2c_2+8c_1c_3-8c_4\right),
    \end{equation}
    with $c_i$ the Chern classes of the Calabi--Yau $X$. Since $c_1=0$, $\int_X X_8=-\chi/24$, yielding the tadpole condition
    \begin{equation}
        N_{\text{M}2}+\frac{1}{4\kappa_{11}^2T_{\text{M}2}}\int_X G_4\wedge G_4 = \frac{\chi}{24}\,.
    \end{equation}
    This restricts the choices of possible flux configurations.

\section{Quotients of Calabi--Yau hypersurfaces in toric ambient spaces}\label{sec:quotientbuilding}

In this section, we will review the description of a family of $n$-dimensional Calabi--Yau hypersurfaces $\mathcal{X}_\Delta$ over its moduli space\footnote{To be precise, we will restrict the following analysis to the moduli space of \textit{polynomial} complex structure deformations.} $\mathcal{M}_{\mathcal{X}_\Delta}$
\begin{equation}\label{eq:CYoverM}
    \pi:\ \mathcal{X}_\Delta\rightarrow\mathcal{M}_{\mathcal{X}_\Delta}
\end{equation}
inside a toric ambient space $\P_\Delta$ \cite{Cox:2000vi}. Starting from the construction of $\P_\Delta$, we will review Batyrev's mirror construction \cite{Batyrev:1993oya} and the hypersurface's complex structure moduli space parametrising deformations of its defining polynomial modulo automorphisms of the ambient space. We then introduce the notion of a quotient family used in the rest of this article. In the second subsection, we introduce the rational middle cohomology following the work of Griffiths \cite{Griffithsarticle}.
\subsection{Complex structure moduli spaces}\label{sec:Mcs}
\par Let $(\Delta, \Delta^*)$ be a pair of dual reflexive lattice polytopes in the integral lattices $N$ and $N^*$, where
\begin{equation}
    \Delta^* = \left\{x\in N^*\,\big|\, \braket{x | y}\geq -1\ \forall y\in\Delta\right\}.
\end{equation}
The ambient space $\P_\Delta$ is constructed from a reflexive polytope $\Delta$ with a triangulation that we demand to be fine, star and regular. This is to say that all points appear in the triangulation, all maximal simplices contain the inner point and all cones have unit simplicial volume.
We denote the resulting fan by $\Sigma(\Delta)$. 
The generators $\{\nu_i\,|\,i\in\{1,\ldots ,s\}\}$ of the fan correspond to the coordinates $x_i$ of $\P_\Delta$. Here, one can ignore those ending at a point inside a face of codimension one of $\P_\Delta$, as we will show later in this section. The Stanley--Reisner ideal $Z(\Sigma)$ is generated by the divisors 
\begin{equation}
    x_{i_1}=\ldots =x_{i_k}=0
\end{equation}
for index sets $I= \{i_1,\ldots ,i_k\}$ where $\{\nu_i\}_{i\in I}$ does not generate a cone in $\Sigma(\Delta)$. Let us assume that there are $m$ relations among the generators $\nu_i$ given by
\begin{equation}
    0 = \sum_{i=1}^{s}Q_{ij}\nu_i,\quad Q\in \mathrm{Mat}_{s\times m}(\Z)\,.
\end{equation}
Then, the coordinates $x_i$ obey the scaling relations
\begin{equation}
    Q_{*,j}(\lambda):\ (x_1,\ldots ,x_{s})\mapsto \left(\lambda^{Q_{1,j}}x_1,\ldots ,\lambda^{Q_{s,j}}x_s\right),\quad\lambda\in\C^*.
\end{equation}
The matrix $Q$ is also called charge matrix and gives the continuous part of the toric group. For non-regular fans, it contains a discrete subgroup given by $N'/N$, where $N$ is the ambient lattice of $\Delta$ and the sublattice $N'\subset N$ is the $\Z$-span of the generators $\nu_i$.

For the regular triangulations considered here, we may write the toric variety corresponding to the fan $\Sigma(\Delta)$ as 
\begin{equation}
    \P_\Delta = \frac{\C^s\setminus Z(\Sigma)}{Q}\,.
\end{equation}
Batyrev showed that the family of hypersurfaces $\mathcal{X}_\Delta$ in $\P_\Delta$ given by the vanishing loci of the polynomials
\begin{equation}\label{eq:defpolyDelta}
    P_{\mathcal{X}_\Delta} = \sum_{\nu\in\Delta}a_{\nu} \prod_{\nu^*\in\Delta^*\setminus\{0\}} x_{\nu^*}^{\braket{\nu | \nu^*}+1},\quad a_{\nu}\in\C\,,
\end{equation}
consists of Calabi--Yau manifolds. One can consider the Calabi--Yau family in a patch where coordinates not corresponding to vertices are non-zero and use the scaling relations to set them equal to one. For the hypersurfaces in weighted projective space considered in this work, this implies that the defining polynomial \eqref{eq:defpolyDelta} becomes a function of $n+2$ variables $x_i$. The moduli $a_i\equiv a_{\nu_i}$ are related to the Batyrev coordinates $z_j$ via
\begin{equation}
    z_j = (-a_0)^{l_{j,0}}\prod_{i=1}^\infty a_i^{l_{j,i}},
\label{eq:batyrevcoordinates}
\end{equation}
where the generators of the Mori cone $l_j$ are defined by the relations
\begin{equation}
    0=\sum_{i=0}^\infty l_{j,i} \bar{\nu}_i\,,
\end{equation}
with $\bar{\nu}_i=(1,\nu_i)$ and $\nu_0=\vec{0}$ corresponding to the inner point of $\Delta$. 
We will denote the fibres of the map \eqref{eq:CYoverM} by
\begin{equation}
    X_{\vec{a}} = \pi^{-1}(\vec{a})\,.
\end{equation}
\par The moduli space $\mathcal{M}_{\mathcal{X}_\Delta}$ of the Calabi--Yau family is given by the deformations $a_\nu$ of $P_{\mathcal{X}_\Delta}$ modulo the automorphisms of the ambient space $\P_\Delta$
\begin{equation}
    \mathcal{M}_{\mathcal{X}_\Delta}=\frac{\text{Def}(\mathcal{X}_\Delta)}{\text{Aut}(\P_\Delta)}\,.
\end{equation}
The latter consists\footnote{This includes only the identity component of $\text{Aut}(\P_\Delta)$. The full group of automorphisms also includes suitable permutations between the homogeneous coordinates.} of coordinate scalings and so-called roots \cite{Cox:2000vi}.  The automorphisms of $\P_\Delta$ are then those of $\C^s\setminus Z(\Delta)$ which commute with the action of $Q$. This shows that the five coordinate scalings are part of $\mathcal{M}_{\mathcal{X}}$. Roots are the generalisation of the off-diagonal elements of $\mathrm{PGL}(5,\C)$ in $\text{Aut}(\P^4)$, where the diagonal elements represent the scalings. Since commutativity with $Q$ means that transformations between the variables $x_i$ must leave the action of $Q$ invariant, roots are described by 
\begin{equation}
    x_i\mapsto x_i + \mu x^D,\ \text{ for } \deg_Q(x^D)=\deg_Q(x_i),\ \quad \mu\in \C.
\end{equation}
\paragraph{\textbf{Example.}}Consider the family
$\mathcal{X}=\P_{2,1,1,1,1}[6]$. The torus action is given by
\begin{equation}
    Q:\,(x_1,x_2,x_3,x_4,x_5)\mapsto (\lambda^2x_1,\lambda x_2,\lambda x_3,\lambda x_4,\lambda x_5),\quad \lambda\in \C^\star.
\end{equation}
Beside the five scalings, we have the roots
\begin{align}
    x_1&\mapsto x_1 + \sum_{j,k=2}^5 A_{jk}^{(1)}x_jx_k,\\
    x_i&\mapsto x_i+\sum_{j=2,j\neq i}^5 A_{j}^{(i)}x_j,\quad i\in\{2,3,4,5\}\,.
\end{align}
Since $A_{jk}^{(1)}$ is symmetric, it has ten degrees of freedom. Adding these to the $3\cdot 4=12$ parameters of $A^{(i)}_j$ and the five scaling relations, we have 
\begin{equation}
    \text{dim}(\text{Aut}(\mathcal{X}))=22+5 =27\,.
\end{equation}
A simple counting of monomials with $Q$-degree six tells us that there are 130 deformations of $X$. In this way, we find
\begin{equation}
    h^{2,1}(X)=\text{dim}(\mathcal{M}_{\mathcal{X}})=130-27=103\,.
\end{equation}\hfill $\blacktriangle$
\par In practise, the automorphisms of the ambient space allow us to transform $\vec{x}$ in a way such that the defining polynomial becomes of the form
\begin{equation}\label{eq:defpolymodAut}
    P_{\mathcal{X}_\Delta}(\vec{a},\vec{x})= P_0(\vec{x})+\sum_{j}a_j\text{def}_j(\vec{x})
\end{equation}
with $\dim{\mathcal{M}_{\mathcal{X}}}$ monomials $\text{def}_j(\vec{x})$. Here, all monomials of $P_{\mathcal{X}_\Delta}$ are inequivalent over the Jacobian ideal $\text{Jac}_{\vec{a}}(P_{\mathcal{X}_\Delta})$ generated by $\{\partial_{x_i}P_{\mathcal{X}_\Delta}(\vec{a},\vec{x})\}_i$, which, for a generic hypersurface with all deformations present, generates the same equivalences as $\text{Aut}(\P_\Delta)$. 
\par Another way of taking into account the reparametrisations of the ambient space is the removal of monomials in \eqref{eq:defpolyDelta} that correspond to points inside codimension one faces of $\Delta^*$. To see this, let us assume that $\hat{\nu}^*$ lies in a face of codimension one of $\Delta^*$. The unique vertex $\nu_v\in\Delta$ with the property $\braket{\nu_v|\nu^*}=-1$ for all points in said face of $\Delta^*$ will then give rise to the only term in \eqref{eq:defpolyDelta} with $x_{\hat{\nu}^*}$-exponent $\braket{\nu_v|\hat{\nu}^*}+1=0$. This means that the divisor $x_{\hat{\nu}^*}=0$ intersects the hypersurface for 
\begin{equation}
    0=P_{\mathcal{X}_\Delta}\Big|_{x_{\hat{\nu}^*}=0} = c_{v_j}\prod_{\nu^*\in \Delta^*\setminus \{0\}} x_{\nu^*}^{\braket{\nu_v|\nu^*}+1}.
\end{equation}
The inner product of $\nu_v$ with any element in the face dual to $\nu_v$ is $-1$. Denoting the vertex opposite to this face by $\nu_v^*$, the condition simplifies to $0=x_{\nu_v^*}$. But since vertices cannot share a simplex with points in their opposite face in a regular triangulation, the Stanley--Reisner ideal contains the set $\{x_{\nu_v^*}=x_{\hat{\nu}^*}=0\}$ and the intersection is not part of the ambient space. We conclude that the hypersurface can be described completely in the patch $x_{\hat{\nu}^*}\neq 0$ and we may use a scaling relation to fix the coordinate to one, effectively ignoring its appearance in \eqref{eq:defpolyDelta}.
\par Note that a general hypersurface in a toric variety can also have non-polynomial deformations, which, as the name suggests, are not described by monomials in the defining polynomial. The following formula by Batyrev \cite{Batyrev:1993oya} counts both polynomial and non-polynomial deformations:
\begin{equation}
    h^{n-1,1} = l(\Delta) - (n+2) -\sum_{\text{codim}(\theta)=1}\hat{l}(\theta) + \sum_{\text{codim}(\theta)=2}\hat{l}(\theta)\cdot \hat{l}(\theta^*)\,,
\end{equation}
where $\theta$ denotes faces of $\Delta$ while $l$ and $\hat{l}$ are the number of points inside the face with and without vertices, respectively. We also introduced the dual face $\theta^*$ to $\theta$, defined by
\begin{equation}
    \theta^*=\left\{ x\in N^* \,\big|\, \braket{x | y}=-1\ \forall y\in \theta \right\}.
\end{equation}
Exchanging $\Delta$ for its dual $\Delta^*$ gives a formula for $h^{1,1}$.
\par
From now on, we will restrict ourselves to hypersurfaces $\P_{w_1,\ldots ,w_{n+2}}[d]$ in weighted projective spaces with weights $w_i$ and $w_{n+1}=1$ where the moduli independent term of the defining polynomial is Fermat, i.e.\ given by 
\begin{equation}
    P_0(\vec{x}) = \sum_{i=1}^{n+2}x_i^{d/w_i}\quad\text{with}\quad d=\sum_{i=1}^{n+2}w_i\,.
\end{equation}
The family $\mathcal{X}_\Delta$ has a residual symmetry given by a group $G$ consisting of discrete phase symmetries that act on the variables $\vec{x}$. It is useful to consider the extension $\hat{G}$, which also acts on the moduli $\vec{a}$ such that the defining polynomial is invariant under its action. We denote an element $g$ in $G$ (or $\hat{g}$ in $\hat{G}$) by $\Z_p: (\beta_1,\ldots ,\beta_{n+2})$, which then acts with $\alpha=e^{2\pi\ii/p}$ as
\begin{equation}
\def\arraystretch{1.5}
    \hat{g}:\ \begin{array}{l}
         x_i\mapsto g(x_i)=\alpha^{\beta_i}x_i\,,\quad \beta_i\in\{0,\ldots ,d-1\}\,, \\
         a_j\mapsto \text{def}_j(\vec{x})/g(\text{def}_j(\vec{x}))a_j\,,
    \end{array}
\def\arraystretch{1}
\end{equation}
where $g$ acts on the deformation monomials multiplicatively. The action of $\hat{G}$ on the moduli simply cancels the phase obtained by $\text{def}_i(\vec{x})$ and restores the invariance of $\mathcal{X}_\Delta$. Since such symmetries identify points in the moduli space that correspond to equivalent hypersurfaces, it is sensible to consider only the moduli space modulo such symmetries. Then, these elements describe closed paths in the moduli space, which, due to the flatness of the Gauss--Manin connection, give rise to path independent monodromy actions. We will come back to this in \cref{sec:hodgesplitting}.
\par
Let $\hat{S}\subset \hat{G}$ be a subgroup. We refer to the part of the moduli space that is invariant under the induced action of $\hat{S}$ as $\text{Inv}_{\hat{S}}(\mathcal{M}_{\mathcal{X}})$. Then, we define the quotient of $\mathcal{X}_\Delta$ by $\hat{S}$ as
\begin{equation}
    \frac{\mathcal{X}_\Delta}{\hat{S}} \coloneqq \pi^{-1}\left(\text{Inv}_{\hat{S}}(\mathcal{M}_{\mathcal{X}_\Delta})\right).
\end{equation}
The invariant slice $\text{Inv}_{\hat{S}}(\mathcal{M}_{\mathcal{X}_\Delta})$ is just $\mathcal{M}_{\mathcal{X}_\Delta}$ with all $a_i$ describing deformations $\text{def}_i(\vec{x})$ that are not invariant under $\hat{S}$ set to zero. Comparing to \eqref{eq:defpolymodAut}, only the invariant moduli appear in the defining polynomial of $\mathcal{X}_\Delta/\hat{S}$.
\par
The abelian symmetry group $\hat{G}$ is isomorphic to
\begin{equation}
    \hat{G}\cong \Z_{d/w_1}\times\ldots \times \Z_{d/w_{n+2}}\,.
\end{equation}
The subgroup descending from the projective scaling of the ambient space is called the quantum symmetry $\hat{Q}$. The mirror group $\hat{H}$ used in the mirror construction by Greene and Plesser \cite{Greene:1989cf} is yet another subgroup of $\hat{G}/\hat{Q}$ which leaves the symmetric deformation $\psi$ belonging to $\prod_{i=1}^{n+2}x_i$ invariant. 
\paragraph{\textbf{Example.}}
    Consider the family $\mathcal{X}=\P_{4,1,1,1,1}[8]$. The defining polynomial has phase symmetries $\hat{G}\cong\Z_2\times \Z_8^4$ generated by (for example)
    \begin{align*}
        &\begin{rcases*}
        \begin{rcases*}
        \Z_2: (1,0,0,0,1)\\
        \Z_8: (0,1,0,0,7)\\
        \Z_8: (0,0,1,0,7)
        \end{rcases*}\hat{H}\\
        \, \Z_8: (0,0,0,-1,0)\\
        \end{rcases*}\hat{G}/\hat{Q}\,,\\
        &\begin{rcases*}
        \, \Z_8: (4,1,1,1,1)
        \end{rcases*}\hat{Q}\,.
    \end{align*}
    The last generator corresponds to (a subset of) the scaling symmetry of the ambient space while the second to last $\Z_8$ does not act trivially on $\prod_ix_i$. Its action on the modulus $\psi$ is given by $\psi\mapsto \exp(2\pi\ii/8)\,\psi$. The quotient family $\mathcal{X}/\hat{H}$ is described by the polynomial
    \begin{equation}
        P_{\mathcal{X}/\hat{H}}(\psi,\vec{x}) = x_1^2+x_2^8+x_3^8+x_4^8+x_5^8 - 8 \psi \prod_{i=1}^5x_i\,.
    \end{equation}\hfill $\blacktriangle$
\par For three different three-dimensional hypersurface families, we will use subgroups $\hat{S}$ of their mirror groups $\hat{H}$ to construct three-parameter models. The resulting quotient families will be described by a polynomial $P$ defining their mirrors with two additional deformations. The fact that the families have a residual symmetry $\hat{H}/\hat{S}$, implies that supersymmetric flux vacua of $\mathcal{X}/\hat{H}$ are also supersymmetric flux vacua of $\mathcal{X}/\hat{S}$, where the moduli transversal to $\text{Inv}_{\hat{H}/\hat{S}}(\mathcal{M}_{\mathcal{X}/\hat{S}})$ are stabilised automatically due to the algebraic form of the periods on this slice of the moduli space.
\par
Taking the quotient associated to such symmetries commutes with removing the redundancies of $\mathcal{M}$ due to the automorphisms of the ambient space. Although the roots themselves do not commute with the action of $\hat{G}$, they are mapped onto themselves with different values of $A^{(i)}_\mathbf{j}$, possibly combined with a scaling. In practice, it might not be obvious which deformations of a quotient family are inequivalent over these automorphisms. One can resolve this issue by either (i) computing a Gröbner basis of the Jacobian ideal and use it to reduce each deformation or (ii) considering the toric description of the quotient and remove the deformations corresponding to points in codimension one faces.
\subsection{Rational cohomology}
\label{sec:rationalcohomology}
\par We may also ask how cyclic groups act on elements of the middle cohomology. We give a review of Griffiths' description  \cite{Griffithsarticle,Doran:2007jw} of cohomology on a Calabi--Yau $n$-fold $X_{\vec{a}}$ in $\P_\Delta$ defined by $P=0$. While the construction holds for toric ambient spaces, we will restrict the discussion to weighted projective spaces. This section concludes with the statement that the Kähler potential is flat in the normal directions of a locus that is invariant under a cyclic symmetry. The covariant derivatives of the F-term equations thus simplify to partial derivatives for flux configurations that stabilise moduli on their fix point locus under such symmetry. 
\par
Rational $(n+1)$-forms on $\P_{\Delta}\setminus X_{\vec{a}}$, denoted by $\mathcal{A}^{n+1}(X_{\vec{a}})$, give rise to $n$-forms on $X_{\vec{a}}$ via the residue map with the property
\begin{equation}
    \frac{1}{2\pi\ii}\int_{T(\gamma)}\phi = \int_\gamma \text{Res}(\phi)
\end{equation}
for an $n$-cycle $\gamma$ in $X_{\vec{a}}$ and $T(\gamma)$ a tubular neighbourhood of $\gamma$ in $\P_{\Delta} \setminus X_{\vec{a}}$. It turns out that this constitutes a map from the cohomology group   $\mathcal{H}(X_{\vec{a}})=\frac{\mathcal{A}^{n+1}(X_{\vec{a}})}{\dd\mathcal{A}^n(X_{\vec{a}})}$ of the ambient space to the primitive middle cohomology group of the hypersurface
\begin{equation}
    \text{Res}:\ \mathcal{H}(X_{\vec{a}})\rightarrow H^n(X_{\vec{a}},\C)\,.
\end{equation}
For three-folds, all forms in the middle cohomology are primitive  since $h^{1,0}=0$. For generic four-folds, the horizontal cohomology (generated by the derivatives of $\Omega$) is just a subset of the primitive cohomology, which itself is just a subset of the middle cohomology.
Splitting the cohomology class above into those of forms with poles of order $k$
\begin{equation}
    \mathcal{H}_k(X_{\vec{a}}) = \frac{\mathcal{A}^{n+1}_k(X_{\vec{a}})}{\dd \mathcal{A}^n_{k-1}(X_{\vec{a}})}\,,
\end{equation}
there exists a filtration 
\begin{equation}
     \mathcal{H}_1(X_{\vec{a}})\subset\ldots\subset  \mathcal{H}_{n+1}(X_{\vec{a}})=\mathcal{H}(X_{\vec{a}})\,.
\end{equation}
Griffiths' theorem then states that the residue maps the $i$-th filtrant of the above into the $i$-th Hodge filtrant in
\begin{equation}
    F^0(X_{\vec{a}})\subset\ldots\subset F^n(X_{\vec{a}})\equiv H^n(X_{\vec{a}},\C)\,,
\end{equation}
where we wrote 
\begin{equation}
    F^i(X_{\vec{a}}) = \bigoplus_{j=0}^i H^{n-j,j}(X_{\vec{a}},\C)\,.
\end{equation}
\par
Since $\varphi$ must be invariant under the scaling relations of the ambient  weighted projective space, homogeneous polynomials of degree $kd$  correspond under the residue map to an element in $F^k(X_{\vec{a}},\C)$. To obtain a map that is injective, one must quotient out by the  Jacobian ideal and instead consider the ring
\begin{equation}\label{eq:polyring}
    \mathcal{R}_{\vec{a}} = \bigoplus_{k=0}^n\frac{\C\left[x_1,\ldots ,x_{n+2}\right]_{kd}}{\text{Jac}_{\vec{a}}(P)}\,,
\end{equation}
where $\C\left[x_1,\ldots ,x_{n+2}\right]_{kd}$ denotes the polynomial ring of (weighted) degree $k d$. Each summand is generated by $h^{n-k,k}$ monomials. The formulation of the residue map most useful for us is then
\begin{equation}
\begin{split}\label{eq:resmap}
    \frac{\C\left[x_1,\ldots ,x_{n+2}\right]_{kd}}{\text{Jac}(P,\vec{a})}&\rightarrow H^{n-k,k}(X_{\vec{a}},\C)\,,\\
    Q&\mapsto\ \mathcal{P}^{n-k,k}\,\text{Res}\left[\frac{Q\mu}{P^{k+1}}\right]
\end{split}
\end{equation}
with the projection $\mathcal{P}^{n-k,k}$ into $H^{n-k,k}(X_{\vec{a}},\C)$ and the volume form
\begin{equation}
    \mu \coloneqq \sum_{i=1}^{n+2} (-1)^i w_ix_i\,\dd x_1\wedge\ldots \wedge \widehat{\dd x_i}\wedge \ldots \wedge \dd x_{n+2}\,,\quad \text{omit }\hat{k}\,.
\end{equation}
\par One can give an account of the parts of the residue contained in a lower filtrant using the covariant derivative. The middle cohomology is generated from the holomorphic $(n,0)$-form $\text{Res}\left[\frac{\mu}{P}\right]$ by acting on it with derivatives w.r.t.\ the complex structure moduli $a_i$ using the Gauss--Manin connection.

The derivative increases the order of the pole and consequently maps the residue into the next filtrant. To map into the next cohomology class only, which is to say that the part contained in the old filtration vanishes, we may use the covariant derivative 
\begin{equation}
\label{eq:GM-connection}
\begin{split}
    D_{a_i}:\ H^{n-k,k}\left(X_{\vec{a}},\C\right)&\rightarrow H^{n-k-1,k+1}\left(X_{\vec{a}},\C\right)\,,\\
    \varphi&\mapsto \partial_{a_i}\varphi+(\partial_{a_i}K_{\text{cs}})\,\varphi\,.
\end{split}
\end{equation}
\par
The description of cohomology classes in terms of monomials allows us to describe their action under the cyclic symmetries discussed before. The period vector is given by integrals of a nowhere-vanishing element in $H^{n,0}\left(X_{\vec{a}},\C\right)$ over an integral basis of the middle homology, cf.\ \eqref{eq:periodvector}. To obtain a form that is invariant under the symmetry group $\hat{G}$, we must multiply the form $\text{Res}\left[\frac{\mu}{P}\right]$ by the modulus that parametrises the symmetric deformation

\begin{align}
    \Omega(\vec{a}) &= a_0\text{Res}\left[\frac{\mu}{P}\right]=\frac{a_0}{2\pi\ii}\oint_\gamma \frac{\mu}{P}\label{eq:OmegaRes}\\
    &= \frac{a_0}{2\pi\ii}\sum_{k=1}^{n+2}(-1)^kw_k\oint_\gamma \frac{x_k\,\dd x_1\wedge \ldots \wedge\widehat{\dd x_{k}}\wedge \ldots \wedge\dd x_{n+2}}{P}\nonumber\\
    &= -a_0 w_1 \frac{x_1\,\dd x_3\wedge \ldots\wedge\dd x_{n+2}}{\frac{\partial P}{\partial x_2}}+a_0\sum_{k=2}^{n+2}(-1)^kw_k \frac{x_k\,\dd x_2\wedge \ldots \wedge\widehat{\dd x_{k}}\wedge \ldots \wedge\dd x_{n+2}}{\frac{\partial P}{\partial x_1}}\,,\nonumber
\end{align}
where we made the coordinate transformation $x_1\mapsto P$ (for $k=1$ instead $x_2\mapsto P$) and used the residue formula with $\gamma$ encircling the hypersurface $P=0$. This rescaling by $a_0$ corresponds to a choice of Kähler gauge, cf.\ \eqref{eq:Kaehlergauge}. 
\par
The invariance of $\Omega(\vec{a})$ under a symmetry group $\hat{S}\subset \hat{G}$ implies that the derivatives of the Kähler potential $K_\text{cs}$ normal to its fixed point locus in the moduli space vanish. This can be seen from the invariance of 
\begin{equation}\label{eq:eKcs}
    e^{-K_{\text{cs}}(\vec{a})}= \ii \int_{X_{\vec{a}}} \Omega(\vec{a})\wedge \bar{\Omega}(\vec{a})
\end{equation}
under $\hat{S}$. Therefore, the expression in \eqref{eq:eKcs} contains the moduli not invariant under $\hat{S}$ only in invariant combinations, which must be of order greater than one. Since the derivative of $K_{\text{cs}}$ is proportional to that of \eqref{eq:eKcs}, we deduce that if $\text{Inv}_{\hat{S}}$ is given by $a_i=0$ for $i\in I\subset\{1,\ldots ,h^{n-1,1}\}$, then
\begin{equation}\label{eq:vanishingdelK2}
    \partial_{a_i}K_{\text{cs}}\Big|_{\text{Inv}_{\hat{S}}}=0\,.
\end{equation}
\par
For the other classes in the middle cohomology, above analysis implies that the transformation under an element in $\hat{G}$ is given by that of its monomial $Q$ assigned by the residue map times $\prod_{i=1}^{n+2}x_i$. With \eqref{eq:vanishingdelK2}, this observation will help us to make a quantitative analysis of the vanishing of certain periods and/or their derivatives on subloci $\text{Inv}_{\hat{S}}(\mathcal{M}_{\mathcal{X}})$.

\subsection{Splitting of Hodge structure}\label{sec:hodgesplitting}
For one-parameter families $X_\psi$ of Calabi--Yau three-folds, a rank-two attractor point \cite{Ferrara:1995ih,Moore:1998pn} is a point in the moduli space $\psi_0\in \mathcal{M}_{X_\psi}$ where the cohomology group
    \begin{equation}\label{eq:attractorpoint}
        H^c(X_{\psi_0},\Z)\coloneqq \left(H^{2,1}(X_{\psi_0},\C)\oplus H^{1,2}(X_{\psi_0},\C)\right)\cap H^3(X_{\psi_0},\Z)
    \end{equation}
    contains two independent elements and is therefore of rank two. This splitting is a special case of the vacuum conditions we discussed in \cref{sec:vacuum_criteria} where the lattice $\Lambda^\perp$ in \eqref{eq:hodgesplitting} is now contained in $H^{3,0}\oplus H^{0,3}$. It follows that, on one-parameter families, these special points give rise to supersymmetric flux vacua. The name attractor point arises from the fact that in the context of extremal black hole solutions of $\mathcal{N}=2$ supergravity the value of the moduli field $\psi$ flows to a constant value $\psi_0$ at the horizon of the black hole where such a splitting occurs. 
    
    Number theoretic methods have proven to be fruitful to study such points. The splitting of Hodge structure induces a factorisation of the numerator of the local zeta function of the Calabi--Yau \cite{Candelas:2019llw}, which can be calculated from the periods as reviewed in \cref{sec:app7}. In this section, we will discuss examples of such splittings of the Hodge structure by instead studying the monodromy action on the middle cohomology. 
    \par
    At the end of the previous section, we gave our choice of the Kähler gauge, which renders $\Omega(z)$ invariant under the symmetry group $\hat{G}$. It follows that the monodromies of the period vector describes the behaviour of the middle homology group under closed paths in the complex structure moduli space. However, due to the duality of the middle homology and cohomology as vector spaces, transformation properties and linear relations in one apply also to the other. This is also why \eqref{eq:attractorpoint} can be seen as a relation among the periods. While this section is concerned with representations of the symmetry group on the cohomology, this lets us prove important relations of the homology along invariant loci in the moduli space. 
    \par
    We consider a hypersurface-family $\P_{w_1,\ldots,w_5}[d]$ with $w_5=1$. As in \cref{sec:Mcs}, we denote by $\hat{G}=\bigtimes_{i=1}^{5}\Z_{d/w_i}$ the full symmetry group of the defining polynomial, generated by the five elements
    \begin{gather}
    \begin{gathered}
        g_1 = (w_1,0,0,0,-w_1 )\,,\ 
        g_2 = (0,w_2,0,0,-w_2)\,,\
        g_3 = (0,0,w_3,0,-w_3)\,,\\
        g_4 = (0,0,0,w_4,-w_4)\,,\
        g_5 = (0,0,0,0,1)\,,
    \end{gathered}
    \end{gather}
    where $g_i$ is an element of $\Z_{d/w_i}$ and induces an action $\frak{M}_j$, $j\in\{1,\ldots,5\}$, on the complex structure moduli via their deformation monomials.
    \par
    From \cref{sec:rationalcohomology}, we know that the middle cohomology is spanned by derivatives of the holomorphic $(n,0)$-form $\text{Res}\left(\frac{\mu}{P}\right)$ w.r.t.\ the complex structure moduli. However, there is a choice to be made which moduli or, in other words, which representatives in the polynomial ring quotient defined in \eqref{eq:polyring}, are used.  We claim that the correct choice consists of the monomials for which the exponents of each variable $x_i$ are less than $d/w_i-1$. This choice is possible since the Jacobian ideal is generated by elements of the form
    \begin{equation}\label{eq:Jacobi}
        \partial_{x_i}P = w_i x_i^{d/w_i-1}- ...
    \end{equation}
    where the ellipsis stands for a polynomial with $x_i$-degree lower than $d/w_i-1$ with which we may replace each appearance of $x_i^{d/w_i-1}$. 
    \paragraph{\textbf{Example.}}For the octic one-parameter family inside the ambient space $\P_{4,1,1,1,1}$, the identity \eqref{eq:Jacobi} for $x_1$ is used to replace the symmetric monomial $\psi\prod_{i=1}^5x_i$ with $\psi^2\prod_{i=2}^5 x_i^2$.~\text{}~\hfill $\blacktriangle$
    \par
    The variables $a_i$ parametrise the above choices for representatives of the quotient ring. We let the middle cohomology be generated by
    \begin{equation}\label{eq:basismc}
        H^3(X,\C) = \text{span}_{\C}\left(\left\{\text{Res}\left(\frac{\mu}{P}\right)\right\}\cup\left\{\text{Res}\left(\partial_{a_i}\frac{\mu}{P}\right)\ \Big|\ i\in\{0,\ldots ,h^{2,1}-1\}\right\}\cup \text{ c.c.}\right),
    \end{equation}
    The first derivatives correspond to classes in $H^{3,0}(X,\C)\oplus H^{2,1}(X,\C)$ and obey with \eqref{eq:GM-connection} and \eqref{eq:OmegaRes}
    \begin{equation} 
    \begin{split}
        -\text{Res}\left(\frac{\text{def}_i\,\mu}{P^2}\right) &= \partial_i \text{Res}\left(\frac{\mu}{P}\right) \\
        &= \underbrace{D_i \text{Res}\left(\frac{\mu}{P}\right)}_{\in H^{2,1}} - \underbrace{(\partial_i K_{\text{cs}})\,\text{Res}\left(\frac{\mu}{P}\right)}_{\in H^{3,0}}.
    \end{split}
    \end{equation}
    This implies that at points in the moduli space where the Kähler potential is flat in the direction of $a_i$, the derivative of $\text{Res}\left(\frac{\mu}{P}\right)$ w.r.t.\ $a_i$ is a form in $H^{2,1}(X,\C)$.
    \par
    The middle cohomology is spanned by $\text{Res}\left(\frac{\mu}{P}\right)$ and its complex conjugate transforming just as $\prod_{i=1}^5 x_i$ resp.\ $\prod_{i=1}^5 \overline{x_i}$ and the derivatives of $\text{Res}\left(\frac{\mu}{P}\right)$ with its complex conjugate partners. The latter set transforms as the product of $\prod_{i=1}^5 x_i$ with the corresponding deformation (or the complex conjugate of it).
    \paragraph{\textbf{Example.}}
    For the sextic one-parameter family in $\P_{2,1,1,1,1}$, the middle cohomology is generated by the set
    \begin{equation}
        \mathcal{S}=\left\{\text{Res}\left(\frac{\mu}{P}\right),\ \text{Res}\left(\frac{\mu\prod_{i=1}^5 x_i}{P^2}\right),\ \overline{\text{Res}\left(\frac{\mu\prod_{i=1}^5 x_i}{P^2}\right)},\ \overline{\text{Res}\left(\frac{\mu}{P}\right)}\right\}.
    \end{equation}
    As a quotient of the mirror group, its middle cohomology is invariant under $g_1,\ldots ,g_4$. Under $g_5$, it undergoes the cyclic transformation
    \begin{equation*}
        \{\alpha,\ \alpha^2,\ \alpha^4,\ \alpha^5\}\,.
    \end{equation*}\hfill $\blacktriangle$
    \par
    We now want to show that certain cyclic symmetries of the moduli space imply the existence of flux vacua on their fixed point loci. Suppose there exists a symmetry $a_i\mapsto e^{2\pi\ii/n}a_i$\,, $i\neq 0$ with all other moduli invariant. In the fundamental domain of the moduli space, this action induces a transport of the integral basis of periods $\vec{\Pi}$ along a closed cycle. By definition, the monodromy action 
    \begin{equation}
        \vec{\Pi}\left(e^{2\pi\ii/n} a_i\right) = \mathfrak{M}_i \vec{\Pi}(a_i)
    \end{equation}
    has integral coefficients. On the other hand, we may consider a basis $\tilde{\vec{\varpi}}$ given in \eqref{eq:basismc}. We sort the basis, such that the above monodromy action takes the form
    \begin{equation}
        \begin{pmatrix}
            \mathbb{1} & 0 & 0\\
            0 & e^{2\pi\ii/n} & 0\\
            0 & 0 & e^{-2\pi\ii/n}
        \end{pmatrix}.
    \end{equation}
    We emphasise that the symmetries we consider here leave the modulus $a_0$ parametrising the symmetric deformation invariant. Since the elements in the basis $\tilde{\vec{\varpi}}$ transform as $\mathrm{def}_i\prod_j x_j$ (or its conjugate), symmetries that let $\prod_j x_j$ transform would not be of the above form. 
    \par
    If now $n\in\{2,3,4,6\}$, we may transition to a basis $\vec{\varpi}$ on which the symmetry action is integral. Here, one uses for $n\in\{3,4,6\}$
    \begin{equation}\label{eq:integral2x2}
        \begin{pmatrix}
            1 & 1\\ e^{2\pi\ii/n} & e^{-2\pi\ii/n}
        \end{pmatrix}\begin{pmatrix}
            e^{2\pi\ii/n} & 0\\
            0 & e^{-2\pi\ii/n}
        \end{pmatrix}\begin{pmatrix}
            1 & 1\\ e^{2\pi\ii/n} & e^{-2\pi\ii/n}
        \end{pmatrix}^{-1}\in M_{2\times 2}(\Q)\,.
    \end{equation}
    We call the matrix representation in this basis $\mathfrak{N}_i$. Since both $\mathfrak{N}_i$ and $\mathfrak{M}_i$ are rational and of finite order, we find a \textit{rational} \cite{Koo} conjugation matrix $A$ with
    \begin{equation}
        \mathfrak{M}_i=A\mathfrak{N}_i A^{-1}. 
    \end{equation}
    This also implies that the transition matrix from $\vec{\Pi}$ to $\vec{\varpi}$ is restricted to the general form
    \begin{equation}
        T_{\vec{\varpi}} = AC\,,
    \end{equation}
    where the matrix $C$, which in general is complex, commutes with $\mathfrak{N}_i$, which we can write as
    \begin{equation}
        \mathfrak{N}_i = \begin{pmatrix}
            n_1 & 0 \\ 0 & n_2
        \end{pmatrix},
    \end{equation}
    where $n_1=\mathbb{1}_{(b_3-2)\times (b_3-2)}$ and $n_2$ is the integral $2\times 2$-matrix from \eqref{eq:integral2x2}. Expressing the matrix $C$ in a similar block form, commutativity gives
    \begin{equation}
        0=\mathfrak{N}_i\begin{pmatrix}
            C_{11} & C_{12}\\ C_{21}&C_{22}
        \end{pmatrix}-\begin{pmatrix}
            C_{11} & C_{12}\\ C_{21}&C_{22}
        \end{pmatrix}\mathfrak{N}_i=\begin{pmatrix}
            n_1 C_{11} -C_{11}n_1&n_1 C_{12} - C_{12}n_2 \\ 
            n_2 C_{21} - C_{21}n_1 & n_2 C_{22} - C_{22}n_2
        \end{pmatrix}.
    \end{equation}
    Since $n_1$ and $n_2$ do not share any eigenvalues, the off-diagonal equations imply that $C_{12}=C_{21}=0$ by the properties of Sylvester equations \cite{sylvester1884equation}. 
    \par
    We conclude that in the neighbourhoods of fixed point loci of such symmetries of order $n\in\{2,3,4,6\}$, a rational basis change can be performed that splits the periods into invariant and non-invariant parts
    \begin{equation}\label{eq:splittingtheory}
        A^{-1}\,\vec{\Pi} =\begin{pmatrix}
            \vec{\Pi}_{\text{inv}} \\ \vec{\Pi}_{\text{n-inv}}
        \end{pmatrix}, \quad A\in M_{b_3\times b_3}(\Q)\,.
    \end{equation}
    Since $\vec{\Pi}_{\text{n-inv}}$ vanishes at the fixed point loci, the last two rows of $A^{-1}$ can be seen as fluxes $f^T\Sigma$ and $g^T\Sigma$ that together with a suitable value for the axio-dilaton define a supersymmetric flux configuration. 
    \par
    On the other hand, invariance of $\vec{\Pi}_{\text{inv}}$ under the transformation in $a_i$ forces its derivative w.r.t.\ $a_i$ at $a_i=0$ to vanish. Together with \eqref{eq:vanishingdelK2}, this implies that the F-term for $a_i$ is satisfied automatically for superpotentials that are linear combinations of $\vec{\Pi}_\text{inv}$. We will encounter such a splitting in \cref{sec:FVsym}.
    \par
    For general hypersurfaces, we find a basis for the middle cohomology spanned by \eqref{eq:basismc} in the patch parametrised by the $a_i$ that transforms among itself with integral coefficients under the monodromy actions. More significantly, in this basis, the matrix representation is in block-diagonal form, each block being an irreducible integral representation of a subgroup of $\Z_{d/w_i}$. The form of these representations is given at the end of this section. We observe that the cohomology of the mirror and the remaining elements transform in separate representations. 
    As before, we will denote the matrix representation of the monodromy in the integral symplectic basis by $\mathfrak{M}_i$ and that in the new basis $\mathfrak{N}_i$. We thus have
    \begin{equation}
        \mathfrak{M}_i = A\,\mathfrak{N}_iA^{-1},
    \end{equation}
    where $A$ is again rational due to the rationality of the monodromy matrices. With the same argument used above,
    it follows that the transition matrix to the integral symplectic basis $\vec{\Pi}$ is of the form $A C$, where now $C$ commutes with all of $\mathfrak{N}_j$. For each sub-representation $i$ not part of the mirror cohomology, there exists a monodromy $j\in\{1,2,3,4\}$ such that the block $n_j^{(i)}$ in $\mathfrak{N}_j$ has no eigenvalue one. If this where the case, the representation either would not be irreducible or the block would transform trivially under the actions of the mirror group, making it part of the mirror cohomology. The commutativity of $C$ with $\mathfrak{N}_j$ again gives rise to Sylvester equations which now prohibit blocks in $C$ that mix the representations of the mirror cohomology with the rest. Therefore, even in the many-parameter cases, on the mirror locus of the moduli space a rational splitting of the periods into vanishing and non-vanishing parts occurs. 
    \par
    We saw in the previous section that derivatives w.r.t.\ complex structure moduli map the middle cohomology onto itself. It follows that the derivative of an element in the mirror cohomology in a direction normal to the mirror locus is given by a linear combination of elements in \eqref{eq:basismc}. Due to the non-trivial transformation properties under the mirror group of this element, its summands must all vanish on the mirror locus. Therefore, supersymmetric flux vacua on the mirror locus of a Calabi--Yau family are still present from the perspective of the whole moduli space. Without this knowledge, a tedious (if even possible) analytical continuation would have had to be performed to a patch where all moduli broken by the mirror group can be set to zero.
    \par
    \par
    We conclude this section with the description of the monodromy representation on the basis given in \eqref{eq:basismc}.
    One observes that the set $\mathcal{S}$ of generators of $H^3(X,\C)$ has a partition 
    \begin{equation}
        \mathcal{S}=\bigcup_i s_i\,,\quad s_i\cap s_j=\emptyset\,,\ i\neq j\,,
    \end{equation}
    such that for all $s_i = (m_1,\ldots ,m_{n_i})$ there exist numbers $l(k)$ that relate the action of the monodromies on $s_i$ in the following way
    \begin{equation}
        \frak{M}_j\big|_{m_k} = \frak{M}_j^{l(k)}\big|_{m_1}.
    \end{equation}
    Here, the slashed action represents the phase the deformation obtains from the action of $\frak{M}_j$. This just means that the phase factor obtained by $m_k$ under the monodromy $j$ is a power of the phase obtained by $m_1$.
    Recalling that $d$ denotes the hypersurface-degree, the possible images of $l$ are the sets of coprime numbers in $\frac{d}{\delta}\{1,\ldots ,\delta-1\}$ for $\delta$ a divisor of $d$.
    \paragraph{\textbf{Example.}}For the hypersurface $\P_{2,1,1,1,1}[6]$, the image of $l$ is either
    \begin{equation*}
        \{1,5\},\ \{2,4\}\text{, or }\{3\}
    \end{equation*}
    corresponding to $\delta=6,3,2$, respectively.\hfill $\blacktriangle$
    \par
    In other words, for all sets $s_i$, we find a ``multi-phase'' $\vec{\alpha}=(\alpha_1,\ldots,\alpha_5)$, where $\alpha_j^d=1$ for all $j$, such that the monodromy actions are given by
    \begin{equation}
        \frak{M}_j m_k=\alpha_j^ {l(k)}m_k\,.
    \end{equation}
    For each such set $s_i$, we use $\beta^\delta=1$ to define new generators
    \begin{equation}\label{eq:repbasis}
    \begin{split}
        \mu_1 &= m_1+\ldots +m_{n_i}\,,\\
        \mu_2 &= \beta^{l(1)}m_1+\ldots +\beta^{l(n_i)}m_{n_i}\,,\\
        &\vdots\\
        \mu_{n_i}&=\beta^{l(1)\,(n_i-1)}m_1+\ldots +\beta^{l(n_i)\,(n_i-1)}m_{n_i}\,.
    \end{split}
    \end{equation}
    We verify case-wise that such groups transform amongst each other over the integers under the monodromies $\frak{M}_j$. We note that, for the mirror cohomology representations, the first four entries of the multi-phase must be one. 
    \par 
    Sometimes, the representation on the mirror cohomology is not irreducible, which indicates the existence of continuous flux vacua in the moduli space. This has already been observed in \cite{DeWolfe:2004ns,DeWolfe:2005gy}. Here, we will briefly exemplify how this follows from the above analysis. 
    \par
    For the complex part in the transition matrix to again be of block-diagonal form, it is important that all elements in the mirror group have unique eigenvalues under the monodromies $\mathfrak{M}_j$. This is guaranteed, since, in a suitable basis, $\hat{G}$ just counts the exponents of the coordinates $x_i$ of the deformation.  Therefore, if the mirror representation decomposes, the spectra of the block matrices in the monodromies $\mathfrak{N}_j$ are disjoint and the above matrix $C$ does not mix the sub-representations. 
    \paragraph{\textbf{Example.}}As discussed in the previous example, for the mirror sextic, the phases under $g_5$ have the $\Z_6$-weight
    \begin{equation*}
        \{1,2,4,5\}\,,
    \end{equation*}
    which splits in the sense above into $\{1,5\}$ and $\{2,4\}$. To obtain a finite Kähler potential at the Fermat point, we must pick a gauge that divides the periods by a factor $\psi$. Then, the above analysis shows that there are two rational linear combinations of the integral symplectic basis $\vec{\Pi}$ that are complex linear combinations of the periods corresponding to the representation $\{2,4\}$. But, since in the chosen gauge the two generators are of form $\psi+\mathcal{O}(\psi)^3$, they vanish at $\psi=0$.
    At the Fermat-point where $\partial_\psi K_\text{cs}=0$, this consitutes a splitting of cohomology in the sense of \eqref{eq:attractorpoint}.\hfill $\blacktriangle$
    \paragraph{\textbf{Example.}}
    For the full octic hypersurface $\P_{4,1,1,1,1}[8]$, there are $70$ representations of order eight and ten of order four. Since the order eight representation consists of four elements (with weights \{1,3,5,7\}) and the one of order four of two (\{2,6\}), this adds up to
    \begin{equation}
        70\cdot 4+10\cdot 2 = 300 = b^3 = 2\cdot (h^{21}+1)
    \end{equation}
    with $h^{2,1}=149$. Furthermore, since the one-parameter deformation $\prod_{i=2}^5x_i^2$ appears in the same representation as $\text{Res}\left(\frac{\eta}{P}\right)$, there is no attractor point at the Fermat point in the mirror quotient.\hfill $\blacktriangle$
    \paragraph{\textbf{Example.}}For the mirror-quotient of the hypersurface $\P_{6,2,2,1,1}[12]$ with two deformations, we obtain
    one order twelve and one order four representation, where the latter consists of the form corresponding to $\psi^2\prod_{i=2}^5x_i^2$ and its complex conjugate. At $\psi^2=0$, the associated periods vanish for the same reasons as for the mirror sextic above and one finds a set of continuous vacua along this locus.\hfill $\blacktriangle$
    \par It is possible to generalise the above analysis of type IIB flux vacua at orbifold points of Calabi--Yau three-folds to F-theory compactifications on Calabi--Yau four-folds. For hypersurfaces $\P_{w_i}[d]$ in weighted projective spaces, the symmetry group $\hat{G}=\bigtimes_{i=1}^{6}\Z_{d/w_i}$ then has the generators
    \begin{gather}
    \begin{gathered}
        g_1 = (w_1,0,0,0,0,-w_1 )\,,\ 
        g_2 = (0,w_2,0,0,0,-w_2)\,,\
        g_3 = (0,0,w_3,0,0,-w_3)\,,\\
        g_4 = (0,0,0,w_4,0,-w_4)\,,\
        g_5 = (0,0,0,0,w_5,-w_5)\,,\
        g_6 = (0,0,0,0,0,1)\,.
    \end{gathered}
    \end{gather}
    At orbifold points of such hypersurfaces, we observe a similar splitting as for Calabi--Yau three-folds when considering the \textit{primitive} middle cohomology. While the horizontal middle cohomology is generated by $\Omega(\vec{z})$ and its derivatives, the primitive cohomology is given by the image of the residue map (cf.\ \eqref{eq:resmap} for Calabi--Yau three-folds).
    \paragraph{\textbf{Example.}}Consider the mirror of the hypersurface $\P_{2,2,1,1,1,1}[8]$. The monomials that give rise to generators of $H^{4,0}\oplus H^{3,1}$ are
    \begin{equation}
        1\quad\text{ and }\quad \prod_{i=1}^6x_i\,.
    \end{equation}
    Then, $H^{1,3}\oplus H^{0,4}$ is spanned by their complex conjugate forms. The elements in $H^{2,2}_\text{prim}$ are obtained by monomials of degree $2\cdot 8$ that are invariant under all symmetries $g_i$, i.e.
    \begin{equation}\label{eq:H22prim8}
        \prod_{i=1}^6 x_i^2\quad\text{ and }\quad \prod_{i=3}^6 x_i^4\,.
    \end{equation}
    Since the second monomial in \eqref{eq:H22prim8} is not the product of deformations in the defining polynomial and therefore cannot be obtained by its derivatives w.r.t.\ the complex structure moduli, its corresponding form is not part of the horizontal cohomology. One finds that the primitive middle cohomology splits into the following two representations:
    \begin{alignat}{2}
        \Z_8:\qquad &\left\langle 1,\ \prod_{i=1}^6 x_i^2,\ \prod_{i=3}^6 x_i^4, \overline{1}\right\rangle\quad &&\text{ with } \Z_8\text{-weights } \left\{1,3,5,7\right\},\\ \label{eq:rep221111}
        \Z_4:\qquad &\left\langle \prod_{i=1}^6x_i,\ \overline{\prod_{i=1}^6x_i}\right\rangle && \text{ with } \Z_4\text{-weights } \left\{1,3\right\},
    \end{alignat}
    where the line over a monomial indicates the complex conjugate of the corresponding form. The $\Z_n$-weights are the exponents of the $n$-th root of unity that are obtained after acting with $g_6$ on the form. Accounting for the measure $\eta$, the $\Z_d$ weight is simply one plus the exponent of $x_6$ of the monomial. It follows, for example, that the representation in \eqref{eq:rep221111} has $\Z_8$ weights $2$ and $8-2=6$.
    \hfill $\blacktriangle$
    \paragraph{\textbf{Example.}}For the mirror sextic four-fold $\P^5[6]$, the horizontal cohomology coincides with the primitive one and 
    we observe the following splitting of the orbifold monodromy representation
    \begin{alignat}{2}
        \Z_6:\qquad &\left\langle 1, \overline{1}\right\rangle\quad &&\text{ with } \Z_6\text{-weights } \left\{1,5\right\},\\
        \Z_3:\qquad &\left\langle \prod_{i=1}^6x_i,\ \overline{\prod_{i=1}^6x_i}\right\rangle\quad && \text{ with } \Z_3\text{-weights } \left\{1,2\right\},\\
        \Z_2:\qquad &\left\langle \prod_{i=1}^6x_i^2\right\rangle && \text{ with } \Z_2\text{-weight } \left\{1\right\}.
    \end{alignat}
    Since the $\Z_2$ representation is contained in $H^{2,2}_\text{prim}$, the discussion above implies that the orbifold furnishes a supersymmetric flux vacuum.
    \hfill $\blacktriangle$
    \paragraph{\textbf{Example.}}For the two elliptically fibred four-folds discussed in \cref{sec:X24,sec:X36}, the horizontal cohomology groups again coincide with the primitive ones and one obtains the two irreducible representations
    \begin{alignat*}{2}
        \P_{12,8,1,1,1,1}[24]:&\qquad &&\Z_{24}\text{-weights:}\quad \left\{1,5,7,11,13,17,19,23\right\},\\
        \P_{18,12,3,1,1,1}[36]:&\qquad &&\Z_{36}\text{-weights:}\quad \left\{1,5,7,11,13,17,19,23,25,29,31,35\right\}.
    \end{alignat*}
    The absence of a sub-representations in $H^{2,2}_\text{prim}$ again implies that there do not exist supersymmetric flux vacua at their orbifold points.
    \hfill $\blacktriangle$
\subsection{Cyclic quotient Calabi--Yau}
\label{sec:sextic}
To study the relation between symmetries in the complex structure moduli space and flux vacua of a Calabi--Yau family in later sections, we begin by introducing a three-parameter model with a cyclic symmetry. Here, we consider one of the four hypersurfaces with a single Kähler modulus in weighted projective space, $\P_{2,1,1,1,1}[6]$. The analogue analysis for the hypersurfaces $\P_{4,1,1,1,1}[8]$ and $\P_{5,2,1,1,1}[10]$ are collected in \cref{sec:app81} and \cref{sec:app101}. Even though the family of quintics in $\P^4$ contains a quotient family with five complex structure moduli, we will not deal with here, due to its Mori cone being non-simplicial. The remainder of this section deals with the computation of the quotient's topological data, the Picard--Fuchs ideal and an integral symplectic basis for the periods.
\par
We consider the family of Calabi--Yau manifolds $\P_{2,1,1,1,1}[6]$ given by the zero locus of
\begin{align}\label{eq:P211116}
    P_{\mathcal{X}_6}=\sum_{\substack{\vec{\nu}\in \N^5 \\ 2\nu_1+\sum_{i=2}^5\nu_i=6}}a_{\vec{\nu}} \vec{x}^{\vec{\nu}}
\end{align}
inside the ambient space $\P_{2,1,1,1,1}$. Following the lines of \cref{sec:quotientbuilding}, we define a quotient of this family $\mathcal{X}_6^{(3)}\coloneqq\P_{2,1,1,1,1}[6]/\hat{S}$, where\footnote{One might also consider the quotient of the family by $\Z_d^2$ as we did for the other models $\mathcal{X}_8$ and $\mathcal{X}_{10}$ described in the appendix. However, for the family of the sextic, the corresponding quotient has a non-simplicial  Mori cone, i.e.\ it is a three-dimensional cone generated by four vectors.} $\hat{S}=\Z_3^2\times \Z_6$ is generated by 
\begin{equation} 
\begin{split}
g_1=\Z_3:(1,0,0,0,2)\,,\
g_2=\Z_3:(1,1,1,0,0)\,,\
g_3=\Z_6:(2,1,0,0,3)\,.
\end{split}
\end{equation}
The invariant monomials form the defining polynomial 
\begin{equation}\label{eq:Psextic}
P_{\mathcal{X}_6^{(3)}}=x_1^3+x_2^6+x_3^6+x_4^6+x_5^6-a_0\prod_{i=1}^5x_i-a_6x_2^3x_5^3-a_7x_3^3x_4^3\,,
\end{equation}
where we used the scaling relations of the ambient spaces' automorphism group to set the moduli parametrising the univariate monomials to one.
To obtain a toric description of the ambient space, we must find a reflexive polytope with (including the inner point) nine points whose vertices map linearly to those of the Newton polytope of $P_{\mathcal{X}_6}$. By scanning the list of Kreuzer and Skarke \cite{kreuzerskarke}, we identify this quotient with the polytope containing the points given in \Cref{tab:X6}.
\begin{table}
\renewcommand{\arraystretch}{1.2}
\centering
\begin{tabular}{| c | c c c c|c c c|}
    \hline \multicolumn{5}{|c |}{points} & \multicolumn{3}{| c|}{$l$-vectors} \\ \hline \hline
    (1 & 0&0&0&0)& 0 & 0 & -3\\ \hline
    (1 &1& 0& 0& 0)& 0 & 0 & 1\\
    (1 &0& 1& 0& 0)& 0 & 1 & 0\\
    (1 &0& 0& 1& 0)& 1 & 0 & 0\\
    (1 &0& 0& 1& 2)& 1 & 0 & 0\\
    (1 &-2& -1& -2& -2)& 0 & 1 & 0\\
    (1 &0& 0& 1& 1)& -2 & 0 & 1\\
    (1 &-1& 0& -1& -1)& 0 & -2 & 1\\ \hline
\end{tabular}
\caption{Integral points and their scaling relations of the polytope describing $\P_{2,1,1,1,1}[6]/\hat{S}$.}
\label{tab:X6}
\end{table}

Using \texttt{SAGEMATH}, we determine the unique fine star triangulation and the generators of the Mori cone listed in the right column of \Cref{tab:X6}.
The intersection ring\footnote{The coefficients of the terms $J_iJ_jJ_k$ are the classical triple intersections $C^{(0)}_{ijk}$.} 
of the Kähler forms dual to them is given by
\begin{equation}\label{eq:intringX6}
R=3J_1J_2J_3+6J_1J_3^2 +6J_2J_3^2+12J_3^3\,.
\end{equation}
The topological invariants of the model are
\begin{align}\label{eq:topdataX6}
	c_2\cdot J_1=24\,,\;c_2\cdot J_2=24\,,\;c_2\cdot J_3=60\,,\;\chi=-120\,.
\end{align}
The first instanton numbers together with the Yukawa couplings are listed in \cref{sec:app6}. The Batyrev coordinates read
\begin{align}
z_1 = \frac{1}{a_6^2}\,,\quad z_2 = \frac{1}{a_7^2}\,,\quad z_3 = \frac{a_6 a_7}{a_0^3}\,.
\end{align}
The Picard--Fuchs ideal for this model is generated by
\begin{align}
{\cal L}_1^{(2)}(\uz)&= \theta_{1}^{2}-z_{1}(2\theta_{1} - \theta_{3} + 1)(2\theta_{1} - \theta_{3})\,,\\
{\cal L}_2^{(2)}(\uz)&=  \theta_{2}^{2}-z_{2}(2\theta_{2} - \theta_{3} + 1)(2\theta_{2} - \theta_{3})\,,\\
{\cal L}_3^{(2)}(\uz) &= \theta_3\left(\theta_{3} - 2\theta_{1} - 2\theta_{2}\right) + 4\theta_{1}\theta_{2}-z_{3}\left(27\theta_{3}(\theta_3+1) + 6\right).
\end{align}
To study the one-parameter limit $a_6=a_7=0$, we consider two patches in the moduli space of the three-parameter quotient which contain the orbifold point and MUM point of the one-parameter model, respectively. The former is parametrised by the polynomial deformations
\begin{equation}
    a_6=\frac{1}{\sqrt{z_1}}\,,\quad a_7=\frac{1}{\sqrt{z_2}}\,,\quad a_0=\frac{1}{(\sqrt{z_1z_2}\,z_3)^{1/3}}\,,
\end{equation}
and the coordinates in the second patch are given by
\begin{equation}
    \tilde{a}_6=a_6=\frac{1}{\sqrt{z_1}}\,,\quad \tilde{a}_7=a_7=\frac{1}{\sqrt{z_2}}\,,\quad \tilde{a}_0=\frac{1}{a_0^3}=\sqrt{z_1z_2}\,z_3\,.
\end{equation}
In order to give an expression for a superpotential in a patch near $b_i=0$, we need to find the transition matrix $T_{b}$ transforming the Frobenius basis $\vec{\varpi}^b$ near $b_i=0$ into the integral symplectic basis $\vec{\Pi}(b)$. The strategy for finding $T_{b}$ is to first construct $\vec{\Pi}(z)$ at the MUM point by matching the leading logarithms to the $A$-side periods coming from the prepotential, cf.\ \eqref{eq:prepotential}.
Then, we need to find a set of points in the moduli space whose combined area of convergence allows for an analytical continuation of $\vec{\Pi}(z)$ to the patch parametrised by $b_i$. As an example we describe the process of analytic continuation of the periods together with the path, Frobenius bases, monodromies and transition matrices for $b_i=\tilde{a}_i$ in \cref{sec:app6}.

In the patch around $a_i = 0$, the analytical continuation combines the Frobenius basis (given up to $\mathcal{O}(a^4)$)
\begin{equation}
\begin{alignedat}{2}
&\varpi_1^a = a_0 \left(1+\frac{a_6^2}{72}+\frac{a_7^2}{72}+\dots\right),\quad
&&\varpi_2^a =  a_0 \left(18 a_6 a_7+a_0^3+\dots\right), \\
&\varpi_3^a =  a_0 \left(a_7+\frac{1}{72} a_6^2 a_7+\frac{2 a_7^3}{27}+\dots\right),
&&\varpi_4^a =  a_0 \left(a_6+\frac{2 a_6^3}{27}+\frac{1}{72}  a_6 a_7^2+\dots\right), \\
&\varpi_5^a = a_0^2 \left(1+\frac{a_6^2}{18}+\frac{a_7^2}{18}+\dots\right),
&&\varpi_6^a = a_0^2 \left(36 a_6 a_7+a_0^3+\dots\right), \\
&\varpi_7^a = a_0^2 \left(a_7+\frac{1}{18} a_6^2 a_7+\frac{25 a_7^3}{216}+\dots\right), \quad
&&\varpi_8^a = a_0^2 \left(a_6+\frac{1}{18}  a_6 a_7^2+\frac{25 a_6^3}{216}+\dots\right). 
\end{alignedat}
\end{equation}
to the integral symplectic basis via
\begin{equation}
    \vec{\Pi}(a) = T_a\,\vec{\varpi}^a,
\end{equation}
where the transition matrix was determined as
\begin{align}
    T_a=\resizebox{0.8\textwidth}{!}{%
    $\left(
\begin{array}{cccccccc}
 -\frac{2\pi^3 2^{2/3} \left(-\sqrt{3}+i\right)}{27 \Gamma \left(\frac{2}{3}\right)^9} & -\frac{\left(\sqrt{3}+3 i\right) \Gamma \left(\frac{2}{3}\right)^3}{576 \pi ^3} & 0 & 0 & \frac{-1+i \sqrt{3}}{12 \Gamma \left(\frac{2}{3}\right)^3} & \frac{\left(3+i \sqrt{3}\right) \Gamma \left(\frac{2}{3}\right)^9}{3072\cdot 2^{2/3} \pi ^6} & 0 & 0 \\
 \frac{2^{2/3} \pi ^3 \left(-\sqrt{3}+i\right)}{27 \Gamma \left(\frac{2}{3}\right)^9} & -\frac{\Gamma \left(\frac{2}{3}\right)^3}{96 \left(-\sqrt{3}+3 i\right) \pi ^3} & -\frac{\sqrt{3}+3 i}{36\cdot 2^{2/3} \Gamma \left(\frac{2}{3}\right)^3} & -\frac{-\sqrt{3}+i}{12\cdot 2^{2/3} \Gamma \left(\frac{2}{3}\right)^3} & -\frac{-3+i \sqrt{3}}{12 \sqrt{3} \left(\sqrt{3}+i\right) \Gamma \left(\frac{2}{3}\right)^3} & -\frac{\left(\sqrt{3}+i\right) \Gamma \left(\frac{2}{3}\right)^9}{2048\cdot 2^{2/3} \sqrt{3} \pi ^6} & \frac{\sqrt{3} \left(\sqrt{3}+i\right) \Gamma \left(\frac{2}{3}\right)^3}{64\cdot2^{1/3} \pi ^3} & \frac{\left(-\sqrt{3}+3 i\right) \Gamma \left(\frac{2}{3}\right)^3}{64 \cdot2^{1/3} \sqrt{3} \pi ^3} \\
 \frac{2^{2/3}\pi^3 \left(-\sqrt{3}+i\right) }{27 \Gamma \left(\frac{2}{3}\right)^9} & -\frac{\Gamma \left(\frac{2}{3}\right)^3}{96 \left(-\sqrt{3}+3 i\right) \pi ^3} & -\frac{-\sqrt{3}+i}{12\cdot 2^{2/3} \Gamma \left(\frac{2}{3}\right)^3} & -\frac{\sqrt{3}+3 i}{36\cdot 2^{2/3} \Gamma \left(\frac{2}{3}\right)^3} & -\frac{-3+i \sqrt{3}}{12 \sqrt{3} \left(\sqrt{3}+i\right) \Gamma \left(\frac{2}{3}\right)^3} & -\frac{\left(\sqrt{3}+i\right) \Gamma \left(\frac{2}{3}\right)^9}{2048\cdot 2^{2/3} \sqrt{3} \pi ^6} & \frac{\left(-\sqrt{3}+3 i\right) \Gamma \left(\frac{2}{3}\right)^3}{64 \cdot2^{1/3} \sqrt{3} \pi ^3} & \frac{\sqrt{3} \left(\sqrt{3}+i\right) \Gamma \left(\frac{2}{3}\right)^3}{64\cdot2^{1/3} \pi ^3} \\
 \frac{2^{2/3}\pi ^3 \left(\sqrt{3}+i\right) }{27 \Gamma \left(\frac{2}{3}\right)^9} & -\frac{\left(-\sqrt{3}+i\right) \Gamma \left(\frac{2}{3}\right)^3}{192 \sqrt{3} \left(-\sqrt{3}+3 i\right) \pi ^3} & \frac{-\sqrt{3}+3 i}{36\cdot 2^{2/3} \Gamma \left(\frac{2}{3}\right)^3} & \frac{-\sqrt{3}+3 i}{36\cdot 2^{2/3} \Gamma \left(\frac{2}{3}\right)^3} & -\frac{1}{6 \sqrt{3} \left(\sqrt{3}+i\right) \Gamma \left(\frac{2}{3}\right)^3} & -\frac{\left(-\sqrt{3}+i\right) \Gamma \left(\frac{2}{3}\right)^9}{2048\cdot2^{2/3} \sqrt{3} \pi ^6} & -\frac{\left(\sqrt{3}+3 i\right) \Gamma \left(\frac{2}{3}\right)^3}{64 \cdot2^{1/3} \sqrt{3} \pi ^3} & -\frac{\left(\sqrt{3}+3 i\right) \Gamma \left(\frac{2}{3}\right)^3}{64\cdot2^{1/3} \sqrt{3} \pi ^3} \\
 -\frac{2^{2/3} \pi ^3\left(-3 \sqrt{3}+i\right) }{9 \Gamma \left(\frac{2}{3}\right)^9} & -\frac{\left(-7 \sqrt{3}+3 i\right) \Gamma \left(\frac{2}{3}\right)^3}{192 \sqrt{3} \left(-\sqrt{3}+3 i\right) \pi ^3} & \frac{-\sqrt{3}+3 i}{12\cdot 2^{2/3} \Gamma \left(\frac{2}{3}\right)^3} & \frac{-\sqrt{3}+3 i}{12\cdot 2^{2/3} \Gamma \left(\frac{2}{3}\right)^3} & \frac{-6+i \sqrt{3}}{6 \sqrt{3} \left(\sqrt{3}+i\right) \Gamma \left(\frac{2}{3}\right)^3} & \frac{\sqrt{3} \left(3 \sqrt{3}+i\right) \Gamma \left(\frac{2}{3}\right)^9}{2048\cdot 2^{2/3} \pi ^6} & -\frac{\sqrt{3} \left(\sqrt{3}+3 i\right) \Gamma \left(\frac{2}{3}\right)^3}{64 \cdot2^{1/3} \pi ^3} & -\frac{\sqrt{3} \left(\sqrt{3}+3 i\right) \Gamma \left(\frac{2}{3}\right)^3}{64\cdot2^{1/3} \pi ^3} \\
 -\frac{2^{2/3}\pi ^3 \left(-\sqrt{3}+i\right) }{9 \Gamma \left(\frac{2}{3}\right)^9} & \frac{\Gamma \left(\frac{2}{3}\right)^3}{96 \left(-\sqrt{3}+3 i\right) \pi ^3} & \frac{-\sqrt{3}+i}{12\cdot 2^{2/3} \Gamma \left(\frac{2}{3}\right)^3} & \frac{\sqrt{3}+3 i}{12\cdot 2^{2/3} \Gamma \left(\frac{2}{3}\right)^3} & \frac{-3+i \sqrt{3}}{12 \sqrt{3} \left(\sqrt{3}+i\right) \Gamma \left(\frac{2}{3}\right)^3} & \frac{\sqrt{3} \left(\sqrt{3}+i\right) \Gamma \left(\frac{2}{3}\right)^9}{2048\cdot 2^{2/3} \pi ^6} & -\frac{\sqrt{3} \left(\sqrt{3}+i\right) \Gamma \left(\frac{2}{3}\right)^3}{64\cdot2^{1/3} \pi ^3} & -\frac{\sqrt{3} \left(-\sqrt{3}+3 i\right) \Gamma \left(\frac{2}{3}\right)^3}{64\cdot2^{1/3} \pi ^3} \\
 -\frac{2^{2/3}\pi ^3 \left(-\sqrt{3}+i\right) }{9 \Gamma \left(\frac{2}{3}\right)^9} & \frac{\Gamma \left(\frac{2}{3}\right)^3}{96 \left(-\sqrt{3}+3 i\right) \pi ^3} & \frac{-\sqrt{3}+i}{12\cdot2^{2/3} \Gamma \left(\frac{2}{3}\right)^3} & \frac{\sqrt{3}+3 i}{12\cdot2^{2/3} \Gamma \left(\frac{2}{3}\right)^3} & \frac{-3+i \sqrt{3}}{12 \sqrt{3} \left(\sqrt{3}+i\right) \Gamma \left(\frac{2}{3}\right)^3} & \frac{\sqrt{3} \left(\sqrt{3}+i\right) \Gamma \left(\frac{2}{3}\right)^9}{2048\cdot2^{2/3} \pi ^6} & -\frac{\sqrt{3} \left(\sqrt{3}+i\right) \Gamma \left(\frac{2}{3}\right)^3}{64\cdot2^{1/3} \pi^3} & -\frac{\sqrt{3} \left(-\sqrt{3}+3 i\right) \Gamma \left(\frac{2}{3}\right)^3}{64\cdot2^{1/3} \pi ^3} \\
 \frac{i\,2^{1/3}}{27 \pi ^3 \Gamma \left(\frac{2}{3}\right)^9} & \frac{i \Gamma \left(\frac{2}{3}\right)^3}{192 \pi ^3} & \frac{i}{6\cdot2^{2/3} \sqrt{3} \Gamma \left(\frac{2}{3}\right)^3} & \frac{i}{6\cdot2^{2/3} \sqrt{3} \Gamma \left(\frac{2}{3}\right)^3} & -\frac{i}{12 \Gamma \left(\frac{2}{3}\right)^3} & -\frac{i \Gamma \left(\frac{2}{3}\right)^9}{1024\cdot2^{1/3} \sqrt{3} \pi ^6} & -\frac{i \sqrt{3} \Gamma \left(\frac{2}{3}\right)^3}{32\cdot2^{1/3} \pi ^3} & -\frac{i \sqrt{3} \Gamma \left(\frac{2}{3}\right)^3}{32\cdot2^{1/3} \pi ^3} \\
\end{array}
\right)
        $%
    }.
\end{align}
\par In the patch near $\tilde{a}_i=0$, the Frobenius basis reads
\begin{equation}
\begin{alignedat}{2}
&\varpi^{\tilde{a}}_1=\sigma_1\,,
&&\varpi^{\tilde{a}}_5=\sigma_5\,,\\
&\varpi^{\tilde{a}}_2=\sigma_1\log(\tilde{a}_0)+\sigma_2\,,
&&\varpi^{\tilde{a}}_6=\sigma_5\log(\tilde{a}_0)+\sigma_6\,,\\
&\varpi^{\tilde{a}}_3=\sigma_1\log(\tilde{a}_0)^2+2\sigma_2\log(\tilde{a}_0)+\sigma_3\,,
&&\varpi^{\tilde{a}}_7=\sigma_7\,,\\
&\varpi^{\tilde{a}}_4=\sigma_1\log(\tilde{a}_0)^3+3\sigma_2\log(\tilde{a}_0)^2+3\sigma_3\log(\tilde{a}_0)+\sigma_4\,,\quad 
&&\varpi^{\tilde{a}}_8=\sigma_7\log(\tilde{a}_0)+\sigma_8\,,
\end{alignedat}
\end{equation}
with (up to terms in $\mathcal{O}(\tilde{a}^4)$)
\begin{equation}
\begin{alignedat}{2}
&\sigma_1=1 + 360\tilde{a}_0^2 + 6\tilde{a}_6\tilde{a}_7\tilde{a}_0+\dots\,,
&&\sigma_2=1386\tilde{a}_0^2 + 27\tilde{a}_6\tilde{a}_7\tilde{a}_0+\dots\,,\\
&\sigma_3=1314\tilde{a}_0^2 + \frac{1}{4}\tilde{a}_7^2 + \frac{1}{4}\tilde{a}_6^2 + 54\tilde{a}_6\tilde{a}_7\tilde{a}_0+\dots\,,\quad
&&\sigma_4=-3942\tilde{a}_0^2+\dots\,,\\
&\sigma_5=\tilde{a}_7 + 24\tilde{a}_6\tilde{a}_0 + 1440\tilde{a}_7\tilde{a}_0^2 + \frac{1}{24}\tilde{a}_7^3+\dots\,,
&&\sigma_6=60\tilde{a}_6\tilde{a}_0 + 4104\tilde{a}_7\tilde{a}_0^2 - \frac{1}{12}\tilde{a}_7^3+\dots\,,\\
&\sigma_7=\tilde{a}_6 + 24\tilde{a}_7\tilde{a}_0 + 1440\tilde{a}_6\tilde{a}_0^2 + \frac{1}{24}\tilde{a}_6^3+\dots\,,
&&\sigma_8=60\tilde{a}_7\tilde{a}_0 + 4104\tilde{a}_6\tilde{a}_0^2 - \frac{1}{12}\tilde{a}_6^3+\dots.
\end{alignedat}
\end{equation}
Here, the expression of the transition matrix is
\begin{align}
T_{\tilde{a}}=\left( \begin {array}{cccccccc} 1&0&0&0&0&0&0&0\\ \noalign{\medskip}-
{\frac{1}{2}}&0&0&0&0&0&{\frac {1}{2\,\pi}}&0\\ \noalign{\medskip}-{
\frac{1}{2}}&0&0&0&{\frac {1}{2\,\pi}}&0&0&0\\ \noalign{\medskip}{
\frac{1}{2}}& -{\frac {\ii}{2\pi}}&0&0&-{\frac {1}{4\,\pi}}&0
&-{\frac {1}{4\,\pi}}&0\\ \noalign{\medskip}{\frac{13}{4}}&0&{\frac {3
}{2\,{\pi}^{2}}}&0&-{\frac {3}{4\,\pi}}&0&-{\frac {3}{4\,\pi}}&0
\\ \noalign{\medskip}{\frac{7}{8}}&{\frac {3\,\ii}{4\pi}}&{
\frac {3}{4\,{\pi}^{2}}}&0&{\frac {3}{8\,\pi}}-{\frac {3\,\ii\ln  \left( 2 \right) }{2\pi^{2}}}&-\frac {3\,\ii}{4\pi^2}&-{\frac {3}{8\,\pi}}&0\\ \noalign{\medskip}{\frac{7}{8}}&{
\frac {3\,\ii}{4\pi}}&{\frac {3}{4\,{\pi}^{2}}}&0&-{\frac {3
}{8\,\pi}}&0&{\frac {3}{8\,\pi}}-{\frac {3\,\ii\ln  \left( 
2 \right) }{2\pi^{2}}}&-{\frac {3\,\ii}{4\pi^2}}
\\ \noalign{\medskip}{\frac{\chi\left(\mathcal{X}^{(1)}_6\right)\,\zeta(3)}{2(2\pi \ii)^3}}&{\frac{\tilde{c}_2\left(\mathcal{X}^{(1)}_6\right)}{24\cdot 2\pi \ii}}&0&
\frac {\ii}{4\pi^3}&{-\frac {3\,\ii\ln  \left( 2
 \right) }{4\pi^{2}}}& -\frac {3\,\ii}{8\pi^2}&{-\frac 
{3\,\ii\ln  \left( 2 \right) }{4\pi^{2}}}&-{
\frac {3\,\ii}{8\pi^2}}\end {array} \right),
\end{align}
where we identified $\chi\left(\mathcal{X}^{(1)}_6\right)=-204$ and $\tilde{c}_2\left(\mathcal{X}^{(1)}_6\right) = \int c_2\cdot J = 42$ as the Euler characteristic and second Chern class of the mirror quotient $\mathcal{X}_6^{(1)}=\P_{2,1,1,1,1}[6]/\hat{H}$.
\section{Flux vacua on symmetric loci}\label{sec:FVsym}

It is generally assumed that flux compactifications on Calabi--Yau manifolds with many complex structure moduli can be simplified by dividing out cyclic symmetries.

We showed in \cref{sec:hodgesplitting} that certain symmetries in the moduli space induce a rational splitting of the periods into invariant and non-invariant parts. On the fixed point locus of such a symmetry, the non-invariant periods and the partial derivatives of the invariant periods normal to the locus vanish. This implies on the one hand that superpotentials built out of linear combinations of the invariant periods automatically satisfy the F-term equations for the transforming moduli, which we will discuss for the family $\mathcal{X}_6^{(3)}$ in \cref{sec:symstab}. On the other hand, the two non-invariant periods give rise to a family of flux vacua along the invariant locus. This will be the topic of \cref{sec:CYcompSUSY}. In \cref{sec:subhodge}, we describe the Hodge substructures arising along these loci in codimension one. In agreement with \cite{Grimm:2024fip}, we find that the F-terms can be expressed in terms of periods of elliptic curves. While the fixed point locus of each symmetry transformation yields supersymmetric flux vacua, we observe that this does not need to be the case for their intersection. This is because the axio-dilaton values of the flux configurations may differ if there is not a symmetry between the sets of vacua. In \cref{sec:symexample}, we consider a three-parameter model with complete symmetry between the moduli. We show that this symmetry renders the vacua in codimension one compatible. Its symmetric slice contains an attractor point of rank two, whose flux vacuum requires a value for the axio-dilaton which is in agreement with that dictated by the fluxes that restrict to the symmetric locus.
\subsection{Stabilisation of symmetric moduli}\label{sec:symstab}
To illustrate the discussion of \cref{sec:hodgesplitting}, we will give its description for the quotient $\mathcal{X}_6^{(3)}$ introduced in \cref{sec:sextic}, for which we will make the spitting of periods explicit. The discussion for the quotients $\mathcal{X}_8^{(3)}$ and $\mathcal{X}_{10}^{(3)}$ is completely analogous, with the well-known exception that the sextic has a supersymmetric vacuum at its Fermat point.
\par 
The defining polynomial $P_{\mathcal{X}_6^{(3)}}$ was given in \eqref{eq:Psextic}. It turns out that the $a_6$-$a_7$-direction of the moduli space is parametrised more suitably in the coordinates
\begin{equation}
    a_\pm\coloneqq a_6\pm a_7\,.
\end{equation}
This is because the moduli space is symmetric under coordinate permutations in the ambient space
\begin{equation}\label{eq:symX6}
    \sigma_\pm:\begin{array}{c}
     \phantom{\pm} x_2\leftrightarrow x_3  \\
     \pm x_4\leftrightarrow x_5
\end{array},
\end{equation}
where $\hat{\sigma}_\pm$ has the fixed point locus $a_\mp=0$. These two $\Z_2$ symmetries will cause a splitting of the periods as we will see below.
In \cref{sec:sextic}, we found an integral symplectic basis $\vec{\Pi}$. As derived in \cref{sec:hodgesplitting}, there exist (non-symplectic) basis changes, for example
\begin{equation}\label{eq:splittingPi6}
    \mat{\vec{\Pi}_{\text{inv}}\\\vec{\Pi}_{+}\\\vec{\Pi}_{-}}(\vec{a})= 
\left(
\begin{array}{cccccccc}
 0 & 0 & 0 & -3 & 1 & -1 & -1 & 2 \\
 1 & 1 & 1 & -1 & 1 & 0 & 0 & 0 \\
 1 & 1 & 1 & 11 & -3 & 0 & 0 & 0 \\
 1 & 0 & 0 & 0 & 0 & 0 & 0 & 0 \\
 1 & 1 & 1 & 3 & -1 & 1 & 1 & 0 \\
 1 & 1 & 1 & 0 & 0 & 0 & 0 & 0 \\
 0 & 1 & -1 & 0 & 0 & 0 & 0 & 0 \\
 0 & 1 & -1 & 0 & 0 & 1 & -1 & 0 \\
\end{array}
\right)
\, \vec{\Pi}(\vec{a})\,,
\end{equation}
where under $a_\pm\mapsto -a_\pm$ only $\vec{\Pi}_\pm$ changes by a sign while $\vec{\Pi}_\mp$ and $\vec{\Pi}_\text{inv}$ remain invariant. The matrix in \eqref{eq:splittingPi6} plays the role of the matrix $A^{-1}$ in \eqref{eq:splittingtheory}.

\par We verify that, for both loci $a_\pm=0$, the partial derivatives normal to the plane of the Kähler potential vanish, cf.\ \eqref{eq:vanishingdelK2}.
This implies for their intersection $a_+=a_-=0$ that a superpotential $W$ given by a linear combination of $\vec{\Pi}_\text{inv}$ satisfies the F-term equations for the moduli $a_+$ and $a_-$. It is important to note that this does not define a supersymmetric vacuum yet, since it is not possible to choose such $W$ that also vanishes identically on the symmetric locus $a_+=a_-=0$. This implies that the F-term for the radial Kähler modulus $T$, cf.\ \eqref{eq:Wequal0}, does not vanish and supersymmetry is broken for generic $a_0$. 

Nevertheless, on the symmetric slice $a_+=a_-=0$, the periods $\vec{\Pi}_\text{inv}$ are rational linear combinations of those of the mirror of $\P_{2,1,1,1,1}[6]$. This implies that the condition for supersymmetric vacua on the symmetric slice of the quotient under consideration, i.e.\ 
\begin{equation}
    g^T\Sigma\vec{\Pi}_\text{inv}=0\,,
\end{equation}
are identical to the one for supersymmetric vacua in the one-parameter family. In the case of the sextic this implies that there is a supersymmetric vacuum located at $a_0=a_6=a_7=0$. In general, one therefore expects that the search of flux vacua in Calabi--Yau families with moduli subject to such symmetries can be performed in the sub-family over the fixed point loci of these symmetries.

\subsection{Calabi--Yau compactifications as supersymmetric flux vacua}\label{sec:CYcompSUSY}

While the previous subsection explained why one can reduce the search for flux vacua in many-moduli Calabi--Yau families to symmetric loci, here, we will describe families of flux vacua in codimension one of the moduli space. We will start with a review of known flux vacua at orbifold points of hypergeometric one-parameter families. Returning to the model $\mathcal{X}_6^{(3)}$, we will then show that two continuous sets of flux vacua exist which restrict the moduli to $a_+=0$ and $a_-=0$, respectively. However, since their value for the axio-dilaton disagree, there is no vacuum configuration restricting to the mirror locus $a_+=a_-=0$.

\paragraph{One-parameter models with supersymmetric flux vacua.}
We briefly review the explicit construction of supersymmetric vacua using as an example the one-parameter sextic three-fold $\P_{2,1,1,1,1}[6]/\hat{H}$ with $\hat{H}$ the mirror group $\Z_3\times\Z_6^2$. This family is given by the vanishing locus of 
\begin{equation}
    P_{\mathcal{X}_6/\hat{H}}(\vec{x},\psi)=x_1^3+x_2^6+x_3^6+x_4^6+x_5^6-\psi\prod_{i=1}^5x_i\,,
\end{equation}
inside $\P_{2,1,1,1,1}$.

For a hypergeometric Calabi--Yau operator with indicials $\{a_1,a_2,a_3,a_4\}$, $a_i\neq a_j$ for $i\neq j$, at infinity and a conifold at $\mu$, the solutions are given by \cite{bateman1953higher}
\begin{equation}
    \widetilde{\varpi}_k=\frac{D\,\prod_{i=1}^4\Gamma(a_i)}{\pi^4e^{\pi\ii a_{\sigma_k(1)}}}\,\frac{1}{2\pi\ii}\int_{C}\frac{\Gamma(s)^4\Gamma(a_{\sigma_k(1)}-s)(-\mu/z)^s}{\prod_{j=2}^4\Gamma(1-a_{\sigma_k(j)}+s)}\dd s\,,
\end{equation}
where $\sigma_k$ is one of the sets $\left\{\{1,2,3,4\},\,\{2,3,4,1\},\,\{3,4,1,2\},\,\{4,1,2,3\}\right\}$ and $D=\mathrm{gcd}(\{a_i\})$. The integration contour $C$ is taken along the right side of the imaginary axis and for $z<\mu$ the contour is closed to the left and for $z>\mu$ to the right side.
\par 
With $\beta^D=1$, we introduce the basis (cf.\ \eqref{eq:repbasis})
\begin{equation}\label{eq:wX6}
    \vec{\varpi} = \frac{3}{32}\begin{pmatrix}\widetilde{\varpi}_1+\widetilde{\varpi}_4\\
    \beta\,\widetilde{\varpi}_1+\beta^5\widetilde{\varpi}_4\\
    \widetilde{\varpi}_2+\widetilde{\varpi}_3\\
    \beta^2\widetilde{\varpi}_2+\beta^4\widetilde{\varpi}_3
    \end{pmatrix}.
\end{equation}
Then, the inverse of the transition matrix translating to the integral symplectic basis via $\vec{\Pi}=T_{\psi}\,\vec{\varpi}$ is given by
\begin{equation}
    T_\psi^{-1}=\begin{pmatrix}-6&-3&6&3\\ -3&12&3&-1\\-6&-1&2&3\\3&2&-1&-3
    \end{pmatrix}.
\end{equation}

Up to an overall factor, the leading order in $\psi$ of $\vec{\Pi}$ is 
\begin{equation}
    \vec{\Pi} = \left(1-i \sqrt{3},\,2,\,1-3 i \sqrt{3},\,2\right)\,\psi+\mathcal{O}(\psi^2)\,.
\end{equation}
The fact that these four entries are twice $\Q$-linear dependent makes $\psi_0=0$ a rank-two attractor point and therefore allows for a supersymmetric vacuum at $\psi_0$.
\par As we explained in \cref{sec:vacuum_criteria}, the vacuum criteria $W=D_iW=0$ are valid only for a regular Kähler potential $K$ and generally must be phrased in the gauge-independent way $e^{K/2}W=e^{K/2}D_iW=0$. In our gauge, this factor behaves as $e^{K/2}\sim \psi^{-1}$. 
\par
From the period structure in \eqref{eq:wX6} and the discussion of \cref{sec:hodgesplitting}, we know that the flux vacua lie on the $\Q$-lattice spanned by the last two rows of $T_\psi^{-1}$, since these give linear combinations of the integral period vector that lie in $H^c(X_0,\Z)$.
Here, we consider the configuration 
\begin{align}
\begin{split}\label{eq:fluxX6}
    f^T\Sigma &= (3,0,-1,-1)\,,\\
    h^T\Sigma &= (0,1,0,-1)\,.
    \end{split}
\end{align}
To guarantee that the resulting superpotential flux $G_3=F_3-\tau H_3$ is contained in $H^{2,1}(X,\C)$ -- which is equivalent to $D_\psi W=0$ -- the axio-dilaton $\tau$ must be chosen appropriately:
\begin{align}
    (f-\tau h)^T\Sigma D_\psi\,\vec{\Pi}\big|_{\psi=0} \overset{!}{=}0 \quad \Rightarrow \quad\tau = -\frac{1}{2}+\frac{\ii \sqrt{3}}{2}\,.
\end{align}
One thus finds a superpotential that stabilises the axio-dilaton and the modulus $\psi$. We emphasise that a different choice of fluxes
\begin{equation}
\begin{pmatrix}
    \tilde f\\ \tilde h 
\end{pmatrix}=\begin{pmatrix}
    a & b \\ c & d 
\end{pmatrix}\begin{pmatrix}
     f\\  h 
\end{pmatrix}\quad\text{with}\quad\begin{pmatrix}
    a & b \\ c & d 
\end{pmatrix}\in\mathrm{GL}(2,\Z)
\end{equation}
gives rise to a vacuum with $\tilde{\tau}=\frac{a\tau + b}{c\tau +d}$. The configuration \eqref{eq:fluxX6} yields the representant of $\tau$ inside the canonical fundamental domain of $\mathrm{SL}(2,\Z)$. Since the string coupling $g_s$ appears in the axio-dilaton as $\tau = C_0 + \frac{\ii}{g_s}$, higher order corrections in $g_s$ cannot be neglected. This holds also for the two attractor points we discuss next. 
\par
In a similar way, the authors of \cite{Gu:2023mgf} identified attractor points at the orbifold point in the models
\begin{equation}
    X_{4,3}=\left(\begin{tabular}{c | c c}
         $\P_{2,1^5}$ & 4&3
        \end{tabular}\right)_{-156}^{1,79}\quad\text{and}\quad X_{6,4}=\left(\begin{tabular}{c | c c}
         $\P_{3,2^2,1^3}$ & 6&4
        \end{tabular}\right)_{-156}^{1,79}.
\end{equation}
The local exponents at infinity are $\{3,4,8,9\}/12$ and $\{2,3,9,10\}/12$, respectively. In the bases
\begin{equation}
    \vec{\varpi}_{X_{4,3}} = \frac{3}{32}\begin{pmatrix}\widetilde{\varpi}_1+\widetilde{\varpi}_4\\
    \beta^3\,\widetilde{\varpi}_1+\beta^9\widetilde{\varpi}_4\\
    \widetilde{\varpi}_2+\widetilde{\varpi}_3\\
    \beta^4\widetilde{\varpi}_2+\beta^8\widetilde{\varpi}_3
    \end{pmatrix}\quad \text{and}\quad \vec{\varpi}_{X_{6,4}} = \frac{1}{48}\begin{pmatrix}\widetilde{\varpi}_1+\widetilde{\varpi}_4\\
    \beta^2\,\widetilde{\varpi}_1+\beta^{10}\widetilde{\varpi}_4\\
    \widetilde{\varpi}_2+\widetilde{\varpi}_3\\
    \beta^3\widetilde{\varpi}_2+\beta^9\widetilde{\varpi}_3
    \end{pmatrix}
\end{equation}
we find the transition matrices
\begin{equation}
    T_{X_{4,3},\psi}^{-1}=\begin{pmatrix}-6&0&3&3\\ 0&9&0&-3\\-6&0&2&3\\3&3&-1&-3
    \end{pmatrix}\quad\text{and}\quad T_{X_{6,4},\psi}^{-1}=\begin{pmatrix}-2&0&2&1\\ -1&3&1&-1\\-2&0&1&1\\0&1&0&-1
    \end{pmatrix}.
\end{equation}
Flux vacua are again generated by the last two rows of the transition matrix. For the model $X_{4,3}$ we can choose the elements
\begin{equation}
    f^T\Sigma=(6, 0, -2, -3)\quad\text{and}\quad h^T\Sigma=(0, 2, 0, -1)
\end{equation}
which require a value for the axio-dilaton of $\tau=\ii\sqrt{3}$. For $X_{6,4}$, the fluxes
\begin{equation}
    f^T\Sigma=(2, 0, -1, -1)\quad\text{and}\quad h^T\Sigma=({0, 1, 0, -1})
\end{equation}
give rise to a vacuum with $\tau=\ii$.
\paragraph{Three-parameter models and their supersymmetric flux vacua.} In the following, we will argue that the three-parameter quotient $\mathcal{X}_6^{(3)}$ of $\P_{2,1,1,1,1}[6]$ described in the previous section allows for supersymmetric vacua along two codimension one loci in its moduli space. We will make the analysis in the neighbourhood of the orbifold point $a_i=0$, allowing us to connect the findings to the vacuum of the one-parameter model found above.
\par
To phrase the analysis in the language of \cref{sec:hodgesplitting}, we first define the differential forms
\begin{equation}
    \label{eq:omegaresidue}
    \omega_i = \text{Res}\left(\frac{\text{def}_i\mu}{P^2}\right),
\end{equation}
with
\begin{equation}
    \label{eq:3paramdefs}
    \text{def}_0 = \prod_{i=1}^5 x_i\,,\quad \text{def}_6=x_2^3x_5^3\quad \text{and}\quad  \text{def}_7=x_3^3x_4^3\,.
\end{equation}
Then, the cohomology group $H^{3}(X_{\vec{a}},\Z)$ splits into the four integral monodromy representations

\begin{gather}
\begin{split}
\label{eq:repssextic}
    H^{3}(X_{\vec{a}},\Z)=&\left<\text{Res}\left(\frac{\mu}{P}\right),\overline{\text{Res}\left(\frac{\mu}{P}\right)}\right>\oplus 
    \left<\text{Res}\left(\frac{\text{def}_0\mu}{P^2}\right),\overline{\text{Res}\left(\frac{\text{def}_0\mu}{P^2}\right)}\right>\\
    &\oplus \left<\text{Res}\left(\frac{\text{def}_6\mu}{P^2}\right),\overline{\text{Res}\left(\frac{\text{def}_6\mu}{P^2}\right)}\right>\oplus 
    \left<\text{Res}\left(\frac{\text{def}_7\mu}{P^2}\right),\overline{\text{Res}\left(\frac{\text{def}_7\mu}{P^2}\right)}\right>,
\end{split}
\end{gather}
meaning that, for each of these two-dimensional subspaces, there exists a basis that transforms integrally under the group $\hat{G}$.
The two representations in the first line are those we encountered before for the mirror quotient. As discussed in \cref{sec:hodgesplitting}, we define a basis for a sub-lattice of the integral cohomology 
\begin{equation}
\begin{split}\label{eq:basisHcsextic}
    \alpha_1 = \omega_6 + \overline{\omega_6}\,,\quad \alpha_2 = \alpha^4\, \omega_6 + \alpha^2\, \overline{\omega_6}\,,\\
    \beta_1 = \omega_7 + \overline{\omega_7}\,,\quad \beta_2 = \alpha\,\omega_7 +\alpha^5\, \overline{\omega_7}\,,
\end{split}
\end{equation}
where we used $\frak{M}_5\omega_6 = \alpha^4 \omega_6$ and $\frak{M}_5\omega_7 = \alpha\, \omega_7$ to generate the basis elements. In this basis $(\alpha_1,\alpha_2,\beta_1,\beta_2)$ the monodromy $\frak{M}_5$ acts as
\begin{equation}
    \frak{M}_5 = \begin{pmatrix}
    0 & 1 & 0 & 0\\ 
    -1 & -1 & 0 & 0\\
    0 & 0 & 0 & 1\\
    0 & 0 & -1 & 1\\
    \end{pmatrix}.
\end{equation}
Defining $\omega_\pm\coloneqq \omega_+\pm\omega_-$, the integral lattice generated by these elements contains a sub-lattice spanned by
\begin{align}
    \alpha_1 + \beta_1 &= \omega_+ + \overline{\omega_+}\\
    \text{and } \alpha_2 - \beta_2 &= -\alpha\,\omega_+ -\alpha^5\overline{\omega_+}
\end{align}
and similarly for the locus $a_-=0$. It turns out that the derivative w.r.t.\ $a_\pm$ of the Kähler potential vanishes along the loci $a_\pm=0$, where thus $\omega_\pm\in H^{2,1}(X_{a_{\pm}=0},\C)$. Before we make the vacua description explicit, we conclude that the existence of vacua can be deduced from a purely group-theoretic analysis of the monodromy action.
\par
It follows from the defining polynomial in \eqref{eq:Psextic} that the family has a residual symmetry
\begin{equation}
    b=\Z_6:(0,0,0,0,5)\,,
\end{equation}
which acts on the deformation parameters as
\begin{equation}\label{eq:bsymm}
    b:\ \begin{array}{l}
        a_0\mapsto \alpha\,a_0\\
        a_-\leftrightarrow a_+
    \end{array},\quad \alpha^6=1\,.
\end{equation}
For both moduli $a_\pm$, we will describe the splitting of cohomology on their vanishing locus and compute the value of the axio-dilaton $\tau$ necessary for supersymmetric vacua. We find that the discussion respects the symmetry generated by $b$. 
\par 
As discussed in \cref{sec:Type IIB flux compactifications} we want to perform an orientifold projection to break $\mathcal{N}=2$ to $\mathcal{N}=1$ supersymmetry. For $\mathcal{X}_6^{(3)}$ we may choose the $\mathbb{Z}_2$ permutation symmetry $x_2\leftrightarrow x_5$ under which the defining polynomial \eqref{eq:Psextic} is invariant and according to \eqref{eq:omegaresidue} $h^{2,1}=h^{2,1}_{-}$ holds. This choice of $\sigma$ corresponds to the presence of an O$7$--plane. The same permutation symmetry is present for the model $\mathcal{X}_8^{(3)}$ discussed in \cref{sec:app8}.
\par
To stabilise $a_-$, we introduce the rank two lattice in $H^c(X_{a_-=0},\Z)$ spanned by the last two rows of the splitting matrix of \eqref{eq:splittingPi6}
\begin{align}
    f_-^T\Sigma &=(0, 1, -1, 0, 0, 0, 0, 0)\,,\\
    h_-^T\Sigma &=(0, 1, -1, 0, 0, 1, -1, 0)
\end{align}
with 
\begin{equation}
    g^T\Sigma\cdot \vec{\Pi}\big|_{a_-=0}=0\,,\quad g\in\{f_-,h_-\}\,.
\end{equation}
We emphasise the importance of the splitting matrix in \eqref{eq:splittingPi6} to be rational for these vacua to exist.
The resulting superpotential is given by $W=F_3^- -\tau H_3^-$ with
\begin{align}
    F_3^-(a_\pm,a_0)&=f_-^T\Sigma\vec{\Pi}(a_\pm,a_0)\,,\\
    H_3^-(a_\pm,a_0)&=h_-^T\Sigma\vec{\Pi}(a_\pm,a_0)\,.
\end{align}
The F-term equation $D_{a_-}W\big|_{a_-=0}=0$ fixes the axio-dilaton to be a non-trivial function of $a_0$ and $a_+$. The first couple of terms in its expansion are given by
\begin{align}
\tau_-&=-\frac{1}{2}+\frac{\ii\sqrt{3}}{2}+a_0 \left(\frac{9 \left(\sqrt{3}-i\right) \Gamma \left(\frac{2}{3}\right)^6}{16\ 2^{2/3} \pi ^3}+\mathcal{O}\left(a_+^2\right)\right)\\
&\qquad +a_0^2 \left(-\frac{27 \left(\left(\sqrt{3}-i\right) \left(3 i+\sqrt{3}\right) \Gamma \left(\frac{2}{3}\right)^{12}\right)}{512\cdot 2^{1/3} \pi ^6}+\mathcal{O}\left(a_+^2\right)\right)+\mathcal{O}\left(a_0^3\right).
\end{align}
We note that, with this choice of generators $f_-$ and $h_-$, the value of $\tau_-$ at the orbifold point in the one-parameter model coincides with the critical value found above
\begin{equation}
    \tau_-(a_0=0)=\tau= -\frac{1}{2}+\frac{\ii\sqrt{3}}{2}.
\end{equation}
Now, to stabilise $a_+$, we utilise the two rows of the splitting matrix in \eqref{eq:splittingPi6} that give $\vec{\Pi}_+$
\begin{align}
    f_+^T\Sigma &=({1, 1, 1, 3, -1, 1, 1, 0})\,,\\
    h_+^T\Sigma &=(1, 1, 1, 0, 0, 0, 0, 0)
\end{align}
satisfying
\begin{equation}
    g^T\Sigma \vec{\Pi}\big|_{a_+=0}=0\,,\quad g\in\{f_+,h_+\}\,.
\end{equation}
Here, the superpotential $W=F_3^- -\tau H_3^-$ is formed by
\begin{align}
    F_3^+(a_\pm,a_0)&=f_+^T\Sigma \vec{\Pi}(a_\pm,a_0)\,,\\
    H_3^+(a_\pm,a_0)&=h_+^T\Sigma \vec{\Pi}(a_\pm,a_0)\,.
\end{align}
The first terms of the critical dependence of $\tau$ on $a_0$ and $a_-$ are
\begin{align}
\tau_{+} &=-\frac{1}{2}+\frac{\ii\sqrt{3}}{2}+a_0 \left(\frac{9 \Gamma \left(\frac{2}{3}\right)^6}{4\ 2^{2/3} \left(-i+\sqrt{3}\right) \pi ^3}+\mathcal{O}\left(a_-^2\right)\right)\\
&\qquad +a_0^2 \left(\frac{27 \left(3-i \sqrt{3}\right) \Gamma \left(\frac{2}{3}\right)^{12}}{256 \sqrt[3]{2} \pi ^6}+\mathcal{O}\left(a_-^2\right)\right)+\mathcal{O}\left(a_0^3\right).
\end{align}
Again, its value at $a_0=0$ is in agreement with the one-parameter case. Furthermore, we recover $b$-symmetry
\begin{equation}\label{eq:tausym}
    \tau_-( \alpha\,a_0,a_\mp)=\tau_+(a_0,a_\pm)\,.
\end{equation}
In fact, the symmetry is present already in the superpotential
\begin{align}
    F_3^-( \alpha\,a_0,a_\mp)&=F_3^+(a_0,a_\pm)\,,\\
    H_3^-( \alpha\,a_0,a_\mp)&=H_3^+(a_0,a_\pm)\,.
\end{align}
The two sets of vacua are not compatible in the sense that there is no continuous supersymmetric vacuum at their intersection $a_6=a_7=0$. For generic values of the one-parameter modulus $a_0$, the two sets of vacua demand contradicting values for the axio-dilaton. Only at $a_0=0$, the solutions coincide and allow for the supersymmetric vacuum discussed at the beginning of this subsection.

\par For the three-parameter quotient of $\P_{4,1,1,1,1}[8]$ discussed in \cref{sec:app8}, a similar analysis can be done. Instead of two representations for the broken deformations in \eqref{eq:repssextic}, there is one (order eight) representation besides that of the mirror quotient cohomology (cf.\ \cref{sec:hodgesplitting}) given by $\left<\omega_6,\,\overline{\omega_7},\,\omega_7,\,\overline{\omega_6}\right>$. Analogous to \eqref{eq:basisHcsextic}, we introduce the basis
\begin{align}
    \alpha_1 &= \omega_6 + \overline{\omega_7}+\omega_7+\overline{\omega_6}=\omega_+ +\overline{\omega_+}\,,\\
    \alpha_2 &= \alpha^5\,\omega_6 + \alpha^7\,\overline{\omega_7}+\alpha\,\omega_7+\alpha^3\,\overline{\omega_6}=\alpha^5\,\omega_- +\alpha^3\,\overline{\omega_-}\,,\\
    \alpha_3 &= \alpha^2\,\omega_6 + \alpha^6\,\overline{\omega_7}+\alpha^2\,\omega_7+\alpha^6\,\overline{\omega_6}=\alpha^2\,\omega_+ +\alpha^6\,\overline{\omega_+}\,,\\
    \alpha_4 &= \alpha^7\,\omega_6 + \alpha^5\,\overline{\omega_7}+\alpha^3\,\omega_7+\alpha^1\,\overline{\omega_6}=\alpha^7\,\omega_- +\alpha\,\overline{\omega_-}\,.
\end{align}
Under monodromy, these elements transform as $\frak{M}_5\,\alpha_i = \alpha_{i+1}$ with $\alpha_5 \equiv -\alpha_1$. We again have $\partial_{a_\pm}K_{\text{cs}}|_{a_\pm=0}=0$ due to the analogous symmetries to \eqref{eq:symX6}. Thus there are rank-two sub-lattices in $H^c(X_{a_+=0},\Z)$ spanned by $\alpha_{1}$ and $\alpha_3$ and in $H^c(X_{a_-=0},\Z)$ spanned by $\alpha_{2}$ and $\alpha_4$. As expected, one again finds supersymmetric vacua along these loci in codimension one. Here, the splitting of \eqref{eq:splittingPi6} can be written as
\begin{equation}
    \mat{\vec{\Pi}_{\text{inv}}\\\vec{\Pi}_{+}\\\vec{\Pi}_{-}}(\vec{a})= 
\left(
\begin{array}{cccccccc}
 0 & 0 & 0 & -2 & 1 & -1 & -1 & 2 \\
 0 & 0 & 0 & -2 & 1 & 0 & 0 & 0 \\
 1 & 1 & 1 & 0 & 1 & 0 & 0 & 0 \\
 -1 & 0 & 0 & 0 & 0 & 0 & 0 & 0 \\
 1 & 1 & 1 & 0 & 0 & 0 & 0 & 0 \\
 0 & 0 & 0 & 2 & -1 & 1 & 1 & 0 \\
 0 & 1 & -1 & 0 & 0 & 0 & 0 & 0 \\
 0 & 0 & 0 & 0 & 0 & 1 & -1 & 0 \\
\end{array}
\right)
\, \vec{\Pi}(\vec{a})\,.
\end{equation}
The two sets of vacua are also incompatible, where, now, the symmetry of \eqref{eq:bsymm} and \eqref{eq:tausym} is with an eighth root of unity.
\par
Things change for the quotient of  $\P_{5,2,1,1,1}[10]$ introduced in \cref{sec:app10}. Due to the weights of the ambient space, there does not exist a symmetry such as \eqref{eq:symX6}. Therefore, one does not find flux vacua along loci of codimension one in the moduli space. We nevertheless give the splitting as
\begin{equation}
    \mat{\vec{\Pi}_{\text{inv}}\\\vec{\Pi}_{\text{n-inv}}}(\vec{a})= 
\left(
\begin{array}{cccccccc}
 0 & 0 & -1 & -3 & 3 & -1 & -1 & 2 \\
 0 & 0 & 0 & -1 & 1 & 0 & 0 & 0 \\
 1 & 0 & 0 & -1 & 1 & 0 & 0 & 0 \\
 3 & 5 & 1 & 0 & 2 & 0 & 0 & 0 \\
 0 & 0 & 1 & 3 & -3 & 1 & 1 & 0 \\
 0 & 0 & 0 & 5 & -5 & 0 & 2 & 0 \\
 -1 & 0 & 0 & 1 & -1 & 2 & 0 & 0 \\
 -3 & -5 & -1 & 0 & 0 & 0 & 0 & 0 \\
\end{array}
\right)
\, \vec{\Pi}(\vec{a})\,,
\end{equation}
where the four periods in $\vec{\Pi}_{\text{n-inv}}$ obtain a sign from the monodromy $(a_6,a_7)\mapsto (-a_6,-a_7)$ induced by the symmetry action of the ambient space $(x_1,x_5)\mapsto (-x_1,-x_5)$, cf.\ \eqref{eq:PX10}.
\par
In \cref{sec:symexample}, we will discuss a three-parameter model with complete symmetry between the moduli where the intersection of two sets of vacua in codimension one yields concurring axio-dilaton values. A similar model with such vacua, the Hulek--Verrill Calabi--Yau (HV CY) \cite{hulek2005modularity}, was studied in \cite{Candelas:2023yrg}. The important difference between these models and ours is the symmetry between the involutions, whose invariant slices allow for vacua in codimension one. For the model in \cref{sec:symexample}, the symmetry between the three $z_i$'s implies that there are vacua on, for example, $z_1=z_2$ and $z_2=z_3$. The fact that these two loci themselves are symmetric, implies that the values of the axio--dilaton are compatible on the intersection. In contrast, the two loci $a_+=0$ and $a_-=0$ are not symmetric since $a_0$ is rotated by $e^{2\pi\ii/6}$ under an exchange of the two, cf.\ \eqref{eq:bsymm}.

\subsection{Hodge substructure}\label{sec:subhodge}
The symmetry also manifests itself in the factorisation of the local zeta function, for a brief introduction of the latter see \cref{sec:app7}. As conjectured in \cite{Kachru:2020sio}, supersymmetric flux vacua are expected to be modular in the sense that the local zeta function contains a quadratic factor belonging to an elliptic curve, whose complex structure modulus is given by the axio-dilaton \cite{Kachru:2020abh}.
This was generalised to F-theory flux vacua on Calabi--Yau four-folds in \cite{Grimm:2024fip}, where it was argued that the derivatives of the vanishing periods are described by families of K3 surfaces. 
Finding flux vacua on the fixed point locus is then equivalent to identifying the attractive K3s inside this family. 
The fact that periods on elliptic curves and K3 surfaces are exact when expressed in terms of the mirror coordinates allowed the authors to perform an exact analysis of F-theory flux vacua on the Hulek--Verrill Calabi--Yau four-fold. 

For the HV CY, the factorisation of the local zeta function on the symmetric locus $\varphi_j=\varphi\,,\ \forall\, j$ was found in \cite{Candelas:2023yrg}. The numerator of the local zeta function consists of four equal factors of degree two, which give rise to a (compatible) value for the axio-dilaton on the symmetric locus. For the family $\mathcal{X}_6^{(3)}$, we would expect to see a factorisation into two \textit{different} quadratic and one quartic factor along the symmetric locus $a_+=a_-=0$. We analyse their corresponding elliptic curves in the following. 

The relevant objects in the F-term equations for the non-invariant moduli can be written as periods of elliptic curves. We will focus on the vacua along $a_+=0$. On the locus $a_+=a_-=0$, one finds that the F-terms
\begin{equation}
    \begin{pmatrix}f_+^T\Sigma\partial_{a_+}\vec{\Pi}\\ h_+^T\Sigma\partial_{a_+}\vec{\Pi}\end{pmatrix}\Bigg|_{a_\pm=0}(a_0)
\end{equation}
are linear combinations of periods $\vec{\Pi}^{\text{el},+}(z)$ of the elliptic curve family with Picard--Fuchs ideal
\begin{equation}
    \mathcal{L}^{(2)}_{\text{el},+}(z) = \theta^2-12 z(3\theta+1)(3\theta+2)\,,
\end{equation}
where $z=1/a_0^3$ and $\theta=z\partial_z$. This family is given by a hypersurface of degree three in $\P^2$. The Riemann symbol of the above operator reads
\begin{equation}
    \mathcal{P}_{\mathcal{L}^{(2)}_{\text{el},+}}\left\{\begin{array}{ccc}  0 & \frac{1}{108} & \infty\\\hline
     0 & 0 & \frac{1}{3}\\
     0 & 0 & \frac{2}{3}
   \end{array},\ z\right\}.
\end{equation}
For sufficiently small values of $a_-\neq 0$, we find that the F-terms are again described by the above operator, where $z$ is then rescaled. 
We choose the integral symplectic basis at the MUM point of $\mathcal{L}^{(2)}_{\text{el},+}$ 
\begin{equation}
    \vec{\Pi}^{\text{el},+}(z) = \begin{pmatrix}
        1 & 0 \\ -\frac{\ii \log 2}{\pi} & \frac{1}{2\pi \ii}
    \end{pmatrix}\vec{\varpi}^{\text{el},+}(z)\,,
\end{equation}
with the Frobenius basis $\vec{\varpi}^{\text{el},+}(z)= (1,\log(z)) + \mathcal{O}(z)$. Then, we find that, in the patch $z>1/108$,
\begin{equation}
    \begin{pmatrix}f_+^T\Sigma\partial_{a_+}\vec{\Pi}\\ h_+^T\Sigma\partial_{a_+}\vec{\Pi}\end{pmatrix}\Bigg|_{a_\pm=0}(a_0)=\frac{1}{\pi}\begin{pmatrix}
        1 & 3\\ 1 & 0 
    \end{pmatrix}\vec{\Pi}^{\text{el},+}(a_0)\,.
\end{equation}\par
The F-terms for the vacua on $a_-=0$ are described by periods of a family with operator $\mathcal{L}^{(2)}_{\text{el},-}(z)=\mathcal{L}^{(2)}_{\text{el},+}(-z)$, as expected from the symmetry \eqref{eq:bsymm}. Here, the matrix transforming the Frobenius basis into an integral symplectic one is given by\footnote{If one identifies the F-terms with a basis of periods on the elliptic curve, the monodromy group is merely rational.}
\begin{equation}
    \vec{\Pi}^{\text{el},-}(z) = \begin{pmatrix}
    1 & 0 \\ \frac{1}{2}-\frac{\ii \log 2}{\pi} & \frac{1}{2\pi \ii}
\end{pmatrix}\vec{\varpi}^{\text{el},-}(z)\,,
\end{equation}
and the F-terms satisfy
\begin{equation}
    \begin{pmatrix}f_-^T\Sigma\partial_{a_-}\vec{\Pi}\\ h_-^T\Sigma\partial_{a_-}\vec{\Pi}\end{pmatrix}\Bigg|_{a_\pm=0}(a_0)=\frac{1}{\pi}\begin{pmatrix}
        1 & 0\\ 4 & -3
    \end{pmatrix}\vec{\Pi}^{\text{el},-}(a_0)\,.
\end{equation}
We may also compute the $j$-invariant of the elliptic curves as a function of the complex structure modulus $\tau^{\text{el},\pm} = \Pi^{\text{el},\pm}_2/\Pi^{\text{el},\pm}_1$, which reads
\begin{equation}
    j(\tau^{\text{el},\pm}(z)) = \pm\frac{(1\pm 864 z)^3}{4z (1 \mp 108 z)^3}\,.
\end{equation}

\subsection{A symmetric example: CICY in \texorpdfstring{$\P^2\times \P^2\times \P^2$}{P2P2P2}}\label{sec:symexample}
In the previous subsections, we showed that the symmetric sub-family of $\mathcal{X}_6^{(3)}$ cannot be described as a supersymmetric flux vacuum since the two vacua in codimension one are not identical from the point of view of the one-parameter model. Here, we give a counterexample where all three moduli appear in a symmetric manner and the symmetric sub-family can be obtained by a supersymmetric flux configuration. Furthermore, we show that the flux vacuum at the attractor point of the one-parameter model is compatible with the value of the axio-dilaton dictated by this configuration. \par
The model under consideration is given by the mirror of the CICY
\begin{equation}\label{eq:CYsym}
    \mathcal{X}_\text{sym}^{(3)}=\left(\begin{tabular}{c | c c c}
         $\P^2$ & 1&1&1 \\
         $\P^2$ & 1&1&1 \\
         $\P^2$ & 1&1&1
        \end{tabular}\right)^{3,48}_{-90},
\end{equation}
with a Mori cone generated by
\begin{align}\begin{split}\label{eq:lsym}
    l_1 &= (-1,-1,-1;1,1,1,0,0,0,0,0,0)\,,\\
    l_2 &= (-1,-1,-1;0,0,0,1,1,1,0,0,0)\,,\\
    l_3 &= (-1,-1,-1;0,0,0,0,0,0,1,1,1)\,.\end{split}
\end{align}
For the computation of topological invariants on CICYs, we refer to \cite{Hosono:1994ax}. Here,and in the following, we always mean the mirror manifold of a CICY when writing about it in the form of \eqref{eq:CYsym}. 
The intersection ring of the Kähler forms is given by
\begin{equation}
    R= 3 J_1^2J_2+3J_1J_2^2+3 J_1^2J_3+6J_1J_2J_3+3J_2^2J_3+3J_1J_3^2+3J_2J_3^2\,.
\end{equation}
Their intersection with the second Chern class are
\begin{equation}
    c_2\cdot J_i= 36\,,\quad i\in\{1,\ldots ,3\}\,.
\end{equation}
The Picard--Fuchs ideal is generated by, for example,\footnote{At the cost of lengthier expressions, the ideal can of course be generated by a set of operators that is invariant under the permutation of any two moduli.}
\begin{align}
    \mathcal{L}_1^{(2)}(\vec{z}) &= \left(z_3-1\right)\left(\theta _1^2-\theta _2 \theta _1+\theta _2^2\right)+ z_1\left(\left(\theta _1+\theta _2+1\right)^2+3 \theta _3 \left(\theta _1+\theta _2+1\right)+3 \theta _3^2\right)\nonumber\\
    &\quad + z_2\left(\left(\theta _1+\theta _2+1\right)^2+3 \theta _3 \left(\theta _1+\theta _2+1\right)+3 \theta _3^2\right),\\
    \mathcal{L}_2^{(2)}(\vec{z}) &= \left(z_1-1\right)\left(\theta _2^2-\theta _3 \theta _2+\theta _3^2\right)+z_2\left(3 \theta _1^2+3 \left(\theta _2+\theta _3+1\right) \theta _1+\left(\theta _2+\theta _3+1\right)^2\right)\nonumber\\
    &\quad + z_3\left(3 \theta _1^2+3 \left(\theta _2+\theta _3+1\right) \theta _1+\left(\theta _2+\theta _3+1\right)^2\right),\\
    \mathcal{L}_3^{(2)}(\vec{z}) &=\left(\theta _1^2-\theta _3 \theta _1+\theta _3^2\right) \left(z_2-1\right)\nonumber\\
    &\quad + z_1\left(\theta _1^2+\left(3 \theta _2+2 \theta _3+2\right) \theta _1+3 \theta _2^2+\left(\theta _3+1\right)^2+3 \theta _2 \left(\theta _3+1\right)\right)\nonumber\\
    &\quad +z_3\left(\theta _1^2+\left(3 \theta _2+2 \theta _3+2\right) \theta _1+3 \theta _2^2+\left(\theta _3+1\right)^2+3 \theta _2 \left(\theta _3+1\right)\right) ,
\end{align}
where the $z_i$ are again the Batyrev coordinates belonging to the $l$-vectors in \eqref{eq:lsym}. 
\par
Since we would like to study the period structure on the symmetric locus $z_1=z_2=z_3=z$, we express the integral symplectic periods $\vec{\Pi}$ in the coordinates $u_1=z_1$, $u_2=z_1-z_2$ and $u_3=z_1-z_3$. At the two loci $u_2=0$ and $u_3=0$, whose intersection is the symmetric locus, we find the relations
\begin{align}
\begin{split}
    h_2^T\Sigma \vec{\Pi}(u_1,u_2=0,u_3)=0\,,\\
    f_2^T\Sigma \vec{\Pi}(u_1,u_2=0,u_3)=0
\end{split}
\end{align}
with $h_2^T\Sigma=(0,1,-1,0,0,0,0,0)$ and $f_2^T\Sigma=(0,0,0,0,0,1,-1,0)$ and 
\begin{align}
\begin{split}
    h_3^T\Sigma \vec{\Pi}(u_1,u_2,u_3=0)=0\,,\\
    f_3^T\Sigma \vec{\Pi}(u_1,u_2,u_3=0)=0
\end{split}
\end{align}
with $h_3^T\Sigma=(0,1,0,-1,0,0,0,0)$ and $f_3^T\Sigma=(0,0,0,0,1,0,-1,0)$. As we discussed in the previous subsections, the axio-dilaton of the configurations is given by
\begin{equation}
    \tau_i = \frac{f_i^T\Sigma\partial_{u_i}\vec{\Pi}|_{u_i=0}}{h_i^T\Sigma\partial_{u_i}\vec{\Pi}|_{u_i=0}}\,.
\end{equation}
The two functions for the axio-dilaton coincide on the symmetric locus, giving
\begin{equation}\label{eq:tausym222}
    \tau = -\frac{3 i \log \left(z\right)}{\pi }+\frac{45 i z^2}{2 \pi }-\frac{999 i z^4}{4 \pi }+\mathcal{O}\left(z^5\right).
\end{equation}
In fact, the equality holds already for the numerators and denominators of $\tau_2$ and $\tau_3$. This implies that they are periods of the same family of elliptic curves, whose Picard--Fuchs operator and Riemann symbol are given by
\begin{align}
\begin{split}\label{eq:symel}
    \mathcal{L}_\text{el}^{(2)}(z) &= \left(\theta+1\right)^2+3 z^2 \left(3 \theta +5\right) \left(3 \theta +7\right),\\
    \mathcal{P}&_{\mathcal{L}^{(2)}_{\text{el}}}\left\{\begin{array}{cccc}  0 & \frac{\ii}{3\sqrt{3}} & -\frac{\ii}{3\sqrt{3}} & \infty\\\hline
     -1 & 0 & 0 & \frac{5}{3}\\
     -1 & 0 & 0 & \frac{7}{3}
   \end{array},\ z\right\}.
   \end{split}
\end{align}
\par
We will now show that the axio-dilaton in \eqref{eq:tausym222} is compatible with that of an attractor point of the one-parameter model fibred over $z_1=z_2=z_3=z$. Similar to the symmetric locus of the HV three-fold discussed in \cite{Candelas:2021lkc}, this model has integral monodromies only when quotiented by the $\Z_3$ symmetry between the coordinates of the three $\P^2$s. This quotient has the effect of dividing the prepotential --- consisting of integrals over the fibre given by three equal pieces --- by $\kappa=3$. Hence, with the topological data of the three-parameter model, we obtain for the prepotential $\mathcal{F}$
\begin{equation}
    \kappa\,\mathcal{F}=-\frac{90}{3!}t^3 + \frac{108}{24}t+\frac{(-90)\,\zeta_3}{2(2\pi\ii)^3} + \mathcal{O}\left(e^{2\pi\ii t}\right).
\end{equation}
Here, we performed implicitly a symplectic basis change of the periods that removed the quadratic terms in $t$ that are necessary for the integrality of the three-parameter monodromies. The holomorphic three-parameter period at the MUM point with $z_1=z_2=z_3=z$ is annihilated by the operator listed as AESZ 17 in \cite{van2018calabi}
\begin{align}
\begin{split}
    \mathcal{L}^{(4)}_\text{AESZ17}(z) &= 5^{2} \theta^4-3 5 z\left(51\theta^4+84\theta^3+72\theta^2+30\theta+5\right)\\
    &\quad +23 z^{2}\left(531\theta^4+828\theta^3+541\theta^2+155\theta+15\right)\\
    &\quad -23^{3} z^{3}\left(423\theta^4+2160\theta^3+4399\theta^2+3795\theta+1170\right)\\
    &\quad +3^{5} z^{4}\left(279\theta^4+1368\theta^3+2270\theta^2+1586\theta+402\right)-3^{10} z^{5}\left((\theta+1)^4\right)
    \end{split}
\end{align}
with Riemann symbol
\begin{equation}
    \mathcal{P}_{\mathcal{L}^{(4)}_\text{AESZ17}}\left\{\begin{array}{cccccc}  0 & \frac{\ii}{3\sqrt{3}}& -\frac{\ii}{3\sqrt{3}} & \frac{1}{27} & \frac{5}{9} & \infty\\\hline
     0 & 0 & 0 & 0 & 0 & 1 \\
     0 & 1 & 1 & 1 & 1 & 1 \\
     0 & 1 & 1 & 1 & 3 & 1 \\
     0 & 2 & 2 & 2 & 4 & 1
   \end{array},\ z\right\}.
\end{equation}
The monodromy transformations of the periods are given in \Cref{tab:mon17}.
\begin{table}
\centering
\setlength{\arraycolsep}{7pt}
\begin{tabular}{ l l }
        $\mathfrak{M}_0 =
\left(\begin{array}{rrrr}
 1 & 0 & 0 & 0 \\
 1 & 1 & 0 & 0 \\
 -\frac{45}{\kappa } & -\frac{90}{\kappa } & 1 & 0 \\
 \frac{24}{\kappa } & \frac{45}{\kappa } & -1 & 1 \\
\end{array}\right)$
 & $\mathfrak{M}_{\frac{1}{27}}=\left(
\begin{array}{rrrr}
 1 & 0 & 0 & -\kappa  \\
 0 & 1 & 0 & 0 \\
 0 & 0 & 1 & 0 \\
 0 & 0 & 0 & 1 \\
\end{array}
\right)$ \vspace{0.3cm}\\
        $\mathfrak{M}_{\frac{\ii}{3\sqrt{3}}}=\left(
\begin{array}{rrrr}
 10 & -9 & -\kappa  & -3 \kappa  \\
 3 & -2 & -\frac{\kappa }{3} & -\kappa  \\
 -\frac{27}{\kappa } & \frac{27}{\kappa } & 4 & 9 \\
 \frac{27}{\kappa } & -\frac{27}{\kappa } & -3 & -8 \\
\end{array}\right)$ & $\mathfrak{M}_{-\frac{\ii}{3\sqrt{3}}}= \left(
\begin{array}{rrrr}
 -8 & -9 & \kappa  & -3 \kappa  \\
 3 & 4 & -\frac{\kappa }{3} & \kappa  \\
 \frac{27}{\kappa } & \frac{27}{\kappa } & -2 & 9 \\
 \frac{27}{\kappa } & \frac{27}{\kappa } & -3 & 10 \\
\end{array}
\right)$ \vspace{0.3cm}\\
$\mathfrak{M}_\infty=\left(
\begin{array}{rrrr}
 -5 & -27 & \kappa  & 7 \kappa  \\
 -4 & -5 & \frac{2 \kappa }{3} & 3 \kappa  \\
 \frac{36}{\kappa } & -\frac{72}{\kappa } & -5 & -9 \\
 -\frac{24}{\kappa } & -\frac{36}{\kappa } & 4 & 19 \\
\end{array}
\right)$&
\end{tabular}
\caption{Monodromy representations $\mathfrak{M}_p$ around $p$, which become integral after performing the quotient with $\kappa=3$.}
\label{tab:mon17}
\end{table}
The attractor point of rank two at $z=-1$ gives rise to supersymmetric flux vacua with fluxes on the lattice generated by\footnote{To restore integrality on the quotient with $\kappa=3$, one can simply multiply both fluxes by $\kappa$, which leaves the axio-dilaton invariant.}
\begin{align}
f^T\Sigma &=\left(-\frac{18}{\kappa }, \frac{45}{\kappa }, 1, 0\right), \\
h^T\Sigma &= \left(\frac{4}{\kappa }, -\frac{14}{\kappa }, 0, 1\right).
\end{align}
These particular fluxes imply the value for the axio-dilaton
\begin{equation}
    \tau = -\frac{1}{2}+\ii\, 1.1210986708\cdots\,.
\end{equation}
Coming back to the three-parameter model, we must evaluate the axio-dilaton at $z_1=z_2=z_3=-1$. This point is outside the region of convergence of the solutions at the MUM point. To avoid having to perform an analytical continuation for the whole period vector, we may construct $\tau$ globally from its description as periods of the elliptic curves described by \eqref{eq:symel}. We find that, up to $\text{Sl}(2,\Z)$-transformations, the two values for $\tau$ indeed agree and that the attractor point is therefore also a supersymmetric flux vacuum from the point of view of the three-parameter model.

\section{Non-pertubative type IIB vacua}
\label{sec:FtheoryVacua}
In this section, we will study supersymmetric F-theory flux vacua at special loci in the moduli space of elliptically fibred four-folds. We start by considering the simplest elliptically fibred four-fold given as a hypersurface in a weighted projective space, the mirror of the model $\P_{12,8,1,1,1,1}[24]$. The analytic continuation of periods and search for supersymmetric flux vacua in this model has been performed in part already in \cite{Cota:2017aal}, where the conifold and orbifold locus was studied, with the result that the former features a family of supersymmetric vacua. It is expected that, in general, the conifold, where an $S^4$ cycle shrinks to zero, features such a family of supersymmetric vacua. In practice, this is due to the presence of a local solution starting as $\Delta^{3/2}$. We review the vacuum found in \cite{Cota:2017aal} and then extend the analysis to the rest of the moduli space. Then, we discuss a three-parameter model, the mirror of $\P_{18,12,3,1,1,1}[36]$, in which we again find a supersymmetric vacuum along a conifold locus. In the following, $\mathcal{Y}$ will always denote a family of Calabi--Yau four-folds.
\subsection{Elliptic fibration over \texorpdfstring{$\P^3$}{P3}}
\label{sec:X24}
The mirror of the model $\P_{12,8,1,1,1,1}[24]$ is a two-parameter model given as the vanishing locus of 
\begin{equation}\label{eq:P1281111}
    P_{\mathcal{Y}_{24}/\hat{H}}(\vec{x},\psi,\phi)=x_1^2+x_2^3+x_3^{24}+x_4^{24}+x_5^{24}+x_6^{24}+\psi\prod_{i=1}^6x_i+\phi\prod_{i=3}^6x_i^6\,.
\end{equation}
We want to study flux vacua in this model in a region of the complex structure moduli space where the (mirror) volume modulus of the elliptic fibre goes to infinity, corresponding to a weakly coupled string. We analyse the global structure of the moduli space and in particular compute all monodromies in this model. 

\par The toric description of the model is given in \Cref{tab:X24}. It describes an elliptic fibration over $\mathbb{P}^3$. The Hodge numbers are
\begin{align}
    h^{1,1}=2\,, \quad h^{2,1}=0\,, \quad h^{2,2}=15564\,, \quad h^{3,1}=3878\,,
\end{align}
with $ h_{\text{prim}}^{2,2}=2$ and for the purpose of constructing an integral basis of periods we further note the following topological invariants:
\begin{gather}
\begin{gathered}
c_2\cdot J_1^2=728\,,\; c_2\cdot J_1 J_2 =182\,, \; c_2\cdot J_2^2 = 48\,,\\
c_3\cdot J_1=-3860\,,\; c_3\cdot J_2=-960\,,\\
\chi=23328\,.
\end{gathered}
\end{gather}
The intersection ring is given by
\begin{equation}
R=64J_1^4 + 16J_1^3J_2 + 4J_1^2J_2^2 + J_1J_2^3\,.
\end{equation}
The differential ideal is generated by the operators
\begin{align}
\begin{split}
     {\cal L}_1^{(2)}(\uz) &=  \left(1-432z_1\right)\theta_1^2-4\left(\theta_2+ 108z_1 \right) \theta_{{1}}-60z_1\,,\\
     {\cal L}_2^{(4)}(\uz)&=
\theta_2^4 - z_2\left(\theta_1 -4\theta_2 \right)  \left( \theta_1 -4\theta_2-1\right)\left( \theta_1 -4\theta_2-2\right)\left( \theta_1 -4\theta_2-3\right).
\end{split}
\end{align}
The four-point couplings are 
\begin{align}
    &C_{1,1,1,1}(\uz)=-\frac{64}{z_1^4\Delta_1}\,,\ C_{1,1,1,2}(\uz)=-\frac{16-6912z_1}{z_1^3z_2\Delta_1}\,,\ C_{1,1,2,2}(\uz)=-\frac{4\left(1-432z_1\right)^2}{z_1^2z_2^2\Delta_1}\,,\\
    &C_{1,2,2,2}(\uz)=-\frac{\left(1-432z_1\right)^3}{z_1z_2^3\Delta_1}\,,\quad C_{2,2,2,2}(\uz)=\frac{64\left(1-864z_1\right)\left(1-864z_1+373248z_1^2 \right)}{z_2^3\Delta_1\Delta_2}\,.
\end{align}

\begin{table}[H]
\renewcommand{\arraystretch}{1.2}
\centering
\begin{tabular}{| c | c c c c c|c c|}
    \hline \multicolumn{6}{|c |}{points} & \multicolumn{2}{| c|}{$l$-vectors} \\ \hline \hline
    (1 & 0&0&0&0&0)& $-$6 & $0$ \\ \hline
    (1 &1& 0& 0& 0 & 0)& 3 & 0 \\
    (1 &0& 1& 0& 0& 0)& 2 & 0 \\
    (1 &0& 0& 1& 0& 0)& 0 & 1 \\
    (1 &0& 0& 0& 1 & 0)& 0 & 1 \\
    (1 &0& 0& 0& 0 & 1)& 0 & 1 \\
    (1 &-12& -8& -1& -1 & -1)& 0 & 1 \\
    (1 &-3& -2& 0& 0 & 0)& 1 & -4 \\\hline 
    (1 &-2& -1& 0& 0 & 0)& - & - \\
    (1 &-1& -1& 0& 0 & 0)& - & - \\
    (1 &-1& 0& 0& 0 & 0)& - & - \\\hline 
\end{tabular}
\caption{Integral points and their scaling relations of polytope describing $\P_{12,8,1,1,1,1}[24]/\hat{H}$. The last three points lie inside a face of codimension one.}
\label{tab:X24}
\end{table}

In order to compute the periods we first discuss the structure of the moduli space. The latter closely resembles that of the elliptically fibred model $\P_{9,6,1,1,1}[18]$. The moduli space and monodromy group of this model was studied extensively in \cite{Candelas:1994hw}, whose analysis we follow for our model at hand. Naively the moduli space is described as $\mathbb{C}^2_{(\psi,\phi)}$. However residual rescalings of the coordinates identify $(\psi,\phi)\leftrightarrow (\alpha\,\psi,\alpha^6\phi)$. Hence we should rather think of the moduli space as $\mathbb{C}^2_{(\psi^6,\phi)}/\mathbb{Z}_4$ where $(\psi^6,\phi)\leftrightarrow (\alpha^6\psi^6,\alpha^6\phi)$. This can be viewed as an affine patch of a $\mathbb{P}_{1,1,4}$ with coordinates $[\psi^6,\phi,1]=[x:y:z]$ which we will use as a compact model of the moduli space. In terms of the homogeneous coordinates the components of the discriminant locus along which the corresponding Calabi--Yau spaces develop conifold singularities read
\begin{align}
\begin{split}
&\Delta_1\equiv2^{24}3^{12}z-(x-2^43^3y)^4=0\,,\\
&\Delta_2\equiv y^4-2^8z=0\,.
\end{split}
\end{align}
\begin{figure}
    \centering
    \includegraphics[width=\textwidth]{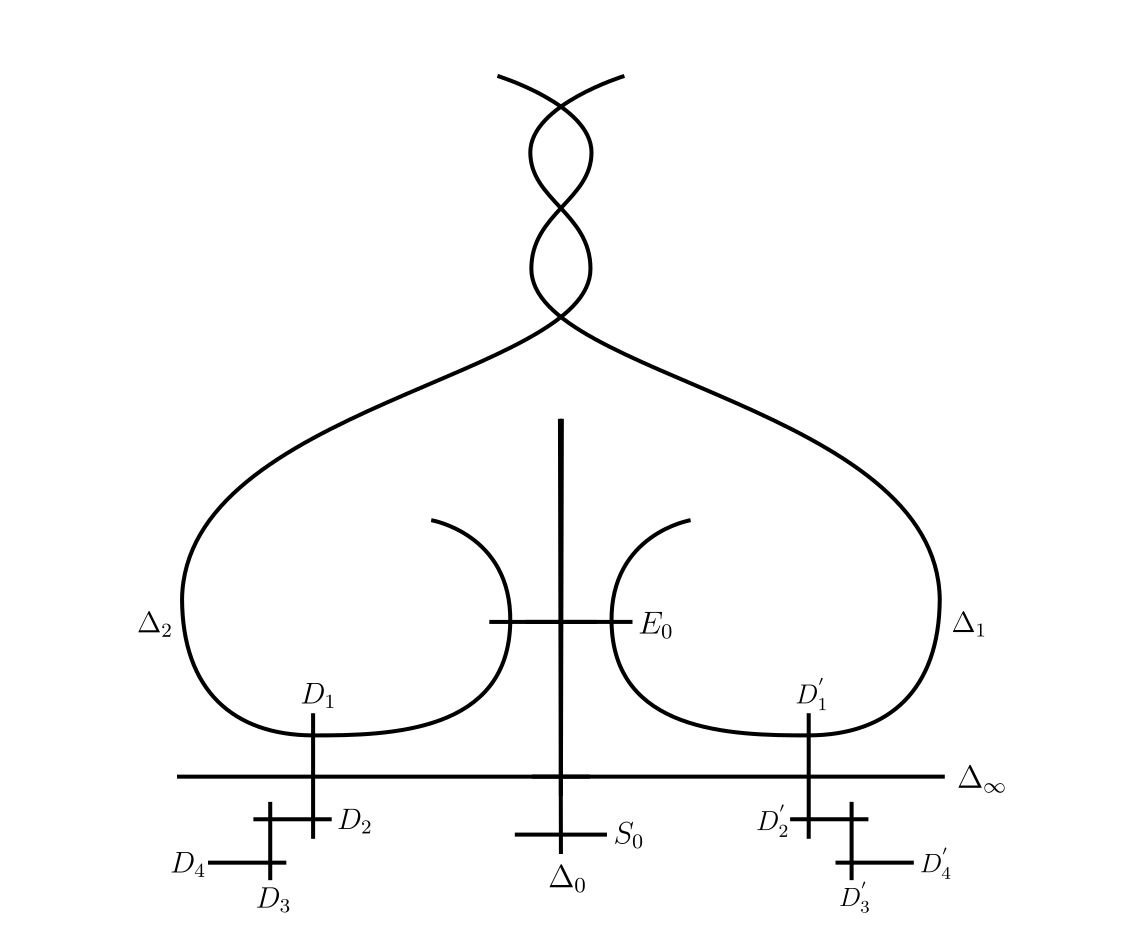}
    \caption{Schematic picture of the moduli space of the model $\P_{12,8,1,1,1,1}[24]$ after resolving singularities.}
    \label{fig:ModuliSpace}
\end{figure}
Further singular loci arise along $\Delta_0\equiv x=0$ and $\Delta_\infty\equiv z=0$. Some of these loci cross at points of tangency such as the four-fold tangency between $\Delta_2$ and $\Delta_\infty$ at \smash{$[x:y:z]=[1:0:0]$}. Resolving these by blowups until the proper transforms of these divisors and all the exceptional divisors are normal crossing results in the schematic picture in \Cref{fig:ModuliSpace}. The local coordinates around the different intersection points are given in \Cref{table:ModuliSpaceCoord}. They are expressed in terms of the Batyrev coordinates
\begin{align}
    z_1 &= \frac{\phi}{\psi^6}\,,\quad z_2 =\frac{1}{\phi^4}\,,
\end{align}
with $z_1$ corresponding to the elliptic fibre.
\begin{table}
	\centering
	\begin{tabular}{|c|c|}
		\hline
		Intersection & Coordinates   \\  
		\hline
		$D_1\cap \Delta_\infty$  & $(z_1,z_2)$  \\ 
     	$D_1\cap D_2$  & $(z_1z_2,z_2^{-1})$  \\ 
		$D_2\cap D_3$  & $(z_1^2z_2,z_1^{-1}z_2^{-1})$  \\ 
		$D_3\cap D_4$  & $(z_1^3z_2,z_1^{-2}z_2^{-1})$  \\ 
	    $D_1\cap \Delta_2$  & $(z_1,\Delta_2)$  \\ 
		$D_1^{'}\cap \Delta_\infty$  & $(\Delta_3,z_2\Delta_3^{-4})$  \\ 
     	$D_1^{'}\cap D_2^{'}$  & $(z_2\Delta_3^{-3},z_2^{-1}\Delta_3^4)$  \\ 
		$D_2^{'}\cap D_3^{'}$  & $(z_2\Delta_3^{-2},z_2^{-1}\Delta_3^3)$  \\ 
		$D_3^{'}\cap D_4^{'}$  & $(z_2\Delta_3^{-1},z_2^{-1}\Delta_3^2)$  \\ 
	    $D_1^{'}\cap \Delta_1$  & $(\Delta_3,\Delta_1\Delta_3^{-4})$  \\ 
	   	$\Delta_0\cap S_0$  & $(z_1^{-1},z_2^{-1})$  \\ 
	   	$E_0\cap \Delta_2$  & $(z_1^{-1},z_1\Delta_2)$  \\ 
	   	$E_0\cap \Delta_0$  & $(\Delta_2,z_1^{-1}\Delta_2^{-1})$  \\ 
	   	$E_0\cap \Delta_1$  & $(z_1^{-1},z_1^{-3}\Delta_1)$  \\ 
	   	$\Delta_0\cap\Delta_\infty$  & $(z_1^{-1},z_2)$  \\ 
	   	$\Delta_1\cap \Delta_2$  & $(1-\alpha_{1,2,3}z_1,\Delta_2)$  \\ 
		\hline
	\end{tabular}
\caption{Coordinates around the intersections in \cref{fig:ModuliSpace}. In the last entry $\alpha_1\equiv2^53^3,\,\alpha_2\equiv2^43^3(1+\ii),\,\alpha_3\equiv2^43^3(1-\ii)$.}
	\label{table:ModuliSpaceCoord}
\end{table}
In this case (as is the case for the model $\P_{9,6,1,1,1}[18]$) the symmetry along the vertical axis stems from a redundancy of the moduli space. There is a symmetry on the moduli space acting as
\begin{align}
(\psi,\phi)\overset{\mathcal{I}}{\longleftrightarrow} (\ii\psi,\phi-\psi^6/(2^43^3))
\end{align}
following from the fact that there are residual transformations that leave the form of \eqref{eq:P1281111} invariant once the above identification is made. The true moduli space is then the quotient by this identification. We note that the two conifold loci $\Delta_1,\Delta_2$ are interchanged by $\mathcal{I}$.

In \cref{sec:appD} we construct an integral basis $\vec{\Pi}$. We analytically continued this integral basis from the region around the MUM point to the other intersections of the discriminant locus in \Cref{fig:ModuliSpace} and computed the monodromy matrices around all divisors. Due to the involution symmetry $\mathcal{I}$, monodromies around divisors that are exchanged when reflecting around $\Delta_0$ in \Cref{fig:ModuliSpace} are related by conjugation. For example, we have 
\begin{align}
\frak{M}_{\Delta_2}=\frak{A}^6 \frak{M}_{\Delta_1} \frak{A}^{-6}
\end{align}
where $A$ is an element of order 24, that can be written as
\begin{align}
    \frak{A}=\frak{M}_{\Delta_1}\frak{M}_{\Delta_2}\frak{M}_{\Delta_\infty}\,.
\end{align}
Similarly, the vanishing cycles $g_1$ and $g_2$ of the two conifold loci are related by $\mathcal{I}$. The cycle $g_1$ whose corresponding period $g_1^{T}\Sigma \vec{\Pi}$ vanishes at $\Delta_1$ is
\begin{align}
    g_1^{T}\Sigma=(1,0,0,0,0,0,0,1)\,.
\end{align}
The vanishing cycle $g_2$ for $\Delta_2$ is then related via

\begin{align}
    g_2^{T}\Sigma=g_1^{T}\Sigma\, \frak{A}^6=(0, -33, -9, -2, 2, -4, 1, 0)\,.
\end{align}
That $g_2$ is indeed a vanishing cycle also follows from the explicit expression of the transition matrix $T$ for the point $z_1=\Delta_2=0$ given in \cref{sec:appD}:
\begin{align}
g_2^T\Sigma T=\bigg(0, 0, 0, 0, 0, 0, 0, \frac{i\sqrt{2}}{6\pi^2}\bigg)\,.
\end{align}
Hence, when aligning the $G_4$-flux with $g_2$ the superpotential is just a multiple of the last element of the Frobenius basis:
\begin{align}\label{eq:1281111W}
W= \frac{i\sqrt{2}}{6\pi^2}\Delta_2^{3/2}\left(1 + \frac{7 \Delta_2}{8} + 6 z_1 \Delta_2 + \frac{677 \Delta_2^2}{896}+\mathcal{O}(\Delta_2)\right)
\end{align}
Since $\vec{\Pi}^\dagger\cdot \Sigma \cdot \vec{\Pi}$ and its derivatives are finite for $\Delta_2=0$ except at $z_1=0$ the F-term equations reduce on this locus to $\partial_i W=0$, which for the superpotential in \eqref{eq:1281111W} are evidently satisfied. The tadpole is given by
\begin{align}
    N_\text{flux}=\frac{1}{2}g_2^T \Sigma  g_2=1\,,
\end{align}
which is of course well below the allowed upper bound of $\chi(Y)/24=972$.

Via the mirror map $z_1\rightarrow 0$ corresponds to $\Im t_E \sim 1/g_s \rightarrow\infty$ where $t_E$ is the complexified Kähler parameter of the fibre and $g_s$ the string coupling. We are thus led to consider vacua close to the locus $z_1=0$ in the moduli space. The above considerations show that there is a continuous family of vacua along $\Delta_2=0$ for all $z_1\neq 0$ and hence with arbitrarily weak coupling.

We further want to mention that there is an $\text{SL}(2,\mathbb{Z})$ action on $t_E$ analogous to the one observed in the model $\P_{9,6,1,1,1}[18]$ \cite{Candelas:1994hw}. The $\text{SL}(2,\mathbb{Z})$ action is induced by monodromies as follows:
\begin{align}
    \frak{M}_{D_1}:\quad &t_E=X^1/X^0\rightarrow t_E+1\,,\\ 
    \frak{M}_{\Delta_0}:\quad &t_E\rightarrow -\frac{1}{t_E+1}\,.
\end{align}
The observations in \cite{Candelas:1994hw} regarding the $q$-expansions carry over to this model as well. In fact the mirror map $z_1(\underline{q})$ at $q_2=0$ agrees with that of the model $\P_{9,6,1,1,1}[18]$ and in particular may be written in terms of the $j$-invariant as follows
\begin{align}
    \frac{1}{z_1(\underline{q})}\bigg\vert_{q_2=0}=864j(t_E)\left\{1+\sqrt{1-\frac{1}{j(t_E)}}\right\}=\frac{1}{q}+312 + 10260 q - 901120 q^2+\dots\,.
\end{align}\
Also we have that 
\begin{align}
    &C_{111}\vert_{q_2=0}=12+4E_4(q_1)\,,
    &C_{112}\vert_{q_2=0}\equiv 4\,,
\end{align}
where $C_{ij\alpha}\equiv C^{(112)}_{ij\alpha}$ are the triple couplings
\begin{align}
    &C_{ij\alpha}\equiv \partial_{t_i}\partial_{t_j}\frac{\tilde{H}_\alpha}{X^0}=\int_{\hat{Y}} J_i \wedge J_j\wedge b_\alpha+\text{instanton corrections}\,,
\end{align}
with $b_\alpha=J_1J_2,J_2^2$ the chosen basis for $H^{2,2}_{\text{prim}}(\hat{Y},\Z)$ and 
 $\tilde{H}_1=H_2, \tilde{H}_2=H_1$ the corresponding double-logarithmic periods.
\par We also searched for supersymmetric vacua around the other intersection points of the discriminant locus. The only further supersymmetric vacua that we find are along $\Delta_1$ when turning on the flux corresponding to the vanishing cycle $g_1$, which is to be expected due to the symmetry $\mathcal{I}$ and along the exceptional divisor $E_0$, which inherits the vanishing cycles from the two conifold loci.

\subsection{Elliptic fibration over \texorpdfstring{$\P_{3,1,1,1}$}{P3}}
\label{sec:X36}

In this section we consider the three-parameter Calabi--Yau four-fold given by the vanishing locus of 
\begin{equation}
    \label{eq:P18123111}
    P_{\mathcal{X}_{36}/\hat{H}}(\vec{x},\psi,\phi,\xi)=x_1^2+x_2^3+x_3^{12}+x_4^{36}+x_5^{36}+x_6^{36}+\psi\prod_{i=1}^6x_i+\phi\prod_{i=3}^6x_i^6+\xi\prod_{i=4}^6 x_i^{12}
\end{equation}
inside the space $\P_{18,12,3,1,1,1}$. The polytope for its toric description together with the generators of the Mori cone is given in \Cref{tab:X36}. It describes an elliptic fibration over the $\mathbb{P}^1$-bundle $\mathbb{P}(\mathcal{O}_{\mathbb{P}^2}\oplus \mathcal{O}(3)_{\mathbb{P}^2})\cong\mathbb{P}_{3,1,1,1}$. The Hodge numbers are
\begin{align}
    h^{1,1}=3\,, \quad h^{2,1}=1\,, \quad h^{2,2}=17486\,, \quad h^{3,1}=4358\,,
\end{align}
with $ h_{\text{prim}}^{2,2}=4$. Integrating the Chern classes yield the invariants
\begin{gather}
\begin{gathered}
c_2\cdot J_1^2=24\,,\; c_2\cdot J_1 J_2 =72\,, \; c_2\cdot J_1 J_3 =138\,,  \; c_2\cdot J_2^2 =216\,, \; c_2\cdot J_2 J_3 =408\,, \; c_2\cdot J_3^2 = 816\,,\\
c_3\cdot J_1=-720\,,\; c_3\cdot J_2=-2160\,,\; c_3\cdot J_3=-4338\,,\\
\chi=26208\,.
\end{gathered}
\end{gather}

\begin{table}[H]
\renewcommand{\arraystretch}{1.2}
\centering
\begin{tabular}{| c | c c c c c|c c c|}
    \hline \multicolumn{6}{|c |}{points} & \multicolumn{3}{| c|}{$l$-vectors} \\ \hline \hline
    (1 & 0&0&0&0&0)& 0 & 0 & $-$6 \\ \hline
    (1 &1& 0& 0& 0 & 0)& 0 & 0 & 3 \\
    (1 &0& 1& 0& 0& 0)& 0 & 0 & 2 \\
    (1 &0& 0& 1& 0& 0)& 0 & 1 & 0 \\
    (1 &0& 0& 0& 1 & 0)& 1 & 0& 0 \\
    (1 &0& 0& 0& 0 & 1)& 1 & 0 & 0\\
    (1 &-18& -12& -3& -1 & -1)& 1 & 0& 0 \\
    (1 &-6& -4& -1& 0 & 0)& -3 & 1& 0 \\
    (1 &-3& -2& 0& 0 & 0)& 0 & -2& 1 \\\hline
    (1 &-2& -1& 0& 0 & 0)& - & - & - \\
    (1 &-1& -1& 0& 0 & 0)& - & - & - \\
    (1 &-1& 0& 0& 0 & 0)& - & - & - \\\hline 
\end{tabular}
\caption{Integral points and their scaling relations of polytope describing $\P_{18,12,3,1,1,1}[36]/\hat{H}$. The last three points lie inside a face of codimension one.}
\label{tab:X36}
\end{table}
\noindent The intersection ring is given by
\begin{equation}
    R = 72 J_3^4+12 J_1 J_3^3+36 J_2 J_3^3+2 J_1^2 J_3^2+18 J_2^2 J_3^2+6 J_1 J_2 J_3^2+9 J_2^3 J_3+3 J_1 J_2^2 J_3+J_1^2 J_2 J_3\,.
\end{equation}
From the Mori cone generators in \Cref{tab:X36}, we infer the Batyrev coordinates 
\begin{equation}
    z_1 = \frac{1}{\xi^3}\,,\quad
    z_2 = \frac{\xi}{\phi^2}\,,\quad 
    z_3 = \frac{\phi}{\psi^6}\,.
\end{equation}
The differential ideal is generated by
\begin{align}
\begin{split}
     {\cal L}_1^{(2)}(\uz) &=\left(1 -4z_1 \right) \theta_1^2+ \left(  z_1\left( 4
\theta_3-2 \right) -3\theta_2 \right) \theta_1-z_
1 \left( \theta_3-1 \right)\theta_3\,,\\
  {\cal L}_2^{(3)}(\uz) &=\theta_2^3- z_2\left( \theta_1-3
\theta_2 \right)\left(\theta_1-3\theta_2 -1\right) 
 \left(\theta_1-3\theta_2 -2\right),\\
 {\cal L}_3^{(2)}(\uz)  &=  \left(1-432z_3\right) \theta_3^{2}+ 2\left(\theta_1 +216\right) \theta_3-60z_3\, .
\end{split}
\end{align}
The discriminant components are
\begin{align}
\begin{aligned}
    &\Delta_1= 1728z_2^3z_1-(1 - 4z_2)^3,\\
    &\Delta_2=1 - 2592z_3 + 2799360z_3^2 - 1612431360z_3^3 - 2239488z_2z_3^2 + 522427760640z_3^4\\
    &+ 3869835264z_2z_3^3 - 90275517038592z_3^5 - 2507653251072z_2z_3^4 + 6499837226778624z_3^6\\
    &+ 722204136308736z_2z_3^5 + 1671768834048z_2^2z_3^4 - 77998046721343488z_2z_3^6\\
    &- 1444408272617472z_2^2z_3^5 + 311992186885373952z_2^2z_3^6 - 415989582513831936z_2^3z_3^6\\
    &- 11231718727873462272z_2^3z_1z_3^6\,,\\
    &\Delta_3=1+27z_1\,.
\end{aligned}
\end{align}
The first two of these are conifold discriminants. For this model we only check for supersymmetric vacua along the simpler of the two conifold loci, i.e.\ component $\Delta_1$. We do so by analytically continuing to the point $z_2=1/4,z_1=z_3=0$. At this point the divisors $\Delta_1$ and ${z_1=0}$ intersect tangentially and so we perform again blowups to reach normal crossing. The resulting local coordinates around the point $z_2=1/4,z_1=z_3=0$ are 
\begin{align}
    w_1=\frac{\Delta_1}{(1-4z_2)^3}\,,\quad w_2=1-4z_2\,,\quad w_3=z_3\,.
\end{align}
The local periods, the global basis and the transition matrix, as well as genus zero and genus one instanton numbers, are given in \cref{sec:appE}. Also here we find that the conifold supports a supersymmetric vacuum by turning on flux along the vanishing cycle. The latter is given by 
\begin{align}
    g_1^{\text{T}}\Sigma=(0, 0, 19, 37, 1, 0, -3, 0, 1, -2, 0, 0) 
\end{align}
with the superpotential 
\begin{align}\label{eq:18123111W}
W= -\frac{1}{9\pi^2}w_1^{3/2}\sqrt{w_2}\left(1-{\frac {71w_{{1}}}{90}}-{
\frac {w_{{1}}w_{{2}}}{15}}+{\frac {14617w_{{1}}^{2}}{22680}}\dots\right).
\end{align}
Again $\vec{\Pi}^\dagger\cdot \Sigma \cdot \vec{\Pi}$ and its derivatives remain finite for $\Delta_1=0$ except at the intersections with $z_2=0$ and $z_3=0$, so that apart from these intersections (and potentially intersections with other divisors) the F-term equations are fulfilled along $\Delta_1$. The tadpole is again
\begin{align}
    N_\text{flux}=\frac{1}{2}g_1^T \Sigma  g_1=1< \chi(Y)/24=1092\,.
\end{align}
\par The same analysis can be performed analogously for the elliptically fibred four-folds over the bases $\mathbb{P}(\mathcal{O}_{\mathbb{P}^2}\oplus \mathcal{O}(1)_{\mathbb{P}^2})$ and $\mathbb{P}(\mathcal{O}_{\mathbb{P}^2}\oplus \mathcal{O}(2)_{\mathbb{P}^2})$, the only major difference being that the discriminant $\Delta_1$ in those cases contains terms of order $\mathcal{O}(z_2^2)$. This complicates the analytic continuation insofar that the transformed Picard--Fuchs operators in the local blow-up coordinates contain square roots. However, the differential equations can still be solved order by order when expanding the square roots and one finds the same structure for the local Frobenius basis. In particular there is again a vanishing cycle corresponding to the local solution starting as $w_1^{3/2}w_2^{1/2}\left(1+\dots\right)$.

\section{Deconstruction of one-parameter four-folds and their flux vacua}
\label{sec:AntisymProducts}
Calabi--Yau operators in different dimensions are related in many ways.
The simplest and most well-known geometric situation is when the Calabi--Yau manifold has a fibration structure over a Fano base.
In particular mirror geometries of K3-fibred Calabi--Yau three-folds with rank $r$, Noether--Lefschetz loci over $\mathbb{P}^1$---as studied first in the context of heterotic/type II duality with $N=2$ 4d supersymmetry~\cite{Kachru:1995wm,Klemm:1995tj}---have the Picard--Fuchs system of the mirror of the K3 with Picard rank $r$ in the limit of the large $\mathbb{P}^1$ limit.
One-parameter Calabi--Yau operators ${\cal L}^{(3)}(z)={\rm Sym}^2\left({\cal L}^{(2)}\right)(z)$ 
describing the $(1,1,1)$ Hodge structure of K3s are always symmetric squares of second order Calabi--Yau operators.
Hadamard products of two second order operators ${\cal L}_1^{(2)}(z)$ and ${\cal L}_2^{(2)}(\phi(z))$ can yield Calabi--Yau three-fold operators of a double elliptically fibred Calabi--Yau three-fold over $\mathbb{P}^1$. 
Further complex parameters contained in $\phi(z)$ and colliding singularities of the universal curves have to be canonical resolvable. 
Suitable choices lead to one-parameter Calabi--Yau three-fold operators ${\cal L}^{(4)}$. 
Most known attractor points arise in Calabi--Yau operators constructed from Hadamard products of elliptic curves as a consequence of an involution symmetry. Following \cite{Elmi2020}, the argument goes as follows. When the elliptic curves are subject to involution symmetries, the resulting Hadamard product also has an involution symmetry. At the fixed points of this symmetry the corresponding automorphism of the underlying variety induces an automorphism on the middle cohomology. This implies a rank two attractor point if the decomposition of the middle cohomology into positive and negative eigenvalues of this automorphism is of Hodge types $(3,0)+ (0,3)$ and $(2,1)+(1,2)$. A list of attractor points can be found in \cite{Bonisch:2022slo}.

In \cite{Almkvist:2004kj}, it was shown that as a consequence of the property \eqref{eq:1111Q} the anti-symmetric square of a  Calabi--Yau three-fold operator yields a Calabi--Yau four-fold operator $\wedge^2 L^{(4)}={\cal L}^{(5)}$. In \cite{2006math.....12215A}, the general relations among the coefficients of the fourth- and fifth-order operators were written. In \cite{Gu:2023mgf}, a formula was given for the fifth-order operators that correspond to the 14 hypergeometric three-fold families. 

Rationality of the basis of periods on the three-fold, i.e.\ the property of having rational monodromy representations, implies that their minors are rational linear combinations of a rational basis for the solutions of the fifth-order operator. As we will show below, this implies that the four-fold flux conditions can be satisfied over a point where the three-fold has an attractor point and thus a type IIB flux vacuum. In particular, the orbifold IIB flux vacua discussed in \cref{sec:CYcompSUSY} can be lifted to four-fold flux vacua. The opposite does not hold: we give examples of four-fold vacua that do not correspond to an attractor point on the three-fold but to a splitting of cohomology over quadratic field extensions. 

We denote the minors of the Wronskian of solutions $\vec{\varpi}$ to a fourth-order Calabi--Yau three-fold operator by
\begin{equation}
    W_{\vec{\varpi}}^{i,j}=2\pi \ii \det{\begin{pmatrix}\varpi_i & \varpi_j\\ \theta \varpi_i & \theta \varpi_j\end{pmatrix}}.
\end{equation}
As already mentioned, these are solutions to a Calabi--Yau operator of order five. We constructed rational bases for the four-fold operators corresponding to the 14 hypergeometric one-parameter families of Calabi--Yau three-folds with periods $\vec{\Pi}^{(3)}$. They can either be obtained directly as the anti-symmetric product of the three-fold periods or by demanding Griffiths transversality together with the leading order behaviour of the highest logarithmic period given by
\begin{equation}
    F_0 = \frac{t^4}{24}+\frac{\tilde{c}_3\, t\, \zeta_3}{2(2\pi\ii)^3}\,,
\end{equation}
where the $i$-th Chern class of the three-fold is given by $c_i=\tilde{c}_i\, H^i$ with $H$ the hyperplane class and $\tilde{c}_i\in\Z$. We normalised the double-logarithmic period such that the intersection form is $\Sigma_{ij}=(-1)^{i+1}\delta_{i,6-j}$. Writing $\vec{\Pi}^{(4)}$ for these periods in ascending logarithmic order, we find the rational map
\begin{equation}\renewcommand{\arraystretch}{1.3}
    \vec{\Pi}^{(4)} = \begin{pmatrix}
    1 & 0 & 0 & 0 & 0\\
    s & -h & 0 & 0 & 0\\
    \frac{\tilde{c}_2}{24} & 0 & h & 0 & 0\\
    0 & 0 & 0 & 2 h & 0\\
    -\frac{1}{2}\left(\frac{\tilde{c}_2}{24}\right)^2 & 0 & -\frac{h\, \tilde{c}_2}{24} & s^2 & -2h^2
    \end{pmatrix}\,
    \begin{pmatrix}
    W^{1,2}_{\vec{\Pi}^{(3)}}\\
    W^{1,3}_{\vec{\Pi}^{(3)}}\\
    W^{2,3}_{\vec{\Pi}^{(3)}}\\
    W^{2,4}_{\vec{\Pi}^{(3)}}\\
    W^{3,4}_{\vec{\Pi}^{(3)}}
    \end{pmatrix},
\end{equation}
where $h=\left(\int_{\hat{X}} H^3\right)^{-1}\in\Q$, $s=h/2$ if $\int_{\hat{X}} H^3$ is odd and $s=0$ if it is even.
Note that Griffiths transversality on the three-fold, more precisely $\vec{\Pi}^{(3)\,T}\Sigma\partial_z \vec{\Pi}^{(3)}=0$, with intersection form as in \eqref{eq:sigmaCY3} implies 
\begin{equation}\label{eq:relminor}
    0=W^{1,4}_{\vec{\Pi}^{(3)}}+W^{2,3}_{\vec{\Pi}^{(3)}}\,.
\end{equation}
\par
With the knowledge that a rational basis for the periods of the four-fold is given by rational linear combinations of minors of the integral period basis of the three-fold, we can deduce that attractor points of Calabi--Yau three-folds imply a flux vacuum on the corresponding four-fold. Recall that the vacuum conditions for type IIB are given by
\begin{gather}
    W=F+\tau H = 0 \quad\text{and}\quad \partial W=\partial F+\tau \partial H = 0\,,\label{eq:IIBred1}\\
    F=H=0\,,\label{eq:IIBred2}
\end{gather}
for some $\tau\in\C$. This redundant phrasing will make the equivalence to M-theory vacua more apparent: \eqref{eq:IIBred1} implies that the minor
\begin{equation}\label{eq:fluxminor}
        G = 2\pi \ii\det{\begin{pmatrix}F & H\\ \theta F & \theta H\end{pmatrix}}.
    \end{equation}
vanishes, due to the linear dependence of the columns. Since the determinant is linear in both columns, $G$ is also a rational linear combination of the periods $\vec{\Pi}^{(4)}$. The second vacuum condition $\partial G=0$ follows from \eqref{eq:IIBred2} and 
\begin{equation}
    \theta G =2\pi \ii \left(F \theta^2 H - H \theta^2 F\right) =0\,.
\end{equation}
We verify that the fifth-order operators belonging to the models $\mathcal{X}_{3,3}$, $\mathcal{X}_{4,6}$, $\mathcal{X}_6$ and $\mathcal{X}_{3,4}$ have rational bases of periods that satisfy the M-theory flux vacuum conditions at points where their three-fold models have attractor points, i.e.\ at $z_{3,3}=-1/2^33^6$,  $z_{4,6}\in\{-1/2^43^3,\infty\}$, $z_6=\infty$ and $z_{3,4}=\infty$.
\par
The orbifold points of the models $\mathcal{X}_5$, $\mathcal{X}_8$, $\mathcal{X}_{10}$ and $\mathcal{X}_{2,12}$ are not attractor points in the sense that their fluxes take values in quadratic field extensions. Nevertheless, we find that their four-fold equivalents have flux vacua at these points over the rationals. The flux lattices of the three-folds are spanned by elements of the form
\begin{align}
    \label{eq:fluxes}
\begin{split}
    f &= \left\{a,0,-b+\sqrt{c},d\right\},\\
    h &= \left\{0,a,0,-b-\sqrt{c}\right\},
\end{split}
\end{align}
for some $a,b,c,d\in\N$. It readily follows that the term multiplying $\sqrt{c}$ of the minor \eqref{eq:fluxminor} with $F=f\cdot \vec{\Pi}^{(3)}$ and $H=h\cdot \vec{\Pi}^{(3)}$ is proportional to the vanishing expression of \eqref{eq:relminor}. The same holds true for the fibres $z=\frac{17\pm12\sqrt{2}}{2^8}$ of the model $\mathcal{X}_{2,2,2,2}$. By studying  factorisations of the local zeta functions over $\mathbb{Z}[\sqrt{2}]$ it was found in \cite{handle:20.500.11811/11048} that these points also feature a splitting of the cohomology over $\mathbb{Q}(\sqrt{2})$. The fluxes again take the form in \cref{eq:fluxes} and the corresponding four-fold points have flux vacua over the rationals.
\par
One might also start from a one-parameter family of Calabi--Yau four-folds and search for an associated fourth-order operator. 
Consider, for example, an integral basis $\vec{\Pi}^{(4)}$ of the four-fold  $\mathcal{Y}_{6}$ given by the mirror of the degree six hypersurface in $\P^5$. The Picard--Fuchs operator is given by
\begin{equation}
    \mathcal{L}^{(5)}_{\mathcal{Y}_{6}} = \theta^5 -6 z \prod_{i=1}^5(6\theta +i)
\end{equation}
and has Riemann symbol
\begin{equation}
    \renewcommand{\arraystretch}{1.3}
    \mathcal{P}_{\mathcal{L}^{(5)}_{\mathcal{Y}_{6}}}\left\{\begin{array}{ccc}  0 & 1/2^63^6 & \infty\\\hline
     0 & 0 & \frac{1}{6}\\
     0 & 1 & \frac{1}{3}\\
     0 & \frac{3}{2} & \frac{1}{2}\\
     0 & 2 & \frac{2}{3}\\
     0 & 3 & \frac{5}{6}
   \end{array},\ z\right\}.
\end{equation}
One can reconstruct functions whose minors in the above sense give linear combinations of $\vec{\Pi}^{(4)}$ and are at the same time solutions to a Calabi--Yau operator of degree four. Up to a Kähler gauge\footnote{The Kähler gauge in which these pullback operators are listed in the AESZ database corresponds to the Yifang--Yang pullback~\cite{2006math.....12215A}, which reduces the degree of the operator.}, this operator is given by AESZ 2.45
\begin{align}
\begin{split}
    \mathcal{L}^{(4)}_{2.45}&= \theta^4+ 2^23^2z (217-18 \theta  (3 \theta -1) (8 \theta  (12 \theta +7)+23))\\
    & \quad +2^4 3^8z^2 (144 \theta  (48 \theta  (9 \theta  (2 \theta +1)+2)-7)+10633)\\
    & \quad -2^{12}3^{14}z^3 (36 \theta  (6 \theta  (24 \theta  (4 \theta +3)+19)+43)-301)\\
    & \quad +2^{22}3^{22}z^4 \theta  (2 \theta +1) (3 \theta +1) (6 \theta +1)
\end{split}
\end{align}
with Riemann symbol
\begin{equation}\renewcommand{\arraystretch}{1.3}
    \mathcal{P}_{\mathcal{L}^{(4)}_{2.45}}\left\{\begin{array}{ccc}  0 & 1/2^63^6 & \infty\\\hline
     0 & \frac{1}{4} & 0\\
     0 & \frac{3}{4} & \frac{1}{6}\\
     0 & \frac{7}{4} & \frac{1}{3}\\
     0 & \frac{9}{4} & \frac{1}{2}\\
   \end{array},\ z\right\}.
\end{equation}
This operator is defined on a branched double cover of $\P^1$ where the branch points are at the singularities $1/2^63^6$ and infinity. This is because there does not exist a rational basis for the periods when considering it over $\P^1$. Only when accounting for the branch cut do the closed paths give rise to rational monodromy representations. We note that one may alternatively use a combined algebraic coordinate and Kähler gauge transformation so as to obtain an equivalent description of the operator AESZ 2.45 with rational monodromies around the singularities. By redefining the periods according to 
\begin{align}
    \Pi(z)\mapsto \frac{1}{\sqrt{1-2^63^6z}}\Pi\left(2z\left(1-2^53^6z\right)\right)
\end{align}
one obtains an operator $\tilde{\mathcal{L}}^{(4)}_{2.45}$ with Riemann symbol
\begin{equation}\renewcommand{\arraystretch}{1.3}
    \mathcal{P}_{\tilde{\mathcal{L}}^{(4)}_{2.45}}\left\{\begin{array}{cccc}  
     0 & 1/2^63^6 & 1/2^53^6 & \infty\\\hline
     0 & 0 & 0 &\frac{1}{2}\\
     0 & 1 & 0 &\frac{5}{6}\\
     0 & 3 & 0 &\frac{7}{6}\\
     0 & 4 & 0 &\frac{3}{2}\\
   \end{array},\ z\right\}.
\end{equation}
The singularity in the interior of the moduli space has been traded for an apparent singularity and the second MUM point is the image of a symmetry acting on the moduli space under which the apparent singularity is a fixed point. In particular the instanton numbers for both MUM points are the same. The symmetry reflects the fact that in this description the two branches are seperated.

The rationality constant $c=\chi/\kappa$ can be obtained by the $p$-adic methods used in \cite{Candelas:2021tqt} to compute the local zeta functions of Calabi--Yau three-folds, which we briefly review in \cref{sec:app7}. For the present discussion, the essential point is that only for the correct value of the constant $c$ it happens that, to any finite $p$-adic order, the matrix representing the Frobenius action $F_p$ on the middle cohomology calculated using the periods converges to an algebraic function. For a lot of operators in the AESZ list the Frobenius is a rational function with the denominator being some $p$-dependent power of the discriminant. However for some operators in the AESZ list, such as the operator $\mathcal{L}^{(4)}_{2.45}$, where the monodromy takes values in field extensions $\mathbb{Q}(\sqrt{D})$, for primes $p$ such that $D$ is a not a quadratic residue mod $p$ the denominator contains square roots of the corresponding irreducible factors of the discriminant. By demanding the Frobenius to converge for finite $p$-adic precision to the expected algebraic expression we then obtain for the operator AESZ 2.45 the value $c=140$.
On the other hand, this is exactly the value for which the matrix relating the integral basis $\vec{\Pi}^{(4)}$ to the basis $W_{\vec{\varpi}_c}^{i,j}$ becomes rational. Here, $\vec{\varpi}_c$ is obtained by a prepotential of the form
\begin{equation}
    \mathcal{F}=\frac{t^3}{3!}-\frac{c\,\zeta_3}{2(2\pi\ii)^3}
\end{equation}
and the usual special geometry structure in \eqref{eq:leadingorder}. This basis has rational monodromy around the MUM point. Around the two remaining singularities, the monodromy representation takes values in $\ii\sqrt{3}\Q$. We observe that the flux vacuum of $\mathcal{Y}_{6}$ at the orbifold point gives rise to an attractor point on the three-fold. The flux vacuum at conifold point of $\mathcal{Y}_{6}$ gives a splitting of Hodge structure on the three-fold over the field extension $\Q\left[\sqrt{3}\right]$. 
\par
A similar pair is given by the four-fold hypersurface of degree ten $\mathcal{Y}_{10}$ inside the ambient space $\P_{5,1,1,1,1,1}$ with
\begin{gather}
    \mathcal{L}^{(5)}_{\mathcal{Y}_{10}} = \theta^5 -2^5 5 z (2 \theta +1) (10 \theta +1) (10 \theta +3) (10 \theta +7) (10 \theta +9),\\
    \renewcommand{\arraystretch}{1.3}
    \mathcal{P}_{\mathcal{L}^{(5)}_{\mathcal{Y}_{10}}}\left\{\begin{array}{ccc}  0 & 1/2^{10}5^5 & \infty\\\hline
     0 & 0 & \frac{1}{10}\\
     0 & 1 & \frac{3}{10}\\
     0 & \frac{3}{2} & \frac{1}{2}\\
     0 & 2 & \frac{7}{10}\\
     0 & 3 & \frac{9}{10}
   \end{array},\ z\right\}\,.
\end{gather}
and the three-fold operator AESZ 2.40
\begin{align}
\begin{split}
    \mathcal{L}^{(4)}_{2.40}&= \theta^4 -2^4 5 z (2000 \theta  (\theta  (20 \theta  (4 \theta +1)+3)-7)-7189)\\
    &\quad +2^{15}5^6 z^2(500 \theta  (\theta  (120 \theta  (2 \theta +1)+23)-5)+10079) \\
    &\quad -2^{25}5^{11}z^3 (250 \theta  (8 \theta  (10 \theta  (4 \theta +3)+7)+19)-1121)\\
    &\quad +2^{32}5^{16}z^4 (20 \theta -1) (20 \theta +3) (20 \theta +7) (20 \theta +11)
\end{split}
\end{align}
whose Riemann symbol reads
\begin{equation}\renewcommand{\arraystretch}{1.3}
    \mathcal{P}_{\mathcal{L}^{(4)}_{2.40}}\left\{\begin{array}{ccc}  0 & 1/2^{10}5^5 & \infty\\\hline
     0 & \frac{1}{4} & -\frac{1}{20}\\
     0 & \frac{3}{4} & \frac{3}{20}\\
     0 & \frac{7}{4} & \frac{7}{20}\\
     0 & \frac{9}{4} & \frac{11}{20}\\
   \end{array},\ z\right\}.
\end{equation}
In the same way as for $\mathcal{Y}_{6}$, we find $c=1740/3$. For this value of $c$, the MUM monodromy is rational while the other two have values in $\ii\Q$. Here, the two vacua of $\mathcal{Y}_{10}$ give an attractor point at the singularity in the interior and a splitting of cohomology over $\Q\left[\sqrt{5}\right]$ at infinity. 
\par
As a last example we studied the Hulek--Verrill one-parameter four-fold defined as the mirror of the CICY with configuration matrix
\begin{equation}
    \text{HV}_4=\left(\begin{tabular}{c | c c}
         $\P^1$ & 1 & 1 \\
         $\P^1$ & 1 & 1 \\
         $\P^1$ & 1 & 1 \\
         $\P^1$ & 1 & 1 \\
         $\P^1$ & 1 & 1 \\
         $\P^1$ & 1 & 1 \\
        \end{tabular}\right)
\end{equation}
restricted to the symmetric locus $z=z_i$ $\forall i\in\{1,\ldots ,6\}$. We constructed a rational basis of periods $\vec{\Pi}^{(4)}$ by setting the leading order of the quadruple-logarithmic period equal to
\begin{equation}
    F_0=\Pi^{\text{asy}}_{\mathcal{O}_{\hat{\text{HV}_4}}}\Big|_{t_i=t}-1
\end{equation}
and demanding Griffiths transversality with the intersection form
\begin{equation}
    \Sigma =\left(
\begin{array}{ccccc}
 0 & 0 & 0 & 0 & 1 \\
 0 & 0 & 0 & -1 & 0 \\
 0 & 0 & \frac{1}{45} & 0 & 0 \\
 0 & -1 & 0 & 0 & 0 \\
 1 & 0 & 0 & 0 & 0 \\
\end{array}
\right).
\end{equation}
The Picard--Fuchs operator for this one-parameter four-fold is given by
\begin{align}
\begin{split}
    \mathcal{L}^{(5)}_{\text{HV}_4} &= \theta^5 -2 z (2 \theta +1) \left(14 \theta  (\theta +1) \left(\theta ^2+\theta +1\right)+3\right) \\
    &\quad + 2^2 z^2 (\theta +1)^3 (196 \theta  (\theta +2)+255) - 2^7 3^2 z^3 (\theta +1)^2 (\theta +2)^2 (2 \theta +3)
\end{split}
\end{align}
and has Riemann symbol
\begin{equation}\renewcommand{\arraystretch}{1}
    \mathcal{P}_{\mathcal{L}^{(5)}_{\text{HV}_4}}\left\{\begin{array}{ccccc}  0 & 1/36 & 1/16 & 1/4 &  \infty\\\hline
     0 & 0 & 0 & 0 & 1\\
     0 & 1 & 1 & 1 & 1\\
     0 & \frac{3}{2} & \frac{3}{2} & \frac{3}{2} & \frac{3}{2}\\
     0 & 2 & 2 & 2 & 2\\
     0 & 3 & 3 & 3 & 2\\
   \end{array},\ z\right\}.
\end{equation}
We find flux vacua at all four singularities away from the MUM point. We note that the vacuum at infinity was found after a Kähler gauge transformation $\Omega(\psi)\mapsto \frac{\Omega(\psi)}{\psi^2\log(\psi)}$ with $\psi=1/z^2$. For the three-fold operator we obtain a Kähler-gauge-transformed version of the operator AESZ 6.1
\begin{align}
\begin{split}
    \mathcal{L}^{(4)}_{6.1}&= \theta^4 + z \left(3-28 \theta ^2 \left(8 \theta ^2+2 \theta +1\right)\right)\\
    &\quad +z^2 \left(2 \theta  (\theta  (5488 \theta  (2 \theta +1)+2999)+510)+261\right)\\
    &\quad -2^2 z^3 \left(2 \theta  (2 \theta  (19352 \theta  (4 \theta +3)+34455)+17475)+7329\right)\\
    &\quad +7 z^4\left(32 \theta  (\theta  (198992 \theta  (\theta +1)+127567)+37569)+181527\right)\\
    &\quad -2^4 z^5 \left(32 \theta  (4 \theta  (130879 \theta  (4 \theta +5)+451039)+572823)+2276415\right)\\
    &\quad +z^6 \left(512 \theta  (\theta  (17159152 \theta  (2 \theta +3)+37992921)+12646716)+699244896\right)\\
    &\quad -2^8 z^7 \left(56 \theta  (2 \theta  (1713176 \theta  (4 \theta +7)+9444349)+6499101)+34797501\right)\\
    &\quad +2^8 z^8 \left(32 \theta  (\theta  (180577544 \theta  (\theta +2)+302563655)+107267295)+350349201\right)\\
    &\quad -2^{13}3^2z^9\left(8 \theta  (2 \theta  (1544264 \theta  (4 \theta +9)+12350901)+9134059)+9376113\right)\\
    &\quad +2^{16}3^4z^{10}\left(224 \theta  (34 \theta  (276 \theta  (2 \theta +5)+1297)+17299)+592815\right)\\
    &\quad -2^{21}3^6 z^{11}(4 \theta +1) (4 \theta  (784 \theta  (2 \theta +5)+3293)+3783)\\
    &\quad +2^{24}3^8z^{12}(4 \theta +1) (4 \theta +3)^2 (4 \theta +5),
\end{split}
\end{align}
whose Riemann symbol reads
\begin{equation}\renewcommand{\arraystretch}{1.3}
    \mathcal{P}_{\mathcal{L}^{(4)}_{6.1}}\left\{\begin{array}{ccccc}  0 & 1/36 & 1/16 & 1/4 &  \infty\\\hline
     0 & \frac{1}{4} & \frac{1}{4} & \frac{1}{4} & \frac{1}{4}\\
     0 & \frac{3}{4} & \frac{3}{4} & \frac{3}{4} & \frac{3}{4}\\
     0 & \frac{7}{4} & \frac{7}{4} & \frac{7}{4} & \frac{3}{4}\\
     0 & \frac{9}{4} & \frac{9}{4} & \frac{9}{4} & \frac{5}{4}
   \end{array},\ z\right\}\,.
\end{equation}
The basis of minors of $\vec{\varpi}_c$ has a rational map to $\vec{\Pi}^{(4)}$ for $c=4/3$. At the first singularity $z=1/36$, we find a splitting of cohomology over $\Q\left[\sqrt{10}\right]$ and the monodromy representation $\frak{M}_{1/36}$ takes values in $\ii\sqrt{10}\Q$. At $z=1/16$, we find a splitting of cohomology over $\Q\left[\sqrt{15}\right]$ and $\frak{M}_{1/16}\in \ii\sqrt{15}\Q$. At $z=1/4$, we find a splitting of cohomology over $\Q\left[\sqrt{6}\right]$ and $\frak{M}_{1/4}\in \ii\sqrt{6}\Q$. At infinity, we find an attractor point and have $\frak{M}_{\infty}\in \ii\Q$. We note that the operator AESZ 6.1 in its original Kähler gauge has real monodromy representations over the same field extensions. In particular, the monodromy around infinity becomes rational for the above value of $c$.

\section{Conifold-transitions}\label{sec:conifoldtrans}
In this section, we study conifold-transitions of the model $\mathcal{X}_6^{(3)}$ introduced in \cref{sec:sextic}. The analogous discussion for the models $\mathcal{X}_8^{(3)}$ and $\mathcal{X}_{10}^{(3)}$ can be found in \cref{sec:appX8contraf,sec:appX10contraf}. After a brief review of conifold transitions, we give a description of $\mathcal{X}_6^{(3)}$ as a CICY in a product of projective spaces. While the moduli spaces are related by a rational map, the holomorphic period changes by an algebraic function. We then discuss its conifold transition to a two-parameter CICY. This second family allows again for a transition to the hypergeometric one-parameter model $\mathcal{X}_{3,2,2}$. 
\par
Locally, the fibre at a conifold point in the moduli space can be described by the quadric \cite{Candelas:1989js}
\begin{equation}\label{eq:quadricconifold}
    \sum_{i=1}^4 w_i^2=0\,.
\end{equation}
There are two ways of regularising the node. The first consists of inserting an $S^3$ via the modification
\begin{equation}
    \sum_{i=1}^4 w_i^2=\epsilon
\end{equation}
with a positive constant $\epsilon$. For the second, one performs a change of variables to bring \eqref{eq:quadricconifold} into the form
\begin{equation}
    X\,Y-U\,V=0\,,
\end{equation}
which can be smoothed by introducing a $\P^1\cong S^2$ with variables $(\lambda_1:\lambda_2)$
\begin{equation}
    \begin{pmatrix}
        X&U\\V&Y
    \end{pmatrix}
    \begin{pmatrix}
        \lambda_1\\ 
        \lambda_2
    \end{pmatrix}=0\,.
\end{equation}
Both regularisations yield smooth Calabi--Yau manifolds and moving from one to the other changes the topology and with it the Hodge numbers $h_{1,1}$ and $h_{2,1}$.
\par 
In type IIB theory, three-cycles can be wrapped by three-branes resulting in black hole hypermultiplets. If the three-cycles are chosen as the vanishing cycles near a conifold point, these hypermultiplets will become massless as one approaches the singularity. It was shown in \cite{Greene:1995hu} for a specific example how the space of vacua solutions for these black holes is parametrised by a single hypermultiplet $v^\alpha$, where $v^\alpha=0$ corresponds to the fibre with shrunken three-cycles. Moving along $v^\alpha$ in this new branch of the moduli space, the v.e.v.\ of the hypermultiplets become finite, breaking the U$(1)$-symmetries corresponding to the vector multiplets associated with the vanishing cycles in $H_{2,1}$. This branch describes a compactification on a Calabi--Yau family with an increased number of hypermultiplets/$(1,1)$-forms but less vector-multiplets/$(2,1)$-forms.
\par
Here, we will consider the inverse process in type IIA theory, which was described in \cite{Klemm:1996kv,Berglund:1996uy}. As the volume of a two-cycle parametrised by the Kähler parameter $t^i$ goes to zero and thus $q^i$ to one, the possibility of a conifold transition can be read off from the prepotential $\mathcal{F}$. If the instanton contributions remain finite as $q^i\rightarrow 1$, a transition is possible to a new Calabi--Yau family, whose Euler number is determined by the resulting constant term of $\mathcal{F}$ (cf.\ the discussion of \eqref{eq:topdataX102}). To verify the conifold transition, a linear map between the Kähler cone generators of both models must be found that renders the prepotential including the instanton corrections equal. In the simplest case considered here and in the following, the instanton numbers of the new model are obtained from that of the old by summing over the degrees belonging to $t^i$. Furthermore, the classical triple intersections must agree for $t_i=0$, the integrated second Chern class must coincide in the remaining Kähler cone generators and the Euler number must transform as mentioned above. Since the Euler number is given by $2(h_{1,1}-h_{2,1})$, this also fixes all Hodge numbers of the new model. 
\par
As mentioned above, we can describe the hypersurface quotient $\mathcal{X}_6^{(3)}$ as the mirror of a complete intersection
\begin{equation}
    \label{eq:X6CICY}
    \mathcal{X}_6^{(3)}=\left(\begin{tabular}{c | c c c}
         $\P^1$ & 0&2&0 \\
         $\P^1$ & 0&0&2 \\
         $\P^4$ & 3&1&1
        \end{tabular}\right)^{3,63}_{-120},
\end{equation}
where we added $(h^{21},h^{11})$ and the Euler number as the super- and subscript, respectively. Its Mori cone generators are given by
\begin{align}\begin{split}
    l_1 &= (0,-2,0;1,1,0,0,0,0,0,0,0)\,,\\
    l_2 &= (0,-2,0;0,0,1,1,0,0,0,0,0)\,,\\
    l_3 &= (-3,-1,-1;0,0,0,0,1,1,1,1,1)\,.\end{split}
\end{align}
Then, the intersection ring, the remaining topological data and the instanton numbers are identical to that of the hypersurface given in \eqref{eq:intringX6}, \eqref{eq:topdataX6} and \Cref{tab:instX6}. 
The holomorphic periods of the two models obey the relation
\begin{equation}\label{eq:X6relationw0}
    \varpi_{0,\text{CICY}}(z_1,z_2,z_3) =\frac{1}{\sqrt{(1-4z_1)(1-4z_2)}}\, \varpi_{0,\text{HS}}\left(z_1,z_2,\frac{z_3}{(1-4z_1)(1-4z_2)}\right) 
\end{equation}
and the discriminant of the CICY can be obtained from that of the hypersurface given in \eqref{eq:discX6} by
\begin{equation}\label{eq:discCICY}
    \Delta_{\text{CICY}}\left(z_1,z_2,z_3\right) = (1-4z_1)^2(1-4z_2)^2\Delta_{\text{HS}}\left(z_1,z_2,\frac{z_3}{(1-4z_1)(1-4z_2)}\right).
\end{equation}
The model undergoes a conifold transition as either $z_1$ or $z_2$ approaches the strong coupling divisors given by $\Delta_i=1-4z_i$, $i=1,2$. For example,
\begin{equation}
    \mathcal{X}_6^{(3)}\xrightarrow{\Delta_1\rightarrow 0} \left(\begin{tabular}{c | c c c}
         $\P^1$ & 0&0&2 \\
         $\P^5$ & 3&2&1
        \end{tabular}\right)^{2,68}_{-132}\eqqcolon \mathcal{X}_6^{(2)}.
\end{equation}
We base this claim on the structure of the instanton numbers, which obey the relations
\begin{equation}
    n_{i_2,i_3}=\sum_{i_1=0}^\infty n_{i_1,i_2,i_3}\,,
\end{equation}
where only finitely many terms contribute to the sum. We differentiate between the instanton numbers of the families by the number of subscripts in $n_{i_1,\ldots,i_k}$. We list these numbers explicitly in \Cref{tab:instX6CICY2} in \cref{sec:appX6contraf}.
The Mori cone of the model $\mathcal{X}_6^{(2)}$ is spanned by 
\begin{align}
    l_2 &= (0,0,-2;1,1,0,0,0,0,0,0)\,,\\
    l_3 &= (-3,-2,-1;0,0,1,1,1,1,1,1)\,,
\end{align}
whose dual Kähler cone generators have the triple intersections
\begin{equation}
    R=6J_2J_3^2+12J_3^3\,.
\end{equation}
Their intersection with the second Chern class are
\begin{equation}
    c_2\cdot J_2 = 24,\ c_2\cdot J_3 = 60\,. 
\end{equation}
As expected, the intersection ring and the integrated second Chern class can be obtained from that of $\mathcal{X}_6^{(3)}$ by setting $J_1=0$.
The Picard--Fuchs ideal is generated by
\begin{align}
    {\cal L}_1^{(2)}(\uz)&=\theta _2^2-z_2\left(2 \theta _2+\theta _3+1\right) \left(2 \theta _2+\theta _3+2\right),\\
    \begin{split}
    {\cal L}_2^{(3)}(\uz)&= \theta _3^2 \left(\theta _3-2 \theta _2\right)+z_2\left(4 \theta _3^3+8
    \theta _2 \theta _3^2+4 \theta _3^2\right)\\
    &\quad -6z_3\left(2 \theta _3+1\right) \left(3 \theta _3+1\right) \left(3 \theta _3+2\right).
    \end{split}
\end{align}
The Picard--Fuchs operator of the K3-fibre can be obtained by taking the limit $z_2\rightarrow 0$, in which ${\cal L}_2^{(3)}(\uz)$ becomes 
\begin{equation}
    \mathcal{L}_{\text{K3}}^{(3)}(z_3) = \theta _3^3-6 z_3\left(2 \theta _3+1\right) \left(3 \theta _3+1\right) \left(3 \theta _3+2\right).
\end{equation}
\par The two-parameter model also has a conifold transition to the hypergeometric one-parameter model $\mathcal{X}_{3,2,2}$:
\begin{equation}
    \mathcal{X}_6^{(2)} \xrightarrow{\Delta_2\rightarrow 0} \left(\begin{tabular}{c | c c c}
         $\P^6$ & 3&2&2
        \end{tabular}\right)_{-144}^{1,73}=\mathcal{X}_{3,2,2}\,,\\
\end{equation}
where now the relations between the instanton numbers read
\begin{equation}
    n_{i_3}=\sum_{i_2=0}^\infty n_{i_2,i_3}\,.
\end{equation}
The first couple of instanton numbers of $\mathcal{X}_{3,2,2}$ are given in \Cref{tab:instX6CICY1} in \cref{sec:appX6contraf}. The Mori cone of $\mathcal{X}_{3,2,2}$ is generated by
\begin{equation}
    l_3=(-3,-2,-2;1,1,1,1,1,1,1)
\end{equation}
and, as expected from the first transition, the triple intersection and the integrated second Chern class are now
\begin{equation}
    R=12J_3^3\,,\ c_2\cdot J_3=60\,.
\end{equation}
The differential ideal is generated by
\begin{equation}
    {\cal L}_1^{(4)}(\uz) = \theta _3^4-12 z_3 \left(3 \theta _3+1\right)\left(2 \theta _3+1\right)^2 \left(3 \theta _3+2\right).
\end{equation}

\section{Conclusions}

The variation of pure and mixed Hodge structures recently found applications in string
compactifications and the calculation of Feynman integrals in perturbative quantum field theories, integrable models and the post-Minkowskian approach to black hole scattering
processes. Our treatment of the Frobenius structure in section \ref{sec:CYPeriodgeometry}
and \ref{sec:periodmatrixnfolds} contains new results with both communities in mind. This 
is, of course, only a beginning of the common application of mixed Hodge structures, in 
particular, relative mixed Hodge structures. 

What is remarkable here is the different 
roles of Calabi--Yau $n$-folds with $n< 3$, $n>3$ and the very special role
of Calabi--Yau three-folds. For $n<3$, we have no Frobenius structure, yet, which involves the three-point functions, which simply occur the first time for the three-folds. On the other hand, at 
least in the one-parameter case, the differential operators for $n=1$ generate the 
one at $n=2$ by symmetric products and many of the Calabi--Yau three-fold operators 
by twisted Hadamard products~\cite{van2018calabi,2021arXiv210308651A}. For example in the 
black hole scattering, all higher dimensional operators that occurred so far can be obtained 
via the aforementioned construction from the $n=1$ Legendre curve operator. 

The Calabi--Yau three-fold case is very special, as, due to Serre duality, the virtual dimension of the 
moduli space of all degree and genera world-sheet instantons is zero and the periods
are, as a consequence of mirror symmetry, transcendental counting functions of these instantons~\cite{MR2003030}.
According to the axioms of twisted (2,2) superconformal string world-sheet theory reflected in the 
Frobenius structure all higher $(n>3)$-point functions factorise into three-point functions. Moreover, 
in many physical problems, again in the one-parameter cases,  the higher dimensional operators 
come from wedge or symmetric products of operators for $n\le 3$~\cite{Duhr:2023eld} and in the genus 
expansion of the topological string on Calabi--Yau three-folds, periods of four dimensional Calabi--Yau 
spaces appear naturally~\cite{Gu:2023mgf}. 

One conundrum in the application of periods integrals 
to Feynman graphs is that higher dimensional Calabi--Yau variations of Hodge structures appear
so frequently in higher loop calculations, where naively higher genus variations of Hodge structure
seem more natural\footnote{Even though these structures can be related~\cite{Jockers:2024uan}.}. One 
reason might be that the topological genus expansion of superstring theory in the critical 
dimensions singles out Calabi--Yau three-folds and the genus expansion of 
topological gravity coupled to Calabi--Yau three-fold matter involves the higher 
dimensional Calabi--Yau periods.

This special role of Calabi--Yau three-folds is also significant in situations 
where the mixed variations of Hodge structures is concerned.  For example, wave function applications
appear also in in the WKB approximation of the Nekrasov--Shatasvili integrable 
system~\cite{Aganagic:2011mi}, which likewise leads to iterated integrals for the 
so-called quantum period, which involve meromorphic forms in the hypercohomology
of (local) Calabi--Yau spaces. Such forms appear also in general in the IBP relations of 
Feynman graphs\footnote{We thank Claude Duhr, Christoph Nega and  Benjamin Sauer for bringing 
this fact to our attention.}. An attempt to generalise the Frobenius structure to 
the closely related open string case was made in~\cite{Lerche:2002yw}.

Using similarity of matrices of finite order over $\Q$, we showed in \cref{sec:quotientbuilding} that the splitting of periods into invariant and non-invariant parts under a symmetry happens over the rational numbers. This implies that flux vacua on the mirror locus existed on the original family with all moduli present.
In \cref{sec:FVsym}, we studied the flux vacua on the three-fold family $\mathcal{X}_6^{(3)}$ and exemplified that simultaneous stabilisations of moduli presumably require an additional symmetry between the vacua for the values of the axio-dilaton to coincide. At the same time, this reveals that mirror loci are not necessarily supersymmetric flux vacua. 

We analysed two examples of elliptically fibred four-folds and their moduli spaces in \cref{sec:FtheoryVacua}. In particular we studied the moduli space of the two-parameter model given by the mirror of $\mathbb{P}_{12,8,1,1,1,1}$ and extended the global analysis of the periods and the search for flux vacua, that was partly performed in \cite{Cota:2017aal}. No further supersymmetric vacua exist at discriminant loci apart from the one given by the flux Poincaré dual to the shrinking $S^4$ at the conifold. It is expected generally that conifold loci feature supersymmetric vacua. We gave further evidence by studying a conifold divisor of the three-parameter model $\P_{18,12,3,1,1,1}[36]$. For the latter we also computed the genus zero and one integer invariants, listed in \cref{sec:appE}. A possibility that deserves further study is the search for isolated supersymmetric vacua, i.e.\ vacua in codimension two in the interior of the moduli spaces. This could be done by searching for persistent factorisations of the local zeta function similar to how attractor points in one-parameters three-fold are found \cite{Candelas:2019llw}.

A connection between one-parameter three-fold and four-fold vacua was established in \cref{sec:AntisymProducts}. We showed that the rationality of the period bases as well as the presence of supersymmetric vacua carries over to the antisymmetric square. We checked that under this operation the attractor points of hypergeometric models lift to supersymmetric vacua on the four-fold side. Even for models with attractor points only over a quadratic field extension of the rationals, on the four-fold we still find a vacuum with rational flux. Conversely, starting with four-fold operators we showed that their F-theory vacua correspond to attractor points, in general over some quadratic field extension, on the respective pullback three-fold. An interesting question is whether the antisymmetric product construction can be generalised to multi-parameter cases. If possible, this might allow to find geometric realisations for some Calabi--Yau type operators which so far lack one. For example, one may speculate that by pulling back the Hulek--Verrill four-fold and studying how the one-parameter symmetric slice maps to the operator AESZ 6.1 on the three-fold side one could find a geometric realisation of the latter as a slice in a multi-parameter three-fold. 

Finally, we identified three pairs of hypergeometric one-parameter families with connected moduli spaces in \cref{sec:conifoldtrans}. The three-parameter families $\mathcal{X}^{(3)}_6$, $\mathcal{X}^{(3)}_8$ and $\mathcal{X}^{(3)}_{10}$ naturally have their corresponding one-parameter family in their moduli space. We found that these models allow for conifold transitions to the models $\mathcal{X}_{3,2,2}$, $\mathcal{X}_{4,2}$ and $\mathcal{X}_{6,2}$, respectively.

\vspace{2cm}
\section*{Acknowledgements}    
The authors thank Pyry Kuusela for clarification on the role of the symmetry interchanging the moduli in the analysis of the three-parameter models and Paul Blesse,  Pierre Lairez, Sara Maggio and Franziska Porkert for useful discussions. We further thank Kilian Bönisch for allowing us to use his \texttt{Pari/GP} implementation of the deformation method for computations of local zeta functions. AK thanks Claude Duhr, Christoph Nega, Benjamin Sauer and Lorenzo Tancredi for discussions on the $\epsilon$-factorised form of the IBP relation and Nicolai Reshetikhin for 
comments on wave form integrals in integrable systems. AK also thanks for support by the Grant 508889767/FOR5372/1 “Modern Foundations of Scattering Amplitudes.”     
\newpage
\appendix
\section{Period geometry identities}
\label{app:periodgeometry}
\subsection{Gauss--Manin connection}
In this appendix, we elaborate on the structure of the Gauss--Manin connection. In \cref{sec:periodmatrixnfolds}, we commented on the natural basis to express the Gauss--Manin connection in and discussed explicitly the cases $n=1,\dots,5$. We find it instructive to give a derivation for the four-fold case in \cref{sec:appCY4} using the Griffiths transversality relations. As mentioned in the main text, the results for $n=4$ hold also in those cases in which the middle cohomology is not horizontal, that is, not spanned by derivatives of the holomorphic $(4,0)$-form. In \cref{sec:appCY6}, we further write down the basis for the case $n=6$. 
\label{sec:gaussmanin}
\subsubsection{Calabi--Yau four-folds}\label{sec:appCY4}
As in \eqref{eq:Omega0}, we expand the holomorphic $(n,0)$-form as
\begin{equation}
    \Omega_0=\alpha_0 + t^i \alpha_i + H^{\alpha} \gamma_{\alpha} + F_i\beta^i + F_0 \beta^0,
\end{equation}
where $H^{\alpha}$, $F_i$ and $F_0$ are functions of $\vec{t}$. We denote the inner product $\eta^{(2,2)}$ on $H^{\text{hor}}_{2,2}(Y,\Z)$ as ``$\cdot$'' and use $\partial_i\equiv \partial_{t_i}$. Griffiths' transversality in \cref{Griffiths} implies the quadratic relations of the period vector
\begin{align}
    0&=2 X\cdot F + H\cdot H\,,\\
    0&=\partial_i F_0 - t^j \partial_i F_j + H\cdot \partial_i H - F_i\,,\\
    0&=\partial_i\partial_j F_0 - t^k\partial_i\partial_j F_k + H\cdot \partial_i\partial_j H\,,\label{eq:GT2}\\
    0&=\partial_i\partial_j\partial_k F_0 - t^l\partial_i\partial_j\partial_k F_l + H\cdot \partial_i\partial_j\partial_k H\,,\label{eq:GT3}\\
    C_{ijkl}&= \partial_i\partial_j\partial_k\partial_l F_0-t^m \partial_i\partial_j\partial_k\partial_l F_m + H\cdot \partial_i\partial_j\partial_k\partial_l H\,.\label{eq:GT4}
\end{align}
Differentiating \eqref{eq:GT2} and \eqref{eq:GT3} gives
\begin{align}
\begin{split}\label{eq:dGT1}
    0&=\partial_i\partial_j\partial_k F_0 -t^l\partial_i\partial_j\partial_k F_l - \partial_i\partial_jF_k + \partial_k H \cdot \partial_i\partial_j H + H \cdot \partial_i\partial_j\partial_kH\\
    &\overset{(\ref{eq:GT3})}{=}-\partial_i\partial_jF_k + \partial_kH\cdot \partial_i\partial_j H\,,\end{split}\\
\begin{split}\label{eq:dGT2}
    0&=\partial_i\partial_j\partial_k\partial_l F_0 - t^m\partial_i\partial_j\partial_k\partial_l F_m - \partial_i\partial_j\partial_k F_l+\partial_l H\cdot \partial_i\partial_j\partial_k H + H\cdot \partial_i\partial_j\partial_k\partial_l H\\
    &\overset{(\ref{eq:GT4})}{=} C_{ijkl} - \partial_i\partial_j\partial_kF_l + \partial_l H\cdot \partial_i\partial_j\partial_k H\,.\end{split}
\end{align}
Taking the derivative of \eqref{eq:dGT1} yields
\begin{equation}
    0 = -\partial_i\partial_j\partial_k F_l +\partial_k\partial_l H \cdot \partial_i\partial_j H + \partial_l H \cdot \partial_i\partial_j\partial_k H\,. 
\end{equation}
Here, we renamed the indices $k\leftrightarrow l$ to make the identification with \eqref{eq:dGT2} apparent, implying
\begin{equation}
    C_{ijkl} = \partial_i\partial_j H\cdot \partial_k\partial_l H = \eta^{(2,2)}_{\alpha\beta} C_{ij}^{\ \ \alpha} C_{kl}^{\ \ \beta},
\end{equation}
where we identified the expressions $\partial_i\partial_j H^\alpha$ as the three-point couplings $C_{ij}^{\ \  \alpha}$. For convenience, we repeat the remaining basis elements given in \cref{sec:CY4}
\begin{align*}
    \chi_i &= \partial_i \Omega_0\,,\tag{\ref{eq:chii}}\\
    h_{\alpha} &= \gamma_{\alpha} + \partial_i H_{\alpha} \beta^i - (H_{\alpha}-t^i \partial_i H_{\alpha})\beta^0,\tag{\ref{eq:halpha}}\\
    \chi^i &= \beta^i +  t^i\beta^0,\tag{\ref{eq:chihi}}\\
    \Omega^0 &= \beta^0.\tag{\ref{eq:Omegah0}}
\end{align*}
We compute
\begin{align}
\begin{split}
    \partial_{i}\chi_j &= \partial_{i}\partial_j H^{\beta}\gamma_{\beta} + \partial_{i}\partial_j F_k \beta^k + \partial_i\partial_j F_0 \beta^0\\
    &= C_{ij}^{\ \ \beta}\gamma_{\beta} +C_{ij}^{\ \ \beta}\partial_k H_\beta\beta^k  -C_{ij}^{\ \ \beta}(H_\beta -t^i \partial_i H_\beta) \beta^0,
    \end{split}
\end{align}
where we used \eqref{eq:dGT1} and, for the $\beta^0$ component, additionally \eqref{eq:GT2}.
The action of the derivative acting on the remaining generators reads
\begin{align}
    \partial_i h_\alpha &= C_{ik\alpha} \beta^k + t^k C_{ik\alpha} \beta^0,\\
    \partial_i \chi^j &= \delta_i^j \beta^0,\\
    \partial_i \Omega^0 &= 0\,.
\end{align}
We can summarise the above equations in the form given already in \cref{sec:CY4}
\begin{equation}\tag{\ref{eq:GMCY4}}
    \partial_i \begin{pmatrix}\Omega_0\\ \chi_j \\ h_\alpha \\ \chi^j \\ \Omega^0
    \end{pmatrix} = \begin{pmatrix}
    0 & \delta_i^k & 0 & 0 & 0 \\
    0 & 0 & C_{ij}^{\ \ \beta} & 0 & 0\\
    0 & 0 & 0 & C_{ik\alpha} & 0\\
    0 & 0 & 0 & 0 & \delta_{i}^j\\
    0 & 0 & 0 & 0 & 0
    \end{pmatrix}
    \begin{pmatrix}\Omega_0\\ \chi_k \\ h_\beta \\ \chi^k \\ \Omega^0
    \end{pmatrix}.
\end{equation}
\subsubsection{Calabi--Yau six-folds}\label{sec:appCY6}
We denote the period basis by $\Pi=(1,t^i,H^\alpha, K_A,L_\Gamma,F_i,F_0)^{T}$ with symmetric intersection pairing  
\begin{equation}
    \Sigma = \begin{pmatrix}
     &        &        &       &  &  & 1  \\
     &        &        & & & \eta^{(1,5)} & \\
     &        &        & & \eta^{(2,4)} & & \\
     &        & &  \eta^{(3,3)} &       & \\
     & & \eta^{(4,2)}&        &       & \\
     & \eta^{(5,1)} & & & \\
   1 &        &        &       & & 
  \end{pmatrix}
\end{equation}
where $\eta^{(5,1)}_{ij}=\eta^{(1,5)}_{ij}=\delta^i_{h^{5,1}-j+1}$ and we set from now on $\eta_{(HL)}\equiv \eta^{(2,4)}=\eta^{(4,2)T}$ as well as $\eta_{(KK)}\equiv \eta^{(3,3)}=\eta^{(3,3)T}$. The three-point functions are as for the five-fold: 
\begin{align}
    &\big(C^{(114)}\big)^{\ \ \alpha}_{ij}=\partial_i\partial_j H^\alpha,\\
    &C^{(123)}_{i \Gamma A}\eta_{(HL)\alpha}^{\ \qquad \Gamma}\equiv C^{(123)}_{i \alpha A}=\partial_i(\partial_j\partial_k K_A \mathfrak{C}^{jk}_{\ \ \alpha})\,.
\end{align}
We again define the inverse three-point coupling $\mathfrak{C}^{ij}_{\ \ \alpha}$ via
\begin{align}
   \mathfrak{C}_{\ \ \alpha}^{ij}\big(C^{(114)}\big)^{\ \ \alpha}_{ij}=\delta^\beta_\alpha\, .
\end{align}
The Gauss--Manin connection reads 
\begin{equation}
    \partial_i \begin{pmatrix}\Omega_0\\ \chi_j \\ h_\alpha  \\ k^A \\ l^\Gamma \\ \chi^j \\ \Omega^0
    \end{pmatrix} = \begin{pmatrix}
    0 & \delta_i^k & 0 & 0 & 0 & 0 & 0\\
    0 & 0 & \big(C^{(114)}\big)^{\ \ \beta}_{ij} &0 & 0 & 0& 0 \\
    0 & 0 & 0 & C_{i\alpha B}^{(123)} & 0  &  0 & 0\\
    0 & 0 & 0 & 0 & \big(C^{(123)}\big)_{i\Delta B} \eta^{AB}_{(KK)} & 0 & 0\\
    0 & 0 & 0 & 0 & 0 & \big(C^{(114)}\big)^{\ \ \gamma}_{ik} \eta^{\qquad \ \Gamma}_{(HL)\gamma} & 0\\
    0 & 0 & 0 & 0 & 0 & 0 & \delta_{i}^j \\
    0 & 0 & 0 & 0 & 0 & 0 & 0\\
    \end{pmatrix}
 \begin{pmatrix}\Omega_0\\ \chi_k \\ h_\beta  \\ k^B \\ l^\Delta \\ \chi^k \\ \Omega^0
    \end{pmatrix}
\end{equation}
in the basis 
\begin{align}
    &\Omega_0=\alpha_0+t^i\alpha_i+H^\alpha \gamma_\alpha +K_A \delta^A+L_\Gamma \epsilon^\Gamma + F_i \beta^i+F_0\beta^0,\\
    &\chi_i=\partial_i \Omega_0\,,\\
    \begin{split}
    &h_\alpha=\gamma_\alpha + \partial_i\partial_j K_A \mathfrak{C}^{ij}_{\ \ \alpha}\delta^A+ \partial_i\partial_j L_
\Gamma \mathfrak{C}^{ij}_{\ \ \alpha}\epsilon^\Gamma \\
&\qquad +\left(-\partial_k L_\Gamma\eta_{(HL)\alpha}^{\qquad \ \Gamma}-\partial_k K_A \eta_{(KK)}^{AB} \partial_i\partial_j K_B\mathfrak{C}^{ij}_{\ \ \alpha} -\partial_k H^\beta \eta_{(HL)\beta}^{\qquad \ \Gamma}\partial_i \partial_j L_\Gamma \mathfrak{C}^{ij}_{\ \ \alpha}\right)\beta^k\\
&\qquad +\left(- L_\Gamma\eta_{(HL)\alpha}^{\qquad \ \Gamma}- K_A \eta_{(KK)}^{AB} \partial_i\partial_j K_B\mathfrak{C}^{ij}_{\ \ \alpha} - H^\beta \eta_{(HL)\beta}^{\qquad \ \Gamma}\partial_i \partial_j L_\Gamma \mathfrak{C}^{ij}_{\ \ \alpha}\right.\\ 
    &\left.\qquad \qquad\! +t^k\left\{\partial_k L_\Gamma\eta_{(HL)\alpha}^{\qquad \ \Gamma}+\partial_k K_A \eta_{(KK)}^{AB} \partial_i\partial_j K_B\mathfrak{C}^{ij}_{\ \ \alpha} +\partial_k H^\beta \eta_{(HL)\beta}^{\qquad \ \Gamma}\partial_i \partial_j L_\Gamma \mathfrak{C}^{ij}_{\ \ \alpha} \right\}\right)\beta^0,\end{split}\\
    \begin{split}
&k^A=\delta^A -\eta_{(KK)}^{AB} \partial_i\partial_j K_B \mathfrak{C}^{ij}_{\ \ \alpha}\big(\eta_{(HL)}^{-1}\big)_\Gamma^{\ \ \alpha}\epsilon^\Gamma  \\
&\hspace{2cm} +\left(-\partial_k K_B\eta_{(KK)}^{AB}+\partial_k H^\alpha \eta_{(KK)}^{AB} \partial_i\partial_j K_B\mathfrak{C}^{ij}_{\ \ \alpha}\right)\beta^k\\
&\hspace{2cm} +\left(-K_B\eta_{(KK)}^{AB}+ H^\alpha \eta_{(KK)}^{AB} \partial_i\partial_j K_B\mathfrak{C}^{ij}_{\ \ \alpha}\right.\\ 
    &\left.\hspace{3cm} \qquad +t^k\left\{\partial_k K_B\eta_{(KK)}^{AB}-\partial_k H^\alpha \eta_{(KK)}^{AB} \partial_i\partial_j K_B\mathfrak{C}^{ij}_{\ \ \alpha} \right\}\right)\beta^0,\end{split}\\
    &l^\Gamma =-\epsilon^\Gamma +\partial_k H^\alpha \eta_{(HL)\alpha}^{\qquad \ \Gamma} \beta^k+\left( H^\alpha\eta_{(HL)\alpha}^{\qquad \ \Gamma}-t^k\partial_k H^\alpha\eta_{(HL)\alpha}^{\qquad \ \Gamma}\right)\beta^0,\\
    &\chi^i=\beta^i-t^i\beta^0,\\
    &\Omega^0=-\beta^0.
\end{align}
\subsection{Differential equations for \texorpdfstring{$n$}\ -point couplings}
In this section, we show that, in Batyrev coordinates, the derivatives of the $n$-point couplings satisfy
\begin{equation}
    \partial_{(i_0}C_{i_1 \ldots i_n)}=\frac{2}{n+1}C_{i_0 i_1,\ldots i_n}\,,\tag{\ref{eq:delYuk}}
\end{equation}
where we extended the definition of the couplings (cf. \eqref{Griffiths}) to any index set
\be
C_{I^{(r)}}(\uz)=\int_{X} \Omega \wedge  {\underline \partial}^r_{I^{(r)}} \Omega = \uPi^T \Sigma {\underline \partial}^r_{I^{(r)}} \uPi\,.
\ee
We omit the $z$ indices: throughout this section, we will use only the couplings in Batyrev coordinates. We also introduce the notation
\begin{equation}
    C_{I^{(r)},J^{(s)}}=\int_{X} {\underline \partial}^r_{I^{(r)}}\Omega \wedge  {\underline \partial}^s_{J^{(s)}} \Omega = {\underline \partial}^r_{I^{(r)}}\uPi^T \Sigma {\underline \partial}^s_{J^{(s)}} \uPi\,
\end{equation}
together with the symmetrised version
\begin{equation}
    C_{r,s}=C_{(I^{(r)},J^{(s)})}\,.
\end{equation}
For better readability, we drop the index set $I^{(r)}\cup J^{(s)}$. In the following, $C_{r,s}$ will appear only in expressions where the indices are clear from the context.

First, we want to show by induction in $k$ that the derivatives of the vanishing Griffiths relations
\begin{equation}
    0 = \partial_{i_0}\ldots \partial_{i_k} C_{i_{k+1}\ldots i_n},\quad 1\leq k< \bigg\lceil{\frac{n+1}{2}}\bigg\rceil
\end{equation}
imply
\begin{equation}
    0 = \sum_{r=0}^{k+1}\binom{k+1}{r}C_{r,n+1-r}\,.\label{eq:diffeqsys}
\end{equation}
For $k=1$, we verify 
\begin{align}
    \begin{split}
    0 &= \partial_{i_0}\partial_{i_1} C_{i_2\ldots i_n}= C_{i_0i_1,i_2\ldots i_n}+C_{i_0,i_1i_2\ldots i_n}+C_{i_1,i_0i_2\ldots i_n} + C_{i_0\ldots i_n}
    \end{split}\\
    \Rightarrow 0 &=C_{2,n-1} + 2 C_{1,n}+C_{n+1}\,.
\end{align}
Assuming that the expression holds for $k$, we consider
\begin{align}
\begin{split}
    0= \partial_{i_0}\ldots \partial_{i_{k+1}} C_{i_{k+2}\ldots i_n} = \partial_{i_0}\ldots \partial_{i_{k}}\left(C_{i_{k+1},i_{k+2}\ldots i_n}+C_{i_{k+1}\ldots i_n}\right),
\end{split}
\end{align}
which implies
\begin{align}
\begin{split}
    0&=\sum_{r=0}^{k+1}\binom{k+1}{r}\left(C_{r+1,n-r}+C_{r,n+1-r}\right) =\sum_{r=1}^{k+2}\binom{k+1}{r-1}C_{r,n+1-r}+\sum_{r=0}^{k+1}\binom{k+1}{r}C_{r,n+1-r}\\
    &=\sum_{r=0}^{k+1}\underbrace{\left(\binom{k+1}{r-1}+\binom{k+1}{r}\right)}_{=\binom{k+2}{r}}C_{r,n+1-r}+C_{k+2,n-k-1}= \sum_{r=0}^{k+2}\binom{k+2}{r}C_{r,n+1-r}\,.
\end{split}
\end{align}
It remains to solve the linear system of equations in \eqref{eq:diffeqsys}. We first consider the odd-dimensional case $n=2m-1$, $m\in\N_{>1}$. Then, \eqref{eq:diffeqsys} can be written in matrix form, where the $m-1$ equations are represented as
\begin{equation}
    0=\begin{pmatrix}
       1 & 2 & 1 & 0 & 0 & \ldots & 0\\
       1 & 3 & 3 & 1 & 0 & \ldots & 0\\
       \vdots & & & & \ddots & & 1\\
       1 & & & \ldots & & &  m
    \end{pmatrix}
    \begin{pmatrix}
       C_{n+1}\\ C_{1,n}\\ \vdots \\ C_{m-1,m+1}
    \end{pmatrix}.
\end{equation}
In the last equation, we used antisymmetry of $\Sigma$ to obtain $C_{m,m}=0$. Starting at the last row, subtracting from each row the one above and repeating this process, one obtains the equivalent equation
\begin{equation}
    0=\begin{pmatrix}
       1 & 2 & 1 & 0 & 0 & \ldots & 0\\
       0 & 1 & 2 & 1 & 0 & \ldots & 0\\
       \vdots & & & & \ddots & & 1\\
       0 & & & \ldots & &1 &  2
    \end{pmatrix}
    \begin{pmatrix}
       C_{n+1}\\ C_{1,n}\\ \vdots \\ C_{m-1,m+1}
    \end{pmatrix}.
\end{equation}
The kernel of this matrix is readily found to be spanned by
\begin{equation}
    k=\left(m,-(m-1),m-2,\ldots ,(-1)^{m+1}\right),
\end{equation}
giving us the solution space
\begin{equation}\label{eq:relCodd}
    C_{r,n+1-r} = (-1)^r\frac{m-r}{m} C_{n+1}=(-1)^r\frac{n+1-2r}{n+1}C_{n+1},\quad n\text{ odd.}
\end{equation}
For even dimensional manifolds $n=2m$, we obtain the system of $m$ equations
\begin{equation}
    0=\begin{pmatrix}
       1 & 2 & 1 & 0 & 0 & \ldots & 0\\
       0 & 1 & 2 & 1 & 0 & \ldots & 0\\
       \vdots & & & & \ddots & & 1\\
       0 & & & \ldots & &1 &  3
    \end{pmatrix}
    \begin{pmatrix}
       C_{n+1}\\ C_{1,n}\\ \vdots \\ C_{m,m+1}
    \end{pmatrix}.
\end{equation}
where we used symmetry of $\Sigma$ to write $W_{m+1,m}=W_{m,m+1}$ in the last equation, resulting in the ``$+1$'' in the last entry of the matrix. 
Here, the kernel is given by the span of
\begin{equation}
    k=\left(2m+1,-(2m-1),2m-3,\ldots ,(-1)^{m}\right)
\end{equation}
yielding the solution space
\begin{equation}
    C_{r,n+1-r} = (-1)^r\frac{2m +1 - 2r}{2m+1}C_{n+1} = (-1)^r\frac{n +1 - 2r}{n+1}C_{n+1}\,,
\end{equation}
which, together with \eqref{eq:relCodd}, now holds for all $n$. Furthermore, for $r=1$, we find the relation
\begin{align}
    \begin{split}
    \partial_{(i_0}C_{i_1 \ldots i_n)}&=C_{(i_0,i_1 \ldots i_n)}+C_{(i_0 \ldots i_n)}
    = \left(\frac{1-n}{1+n}+1\right)C_{n+1}
    = \frac{2}{n+1}C_{n+1}\,.
    \end{split}\label{eq:delYuk}
\end{align}
\section{Data for three-folds}
\subsection{Data for quotient of \texorpdfstring{$\P_{2,1,1,1,1}[6]$}{X6}}
\label{sec:app6}
\subsubsection{Instanton numbers and Yukawa couplings}\label{sec:InstYukX6}
\begin{table}
    \centering
    \begin{tabular}{|c|c|}
    \hline
    $(i_1,i_2,i_3)$ &$n_{i_1,i_2,i_3}$\\ \hline
    (1, 0, 0) & 6\\ \hline 
    (0, 1, 0) & 6\\ \hline 
    (0\, 0, 1) & 180\\ 
    (0, 0, 2) & 180\\  
    (0, 0, 3) & 120\\ 
    (0, 0, 4) & 180\\ 
    (0, 0, 5) & 180\\ \hline 
    (1, 0, 1) & 180\\  
    (2, 0, 2) & 180\\ \hline 
    (0, 1, 1) & 180\\
    (0, 2, 2) & 180\\ \hline 
    (1, 1, 1) & 180\\ \hline 
    \end{tabular}
    \hspace{0.5cm} 
    \begin{tabular}{|c|c|}
    \hline
    $(i_1,i_2,i_3)$ &$n_{i_1,i_2,i_3}$\\ \hline
    (1, 0, 2) & 2322\\ \hline
    (0, 1, 2) & 2322\\ \hline 
    (1, 1, 2) & 12420\\ \hline
    (2, 1, 2) & 2322\\ \hline
    (1, 2, 2) & 2322\\ \hline 
    (1, 0, 3) & 17616\\ \hline
    (0, 1, 3) & 17616\\ \hline
    (2, 0, 3) & 17616\\ \hline
    (1, 1, 3) & 367524\\ \hline 
    (0, 2, 3) & 17616\\ \hline 
    (1, 0, 4) & 94554\\ \hline 
    (0, 1, 4) & 94554\\ \hline
    \end{tabular}
    \caption{Instanton numbers for the family $\mathcal{X}_6$.}
    \label{tab:instX6}
\end{table}
Defining
\begin{align}
\begin{split}\label{eq:discsX6}
   \Delta_1&=1-4 z_1\,,\quad \Delta_2=1-4 z_2\,,\\
   \Delta_c&= 1 - 108 z_3 + 1458\left(3-4 z_1 - 4 z_2-16z_1z_2 \right)  z_3^2 - 78732 \Delta_1\Delta_2 z_3^3\\&\qquad +  531441 \Delta_1^2 \Delta_2^2 z_3^4\,,
\end{split}
\end{align}
we can write the Yukawa couplings as
{\footnotesize
\begin{align}
\begin{split}\label{eq:YukX6Q}
    C_{111}&=\frac{6 \left(1+4 z_1-54 \left(1+12 z_1\right) z_3+729\Delta_2 \left(1+24 z_1+16 z_1^2\right) z_3^2\right)}{z_1^2 \Delta_1^2\Delta_c}\,,\\
    C_{112}&= -\frac{6 \left(1-108 z_3+729 \left(5-4 z_1-4 \left(3+4 z_1\right) z_2\right) z_3^2-39366 \Delta_1 \Delta_2 z_3^3\right)}{z_1 z_2\Delta_1 \Delta_c}\,,\\
    C_{113}&= -\frac{12 \left(1-108 z_3+729 \left(3+4 z_1\right) \Delta_2 z_3^2\right)}{z_1z_3\Delta_1  \Delta_c}\,,\\
    C_{122}&=-\frac{6 \left(1-108 z_3+729 \left(5-12 z_1-4 \left(1+4 z_1\right) z_2\right) z_3^2-39366 \Delta_1\Delta_2 z_3^3\right)}{z_1 z_2 \Delta_2 \Delta_c}\,,\\
    C_{123}&=\frac{3 \left(1-81 z_3+729 \left(3-4 z_1-4 \left(1+4 z_1\right) z_2\right) z_3^2-19683 \Delta_1\Delta_2 z_3^3\right)}{z_1 z_2 z_3 \Delta_c}\,,\\
    C_{133}&=\frac{6 \left(1-54 z_3+729 \left(1+4 z_1\right) \Delta_2 z_3^2\right)}{z_1 z_3^2 \Delta_c}\,,\\
    C_{222}&=\frac{6 \left(1+4 z_2-54 \left(1+12 z_2\right) z_3+729\Delta_1 \left(1+24 z_2+16 z_2^2\right) z_3^2\right)}{z_2^2 \Delta_2\Delta_c}\,,\\
    C_{223}&=-\frac{12 \left(1-108 z_3+729 \Delta_1 \left(3+4 z_2\right) z_3^2\right)}{z_2  z_3 \Delta_2\Delta_c}\,,\\
    C_{233}&=\frac{6 \left(1-54 z_3+729 \left(1+4 z_2\right) \Delta_1 z_3^2\right)}{z_2 z_3^2 \Delta_c}\,,\quad 
    C_{333}=  \frac{12\left(1 - 729\Delta_1\Delta_2z_3^2 \right)}{z_3^3\Delta_c}\,.
  \end{split}
\end{align}}
The first non-vanishing instanton number for  $P_{2,1,1,1,1}[6]$ are listed in \Cref{tab:instX6}. These are subject to the symmetries 
\begin{equation}
    n_{i_1,i_2,i_3}=n_{i_3-i_1,i_2,i_3}\quad\text{and}\quad n_{i_1,i_2,i_3}=n_{i_2,i_1,i_3}\,.
\end{equation}

\subsubsection{Analytical continuation and monodromies}
\par We illustrate how to find an integral symplectic basis in the patch of the moduli space with local coordinates
\begin{equation}
    \tilde{a}_6=a_6=\frac{1}{\sqrt{z_1}}\,,\quad \tilde{a}_7=a_7=\frac{1}{\sqrt{z_2}}\,,\quad \tilde{a}_0=\frac{1}{a_0^3}=\sqrt{z_1z_2}z_3\,.
\end{equation}
The operators in this new patch are written as
\begin{align}
\begin{split}
\widetilde{{\cal L}}_1^{(2)}&=\tilde{\theta}_1\tilde{\theta}_2-3\tilde{a}_0\tilde{a}_6\tilde{a}_7(3\tilde{\theta}_3+1)(3\tilde{\theta}_3+2)\,,\\
    \widetilde{{\cal L}}_2^{(2)}&=4\tilde{\theta}_1(\tilde{\theta}_1-1)-\tilde{a}_6^2(\tilde{\theta}_1-\tilde{\theta}_3)^2,\\
   \widetilde{{\cal L}}_3^{(2)}&= 4\tilde{\theta}_2(\tilde{\theta}_2-1)-\tilde{a}_7^2(\tilde{\theta}_2-\tilde{\theta}_3)^2,
   \end{split}
\end{align}
where
$\tilde{\theta}_1=\tilde{a}_6 \partial_{\tilde{a}_6},\tilde{\theta}_2=\tilde{a}_7 \partial_{\tilde{a}_7}$ and $\tilde{\theta}_3=\tilde{a}_0 \partial_{\tilde{a}_0}$.

We compute a Frobenius basis $\varpi^{\tilde{a}}_i,\; i=1,\dots, 8$ in these coordinates:
\begin{equation}
\begin{alignedat}{2}
&\varpi^{\tilde{a}}_1=\sigma_1\,,
&&\varpi^{\tilde{a}}_5=\sigma_5\,,\\
&\varpi^{\tilde{a}}_2=\sigma_1\log(\tilde{a}_0)+\sigma_2\,,
&&\varpi^{\tilde{a}}_6=\sigma_5\log(\tilde{a}_0)+\sigma_6\,,\\
&\varpi^{\tilde{a}}_3=\sigma_1\log(\tilde{a}_0)^2+2\sigma_2\log(\tilde{a}_0)+\sigma_3\,,
&&\varpi^{\tilde{a}}_7=\sigma_7\,,\\
&\varpi^{\tilde{a}}_4=\sigma_1\log(\tilde{a}_0)^3+3\sigma_2\log(\tilde{a}_0)^2+3\sigma_3\log(\tilde{a}_0)+\sigma_4\,,\quad 
&&\varpi^{\tilde{a}}_8=\sigma_7\log(\tilde{a}_0)+\sigma_8\,,
\end{alignedat}
\end{equation}
with (up to terms in $\mathcal{O}(\tilde{a}^4)$)
\begin{equation}
\begin{alignedat}{2}
&\sigma_1=1 + 360\tilde{a}_0^2 + 6\tilde{a}_6\tilde{a}_7\tilde{a}_0+\dots\,,
&&\sigma_2=1386\tilde{a}_0^2 + 27\tilde{a}_6\tilde{a}_7\tilde{a}_0+\dots\,,\\
&\sigma_3=1314\tilde{a}_0^2 + \frac{1}{4}\tilde{a}_7^2 + \frac{1}{4}\tilde{a}_6^2 + 54\tilde{a}_6\tilde{a}_7\tilde{a}_0+\dots\,,\quad
&&\sigma_4=-3942\tilde{a}_0^2+\dots\,,\\
&\sigma_5=\tilde{a}_7 + 24\tilde{a}_6\tilde{a}_0 + 1440\tilde{a}_7\tilde{a}_0^2 + \frac{1}{24}\tilde{a}_7^3+\dots\,,
&&\sigma_6=60\tilde{a}_6\tilde{a}_0 + 4104\tilde{a}_7\tilde{a}_0^2 - \frac{1}{12}\tilde{a}_7^3+\dots\,,\\
&\sigma_7=\tilde{a}_6 + 24\tilde{a}_7\tilde{a}_0 + 1440\tilde{a}_6\tilde{a}_0^2 + \frac{1}{24}\tilde{a}_6^3+\dots\,,
&&\sigma_8=60\tilde{a}_7\tilde{a}_0 + 4104\tilde{a}_6\tilde{a}_0^2 - \frac{1}{12}\tilde{a}_6^3+\dots.
\end{alignedat}
\end{equation}

For the analytical continuation of the integral symplectic basis to the coordinate patch parametrised by $\tilde{a}$, we must first go to the patch with coordinates $u_1=1-4z_1,u_2=1-4z_2,u_3=z_3$. The first two coordinates describe strong coupling divisors. After computing again a Frobenius basis of periods, we may numerically compute the monodromies and the transition matrix for this point. The exact values of the entries are determined experimentally. Using the periods at this point we can then find the transition matrix for our point of interest. We obtain for the latter
\begin{align}
T_{\tilde{a}}=\left( \begin {array}{cccccccc} 1&0&0&0&0&0&0&0\\ \noalign{\medskip}-
{\frac{1}{2}}&0&0&0&0&0&{\frac {1}{2\,\pi}}&0\\ \noalign{\medskip}-{
\frac{1}{2}}&0&0&0&{\frac {1}{2\,\pi}}&0&0&0\\ \noalign{\medskip}{
\frac{1}{2}}& -{\frac {\ii}{2\pi}}&0&0&-{\frac {1}{4\,\pi}}&0
&-{\frac {1}{4\,\pi}}&0\\ \noalign{\medskip}{\frac{13}{4}}&0&{\frac {3
}{2\,{\pi}^{2}}}&0&-{\frac {3}{4\,\pi}}&0&-{\frac {3}{4\,\pi}}&0
\\ \noalign{\medskip}{\frac{7}{8}}&{\frac {3\,\ii}{4\pi}}&{
\frac {3}{4\,{\pi}^{2}}}&0&{\frac {3}{8\,\pi}}-{\frac {{\frac {3\,\ii}{2
}}\ln  \left( 2 \right) }{{\pi}^{2}}}&-\frac {3\,\ii}{4\pi^2}&-{\frac {3}{8\,\pi}}&0\\ \noalign{\medskip}{\frac{7}{8}}&{
\frac {3\,\ii}{4\pi}}&{\frac {3}{4\,{\pi}^{2}}}&0&-{\frac {3
}{8\,\pi}}&0&{\frac {3}{8\,\pi}}-{\frac {{\frac {3\,\ii}{2}}\ln  \left( 
2 \right) }{{\pi}^{2}}}&-{\frac {3\,\ii}{4\pi^2}}
\\ \noalign{\medskip}{\frac{\chi\left(\mathcal{X}^{(1)}_6\right)\,\zeta(3)}{2(2\pi \ii)^3}}&{\frac{\tilde{c}_2\left(\mathcal{X}^{(1)}_6\right)}{24\cdot 2\pi \ii}}&0&
\frac {\ii}{4\pi^3}&{\frac {-{\frac {3\,\ii}{4}}\ln  \left( 2
 \right) }{{\pi}^{2}}}& -\frac {3\,\ii}{8\pi^2}&{\frac 
{-{\frac {3\,\ii}{4}}\ln  \left( 2 \right) }{{\pi}^{2}}}&-{
\frac {3\,\ii}{8\pi^2}}\end {array} \right),
\end{align}
where we identified $\chi\left(\mathcal{X}_6^{(1)}\right)=-204$ and $\tilde{c}_2\left(\mathcal{X}_6^{(1)}\right)=c_2\left(\mathcal{X}_6^{(1)}\right)\cdot J=42$.

As a consistency check we compute the monodromies around the divisor $\tilde{a}_0=0$ (the other monodromies are trivial):
\begin{align}
&\frak{M}_{\tilde{a}_0}={\footnotesize\left( \begin {array}{cccccccc} 1&0&0&0&0&0&0&0\\ 0
&1&0&0&0&0&0&0\\ 0&0&1&0&0&0&0&0
\\1&0&0&1&0&0&0&0\\ -6&-6&-6&-12&1
&0&0&0\\ -3&-3&0&-6&0&1&0&0\\ -3&0
&-3&-6&0&0&1&0\\ 7&3&3&6&-1&0&0&1\end {array}
 \right)}.
\end{align}
The leading behaviour of $\vec{\Pi}=T_{\tilde{a}}\vec{\varpi}^{\tilde{a}}$ is found to be
\begin{align}
\vec{\Pi}\sim{\small\left( \begin {array}{c} 
1 \\
-\frac{1}{2} + \frac{\tilde{a}_6}{2\pi} \\
-\frac{1}{2} + \frac{\tilde{a}_7}{2\pi} \\
-\frac{\ii\ln(\tilde{a}_0)}{2\pi} - \frac{\tilde{a}_6 + \tilde{a}_7}{4\pi} + \frac{1}{2} \\
\frac{3\ln(\tilde{a}_0)^{2}}{2\pi^{2}} - \frac{3(\tilde{a}_6 + \tilde{a}_7)}{4\pi} +\frac{13}{4} \\
 \frac{3\ln(\tilde{a}_0)^{2}}{4\pi^{2}} + \frac{3\ii\ln(\tilde{a}_0)}{4\pi} - \frac{3(\tilde{a}_6 - \tilde{a}_7)}{8\pi} - \frac{3\ii\ln(2)\tilde{a}_7}{2\pi^{2}} - \frac{3\ii \ln(\tilde{a}_0)\tilde{a}_7}{4\pi^{2}}+\frac{7}{8} \\
 \frac{3\ln(\tilde{a}_0)^{2}}{4\pi^{2}} + \frac{3\ii\ln(\tilde{a}_0)}{4\pi} + \frac{3(\tilde{a}_6 - \tilde{a}_7)}{8\pi} - \frac{3\ii\ln(2)\tilde{a}_6}{2\pi^{2}} - \frac{3\ii \ln(\tilde{a}_0)\tilde{a}_6}{4\pi^{2}}+\frac{7}{8}  \\
-\frac{51\ii\zeta(3)}{4\pi^{3}} - \frac{7\ii \ln(\tilde{a}_0)}{8\pi}  -\frac{\ii\ln(\tilde{a}_0)^{3}}{2\pi^{3}} - \frac{3\ii\ln(2)(\tilde{a}_6 + \tilde{a}_7)}{4\pi^{2}} - \frac{3\ii(\tilde{a}_6 + \tilde{a}_7)\ln(\tilde{a}_0)}{8\pi^{2}}
\end {array} \right)
}. 
\end{align}

For the coordinates $u_1=1-4z_1,u_2=1-4z_2,u_3=z_3$ of the intermediate point the Frobenius basis reads
\begin{equation}
\begin{alignedat}{2}
&\varpi^{u}_1=\sigma_1\,,
&&\hspace{-2cm}\varpi^{u}_2=\sigma_1\log(u_3)+\sigma_2\\
&\varpi^{u}_3=\sigma_1\log(u_3)^2+2\sigma_2\log(u_3)+\sigma_3\,,\\
&\varpi^{u}_4=\sigma_1\log(u_3)^3+3\sigma_2\log(u_3)^2+3\sigma_3\log(u_3)+\sigma_4\,,\\
&\varpi^{u}_5=\sigma_5\,,
&&\hspace{-2cm}\varpi^{u}_6=\sigma_5(\log(u_1)+\log(u_3))+\sigma_6\,,\\
&\varpi^{u}_7=\sigma_7\,,
&&\hspace{-2cm}\varpi^{u}_8=\sigma_7(\log(u_2)+\log(u_3))+\sigma_8\,,\\
\end{alignedat}
\end{equation}
where {\footnotesize
\begin{align}
\begin{split}
&\sigma_1=1 + 6u_3 + \frac{405}{2}u_3^2+\dots\,, \\
&\sigma_2=- \frac{1}{2}u_1 - \frac{1}{2}u_2 +27u_3 - \frac{1}{4}u_1^2 - 3u_1u_3 - \frac{1}{4}u_2^2 - 3u_2u_3 + \frac{8397}{8}u_3^2+\dots\,,\\
&\sigma_3= - u_1- u_2 + 30u_3 - \frac{5}{12}u_1^2 + \frac{1}{2}u_1u_2- 21u_1u_3 - \frac{5}{12}u_2^2- 21u_2u_3   +\frac{17505}{8}u_3^2+\dots\,,\\
&\sigma_4=- 6u_1 - 6u_2 -180u_3   - \frac{17}{6}u_1^2+ 3u_1u_2- \frac{17}{6}u_2^2- \frac{39339}{8}u_3^2+\dots\,,\\
&\sigma_5=\sqrt{u_1}\left(1 + \frac{1}{3}u_1 + \frac{1}{5}u_1^2+ 2u_1u_3 \right)+\dots\,,\\
&\sigma_6=\sqrt{u_1}\left(\frac{5}{18}u_1-\frac{1}{2}u_2 + \frac{61}{300}u_1^2 - \frac{1}{6}u_1u_2 + \frac{23}{3}u_1u_3  - \frac{1}{4}u_2^2\right)+\dots\,,\\
&\sigma_7=\sqrt{u_2}\left(1 + \frac{1}{3}u_2 + \frac{1}{5}u_2^2+2u_2u_3\right)+\dots\,,\\
&\sigma_8=\sqrt{u_2}\left( - \frac{1}{2}u_1 +\frac{5}{18}u_2 - \frac{1}{4}u_1^2- \frac{1}{6}u_1u_2 + \frac{61}{300}u_2^2+\frac{23}{3}u_2u_3  \right)+\dots\,.\\
\end{split}
\end{align}}
The transition matrix in this patch is 
\begin{align}
T_u ={\scriptsize
\left( \begin{array}{cccccccc}
1 & 0 & 0 & 0 & 0 & 0 & 0 & 0 \\
0 & 0 & 0 & 0 & \frac{\ii}{\pi} & 0 & 0 & 0 \\
0 & 0 & 0 & 0 & 0 & 0 & \frac{\ii}{\pi} & 0 \\
\frac{\ii\ln(2)}{\pi} & \frac{-\frac{\ii}{2}}{\pi} & 0 & 0 & \frac{1}{2\pi \ii} & 0 &\frac{1}{2\pi \ii} & 0 \\
\frac{5}{2} + \frac{6\ln(2)^2}{\pi^2} & -\frac{6\ln(2)}{\pi^2} & \frac{3}{2\pi^2} & 0 & 0 & 0 & 0 & 0 \\
\frac{5}{4} + \frac{3\ln(2)^2}{\pi^2} & -\frac{3\ln(2)}{\pi^2} & \frac{3}{4\pi^2} & 0 & 0 & 0 & -\frac{3}{\pi^2} & \frac{3}{2\pi^2} \\
\frac{5}{4} + \frac{3\ln(2)^2}{\pi^2} & -\frac{3\ln(2)}{\pi^2} & \frac{3}{4\pi^2} & 0 & -\frac{3}{\pi^2} & \frac{3}{2\pi^2} & 0 & 0 \\
\frac{-5\ln(2)\pi^2 + 4\ln(2)^3 + 36\zeta(3)}{2\ii\pi^3}  & \frac{5\pi^2 - 12\ln(2)^2}{4\ii\pi^3} & -\frac{3\ii\ln(2)}{2\pi^3} & \frac{\ii}{4\pi^3} & 0 & 0 & 0 & 0
\end{array} \right)}.
\end{align}
The monodromies around the divisors $u_i=0$ are respectively given by
\begin{align}
\begin{split}
&\frak{M}_{u_1} ={\footnotesize \left( \begin{array}{cccccccc}
1 & 0 & 0 & 0 & 0 & 0 & 0 & 0 \\
0 & -1 & 0 & 0 & 0 & 0 & 0 & 0 \\
0 & 0 & 1 & 0 & 0 & 0 & 0 & 0 \\
0 & 1 & 0 & 1 & 0 & 0 & 0 & 0 \\
0 & 0 & 0 & 0 & 1 & 0 & 0 & 0 \\
0 & 0 & 0 & 0 & 0 & 1 & 0 & 0 \\
0 & -3 & 0 & 0 & 1 & 0 & -1 & 0 \\
0 & 0 & 0 & 0 & 0 & 0 & 0 & 1
\end{array} \right)},\quad 
\frak{M}_{u_2} = {\footnotesize \left( \begin{array}{cccccccc}
1 & 0 & 0 & 0 & 0 & 0 & 0 & 0 \\
0 & 1 & 0 & 0 & 0 & 0 & 0 & 0 \\
0 & 0 & -1 & 0 & 0 & 0 & 0 & 0 \\
0 & 0 & 1 & 1 & 0 & 0 & 0 & 0 \\
0 & 0 & 0 & 0 & 1 & 0 & 0 & 0 \\
0 & 0 & -3 & 0 & 1 & -1 & 0 & 0 \\
0 & 0 & 0 & 0 & 0 & 0 & 1 & 0 \\
0 & 0 & 0 & 0 & 0 & 0 & 0 & 1
\end{array} \right)},\\
&\frak{M}_{u_3} = {\footnotesize\left( \begin{array}{cccccccc}
1 & 0 & 0 & 0 & 0 & 0 & 0 & 0 \\
0 & 1 & 0 & 0 & 0 & 0 & 0 & 0 \\
0 & 0 & 1 & 0 & 0 & 0 & 0 & 0 \\
1 & 0 & 0 & 1 & 0 & 0 & 0 & 0 \\
-6 & -6 & -6 & -12 & 1 & 0 & 0 & 0 \\
-3 & -3 & 0 & -6 & 0 & 1 & 0 & 0 \\
-3 & 0 & -3 & -6 & 0 & 0 & 1 & 0 \\
7 & 3 & 3 & 6 & -1 & 0 & 0 & 1
\end{array} \right)}.
\end{split}
\end{align}

\subsubsection{Conifold transitions}\label{sec:appX6contraf}
The instanton numbers for the two-parameter CICY are given in the following table. Since the model appears as the conifold transition of $\mathcal{X}_6^{(3)}$ when $q_1\rightarrow 1$, we named the two Kähler parameters $t_2$ and $t_3$.
\begin{table}[H]
    \centering
    \begin{tabular}{|c|cccccc|}\hline
    &$i_3=0$&$i_3=1$&$i_3=2$&$i_3=3$&$i_3=4$&$i_3=5$ \\\hline
 $i_2=0$&- & 360 & 2682 & 35472 & 606348 & 12210408 \\
 $i_2=1$&6 & 360 & 17064 & 770280 & 33726420 & 1444231296 \\
 $i_2=2$&0 & 0 & 2682 & 770280 & 99533664 & 9382024152 \\
 $i_2=3$&0 & 0 & 0 & 35472 & 33726420 & 9382024152 \\
 $i_2=4$&0 & 0 & 0 & 0 & 606348 & 1444231296 \\
 $i_2=5$&0 & 0 & 0 & 0 & 0 & 12210408 \\\hline
    \end{tabular}
    \caption{Instanton numbers for the model $\mathcal{X}_6^{(2)}$.}
    \label{tab:instX6CICY2}
\end{table}
\begin{table}[H]
    \centering
    \begin{tabular}{|ccccc|}\hline
    $i_3=1$&$i_3=2$&$i_3=3$&$i_3=4$&$i_3=5$ \\\hline
720& 22428& 1611504& 168199200& 21676931712 \\\hline
    \end{tabular}
    \caption{Instanton numbers for the model $\mathcal{X}_6^{(1)}=\mathcal{X}_{3,2,2}$.}
    \label{tab:instX6CICY1}
\end{table}
\subsection{Data for quotient of \texorpdfstring{$\P_{4,1,1,1,1}[8]$}{X8}}\label{sec:app8}
\subsubsection{Toric description}\label{sec:app81}
We consider the family of Calabi--Yau manifolds given by the zero locus of octics
\begin{align}\label{eq:P411118}
    P_{\mathcal{X}_8}=\sum_{\substack{\vec{\nu}\in \N^5 \\ 4\nu_1+\sum_{i=2}^5\nu_i=8}}a_{\vec{\nu}} \vec{x}^{\vec{\nu}}
\end{align}
inside the ambient space $\P_{4,1,1,1,1}$.
Dividing by the group $\hat{S}=\Z_8^2$ generated by the phase symmetries
\begin{align}
    g_1=\Z_8:(4,0,3,1,0)\,\quad g_2=\Z_8:(0,2,1,5,0)\,,
\end{align}
only nine of the 201 monomials in the sum of \eqref{eq:P411118} remain (i.e.\ are invariant under both group actions). 
After taking the quotient w.r.t.\ the Jacobian ideal, the defining polynomial of this quotient reads
\begin{align} \label{eq:X8}
    P_{\mathcal{X}_8/\hat{S}}=x_1^2+x_2^8+x_3^8+x_4^8+x_5^8-a_0\prod_{i=1}^5x_i-a_6x_2^4x_5^4-a_7x_3^4x_4^4
\end{align}
The reflexive polytope corresponding to its ambient space has the integral points and scaling relations listed in \Cref{tab:X8}.
\begin{table}[H]
\renewcommand{\arraystretch}{1.2}
\centering
\begin{tabular}{| c | c c c c|c c c|}
    \hline \multicolumn{5}{|c |}{points} & \multicolumn{3}{| c|}{$l$-vectors} \\ \hline \hline
    (1 & 0&0&0&0)& 0 & 0 & -4\\ \hline
    (1 &1& 0& 0& 0)& 0 & 0 & 2\\
    (1 &0& 1& 0& 0)& 0 & 1 & 0\\
    (1 &0& 0& 1& 0)& 1 & 0 & 0\\
    (1 &0& 0& 1& 2)& 1 & 0 & 0\\
    (1 &-4& -1& -2& -2)& 0 & 1 & 0\\
    (1 &-2& 0& -1& -1)& 0 & -2 & 1\\
    (1 &0& 0& 1& 1)& -2 & 0 & 1\\\hline
    (1 &-1& 0& 0& 0)& - & - & -\\ \hline
\end{tabular}
\caption{Integral points and their scaling relations of polytope describing $\P_{4,1,1,1,1}[8]/\hat{S}$. The last point lies inside a face of codimension one.}
\label{tab:X8}
\end{table}
The intersection ring of the Kähler cone generators $J_i$ on the mirror Calabi--Yau is given by
\begin{equation}\label{eq:intringX8}
    R=2\,J_1 J_2 J_3+4\,J_1 J_3^2+4\,J_2J_3^2+8\,J_3^3\,.
\end{equation}
The topological invariants of this model read
\begin{equation}\label{eq:topdataX8}
    c_2\cdot J_1 = 24,\ c_2\cdot J_2 = 24,\, c_2\cdot J_3 = 56,\ \chi=-160\,.
\end{equation}
The first instanton numbers and the Yukawa couplings are listed at the end of the section.
The Batyrev coordinates are given by
\begin{equation}
    z_1 = \frac{1}{a_7^2}\,,\quad
    z_2 = \frac{1}{a_6^2}\,,\quad
    z_3 = \frac{a_6 a_7}{a_0^4}\,.
\end{equation}
The differential ideal is generated by the operators
\begin{align}
\begin{split}
    {\cal L}_1^{(2)}(\uz) &= (2\theta_{1}-\theta_{3})(2\theta_{2}-\theta_{3})-4z_{3}(3+16\theta_{3}+16\theta_{3}^{2})\,,\\
     {\cal L}_2^{(2)}(\uz) &= \theta_{1}^{2}-z_{1}\left(4\theta_{1}^{2}+\theta_{1}(2-4\theta_{3})+(\theta_{3}-1)\theta_{3}\right),\\
     {\cal L}_3^{(2)}(\uz)&= \theta_{2}^{2}-z_{2}\left(4\theta_{2}^{2}+\theta_{2}(2-4\theta_{3})+(\theta_{3}-1)\theta_{3}\right).
\end{split}
\end{align}

\subsubsection{Instanton numbers and Yukawa couplings}

The first non-vanishing instanton numbers are listed in \Cref{tab:instX8}.
\begin{table}[H]
    \centering
    \begin{tabular}{|c|c|}
    \hline
    $(i_1,i_2,i_3)$ &$n_{i_1,i_2,i_3}$\\ \hline
    (0, 0, 1) & 320\\
    (0, 0, 2) & 160\\
    (0, 0, 3) & 320\\
    (0, 0, 4) & 160\\
    (0, 0, 5) & 320\\ \hline
    (0, 1, 0) & 4\\ \hline
    (0, 1, 1) & 320\\
    (0, 2, 2) & 160\\ \hline 
    (0, 1, 2) & 9712\\ \hline
    (0, 1, 3) & 143872\\ \hline
    (0, 1, 4) & 1243176\\ \hline
    (0, 2, 3) & 143872\\ \hline
    \end{tabular}
    \hspace{0.5cm}
    \begin{tabular}{|c|c|}
    \hline
    $(i_1,i_2,i_3)$ &$n_{i_1,i_2,i_3}$\\ \hline
    (1, 0, 0) & 4\\ \hline
    (1, 0, 1) & 320\\ 
    (2, 0, 2) & 160\\ \hline
    (1, 0, 2) & 9712\\ \hline
    (1, 0, 3) & 143872\\ \hline
    (1, 0, 4) & 1243176\\ \hline
    (1, 1, 1) & 320\\ \hline
    (1, 1, 2) & 52800\\ \hline
    (1, 1, 3) & 3625728\\ \hline
    (1, 2, 2) & 9712\\ \hline
    (2, 0, 3) & 143872\\ \hline
    (2, 1, 2) & 9712\\ \hline
    \end{tabular}
    \caption{Instanton numbers for the family $\mathcal{X}_8$.}
    \label{tab:instX8}
\end{table}
Similar to the sextic model, we have the symmetries
\begin{equation}
    n_{i_1,i_2,i_3}=n_{i_3-i_1,i_2,i_3}\quad\text{and}\quad n_{i_1,i_2,i_3}=n_{i_2,i_1,i_3}\,.
\end{equation}
The couplings are given by{\footnotesize
\begin{align}
\begin{split}
    C_{111}&=\frac{4 \left(1+4 z_1-128 \left(1+12 z_1\right) z_3+4096\Delta_2 \left(1+24 z_1+16 z_1^2\right) z_3^2\right)}{z_1^2 \Delta_1^2\Delta_c}\,,\\
    C_{112}&= -\frac{4 \left(1-256 z_3+4096 \left(5-4 z_1-4 \left(3+4 z_1\right) z_2\right) z_3^2-524288 \Delta_1 \Delta_2 z_3^3\right)}{z_1 z_2\Delta_1 \Delta_c}\,,\\
    C_{113}&= -\frac{8 \left(1-256 z_3+4096 \left(3+4 z_1\right) \Delta_2 z_3^2\right)}{z_1z_3\Delta_1  \Delta_c}\,,\\
    C_{122}&=-\frac{4 \left(1-256 z_3+4096 \left(5-12 z_1-4 \left(1+4 z_1\right) z_2\right) z_3^2-524288  \Delta_1\Delta_2 z_3^3\right)}{z_1 z_2 \Delta_2 \Delta_c}\,,\\
    C_{123}&=\frac{2 \left(1-192 z_3+4096 \left(3-4 z_1-4 \left(1+4 z_1\right) z_2\right) z_3^2-262144 \Delta_1\Delta_2 z_3^3\right)}{z_1 z_2 z_3 \Delta_c}\,,\\
    C_{133}&=\frac{4 \left(1-128 z_3+4096 \left(1+4 z_1\right) \Delta_2 z_3^2\right)}{z_1 z_3^2 \Delta_c}\,,\\
    C_{222}&=\frac{4 \left(1+4 z_2-128 \left(1+12 z_2\right) z_3+4096\Delta_1 \left(1+24 z_2+16 z_2^2\right) z_3^2\right)}{z_2^2 \Delta_2^2\Delta_c}\,,\\
    C_{223}&=-\frac{8 \left(1-256 z_3+4096 \Delta_1 \left(3+4 z_2\right) z_3^2\right)}{z_2  z_3 \Delta_2\Delta_c}\,,\\
    C_{233}&=\frac{4 \left(1-128 z_3+4096 \left(1+4 z_2\right) \Delta_1 z_3^2\right)}{z_2 z_3^2 \Delta_c}\,,\quad C_{3,3,3}=  \frac{8\left(1 - 4096\Delta_1\Delta_2z_3^2 \right)}{z_3^3\Delta_c}\,,\\
  \end{split}
\end{align}}
where we defined
\begin{align}
\begin{split}\label{eq:discX6}
   \Delta_1&=1-4 z_1\,,\quad \Delta_2=1-4 z_2\,,\\
   \Delta_c&= 1 - 256 z_3 + 8192\left(3-4 z_1 - 4 z_2-16z_1z_2 \right)  z_3^2 - 1048576 \Delta_1\Delta_2 z_3^3\\&\qquad +  16777216 \Delta_1^2 \Delta_2^2 z_3^4\,.
\end{split}
\end{align}
We note that these Yukawa couplings are related to those of the family $\mathcal{X}_6^{(3)}$ in \eqref{eq:YukX6Q} via
\begin{equation}
    \frac{1}{6\cdot 3^{3i_3}}C_{i_1,i_2,i_3}^{\mathcal{X}_6^{(3)}}\left(z_1,z_2,\frac{z_3}{3^3}\right)=\frac{1}{4\cdot 4^{3i_3}}C_{i_1,i_2,i_3}^{\mathcal{X}_8^{(3)}}\left(z_1,z_2,\frac{z_3}{4^3}\right).
\end{equation}

\subsubsection{Analytical continuation and monodromies}
The Frobenius basis in the coordinates $u_1=\sqrt{z_1}^{-1}$, $u_2=\sqrt{z_2}^{-1}$, $u_3=\sqrt{z_1 z_2} z_3$ reads:

\begin{equation}
\begin{alignedat}{2}
&\varpi^u_1=\sigma_1\,,\quad
&&\hspace{-2cm}\varpi_2^u=\sigma_1\log(u_3)+\sigma_2\,,\\
&\varpi^u_3=\sigma_1\log(u_3)^2+2\sigma_2\log(u_3)+\sigma_3\,,\\
&\varpi^u_4=\sigma_1\log(u_3)^3+3\sigma_2\log(u_3)^2+3\sigma_3\log(u_3)+\sigma_4\,,\\
&\varpi^u_5=\sigma_5\,,\quad
&&\hspace{-2cm}\varpi_6^u=\sigma_5\log(u_3)+\sigma_6\,,\\
&\varpi_7^u=\sigma_7\,,\quad
&&\hspace{-2cm}\varpi_8^u=\sigma_7\log(u_3)+\sigma_8\,,
\end{alignedat}
\end{equation}
where
\begin{equation}
\begin{alignedat}{2}
&\sigma_1=1 + 1680u_3^2 + 12u_1u_2u_3+\dots\,,\quad
&&\sigma_2=7904u_3^2 + 64u_1u_2u_3+\dots\,,\\
&\sigma_3=7376u_3^2 + \frac{1}{4}u_2^2 + \frac{1}{4}u_1^2 + 128u_1u_2u_3+\dots\,,\quad
&&\sigma_4=-22128u_3^2+\dots\,,\\
&\sigma_5=u_2 + 48u_1u_3 + 6720u_2u_3^2 + \frac{1}{24}u_2^3+\dots\,,\quad
&&\sigma_6=160u_1u_3 + 24896u_2u_3^2 - \frac{1}{12}u_2^3+\dots\,,\\
&\sigma_7=u_1 + 48u_2u_3 + 6720u_1u_3^2 + \frac{1}{24}u_1^3+\dots\,,\quad
&&\sigma_8=160u_2u_3 + 24896u_1u_3^2 - \frac{1}{12}u_1^3+\dots.\\
\end{alignedat}
\end{equation}
The transition matrix is
\begin{align}T_u=\left( \begin{array}{cccccccc} 
1 & 0 & 0 & 0 & 0 & 0 & 0 & 0 \\
-\frac{1}{2} & 0 & 0 & 0 & 0 & 0 & \frac{1}{2\pi} & 0 \\
-\frac{1}{2} & 0 & 0 & 0 & \frac{1}{2\pi} & 0 & 0 & 0 \\
\frac{1}{2} & \frac{1}{2\pi\ii} & 0 & 0 & -\frac{1}{4\pi} & 0 & -\frac{1}{4\pi} & 0 \\
\frac{17}{6} & 0 & \frac{1}{\pi^2} & 0 & -\frac{1}{2\pi} & 0 & -\frac{1}{2\pi} & 0 \\
\frac{11}{12} & \frac{1}{2\pi\ii} & \frac{1}{2\pi^2} & 0 & \frac{-4\ii\ln(2) + \pi}{4\pi^2} & -\frac{1}{2\pi^3} & -\frac{1}{4\pi} & 0 \\
\frac{11}{12} & \frac{1}{2\pi\ii} & \frac{1}{2\pi^2} & 0 & -\frac{1}{4\pi} & 0 & \frac{-4\ii\ln(2) + \pi}{4\pi^2} & -\frac{1}{2\pi^3} \\
\frac{\chi\left(\mathcal{X}_8^{(1)}\right)\zeta(3)}{2(2\pi\ii)^3} & \frac{\tilde{c}_2\left(\mathcal{X}_8^{(1)}\right)}{24\cdot 2\pi\ii} & 0 & \frac{\ii}{6\pi^3} & -\frac{\ii\ln(2)}{2\pi^2} & -\frac{\ii}{4\pi^2} & \frac{-\ii\ln(2)}{2\pi^2} & -\frac{\ii}{4\pi^2}
\end{array} \right),
\end{align}
with $\chi\left(\mathcal{X}_8^{(1)}\right)=-296$ and $\tilde{c}_2\left(\mathcal{X}_8^{(1)}\right)=44$.

\subsubsection{Conifold transition}\label{sec:appX8contraf}
For the hypersurface quotient $\mathcal{X}_8^{(3)}$, we find the complete intersection
\begin{align}
   \mathcal{X}_8^{(3)}&= \left(\begin{tabular}{c | c c c}
         $\P^1$ & 0&2&0 \\
         $\P^1$ & 0&0&2 \\
         $\P_{2,1^4}$ & 4&1&1
        \end{tabular}\right)^{3,83}_{-160},
\end{align}
Its Mori cone generators are given by
\begin{align}
    l_1 &= (0,-2,0;1,1,0,0,0,0,0,0,0)\,,\\
    l_2 &= (0,-2,0;0,0,1,1,0,0,0,0,0)\,,\\
    l_3 &= (-4,-1,-1;0,0,0,0,2,1,1,1,1)\,.
\end{align}
The intersection ring, the remaining topological data and the instanton numbers are again identical to that of the hypersurface given in \eqref{eq:intringX8}, \eqref{eq:topdataX8} and \Cref{tab:instX8}. 
As for $\mathcal{X}_6$, the holomorphic periods obey the relation \eqref{eq:X6relationw0} and the discriminant of the CICY can be obtained from that of the hypersurface given in \eqref{eq:discX8} by the same relation \eqref{eq:discCICY}.
\par
The model undergoes a conifold transition as either $z_1$ or $z_2$ approaches the strong coupling divisors given by $\Delta_i=1-4z_i$, $i=1,2$. For example,
\begin{equation}
    \mathcal{X}_8^{(3)} \xrightarrow{\Delta_1\rightarrow 0} \left(\begin{tabular}{c | c c c}
         $\P^1$ & 0&0&2 \\
         $\P_{2,1^5}$ & 4&2&1
        \end{tabular}\right)^{2,86}_{-168}\eqqcolon \mathcal{X}_8^{(2)}.
\end{equation}
The model $\mathcal{X}_8^{(2)}$ coincides with the hypersurface family $\P_{2,2,2,1,1}[8]$.
The instanton numbers are again obtained from the same summation relation as for $\mathcal{X}_6$ and are listed in the following table.

\begin{table}[H]
    \centering
    \begin{tabular}{|c|cccccc|}\hline
    &$i_3=0$&$i_3=1$&$i_3=2$&$i_3=3$&$i_3=4$&$i_3=5$ \\\hline
 $i_2=0$&- & 640 & 10032 & 288384 & 10979984 & 495269504 \\
 $i_2=1$&4 & 640 & 72224 & 7539200 & 757561520 & 74132328704 \\
 $i_2=2$&0 & 0 & 10032 & 7539200 & 2346819520 & 520834042880 \\
 $i_2=3$&0 & 0 & 0 & 288384 & 757561520 & 520834042880 \\
 $i_2=4$&0 & 0 & 0 & 0 & 10979984 & 74132328704 \\
 $i_2=5$&0 & 0 & 0 & 0 & 0 & 495269504\\\hline
    \end{tabular}
    \caption{Instanton numbers for the model $\mathcal{X}_8^{(2)}$.}
    \label{tab:instX8CICY2}
\end{table}

The Mori cone of the model $\mathcal{X}_8^{(2)}$ is spanned by 
\begin{align}
    l_2 &= (0,0,-2;1,1,0,0,0,0,0,0)\,,\\
    l_3 &= (-4,-2,-1;0,0,2,1,1,1,1,1)\,,
\end{align}
whose dual Kähler cone generators have the triple intersections
\begin{equation}
    R=4J_2J_3^2+8J_3^3\,.
\end{equation}
Their intersection with the second Chern class and the Euler number are
\begin{equation}
    c_2\cdot J_2 = 24,\ c_2\cdot J_3 = 56\,. 
\end{equation}
These are again obtained from $\mathcal{X}_8^{(3)}$ by setting $J_1=0$.
The Picard--Fuchs ideal is generated by
\begin{align}
    {\cal L}_1^{(2)}(\uz)&=\theta _2^2-z_2\left(2 \theta _2+\theta _3+1\right) \left(2 \theta _2+\theta _3+2\right),\\
    {\cal L}_2^{(3)}(\uz)&= \theta _3^2 \left(\theta _3-2 \theta _2\right)+4z_2 \theta _3^2 \left(2 \theta _2+\theta _3+1\right)-8 z_3\left(2 \theta _3+1\right) \left(4 \theta _3+1\right) \left(4 \theta _3+3\right).
\end{align}
\par The one-parameter model now is $\mathcal{X}_{4,2}$
\begin{equation}
    \mathcal{X}_8^{(2)} \xrightarrow{\Delta_2\rightarrow 0} \left(\begin{tabular}{c | c c c}
         $\P_{2,1^6}$ & 4&2&2
        \end{tabular}\right)_{-176}^{1,89}=\left(\begin{tabular}{c | c c c}
         $\P^5$ & 4&2
        \end{tabular}\right)_{-176}^{1,89}=\mathcal{X}_{4,2}\,,
\end{equation}
whose instanton numbers given in the following table can be obtained from the two-parameter model as before.
\begin{table}[H]
    \centering
    \begin{tabular}{|ccccc|}\hline
    $i_3=1$&$i_3=2$&$i_3=3$&$i_3=4$&$i_3=5$ \\\hline
1280& 92288& 15655168& 3883902528& 1190923282176 \\\hline
    \end{tabular}
    \caption{Instanton numbers for the model $\mathcal{X}_8^{(1)}=\mathcal{X}_{4,2}$.}
    \label{tab:instX8CICY1}
\end{table}
The Mori cone of $\mathcal{X}_{4,2}$ is generated by
\begin{equation}
    l_3=(-4,-2;1,1,1,1,1,1)
\end{equation}
with the triple intersection and the integrated second Chern class
\begin{equation}
    R=8J_3^3\,,\ c_2\cdot J_3=56\,.
\end{equation}
Its differential ideal is generated by
\begin{equation}
    {\cal L}_1^{(4)}(\uz) = \theta _3^4-16 z_3\left(2 \theta _3+1\right)^2 \left(4 \theta _3+1\right) \left(4 \theta _3+3\right).
\end{equation}

\subsection{Data for quotient of \texorpdfstring{$\P_{5,2,1,1,1}[10]$}{X10}}\label{sec:app10}
\subsubsection{Toric description}\label{sec:app101}
Next we discuss a quotient of $\P_{5,2,1,1,1}[10]$ by the group $\hat{S}$ generated by 
\begin{align}
\begin{split}
g_1=\Z_5:(0,2,1,1,1)\,,\quad g_2=\Z_{10}:(5,0,3,2,0)\,,\quad g_3=\Z_{10}:(5,0,1,3,1)\,.
\end{split}
\end{align}
The defining polynomial is 
\begin{equation}\label{eq:PX10}
P_{\mathcal{X}_{10}/\hat{S}}=a_1x_1^2+a_2x_2^5+a_3x_3^{10}+a_4x_4^{10}+a_5x_5^{10}-a_0\prod_{i=1}^5x_i-a_6x_1x_3^5-a_7x_4^5x_5^5\,.
\end{equation}
This corresponds to the polytope (excluding two points inside codimension one faces) given in \Cref{tab:X10}.
\begin{table}[H]
\renewcommand{\arraystretch}{1.2}
\centering
\begin{tabular}{| c | c c c c|c c c|}
    \hline \multicolumn{5}{|c |}{points} & \multicolumn{3}{| c|}{$l$-vectors} \\ \hline \hline
    (1 & 0&0&0&0)& 0 & 0 & -5\\ \hline
    (1 &1& 0& 0& 0)& 1 & 0 & 0\\
    (1 &0& 1& 0& 0)& 0 & 0 & 1\\
    (1 &0& 0& 1& 0)& 0 & 1 & 0\\
    (1 &0& 0& 1& 2)& 0 & 1 & 0\\
    (1 &-5& -2& -2& -2)& 1 & 0 & -2\\
    (1 &-2& -1& -1& -1)& -2 & 0 & 5\\
    (1 &0& 0& 1& 1)& 0 & -2 & 1\\\hline
    (1 &-1& 0& 0& 0)& - & - & -\\
    (1 &-3& -1& -1& -1)& - & - & -\\ \hline
\end{tabular}
\caption{Integral points and their scaling relations of polytope describing $\P_{5,2,1,1,1}[10]/\hat{S}$. The last two points lie inside a face of codimension one.}
\label{tab:X10}
\end{table}
The intersection ring is given by
\begin{equation}
R=40J_1^3 + 10J_1^2J_2 + 20J_1^2J_3 + 5J_1J_2J_3 + 10J_1J_3^2 + 2J_2J_3^2 + 4J_3^3\,.
\end{equation}
The topological invariants of the model are
\begin{align}
c_2\cdot J_1=100\,,\;c_2\cdot J_2=24\,,\;c_2\cdot J_3=52\,,\;\chi=-192\,,
\end{align}
and the Batyrev coordinates are now
\begin{align}
z_1 = \frac{1}{a_6^2}\,,\quad z_2 = \frac{1}{a_7^2}\,,\quad z_3 = \frac{a_6^5a_7}{a_0^5}\,.
\end{align}
The differential ideal is generated by
\begin{align}
\begin{split}
 {\cal L}_1^{(2)}(\uz) &= \theta_{2}^2 - z_{2} (2\theta_{2} - \theta_{3})(2\theta_{2} - \theta_{3} + 1)\,, \\
{\cal L}_2^{(2)}(\uz)&= \left( 560z_{1}^{2}z_3 - 1200z_{1}^{3} + 180z_{1}^{2} - 35z_{3} + 18z_{1} + 3 \right) \theta_{1}^{2}\\
&+ \left(-12\theta_{2} + \left( 6000z_{1}^{3} - 4000z_{1}^{2}z_{3} + 40z_{3} \right) \theta_{3}\right)\theta_{1}+ 30\theta_{2}\theta_{3}+ 180z_{1}^{2}z_{3}\\
&  + \left( -7500z_{1}^{3} + 8000z_{1}^{2}z_{3} - 750z_{1}^{2} + 50z_{1}z_{3} + 60z_{3} - 75z_{1} - 15 \right) \theta_{3}^{2}\\
&+\left( - 320z_{1}^{2}z_{3} - 600z_{1}^{3} + 280z_{1}z_{3} - 60z_{1}^{2} - 15z_{3} - 6z_{1}\right)\theta_{1}\\
&+ \left( 1500z_{1}^{3} + 2000z_{1}^{2}z_{3} - 610z_{1}z_{3} + 150z_{1}^{2} + 30z_{3} + 15z_{1} \right) \theta_{3}\,.
\end{split}
\end{align}

\subsubsection{Instanton numbers and Yukawa couplings}
The first non-vanishing instanton numbers are listed in \Cref{tab:instX10}. 
\begin{table}[H]
    \centering
    \begin{tabular}{|c|c|}
    \hline
    $(i_1,i_2,i_3)$ &$n_{i_1,i_2,i_3}$\\ \hline
    (0, 0, 1) & -2\\ \hline
    (0, 1, 0) & 2\\\hline
    (0, 1, 1) & -2\\\hline
    (0, 1, 2) & -4\\\hline
    (0, 1, 3) & -6\\\hline
    (0, 1, 4) & -8\\\hline
    (0, 2, 3) & -6\\\hline
    (1, 0, 0) & 30\\\hline
    (1, 0, 1) & 30\\\hline
    \end{tabular}
    \hspace{0.5cm} 
    \begin{tabular}{|c|c|}
    \hline
    $(i_1,i_2,i_3)$ &$n_{i_1,i_2,i_3}$\\ \hline
    (1, 1, 1) & 30\\\hline
    (1, 1, 2) & 90\\\hline
    (1, 1, 3) & 150\\\hline
    (2, 0, 1) & 1220\\\hline
    (2, 1, 1) & 1220\\\hline
    (2, 1, 2) & -870\\\hline
    (3, 0, 1) & 1220\\\hline
    (3, 0, 2) & 1220\\\hline
    (3, 1, 1) & 1220\\\hline
    (4, 0, 1) & 30\\\hline
    \end{tabular}
    \caption{Instanton numbers for the family $\mathcal{X}_{10}$.}
    \label{tab:instX10}
\end{table}
These obey the following relations
\begin{equation}
    n_{i_1,i_2,i_3}=n_{5i_3-i_1,i_2,i_3}\quad\text{and}\quad n_{i_1,i_2,i_3}=n_{i_1,i_3-i_2,i_3}.
\end{equation}
The Yukawa couplings for this model are too lengthy to write them out here. 
Instead, we give their conifold locus as the vanishing set of 

\begin{equation}
    \begin{split}
    \Delta_c&=1 - 4 \Big[2 - 25 z_1 (1 - 20 z_1) z_3 + 2 \Big(8 - 32 z_2 - 25 z_1 \big(8 - 32 z_2 \\
    &\quad -5 z_1 (67 - 68 z_2 - 100 z_1 (7 - 4 z_2 + 4 z_1 (-13 + 4 z_1 + 4 (7 + 4 z_1) z_2))\big)\Big) z_3^2 \\
    &\quad - 12500 z_1^2 (1 - 4 z_1)^3 (2 - 25 z_1 (1 - 20 z_1)) (1 - 4 z_2) z_3^3 + 9765625 (1 - 4 z_1)^6 z_1^4 (1 - 4 z_2)^2 z_3^4 \Big]
    \end{split}
\end{equation}

\subsubsection{Analytical continuation and monodromies}
The Frobenius basis in the coordinates $u_1=\sqrt{z_1}^{-1},u_2=\sqrt{z_2}^{-1},u_3=\sqrt{z_1}^5\sqrt{z_2}z_3$ reads
\begin{equation}
\begin{alignedat}{2}
&\varpi_1^u=\sigma_1\,,\quad 
&&\hspace{-2cm}\varpi_2^u=\sigma_1\log(u_3)+\sigma_2\,,\\
&\varpi_3^u=\sigma_1\log(u_3)^2+2\sigma_2\log(u_3)+\sigma_3\,,\\
&\varpi_4^u=\sigma_1\log(u_3)^3+3\sigma_2\log(u_3)^2+3\sigma_3\log(u_3)+\sigma_4\,,\\
&\varpi_5^u=\sigma_5\,,\quad
&&\hspace{-2cm}\varpi_6^u=\sigma_5\log(u_3)+\sigma_6\,,\\
&\varpi_7^u=\sigma_7\,,\quad
&&\hspace{-2cm}\varpi_8^u=\sigma_7\log(u_3)+\sigma_8\,,
\end{alignedat}
\end{equation}
with
\begin{equation}
\begin{alignedat}{2}
&\sigma_1 = 1 + 15120u_3^2+\dots\,, \quad
&&\sigma_2 = 89760u_3^2+\dots\,, \\
&\sigma_3 = \frac{{u_1^2}}{4}+ \frac{{u_2^2}}{4} + 115600u_3^2+\dots\,, \quad
&&\sigma_4 = 173400u_3^2+\dots\,, \\
&\sigma_5 = u_2 + 240u_1u_3+\dots\,, \quad
&&\sigma_6 = 1120u_1u_3+\dots\,, \\
&\sigma_7 = u_1 + 128u_2u_3+\dots\,, \quad
&&\sigma_8 = 224u_2u_3+\dots.
\end{alignedat}
\end{equation}
The transition matrix is 
\begin{align}
T_u = \left( \begin{array}{cccccccc}
1 & 0 & 0 & 0 & 0 & 0 & 0 & 0 \\
-\frac{1}{2} & 0 & 0 & 0 & 0 & 0 & \frac{1}{2\pi} & 0 \\
-\frac{1}{2} & 0 & 0 & 0 & \frac{1}{2\pi} & 0 & 0 & 0 \\
\frac{3}{2} & \frac{1}{2\pi \ii} & 0 & 0 & -\frac{1}{4\pi} & 0 & -\frac{5}{4\pi} & 0 \\
\frac{35}{12} & 0 & \frac{1}{2\pi^2} & 0 & -\frac{1}{4\pi} & 0 & -\frac{5}{4\pi} & 0 \\
\frac{29}{24} & \frac{\ii}{4\pi} & \frac{1}{4\pi^2} & 0 & \frac{-4\ii\ln(2) + \pi}{8\pi^2} & \frac{1}{4\ii\pi^2} & -\frac{5}{8\pi} & 0 \\
\frac{85}{24} & \frac{5\ii}{4\pi} & \frac{5}{4\pi^2} & 0 & -\frac{5}{8\pi} & 0 & \frac{-60\ii\ln(2) + 5\pi}{8\pi^2} & \frac{5}{4\ii\pi^2} \\
\frac{\chi\left(\mathcal{X}_{10}^{(1)}\right)\zeta(3)}{2(2\pi\ii)^3} & \frac{\tilde{c}_2\left(\mathcal{X}_{10}^{(1)}\right)}{24\cdot 2\pi\ii} & 0 & \frac{\ii}{12\pi^3} & \frac{\ln(2)}{4\ii\pi^2} & \frac{1}{8\ii\pi^2} & \frac{15\ln(2)}{4\ii\pi^2} & \frac{5}{8\ii\pi^2}
\end{array}\right)
\end{align}
with $\chi\left(\mathcal{X}_{10}^{(1)}\right)=-288$ and $\tilde{c}_2\left(\mathcal{X}_{10}^{(1)}\right)=34$.
\subsubsection{Conifold transition}\label{sec:appX10contraf}
\text{}

While we are currently unable to identify the three-parameter CICY corresponding to the quotient family $\mathcal{X}_{10}^{(3)}$, we found the two- and one-parameter CICY it transitions to for $q_1,q_2\rightarrow 0$. The three-parameter model and a two-parameter model, which is obtained by the conifold transition $\Delta_2\rightarrow 0$ and discussed below, are neither represented in the list of hypersurfaces in toric ambient spaces \cite{kreuzerskarke} nor in that of CICYs in products of projective spaces \cite{CICYlist}. We conclude that these must be of a more general type, such as CICYs in products of weighted projective spaces as is the case for the model $\mathcal{X}_8^{(3)}$ discussed in \cref{sec:app8}. The first transition reads
\begin{equation}
    \mathcal{X}_{10}^{(3)} \xrightarrow{\Delta_1\rightarrow 0} \left(\begin{tabular}{c | c c c}
         $\P^1$ & 0&0&2 \\
         $\P_{3,2,1^4}$ & 6&2&1
        \end{tabular}\right)^{2,128}_{-252}\eqqcolon \mathcal{X}_{10}^{(2)}\,.
\end{equation}
This CICY $\mathcal{X}_{10}^{(2)}$ is equivalent to the hypersurface family  $P_{6,2,2,1,1}[12]$. The instanton numbers are again obtained from the same summation relation as for $\mathcal{X}_6$ and are listed in the following table.

\begin{table}[H]
    \centering
    \begin{tabular}{|c|cccccc|}\hline
    &$i_3=0$&$i_3=1$&$i_3=2$&$i_3=3$&$i_3=4$&$i_3=5$ \\\hline
 $i_2=0$&- & 2496 & 223752 & 38637504 & 9100224984 & 2557481027520 \\
 $i_2=1$&2 & 2496 & 1941264 & 1327392512 & 861202986072 & 540194037151104 \\
 $i_2=2$&0 & 0 & 223752 & 1327392512 & 2859010142112 & 4247105405354496 \\
 $i_2=3$&0 & 0 & 0 & 38637504 & 861202986072 & 4247105405354496 \\
 $i_2=4$&0 & 0 & 0 & 0 & 9100224984 & 540194037151104 \\
 $i_2=5$&0 & 0 & 0 & 0 & 0 & 2557481027520 \\\hline
    \end{tabular}
    \caption{Instanton numbers of $\mathcal{X}_{10}^{(2)}$.}
    \label{tab:instX10CICY2}
\end{table}

The Mori cone of the model $\mathcal{X}_{10}^{(2)}$ is spanned by 
\begin{align}
    l_2 &= (0,0,-2;1,1,0,0,0,0,0,0)\,,\\
    l_3 &= (-6,-2,-1;0,0,3,2,1,1,1,1)\,,
\end{align}
whose dual Kähler cone generators have the triple intersections
\begin{equation}
    R=2J_2J_3^2+4J_3^3\,.
\end{equation}
Their intersection with the second Chern class are
\begin{equation}
    c_2\cdot J_2 = 24\,,\ c_2\cdot J_3 = 52\,. 
\end{equation}
Both $R$ and $c_2\cdot J$ are again obtained from $\mathcal{X}_{10}^{(3)}$ by setting $J_1=0$.
The Picard--Fuchs ideal is generated by
\begin{align}
    {\cal L}_1^{(2)}(\uz)&=\theta _2^2-z_2\left(2 \theta _2+\theta _3+1\right) \left(2 \theta _2+\theta _3+2\right),\\
    {\cal L}_2^{(3)}(\uz)&= \theta _3^2 \left(\theta _3-2 \theta _2\right)+4z_2 \theta _3^2 \left(2 \theta _2+\theta _3+1\right)-24z_3 \left(2 \theta _3+1\right) \left(6 \theta _3+1\right) \left(6 \theta _3+5\right).
\end{align}
\par The one-parameter model now is $\mathcal{X}_{6,2}$
\begin{equation}
    \mathcal{X}_{10}^{(2)} \xrightarrow{\Delta_2\rightarrow 0} \left(\begin{tabular}{c | c c c}
         $\P_{3,2,1^5}$ & 6&2&2
        \end{tabular}\right)_{-256}^{1,129}=\left(\begin{tabular}{c | c c c}
         $\P_{3,1^5}$ & 6&2
        \end{tabular}\right)_{-256}^{1,129}=\mathcal{X}_{6,2}\,,
\end{equation}
whose instanton numbers coincide with the sum over $i_2$ in the two-parameter model before:
\begin{table}[H]
    \centering
    \begin{tabular}{|ccccc|}\hline
    $i_3=1$&$i_3=2$&$i_3=3$&$i_3=4$&$i_3=5$ \\\hline
4992& 2388768& 2732060032& 4599616564224& 9579713847066240 \\\hline
    \end{tabular}
    \caption{Instanton numbers for the model $\mathcal{X}_{10}^{(1)}=\mathcal{X}_{6,2}$.}
    \label{tab:instX10CICY1}
\end{table}
The Mori cone of $\mathcal{X}_{4,2}$ is generated by
\begin{equation}
    l_3=(-6,-2;3,1,1,1,1,1)
\end{equation}
with the triple intersection and the integrated second Chern class
\begin{equation}
    R=4J_3^3\,,\ c_2\cdot J_3=52\,.
\end{equation}
Its differential ideal is generated by
\begin{equation}
    {\cal L}_1^{(4)}(\uz) = \theta _3^4-48 z_3 \left(6 \theta _3+1\right)\left(2 \theta _3+1\right)^2 \left(6 \theta _3+5\right).
\end{equation}

\par
In contrast to the models $\mathcal{X}_6^{(3)}$ and $\mathcal{X}_8^{(3)}$, here, the Kähler moduli space is not symmetric in the first two coordinates. Therefore, there should be another two-parameter model that arises from $\mathcal{X}_{10}^{(3)}$ as $\Delta_2\rightarrow 0$. Although we cannot give its toric description as of the writing of this article, we know that its topological data is given as follows: its intersection ring reads
\begin{equation}
    R = 40 J_1^3+20J_1^2J_3+10J_1J_3^2+4J_3^3\,,
\end{equation}
the second Chern class and Euler number are
\begin{equation}\label{eq:topdataX102}
    c_2\cdot J_1 = 100\,,\ c_2\cdot J_3=52\,\ \chi=\chi\left(\mathcal{X}_{10}^{(3)}\right)-2\cdot n_{0,0}=-196\,,
\end{equation}
and its first couple of instanton numbers are as listed in \Cref{tab:instrelX10_2}. Here, we used that the constant terms in the prepotential (cf.\ \eqref{eq:prepotential}) of $\mathcal{X}_{10}^{(3)}$ for $q_2\rightarrow 1$ appear as
\begin{equation}\label{eq:chifrominstantons}
    \frac{\chi\left(\mathcal{X}_{10}^{(3)}\right)\,\zeta(3)}{2(2\pi\ii)}-\frac{1}{(2\pi\ii)^3}\sum_{i=1}^\infty n_{0,i,0}\mathrm{Li}_3(1)
\end{equation}
with $\mathrm{Li}_3(1)=\zeta(3)$ and $\sum_{i=1}^\infty n_{0,i,0}=n_{0,0}$, resulting in the constant term 
\begin{equation}
    \frac{\left[\left(\mathcal{X}_{10}^{(3)}\right) -2n_{0,0}\right]\zeta(3)}{(2\pi\ii)^3}
\end{equation}
in the prepotential of $\mathcal{X}_{10}^{(2)}$.
\begin{table}[H]
    \centering
    \begin{tabular}{|c|cccccc|}\hline
    &$i_3=0$&$i_3=1$&$i_3=2$&$i_3=3$&$i_3=4$&$i_3=5$ \\\hline
 $i_1=0$&(2) & -4 & -4 & -12 & -48 & -240 \\
 $i_1=1$&30 & 60 & 90 & 300 & 1470 & 8640 \\
 $i_1=2$&0 & 2440 & -870 & -3480 & -21060 & -148080 \\
 $i_1=3$&0 & 2440 & 8940 & 31680 & 211450 & 1715140 \\
 $i_1=4$&0 & 60 & 536290 & -255420 & -1728600 & -15568200 \\
 $i_1=5$&0 & -4 & 1299876 & 3938928 & 15256162 & 128522400 \\ \hline
    \end{tabular}
   \caption{Instanton numbers for the model $\mathcal{X}_{10}^{(2)}$ obtained via $q_2\rightarrow 1$ in $\mathcal{X}_{10}^{(3)}$. The entry $n_{0,0}=2$ merely allows to give the Euler number as in \eqref{eq:topdataX102} and is not considered an instanton number.}
    \label{tab:instrelX10_2}
\end{table}
We obtain the holomorphic period at the MUM point from that of $\mathcal{X}_{10}^{(3)}$ by setting $z_2=1/4$ and rescaling $z_3\rightarrow 2z_3$ for integrality of the coefficients. This allows us to determine the Picard--Fuchs-ideal, which is generated by
\begin{align}
    {\cal L}_1^{(2)}(\uz)&=\theta _1 \left(\theta _1-2 \theta _2\right)-z_1\left(2 \theta _1-5 \theta _2\right) \left(2 \theta _1-5 \theta _2+1\right) ,\\
    \begin{split}
    {\cal L}_2^{(3)}(\uz) &= \left(2 \theta _1-5 \theta _2\right) \theta _2^2 
    +2  z_2\left(\theta _1-2 \theta _2\right) \left(35 \theta _1^2+288 \theta _2 \theta _1-5 \left(2 \theta _2+1\right)^2\right)\\
    &\quad +2 z_1 z_2\left(-450 \theta _1^3+7485 \theta _2^2 \theta _1-7750 \theta _2^3+\left(\theta _1 \left(159-992 \theta _1\right)+30\right) \theta _2\right) \\
    &\quad +20  z_1^2 z_2\left(124 \theta _1^3+6 \left(17-26 \theta _2\right) \theta _1^2-\theta _2 \left(1625 \theta _2+203\right) \theta _1+3200 \theta _2^3+52 \theta _2+6\right).
    \end{split}
\end{align}

\section{Data for four-folds}
\subsection{Data for quotient of \texorpdfstring{$\P_{12,8,1,1,1,1}[24]$}{X24}}\label{sec:appD}

\subsubsection{Integral basis}
We write the period vector as $\vec{\Pi}=(X^0,X^1,X^2,H_1,H_2,F_2,F_1,F_0)^{T}$ and choose for the primitive cohomology $H^{2,2}_{\text{prim}}$ the basis $J_1J_2,J_2^2$. Following the discussion of \cref{sec:CY4}, we perform a basis change from the asymptotic period vector $\vec{\Pi}^{\text{asy}}$ in \eqref{eq:ident} to one with block-anti-diagonal intersection form. This new period vector has the asymptotic behaviour
\begin{align}
\vec{\Pi}\sim\scriptsize 
\left(\begin{array}{c}
 1 \\ -t^1 \\ -t^2 \\ 2 (t^1)^2+t^2 t^1+2 t^1-2 \\ 8 (t^1)^2+4 t^2 t^1+8 t^1+\frac{(t^2)^2}{2}+\frac{t^2}{2}-\frac{91}{12} \\ -\frac{8}{3} (t^1)^3-2 t^2
   (t^1)^2-(t^1)^2-\frac{1}{2} (t^2)^2 t^1-\frac{1}{2} t^2 t^1+\frac{37 t^1}{4}+\frac{\frac{55 \pi ^3}{4}+360 i \zeta (3)}{3 \pi ^3}-\frac{43}{12} \\ -\frac{32}{3} (t^1)^3-8
   t^2 (t^1)^2-16 (t^1)^2-2 (t^2)^2 t^1-8 t^2 t^1-8 t^1-\frac{(t^2)^3}{6}-(t^2)^2-\frac{41 t^2}{4}+\frac{965 i \zeta (3)}{2 \pi ^3}+\frac{91}{6} \\ \frac{i (-3860 t^1-960
   t^2) \zeta (3)}{8 \pi ^3}+\frac{8 (t^1)^4}{3}+\frac{8}{3} t^2 (t^1)^3+(t^2)^2 (t^1)^2+\frac{91 (t^1)^2}{6}+\frac{1}{6} (t^2)^3 t^1+\frac{91}{12} t^2 t^1+(t^2)^2-\frac{43}{6}
   \\
\end{array}\right).
\end{align}

Then, the matrix block in $\Sigma$ (cf.\ \eqref{eq:sigma}) giving the dual intersection form on $H_{2,2}^{\text{prim}}(\hat{Y},\Z)$ reads
 \begin{align}
 \eta^{(2,2)}= \begin{pmatrix}
      -4 & 1 \\
      1 & 0 \\
     \end{pmatrix}.
 \end{align}
In this basis, the monodromies around the two MUM-divisors are 

\begin{align}
\frak{M}_{D_1} = {\tiny \begin{pmatrix}
 1 & 0 & 0 & 0 & 0 & 0 & 0 & 0 \\
 -1 & 1 & 0 & 0 & 0 & 0 & 0 & 0 \\
 0 & 0 & 1 & 0 & 0 & 0 & 0 & 0 \\
 4 & -4 & -1 & 1 & 0 & 0 & 0 & 0 \\
 16 & -16 & -4 & 0 & 1 & 0 & 0 & 0 \\
 -2 & 2 & 2 & 0 & -1 & 1 & 0 & 0 \\
 -65 & 32 & 14 & 0 & -4 & 0 & 1 & 0 \\
 33 & -33 & 0 & 0 & 0 & 0 & -1 & 1 \\
\end{pmatrix}},\quad
\frak{M}_{\Delta_\infty} ={ \tiny \begin{pmatrix}
 1 & 0 & 0 & 0 & 0 & 0 & 0 & 0 \\
 0 & 1 & 0 & 0 & 0 & 0 & 0 & 0 \\
 -1 & 0 & 1 & 0 & 0 & 0 & 0 & 0 \\
 0 & -1 & 0 & 1 & 0 & 0 & 0 & 0 \\
 1 & -4 & -1 & 0 & 1 & 0 & 0 & 0 \\
 -2 & -1 & 0 & -1 & 0 & 1 & 0 & 0 \\
 -19 & 2 & 2 & 0 & -1 & 0 & 1 & 0 \\
 2 & -17 & -2 & 0 & 0 & -1 & 0 & 1 \\
\end{pmatrix} }.
\end{align}
By analytically continuing the above basis to all intersections of the discriminant loci in \Cref{fig:ModuliSpace}, we verified that the above basis has integral monodromy representations which preserve the intersection matrix
\begin{align*}
\frak{M}^{T}\Sigma \frak{M}=\Sigma\,.  
\end{align*}
The remaining monodromies read

\begin{align}
\frak{M}_{D_1} ={\tiny \begin{pmatrix}
 1 & 0 & 0 & 0 & 0 & 0 & 0 & 0 \\
 -1 & 1 & 0 & 0 & 0 & 0 & 0 & 0 \\
 0 & 0 & 1 & 0 & 0 & 0 & 0 & 0 \\
 4 & -4 & -1 & 1 & 0 & 0 & 0 & 0 \\
 16 & -16 & -4 & 0 & 1 & 0 & 0 & 0 \\
 -2 & 2 & 2 & 0 & -1 & 1 & 0 & 0 \\
 -65 & 32 & 14 & 0 & -4 & 0 & 1 & 0 \\
 33 & -33 & 0 & 0 & 0 & 0 & -1 & 1 \\
\end{pmatrix}},\quad 
\frak{M}_{D_1^{\prime}} = { \tiny \begin{pmatrix}
 1 & 1 & 0 & 0 & 0 & 0 & 0 & 0 \\
 0 & 1 & 0 & 0 & 0 & 0 & 0 & 0 \\
 0 & 0 & 1 & 1 & 0 & 0 & 0 & 0 \\
 0 & -4 & 0 & 1 & 0 & 0 & 0 & 0 \\
 0 & 0 & -2 & -2 & 1 & 1 & 0 & 0 \\
 0 & 2 & 0 & -2 & 0 & 1 & 0 & 0 \\
 -1 & -33 & 6 & 22 & -4 & -4 & 1 & 1 \\
 0 & -1 & 0 & 0 & 0 & 0 & 0 & 1 \\
\end{pmatrix} },
\end{align}

\begin{align}
\frak{M}_{\Delta_\infty} = { \tiny \begin{pmatrix}
 1 & 0 & 0 & 0 & 0 & 0 & 0 & 0 \\
 0 & 1 & 0 & 0 & 0 & 0 & 0 & 0 \\
 -1 & 0 & 1 & 0 & 0 & 0 & 0 & 0 \\
 0 & -1 & 0 & 1 & 0 & 0 & 0 & 0 \\
 1 & -4 & -1 & 0 & 1 & 0 & 0 & 0 \\
 -2 & -1 & 0 & -1 & 0 & 1 & 0 & 0 \\
 -19 & 2 & 2 & 0 & -1 & 0 & 1 & 0 \\
 2 & -17 & -2 & 0 & 0 & -1 & 0 & 1 \\
\end{pmatrix} },
\end{align}

\begin{align}
\frak{M}_{\Delta_2} &= { \tiny \begin{pmatrix}
 1 & 0 & 0 & 0 & 0 & 0 & 0 & 0 \\
 0 & -32 & -9 & -2 & 2 & -4 & 1 & 0 \\
 0 & 132 & 37 & 8 & -8 & 16 & -4 & 0 \\
 0 & 66 & 18 & 5 & -4 & 8 & -2 & 0 \\
 0 & 198 & 54 & 12 & -11 & 24 & -6 & 0 \\
 0 & 297 & 81 & 18 & -18 & 37 & -9 & 0 \\
 0 & 1089 & 297 & 66 & -66 & 132 & -32 & 0 \\
 0 & 0 & 0 & 0 & 0 & 0 & 0 & 1 \\
\end{pmatrix} },\quad 
\frak{M}_{\Delta_1} ={ \tiny \begin{pmatrix}
 0 & 0 & 0 & 0 & 0 & 0 & 0 & -1 \\
 0 & 1 & 0 & 0 & 0 & 0 & 0 & 0 \\
 0 & 0 & 1 & 0 & 0 & 0 & 0 & 0 \\
 0 & 0 & 0 & 1 & 0 & 0 & 0 & 0 \\
 0 & 0 & 0 & 0 & 1 & 0 & 0 & 0 \\
 0 & 0 & 0 & 0 & 0 & 1 & 0 & 0 \\
 0 & 0 & 0 & 0 & 0 & 0 & 1 & 0 \\
 -1 & 0 & 0 & 0 & 0 & 0 & 0 & 0 \\
\end{pmatrix} },
\end{align}

\begin{align}
\frak{M}_{D_2} ={ \tiny \begin{pmatrix}
 1 & 0 & 0 & 0 & 0 & 0 & 0 & 0 \\
 -1 & -32 & -9 & -2 & 2 & -4 & 1 & 0 \\
 1 & 132 & 37 & 8 & -8 & 16 & -4 & 0 \\
 3 & 30 & 8 & 3 & -2 & 4 & -1 & 0 \\
 12 & 186 & 51 & 12 & -11 & 24 & -6 & 0 \\
 2 & 297 & 82 & 19 & -19 & 37 & -9 & 0 \\
 -34 & 1107 & 306 & 66 & -69 & 132 & -32 & 0 \\
 18 & -16 & 3 & 1 & -1 & 1 & -1 & 1 \\
\end{pmatrix} },\quad
\frak{M}_{D_2^{\prime}} = { \tiny \begin{pmatrix}
 -2 & -18 & -2 & -1 & 0 & -1 & 0 & -1 \\
 0 & 1 & 0 & 0 & 0 & 0 & 0 & 0 \\
 1 & 1 & 1 & 1 & 0 & 0 & 0 & 0 \\
 0 & -3 & 0 & 1 & 0 & 0 & 0 & 0 \\
 0 & 4 & -1 & -1 & 1 & 1 & 0 & 0 \\
 2 & 2 & 0 & -1 & 0 & 1 & 0 & 0 \\
 16 & -14 & 3 & 19 & -3 & -3 & 1 & 1 \\
 -1 & -1 & 0 & 0 & 0 & 0 & 0 & 0 \\
\end{pmatrix} },
\end{align}

\begin{align}
\frak{M}_{D_3} = { \tiny \begin{pmatrix}
 1 & 0 & 0 & 0 & 0 & 0 & 0 & 0 \\
 -1 & 67 & 19 & 0 & -3 & 8 & -2 & 0 \\
 2 & -264 & -75 & 0 & 12 & -32 & 8 & 0 \\
 2 & -101 & -29 & -1 & 5 & -12 & 3 & 0 \\
 9 & -404 & -116 & 0 & 19 & -48 & 12 & 0 \\
 5 & -628 & -180 & 0 & 28 & -75 & 19 & 0 \\
 -8 & -2170 & -622 & 0 & 97 & -264 & 67 & 0 \\
 8 & -33 & -4 & 1 & 0 & -2 & 0 & 1 \\
\end{pmatrix} },\quad
\frak{M}_{D_3^{\prime}} ={ \tiny \begin{pmatrix}
 1 & 17 & 2 & 0 & 0 & 1 & 0 & 2 \\
 0 & 1 & 0 & 0 & 0 & 0 & 0 & 0 \\
 -1 & -17 & -1 & 0 & 0 & -1 & 0 & -1 \\
 0 & -2 & 0 & 1 & 0 & 0 & 0 & 0 \\
 1 & 9 & 0 & 0 & 1 & 1 & 0 & 0 \\
 -2 & -35 & -4 & -2 & 0 & -1 & 0 & -2 \\
 -19 & -315 & -33 & 0 & -2 & -19 & 1 & -16 \\
 2 & 18 & 2 & 1 & 0 & 1 & 0 & 1 \\
\end{pmatrix} },
\end{align}

\begin{align}
\frak{M}_{D_4} &={ \tiny \begin{pmatrix}
 1 & 0 & 0 & 0 & 0 & 0 & 0 & 0 \\
 0 & -32 & -9 & 2 & 1 & -4 & 1 & 0 \\
 -1 & 132 & 37 & -8 & -4 & 16 & -4 & 0 \\
 0 & 65 & 18 & -3 & -2 & 8 & -2 & 0 \\
 1 & 194 & 53 & -12 & -5 & 24 & -6 & 0 \\
 -2 & 296 & 81 & -19 & -9 & 37 & -9 & 0 \\
 -19 & 1091 & 299 & -66 & -34 & 132 & -32 & 0 \\
 2 & -17 & -2 & 0 & 0 & -1 & 0 & 1 \\
\end{pmatrix} }, \quad
\frak{M}_{D_4^{\prime}} ={ \tiny \begin{pmatrix}
 0 & 0 & 0 & 0 & 0 & 0 & 0 & -1 \\
 0 & 1 & 0 & 0 & 0 & 0 & 0 & 0 \\
 0 & 0 & 1 & 0 & 0 & 0 & 0 & 1 \\
 0 & -1 & 0 & 1 & 0 & 0 & 0 & 0 \\
 0 & -4 & -1 & 0 & 1 & 0 & 0 & -1 \\
 0 & -1 & 0 & -1 & 0 & 1 & 0 & 2 \\
 0 & 2 & 2 & 0 & -1 & 0 & 1 & 19 \\
 -1 & -17 & -2 & 0 & 0 & -1 & 0 & -2 \\
\end{pmatrix} }, 
\end{align}

\begin{align} 
\frak{M}_{\Delta_0} = { \tiny \begin{pmatrix}
 1 & -1 & 0 & 0 & 0 & 0 & 0 & 0 \\
 1 & 0 & 0 & 0 & 0 & 0 & 0 & 0 \\
 -4 & 0 & 1 & -1 & 0 & 0 & 0 & 0 \\
 -4 & 4 & 1 & 0 & 0 & 0 & 0 & 0 \\
 -6 & 0 & 2 & -2 & 1 & -1 & 0 & 0 \\
 -6 & -2 & 0 & -2 & 1 & 0 & 0 & 0 \\
 -32 & 0 & 2 & -16 & 4 & 0 & 1 & -1 \\
 0 & -32 & -6 & -10 & 4 & -4 & 1 & 0 \\
\end{pmatrix} },
\end{align}
\begin{align}
\frak{M}_{S_0} &={ \tiny \begin{pmatrix}
 0 & 0 & 0 & 0 & 0 & 0 & 0 & -1 \\
 0 & -32 & -6 & -10 & 4 & -4 & 1 & -1 \\
 0 & 132 & 25 & 40 & -16 & 16 & -4 & 3 \\
 0 & 100 & 18 & 31 & -12 & 12 & -3 & 3 \\
 0 & 202 & 37 & 60 & -23 & 24 & -6 & 6 \\
 0 & 266 & 48 & 81 & -32 & 33 & -8 & 6 \\
 0 & 1091 & 197 & 330 & -131 & 132 & -32 & 16 \\
 -1 & 19 & 2 & 1 & 0 & 1 & 0 & -2 \\
\end{pmatrix} },
\end{align}
\begin{align}
\frak{M}_{E_0} &={ \tiny \begin{pmatrix}
 0 & -1 & 0 & 0 & 0 & 0 & 0 & -1 \\
 1 & -33 & -6 & -10 & 4 & -4 & 1 & -1 \\
 -4 & 132 & 25 & 39 & -16 & 16 & -4 & 4 \\
 -4 & 136 & 25 & 40 & -16 & 16 & -4 & 4 \\
 -6 & 198 & 38 & 58 & -23 & 23 & -6 & 6 \\
 -6 & 196 & 36 & 58 & -23 & 24 & -6 & 6 \\
 -32 & 1089 & 200 & 314 & -128 & 132 & -32 & 32 \\
 -1 & 1 & 0 & 0 & 0 & 0 & 0 & 0 \\
\end{pmatrix} }.
\end{align}
We have written side by side matrices that, as a consequence of the involution symmetry $\mathcal{I}$ acting on the moduli space, are related by conjugation. For example, we have 
\begin{align}
\frak{M}_{1}=\frak{A}^{6} \frak{M}_{D_1^{\prime}} \frak{A}^{-6},
\end{align}
where again
\begin{align}
    \frak{A}=\frak{M}_{\Delta_1}\frak{M}_{\Delta_2}\frak{M}_2\,.
\end{align}
\subsubsection{Analytical continuation}
For usage in the text we explicitly give the results of the analytic continuation to the region around the point $z_1=\Delta_2=1-256z_2=0$. The Frobenius basis in the coordinates $z_1,\Delta_2$ is
\begin{align}
\begin{split}
&\varpi_1=\sigma_1\,,\quad
\varpi_2=\sigma_2\,,\quad
\varpi_3=\sigma_3\,,\\
&\varpi_4=(\sigma_1+60\sigma_3)\log(z_1)+\sigma_4\,,\\
&\varpi_5=(\sigma_1+60\sigma_3)\log(z_1)^2+\left(-\frac{1}{2}\sigma_2+744\sigma_3+2\sigma_4\right)\log(z_1)+\sigma_5\,,\\
&\varpi_6=(\sigma_1+60\sigma_3)\log(z_1)^3+\left(-\frac{3}{4}\sigma_2+1116\sigma_3+3\sigma_4\right)\log(z_1)^2\\ &\qquad+\left(360\sigma_3+3\sigma_5\right)\log(z_1)+\sigma_6\,,\\
&\varpi_7=(\sigma_1+60\sigma_3)\log(z_1)^4+\left(-\sigma_2+1488\sigma_3+4\sigma_4\right)\log(z_1)^3\\ &\qquad +\left(720\sigma_3+6\sigma_5\right)\log(z_1)^2
+\left(-1440\sigma_3+4\sigma_6\right)\log(z_1)+\sigma_7\,,\\
&\varpi_8=\Delta_2^{3/2}\sigma_8\,,
\end{split}
\end{align}
where
\begin{equation}
\begin{alignedat}{2}
&\sigma_1=1 - \frac{24255 z_1^2}{2} + \frac{\Delta_2^2}{64} + 15 z_1 \Delta_2+\dots\,,\quad
&&\sigma_2=\Delta_2 + 13860 z_1^2 + \frac{9 \Delta_2^2}{16}+\dots\,,\\
&\sigma_3=z_1 - \frac{z_1 \Delta_2}{4} + \frac{3465 z_1^2}{8} - \frac{\Delta_2^2}{3840}+\dots\,,\quad
&&\sigma_4=78 z_1 \Delta_2 - \frac{118431 z_1^2}{2} + \frac{9 \Delta_2^2}{80}+\dots\,,\\
&\sigma_5=-156 z_1 \Delta_2 + 131319 z_1^2 + \frac{3 \Delta_2^2}{32}+\dots\,,\quad
&&\sigma_6=-180 z_1 \Delta_2 - \frac{3 \Delta_2^2}{32} + 155520 z_1^2+\dots\,,\\
&\sigma_7=720 z_1 \Delta_2 - 622080 z_1^2+\dots\,,\quad
&&\sigma_8=1 + \frac{7 \Delta_2}{8} + 6 z_1 \Delta_2 + \frac{677 \Delta_2^2}{896}+\dots\,.
\end{alignedat}
\end{equation}
The transition matrix determined up to 50 digits is given on the following page. The constants $c_1,\dots,c_9$ are undetermined coefficients that are in part related via Legendre relations. Their numerical values are
\begin{align}
\begin{split}
    &c_1=\left(-0.01137020804807324115645436753579786935025280881041\dots\right)\ii,\\
    &c_2=\left(-49.74036836198296197122972210454792141879531840041397\dots\right)\ii,\\
    &c_3=\left(0.86146429753394651653383174024754038629117804246408\dots\right)\ii,\\
    &c_4=\left(4.97048721716722460797722430347979820970974411063181\dots\right)+\\
    &\qquad\quad \left(4.17964065527656611266054563948178518353389612265763\dots\right)\ii,\\
    &c_5=\left(0.14205037010715961297607488910276182427846535064952\dots\right)+\\
    &\qquad\quad \left(-0.01149316430287095333651772362877814091469418669510\dots\right)\ii,\\
    &c_6=\left(179.47983785957501748568545472154643904005732998370000\dots\right)+\\
    &\qquad\quad \left(645.48041853007703143539752889871662356411831978047000\dots\right)\ii,\\
    &c_7=-3.83947524943659277371848520161485985445918253577120\dots,\\
    &c_8=0.45930055964455129790789789016629125905021273110251\dots,\\
    &c_9=-984.69426387610843724544581483308790982392878222589000\dots.\\
    \end{split}
\end{align}

\scalebox{0.8}{
\begin{minipage}[t][40cm][c]{\textwidth}
\rotatebox{90}{%
\parbox{1\textwidth}{%
\begin{align*}
\begin{aligned}
&\hspace{18cm}
   T=\left(\begin{matrix}
 1 & 0 \\[0.4cm]
 \frac{\pi  c_3-4 i \ln (2)}{4 \pi } & -c_1 \\[0.4cm]
 -c_3 & \frac{8 \pi  c_1-i}{2 \pi } \\[0.4cm]
 \frac{\pi ^3 (-51 c_3+24 c_4-182)-2895 i \zeta (3)-64 i \ln ^3(2)-192 \pi  \ln ^2(2)+258 i \pi ^2 \ln (2)}{36 \pi ^3} & \frac{\pi ^3 (272 c_1+32
   c_5)+9 i \pi ^2-32 i \ln ^2(2)-64 \pi  \ln (2)}{48 \pi ^3} \\[0.4cm]
 \frac{\pi ^2 (-18 c_3-91)-96 \ln ^2(2)+96 i \pi  \ln (2)}{12 \pi ^2} & \frac{24 \pi ^2 c_1+i \pi -8 \ln (2)}{4 \pi ^2} \\[0.4cm]
 \frac{\pi ^3 (-69 c_3-12 c_4+91)+5790 i \zeta (3)+128 i \ln ^3(2)+96 \pi  \ln ^2(2)+240 i \pi ^2 \ln (2)}{36 \pi ^3} & \frac{\pi ^3 (184 c_1-8
   c_5)-3 i \pi ^2+32 i \ln ^2(2)+16 \pi  \ln (2)}{24 \pi ^3} \\[0.4cm]
 \frac{\pi ^3 (182-99 c_3)+5790 i \zeta (3)+128 i \ln ^3(2)+192 \pi  \ln ^2(2)-96 i \pi ^2 \ln (2)}{12 \pi ^3} & \frac{264 \pi ^3 c_1-41 i \pi ^2+32 i \ln^2(2)+32 \pi  \ln (2)}{8 \pi ^3} \\[0.4cm]
 c_7 & c_8 
\end{matrix}\right.\ldots\\[2cm]
&\ldots\qquad  \hspace{13cm}
\left.\begin{matrix}
  60 & 0 & 0 & 0 & 0 & 0\\[0.4cm]
 -c_2 & \frac{i}{2 \pi } & 0 & 0 & 0 & -\frac{i}{6 \sqrt{2} \pi ^2} \\[0.4cm]
 \frac{4 \pi  c_2+744 i-240 i \ln (2)}{\pi } & 0 & 0 & 0 & 0 & \frac{i \sqrt{2}}{3 \pi ^2} \\[0.4cm]
 \frac{\pi ^3 (17 c_2+2 c_6-910)-14475 i \zeta (3)-837 i \pi ^2-480 \pi -240 i-320 i \ln ^3(2)+(-960 \pi +2976 i) \ln ^2(2)+\left(-480 i+5952 \pi +270 i \pi^2\right) \ln (2)}{3 \pi ^3} & \frac{2 \ln (2)-i \pi }{\pi ^2} & -\frac{1}{2 \pi ^2} & 0 & 0 & \frac{i}{3 \sqrt{2} \pi ^2} \\[0.4cm]
 \frac{\pi ^2 (6 c_2-455)-372 i \pi -240-480 \ln ^2(2)+(2976+120 i \pi ) \ln (2)}{\pi ^2} & \frac{8 \ln (2)-4 i \pi }{\pi ^2} & -\frac{2}{\pi ^2} & 0 & 0 &
   \frac{i}{\sqrt{2} \pi ^2} \\[0.4cm]
 \frac{\pi ^3 (23 c_2-c_6+455)+28950 i \zeta (3)+558 i \pi ^2+240 \pi +480 i+640 i \ln ^3(2)+(480 \pi -5952 i) \ln ^2(2)+\left(960 i-2976 \pi -180 i \pi
   ^2\right) \ln (2)}{3 \pi ^3} & \frac{-37 i \pi ^2-32 i \ln ^2(2)-8 \pi  \ln (2)}{8 \pi ^3} & \frac{\pi +8 i \ln (2)}{4 \pi ^3} & -\frac{i}{3 \pi ^3} & 0 & \frac{3 i}{2
   \sqrt{2} \pi ^2} \\[0.4cm]
 \frac{\pi ^3 (33 c_2+910)+28950 i \zeta (3)+7626 i \pi ^2+480 \pi +480 i+640 i \ln ^3(2)+(960 \pi -5952 i) \ln ^2(2)+\left(960 i-5952 \pi -2460 i \pi ^2\right) \ln
   (2)}{\pi ^3} & \frac{4 i \pi ^2-16 i \ln ^2(2)-16 \pi  \ln (2)}{\pi ^3} & \frac{4 \pi +8 i \ln (2)}{\pi ^3} & -\frac{4 i}{3 \pi ^3} & 0 & \frac{11 i}{2 \sqrt{2} \pi ^2}
   \\[0.4cm]
 c_9 & \frac{-2895 \zeta (3)-64 \ln ^3(2)+182 \pi ^2 \ln (2)}{12 \pi ^4} & \frac{96 \ln ^2(2)-91 \pi ^2}{24 \pi ^4} & -\frac{4 \ln (2)}{3 \pi ^4} & \frac{1}{6 \pi
   ^4} & 0 \\[0.4cm]
\end{matrix}\right)
\end{aligned}
\end{align*}
}
}
\end{minipage}
}

\subsection{Data for quotient of \texorpdfstring{$\P_{18,12,3,1,1,1}[36]$}{X36}}\label{sec:appE}

\subsubsection{Integral basis}
We write the integral period vector as \begin{equation}
    \vec{\Pi}=(X^0,X^1,X^2,X^3,H_1,H_2,H_3,H_4,F_3,F_2,F_1,F_0)^{T}
\end{equation}
and choose for $H^{2,2}_{\text{prim}}$ the basis $J_1^2,J_1 J_2,J_1 J_3,J_2 J_3$. We find the period vector $\vec{\Pi}$ in the same way as for $\P_{12,8,1,1,1,1}[24]$ in \cref{sec:appD}. Since the expressions become too lengthy, we will not list them explicitly. The $\eta^{(2,2)}$-block of the intersection form \eqref{eq:sigma} is given by
\begin{align}
     \eta^{(2,2)}=\left(
\begin{array}{cccc}
 0 & 0 & 0 & 1 \\
 0 & 0 & 1 & -3 \\
 0 & 1 & -2 & 0 \\
 1 & -3 & 0 & 0 \\
\end{array}
\right).
 \end{align}

\subsubsection{Analytical continuation}
We give the results of the analytic continuation to the region around the point $z_2=1/4,z_1=z_3=0$. The Frobenius basis in the coordinates $w_1=\frac{\Delta_1}{(1-4z_2)^3},w_2=1-4z_2,w_3=z_3$ is
\begin{align}
\begin{split}
&\varpi_1=\sigma_1\,,\quad
\varpi_2=\sigma_1\log(w_2)+\sigma_2\,,\quad
\varpi_3=\sigma_1\log(w_3)+\sigma_3\\
&\varpi_4=\sigma_1\log(w_2)^2+2\sigma_2\log(w_2)+\sigma_4\,,\\
&\varpi_5=\sigma_1\log(w_2)\log(w_3)+\sigma_2\log(w_3)+\sigma_3\log(w_2)+\sigma_5\,,\\
&\varpi_6=\sigma_1\log(w_3)^2+2\sigma_3\log(w_3)+\sigma_6\,,\\
&\varpi_7=\sigma_1\log(w_2)^2\log(w_3)+2\sigma_2\log(w_2)\log(w_3)+\sigma_3\log(w_2)^2+\sigma_4\log(w_3)\\
&\qquad+2\sigma_5\log(w_2)+\sigma_7\,,\\
&\varpi_8=\sigma_1\left(\log(w_2)\log(w_3)^2+\frac{2}{3}\ln(w_3)^3\right)+\sigma_2\log(w_3)^2+2\sigma_3\left(\log(w_2)\log(w_3)+\log(w_3)^2\right)\\
&\qquad+2\sigma_5\log(w_3)+\sigma_6\left(\log(w_2)+2\log(w_3)\right)+\sigma_8\,,\\
&\varpi_9=\sigma_1\left(\log(w_2)^2\log(w_3)^2+\frac{4}{3}\ln(w_2)\log(w_3)^3+\frac{2}{3}\log(w_3)^4\right)\\
&\qquad+2\sigma_2\left(\log(w_2)\log(w_3)^2+\frac{2}{3}\log(w_3)^3\right)\\
&\qquad+2\sigma_3\left(\log(w_2)^2\log(w_3)+2\log(w_2)\log(w_3)^2+\frac{4}{3}\log(w_3)^3\right)\\
&\qquad+\sigma_4\log(w_3)^2+4\sigma_5\left(\log(w_2)\log(w_3)+\log(w_3)^2\right)\\
&\qquad+\sigma_6\left(\log(w_2)^2+4\log(w_2)\log(w_3)+4\log(w_3)^2\right)\\
&\qquad+2\sigma_7\log(w_3)+2\sigma_8\left(\log(w_2)+2\log(w_3)\right)+\sigma_9\,,\\
&\varpi_{10}=\sigma_{10}\,,\quad
\varpi_{11}=\sigma_{11}\,,\quad
\varpi_{12}=\sigma_{12}\,,\\
\end{split}
\end{align}
where
\begin{equation}
\begin{alignedat}{2}
&\sigma_1=1 + 60w_3 + 20790w_3^2+\dots\,,\\
\quad
&\sigma_2=
\frac{w_1}{3} + w_2 + \frac{w_2^2}{2} + 6930w_3^2 - \frac{w_1^2}{6} + 60w_2w_3 + 20w_1w_3+\dots\,,\\
&\sigma_3=
-\frac{w_2}{2} + 372w_3 - \frac{w_2^2}{4} + 140733w_3^2 - 30w_2w_3\dots\,,\\
&\sigma_4=
w_2^2 + \frac{2w_1w_2}{3} + 13860w_3^2 + \frac{w_1^2}{9}
+\dots\,,\\
&\sigma_5=
w_2 + 120w_3 + \frac{w_2^2}{6} - \frac{w_1w_2}{6} + 87336w_3^2 + 312w_2w_3 + 124w_1w_3+\dots\,,\\
&\sigma_6=
-w_2 - \frac{5w_2^2}{12} + 138384w_3^2 - 312w_2w_3+\dots\,,\\
&\sigma_7=
-4w_2 + 480w_3 - \frac{8w_2^2}{9} + \frac{2w_1w_2}{3} + 278622w_3^2 + 80w_1w_3
+\dots\,,\\
&\sigma_8=
2w_2 - 480w_3 + \frac{4w_2^2}{9} - \frac{w_1w_2}{3} - 189072w_3^2
+\dots\,,\\
&\sigma_9=
-8w_2 - \frac{52w_2^2}{27} + \frac{4w_1w_2}{3} - 385920w_3^2 - 320w_1w_3
+\dots\,,\\
&\sigma_{10}=w_2^{\frac{1}{2}}\left(
1 + \frac{10w_2}{27} + \frac{247w_2^2}{1125} + \frac{w_1w_2}{54} + \frac{w_1^2}{72} + \frac{200w_2w_3}{9}
+\dots\right)\,,\\
&\sigma_{11}=w_2^{\frac{1}{2}}\left(w_1 + \frac{2w_2}{3} + \frac{98w_2^2}{125} + \frac{w_1w_2}{3} - \frac{17w_1^2}{36} + 40w_2w_3
+\dots\right)\,,\\
&\sigma_{12}=w_1^{\frac{3}{2}}w_2^{\frac{1}{2}}\left(
1 - \frac{71w_1}{90} - \frac{w_1w_2}{15} + \frac{14617w_1^2}{22680}
+\dots\right).\\
\end{alignedat}
\end{equation}
The transition matrix determined up to 10 digits is given on the next page.

We introduced again constants $c_1,\dots,c_4$ for which we do not know the exact expressions. These are related by three Legendre relations
\begin{align}
\begin{split}
      &4 \pi ^2 c_1^2-(6 c_1+c_3)^2-36=0\,, \quad (6 c_1+c_3) (6 c_2+c_4)-4 \pi ^2 c_1 c_2=0\,,\\
   &4 \pi ^2 c_2^2-(6c_2+c_4)^2+1=0\,. 
\end{split}
\end{align}
and so there is only one independent undetermined constant. Choosing, for example, $c_1$, its numerical value is 
\begin{align}
    c_1=0.9811667786\dots\,.
\end{align}

\scalebox{0.6}{
\begin{minipage}[t][52cm][c]{\textwidth}
\rotatebox{90}{%
\parbox{1\textwidth}{%
\begin{align*}
\begin{aligned}
&\hspace{18cm}
   T=\left(\begin{matrix}
 1 & 0 & 0 \\[0.4cm]
 -\frac{3 i \log (3)}{2 \pi } & \frac{3 i}{2 \pi } & 0 \\[0.3cm]
 0 & 0 & 0 \\[0.4cm]
 -\frac{i \log (2)}{2 \pi } & 0 & \frac{i}{2 \pi } \\
 \frac{-136 \pi ^2-36 \log ^2(2)-9 \log ^2(3)+\log (2) (-36 \log (3)+72 i \pi )+6 i \pi  \log (3)}{8 \pi ^2} & \frac{-3 i \pi +18 \log (2)+9 \log (3)}{4 \pi ^2} & \frac{-18 i
   \pi +18 \log (2)+9 \log (3)}{2 \pi ^2} \\[0.4cm]
 \frac{-23 \pi ^2-6 \log ^2(2)+\log (2) (-6 \log (3)+12 i \pi )}{4 \pi ^2} & \frac{3 \log (2)}{2 \pi ^2} & \frac{-6 i \pi +6 \log (2)+3 \log (3)}{2 \pi ^2} \\[0.4cm]
 \frac{-23 \pi ^2-6 \log ^2(2)+\log (2) (-6 \log (3)+12 i \pi )}{8 \pi ^2} & \frac{3 \log (2)}{4 \pi ^2} & \frac{-6 i \pi +6 \log (2)+3 \log (3)}{4 \pi ^2} \\[0.4cm]
 \frac{-4 \pi ^2-\log ^2(2)+2 i \pi  \log (2)}{4 \pi ^2} & 0 & \frac{\log (2)-i \pi }{2 \pi ^2} \\[0.4cm]
 \frac{4338 i \zeta (3)+136 \pi ^3+12 i \log ^3(2)+\log ^2(2) (36 \pi +18 i \log (3))+9 \pi  \log ^2(3)+\log (2) \left(-36 i \pi ^2+9 i \log ^2(3)+36 \pi  \log (3)\right)+63
   i \pi ^2 \log (3)}{8 \pi ^3} & \frac{-63 i \pi ^2-18 i \log ^2(2)+\log (2) (-36 \pi -18 i \log (3))-18 \pi  \log (3)}{8 \pi ^3} & \frac{36 i \pi ^2-36 i \log ^2(2)-9 i
   \log ^2(3)+\log (2) (-72 \pi -36 i \log (3))-36 \pi  \log (3)}{8 \pi ^3} \\[0.4cm]
 \frac{4338 i \zeta (3)+69 \pi ^3+12 i \log ^3(2)+\log ^2(2) (18 \pi +18 i \log (3))+\log (2) \left(-148 i \pi ^2+9 i \log ^2(3)+18 \pi  \log (3)\right)+69 i \pi ^2 \log
   (3)}{16 \pi ^3} & \frac{-69 i \pi ^2-18 i \log ^2(2)+\log (2) (-18 \pi -18 i \log (3))}{16 \pi ^3} & \frac{148 i \pi ^2-36 i \log ^2(2)-9 i \log ^2(3)+\log (2) (-36 \pi
   -36 i \log (3))-18 \pi  \log (3)}{16 \pi ^3} \\[0.4cm]
 \frac{714 i \zeta (3)+4 \pi ^3+2 i \log ^3(2)+\log ^2(2) (\pi +3 i \log (3))-23 i \pi ^2 \log (2)}{8 \pi ^3} & -\frac{3 i \log ^2(2)}{8 \pi ^3} & \frac{23 i \pi ^2-6 i \log
   ^2(2)+\log (2) (-2 \pi -6 i \log (3))}{8 \pi ^3} \\[0.4cm]
 \frac{9 \zeta (3) (241 \log (2)+119 \log (3))}{8 \pi ^4}+\frac{-513 \pi ^4+12 \log ^4(2)+24 \log ^3(2) \log (3)-272 \pi ^2 \log ^2(2)-72 \pi ^2 \log ^2(3)+18 \log ^2(2) \log
   ^2(3)-276 \pi ^2 \log (2) \log (3)}{64 \pi ^4} & \frac{3 \left(-714 \zeta (3)-2 \log ^3(2)-3 \log ^2(2) \log (3)+23 \pi ^2 \log (2)+12 \pi ^2 \log (3)\right)}{16 \pi ^4} &
   \frac{-4338 \zeta (3)-12 \log ^3(2)-18 \log ^2(2) \log (3)-9 \log (2) \log ^2(3)+136 \pi ^2 \log (2)+69 \pi ^2 \log (3)}{16 \pi ^4}
\end{matrix}\right.\ldots\\[2cm]
&\ldots\qquad  \hspace{20cm}
\left.\begin{matrix}
 0 & 0 & 0 & 0 & 0 & 0 & 0 & 0 & 0 \\[0.3cm]
 0 & 0 & 0 & 0 & 0 & 0 & 0 & 0 & 0 \\[0.3cm]
 0 & 0 & 0 & 0 & 0 & 0 & -\frac{i c_1}{\pi } & -\frac{i c_2}{\pi } & -\frac{1}{9 \pi ^2} \\[0.3cm]
 0 & 0 & 0 & 0 & 0 & 0 & \frac{i c_1}{2 \pi } & \frac{i c_2}{2 \pi } & \frac{1}{18 \pi ^2} \\[0.3cm]
 -\frac{9}{8 \pi ^2} & -\frac{9}{2 \pi ^2} & -\frac{9}{2 \pi ^2} & 0 & 0 & 0 & -\frac{9 i c_1}{2 \pi } & -\frac{9 i c_2}{2 \pi } & -\frac{1}{2 \pi ^2} \\[0.3cm]
 0 & -\frac{3}{2 \pi ^2} & -\frac{3}{2 \pi ^2} & 0 & 0 & 0 & -\frac{3 i c_1}{2 \pi } & -\frac{3 i c_2}{2 \pi } & -\frac{1}{6 \pi ^2} \\[0.3cm]
 0 & -\frac{3}{4 \pi ^2} & -\frac{3}{4 \pi ^2} & 0 & 0 & 0 & \frac{-6 i \pi  c_1-6 c_1-c_3}{4 \pi ^2} & \frac{-6 i \pi  c_2-6
   c_2-c_4}{4 \pi ^2} & -\frac{1}{6 \pi ^2} \\[0.3cm]
 0 & 0 & -\frac{1}{4 \pi ^2} & 0 & 0 & 0 & -\frac{i c_1}{2 \pi } & -\frac{i c_2}{2 \pi } & -\frac{1}{18 \pi ^2} \\[0.3cm]
 \frac{9 \pi +9 i \log (2)}{8 \pi ^3} & \frac{9 (2 \pi +2 i \log (2)+i \log (3))}{4 \pi ^3} & \frac{9 (2 \pi +2 i \log (2)+i \log (3))}{4 \pi ^3} & -\frac{9 i}{8 \pi ^3} &
   -\frac{9 i}{4 \pi ^3} & 0 & \frac{37 i c_1}{2 \pi } & \frac{37 i c_2}{2 \pi } & \frac{37}{18 \pi ^2} \\[0.3cm]
 \frac{9 i \log (2)}{16 \pi ^3} & \frac{9 (\pi +2 i \log (2)+i \log (3))}{8 \pi ^3} & \frac{9 (\pi +2 i \log (2)+i \log (3))}{8 \pi ^3} & -\frac{9 i}{16 \pi ^3} & -\frac{9
   i}{8 \pi ^3} & 0 & \frac{3 (24 i \pi  c_1+6 c_1+c_3)}{8 \pi ^2} & \frac{3 (24 i \pi  c_2+6 c_2+c_4)}{8 \pi ^2} & \frac{19}{18 \pi
   ^2} \\
 0 & \frac{3 i \log (2)}{4 \pi ^3} & \frac{\pi +6 i \log (2)+3 i \log (3)}{8 \pi ^3} & 0 & -\frac{3 i}{8 \pi ^3} & 0 & 0 & 0 & 0 \\[0.3cm]
 \frac{9 \log ^2(2)}{32 \pi ^4}-\frac{9}{8 \pi ^2} & \frac{3 \left(-23 \pi ^2+6 \log ^2(2)+6 \log (2) \log (3)\right)}{16 \pi ^4} & \frac{-136 \pi ^2+36 \log ^2(2)+9 \log
   ^2(3)+36 \log (2) \log (3)}{32 \pi ^4} & -\frac{9 \log (2)}{16 \pi ^4} & -\frac{9 (2 \log (2)+\log (3))}{16 \pi ^4} & \frac{9}{32 \pi ^4} & 0 & 0 & 0 \\
\end{matrix}\right)
\end{aligned}
\end{align*}
}
}
\end{minipage}
}

\subsubsection{Integer Invariants}
In the following, we collect the genus zero and genus one invariants \cite{Klemm:2007in2} for the model $\P_{18,12,3,1,1,1}[36]$. The genus zero invariants are given in \Cref{tab:ffinstantons1,tab:ffinstantons2,tab:ffinstantons3,tab:ffinstantons4}. For the genus one invariants we compute the genus one free energy 
\begin{align}
    F^{(1)}=\left(\frac{\chi}{24}-h^{1,1}-2\right)\log X_0+\log\det\left(\frac{1}{2\pi \ii}\frac{\partial \underline{z}}{\partial \underline{t}}\right)+\sum_i b_{z_i} \log z_i+ \sum_\alpha b_{\Delta_\alpha} \log \Delta_\alpha\,.
\end{align}
The last two terms correspond to a holomorphic ambiguity that has to be fixed by the boundary behaviour at the components of the discriminant. For the divisors meeting at the MUM point the parameters are known to be given by
\begin{align}
  b_{z_i}=-\frac{1}{24}\int_{\hat{Y}} c_3\wedge J_i -1\,.
\end{align}
For conifold components the behaviour is universally $b_{\Delta_\text{con}}=-1/24$. For the model at hand this leaves only the constant $b_{\Delta_3}$ undetermined and consequently we express the genus one integer invariants in \cref{tab:ffgenusone} in terms of $b\equiv b_{\Delta_3}$. 

\begin{table}[H]
    \centering
    \vspace{0.3cm}
    \resizebox{0.8\textwidth}{!}{
    \begin{tabular}{|c|c|c|c|c|c|}
        \hline
        $n^{(0)}_{(d_1, d_2,0)}(H_1)$ & $d_2 = 0$ & 1 & 2 & 3 & 4 \\
        \hline
        $d_1 = 0$ & * &-3&0&0&0 \\
        1 & 0 & 6 & 3 & -102 & -459 \\
        2 &  0 & -15 & -12 & -9 & 912 \\
        3 &  0 & 96 & 63 & 54 & -2250 \\
        4 & 0 & -858 & -540 & -459 & 14304 \\
        \hline
        $n^{(0)}_{(d_1, d_2,1)}(H_1)$ & $d_2 = 0$ & 1 & 2 & 3 & 4 \\
        \hline
        $d_1 = 0$ &  0 & 720 & 0 & 0 & 0\\
        1 &0 & -1440 & 0 & 51840 & 244800  \\
        2 &0 & 3600 & 0 & 0 & -489600 \\
        3 & 0 & -23040 & 0 & 0 & 1224000  \\
        4 &  0 & 205920 & 0 & 0 & -7833600 \\
        \hline
        $n^{(0)}_{(d_1, d_2,2)}(H_1)$ & $d_2 = 0$ & 1 & 2 & 3 & 4 \\
        \hline
        $d_1 = 0$ &0 & 424332 & 1440 & 0 & 0 \\
        1 &  0 & -848664 & -754920 & -15260400 & -69125400 \\
        2 & 0 & 2121660 & 3016800 & 2245320 & 139747680\\
        3 & 0 & -13578624 & -26331480 & -17962560 & -360595800\\
        4 &0 & 121358952 & 300549600 & 195342840 & 2373675840 \\
        \hline
    \end{tabular}}
      \caption{Genus 0 invariants associated to $H_1$ of $\P_{18,12,3,1,1,1}[36]$ for degree $d_3 = 0, 1, 2$ corresponding to the elliptic fibre.}
         \label{tab:ffinstantons1}
\end{table}

\begin{table}[H]
    \centering
    \vspace{0.3cm}
    \resizebox{0.75\textwidth}{!}{
    \begin{tabular}{|c|c|c|c|c|c|}
        \hline
        $n^{(0)}_{(d_1, d_2,0)}(H_2)$ & $d_2 = 0$ & 1 & 2 & 3 & 4 \\
        \hline
        $d_1 = 0$ & * & 0 & 0 & 0 & 0 \\
        1 &0 & 3 & -6 & -39 & -108 \\
        2 & 0 & -6 & 3 & 12 & 372 \\
        3 &  0 & 36 & -18 & -45 & -882 \\
        4 & 0 & -312 & 144 & 318 & 5550  \\
        \hline
        $n^{(0)}_{(d_1, d_2,1)}(H_2)$ & $d_2 = 0$ & 1 & 2 & 3 & 4 \\
        \hline
        $d_1 = 0$ &  0 & 0 & 0 & 0& 0\\
        1 & 0 & -720 & 2160 & 18000 & 55440  \\
        2 &0 & 1440 & -2160 & -2160 & -191520  \\
        3 & 0 & -8640 & 10800 & 8640 & 442800 \\
        4 &  0 & 74880 & -86400 & -62640 & -2774160 \\
        \hline
        $n^{(0)}_{(d_1, d_2,2)}(H_2)$ & $d_2 = 0$ & 1 & 2 & 3 & 4 \\
        \hline
        $d_1 = 0$ & 0 & 0 & 0 & 0 & 0 \\
        1 &  0 & -424332 & -754920 & -4581360 & -14550840\\
        2 &  0 & 848664 & 1508400 & -767880 & 51565680\\
        3 & 0 & -5091984 & -11284920 & 4568400 & -110169720\\
        4 &0 & 44130528 & 120219840 & -46218600 & 654151680 \\
        \hline
    \end{tabular}}
      \caption{Genus 0 invariants associated to $H_2$ of $\P_{18,12,3,1,1,1}[36]$ for degree $d_3 = 0, 1, 2$ corresponding to the elliptic fibre.}
        \label{tab:ffinstantons2}
\end{table}

\begin{table}[H]
    \centering
    \vspace{0.3cm}
    \resizebox{0.75\textwidth}{!}{
    \begin{tabular}{|c|c|c|c|c|c|}
        \hline
        $n^{(0)}_{(d_1, d_2,0)}(H_3)$ & $d_2 = 0$ & 1 & 2 & 3 & 4 \\
        \hline
        $d_1 = 0$ & * & 0 & 0 & 0 & 0 \\
        1 &3 & 0 & 0 & 0 & 0 \\
        2 &  -12 & 0 & 0 & 0 & 0 \\
        3 &   81 & 0 & 0 & 0 & 0 \\
        4 &  -768 & 0 & 0 & 0 & 0 \\
        \hline
        $n^{(0)}_{(d_1, d_2,1)}(H_3)$ & $d_2 = 0$ & 1 & 2 & 3 & 4 \\
        \hline
        $d_1 = 0$ &   720 & 720 & 0 & 0 & 0\\
        1 &  0 & -1440 & 0 & 14400 & 50400 \\
        2 & 0 & 3600 & 0 & 0 & -100800  \\
        3 &  0 & -23040 & 0 & 0 & 252000 \\
        4 & 0 & 205920 & 0 & 0 & -1612800  \\
        \hline
        $n^{(0)}_{(d_1, d_2,2)}(H_3)$ & $d_2 = 0$ & 1 & 2 & 3 & 4 \\
        \hline
        $d_1 = 0$ & 1440 & 848664 & 1440 & 0 & 0\\
        1 & 0 & -2196288 & -6480 & -6156000 & -24591600\\
        2 &  0 & 6738120 & 23040 & 0 & 49183200\\
        3 &0 & -51107328 & -136080 & 0 & -122958000\\
        4 &0 & 528123024 & 1173600 & 0 & 786931200 \\
        \hline
    \end{tabular}}
      \caption{Genus 0 invariants associated to $H_3$ of $\P_{18,12,3,1,1,1}[36]$ for degree $d_3 = 0, 1, 2$ corresponding to the elliptic fibre.}
    \label{tab:ffinstantons3}
\end{table}

\begin{table}[H]
    \centering
    \vspace{0.3cm}
    \resizebox{0.8\textwidth}{!}{
    \begin{tabular}{|c|c|c|c|c|c|}
        \hline
        $n^{(0)}_{(d_1, d_2,0)}(H_4)$ & $d_2 = 0$ & 1 & 2 & 3 & 4 \\
        \hline
        $d_1 = 0$ & * & 0 & 0 & 0 & 0 \\
        1 &0 & 0 & 0 & 0 & 0 \\
        2 &0 & 0 & 0 & 0 & 0  \\
        3 &0 & 0 & 0 & 0 & 0  \\
        4 & 0 & 0 & 0 & 0 & 0 \\
        \hline
        $n^{(0)}_{(d_1, d_2,1)}(H_4)$ & $d_2 = 0$ & 1 & 2 & 3 & 4 \\
        \hline
        $d_1 = 0$ &   2160 & 2160 & 0 & 0 & 0 \\
        1 & 0 & -4320 & 0 & 43200 & 151200 \\
        2 & 0 & 10800 & 0 & 0 & -302400 \\
        3 & 0 & -69120 & 0 & 0 & 756000  \\
        4 & 0 & 617760 & 0 & 0 & -4838400  \\
        \hline
        $n^{(0)}_{(d_1, d_2,2)}(H_4)$ & $d_2 = 0$ & 1 & 2 & 3 & 4 \\
        \hline
        $d_1 = 0$ & 4320 & 2795472 & 4320 & 0 & 0\\
        1 &0 & -5590944 & -19440 & -18468000 & -73774800 \\
        2 & 0 & 13977360 & 69120 & 0 & 147549600\\
        3 &0 & -89455104 & -408240 & 0 & -368874000 \\
        4 &0 & 799504992 & 3520800 & 0 & 2360793600  \\
        \hline
    \end{tabular}}
      \caption{Genus 0 invariants associated to $H_4$ of $\P_{18,12,3,1,1,1}[36]$ for degree $d_3 = 0, 1, 2$ corresponding to the elliptic fibre.}
    \label{tab:ffinstantons4}
\end{table}

\begin{table}[H]
    \centering
    \vspace{0.3cm}
    \resizebox{0.8\textwidth}{!}{
    \begin{tabular}{|c|c|c|c|c|c|}
        \hline
        $n^{(1)}_{(d_1, d_2,0)}$ & $d_2 = 0$ & 1 & 2 & 3 & 4 \\
        \hline
        $d_1 = 0$ & * & 0 & 0 & 0 & 0  \\
        1 &$6+27b$ & 0 & 0 & 0 & 0 \\
        2 &$-45 - 243b$ & 0 & 0 & 0 & 0  \\
        3 &$394 + 2394b$ &9 & 0 & -9 & -18  \\
        4 & $-5115 - 32616b$ &-90 & -18 & 144 & 630\\
        \hline
        $n^{(1)}_{(d_1, d_2,1)}$ & $d_2 = 0$ & 1 & 2 & 3 & 4 \\
        \hline
        $d_1 = 0$ &  -18 & -18 & 0 & 0 & 0  \\
        1 & 0 & 36 & 0 & -360 & -1260  \\
        2 & 0 & -90 & 0 & 0 & 2520\\
        3 & 0 & -1584 & 2160 & 2160 & -4140  \\
        4 & 0 & 16452 & -17280 & -30240 & -223200 \\
        \hline
        $n^{(1)}_{(d_1, d_2,2)}$ & $d_2 = 0$ & 1 & 2 & 3 & 4 \\
        \hline
        $d_1 = 0$ &  0 & 4266 & 0 & 0 & 0\\
        1 &0 &$-507492 - 2245320b$ &  -4212 & 180360 & 742500 \\
        2 & 0 & $5509890 + 31434480b$ &  16848 & 0 & -1485000 \\
        3 &0 & $-95213988 - 570311280b$& -2344572 & 10800 & 5968620\\
        4 &0 & $1626089796 + 10306018800b$ &  32300640 & 1362960 & 6093360  \\
        \hline
    \end{tabular}}
      \caption{Genus 1 invariants of $\P_{18,12,3,1,1,1}[36]$ for degree $d_3 = 0, 1, 2$ corresponding to the elliptic fibre.}
      \label{tab:ffgenusone}
\end{table}

\section{Local Zeta Functions}
\label{sec:app7}
The local zeta function can be seen as a generating function for the number of points of the Calabi--Yau over finite fields. More precisely given a Calabi--Yau variety $X$ defined as a complete intersection in a projective space, we can define the variety $X_p\coloneqq X/\mathbb{F}_p$ by considering the defining equations mod $p$. The local zeta function of $X_p$ is then defined as
\begin{equation}
    \zeta_p(X/\mathbb{F}_p,T)=\exp\left(\sum_{n=1}^{\infty}\#X_p(\mathbb{F}_{p^n})\frac{T^n}{n}\right)
\end{equation}
where $\#X_p(\mathbb{F}_{p^n})$ is the number of solutions to the defining equations of $X_p$ with the variables as elements of $\mathbb{F}_{p^n}$. The Weil conjectures\cite{bams/1183513798} strongly constrain the form of the local zeta function and for Calabi--Yau three-folds result in 
\begin{align}
    \zeta \left(X / \mathbb{F}_p, T \right) = \frac{P_3(X / \mathbb{F}_p, T)}{(1 - T)(1 - pT)^{h^{1,1}}(1 - p^2 T)^{h^{1,1}}(1 - p^3 T)}
\end{align}
reducing its computation to only the degree $b_3$ polynomial $P_3(X / \mathbb{F}_p, T)$. 

For families of Calabi--Yau manifolds powerful methods have been developed to compute $P_3(X / \mathbb{F}_p, T)$ from the periods of the holomorphic $(3,0)$-form of the Calabi--Yau \cite{Candelas:2021tqt,Candelas:2024vzf}, which we now briefly summarize. The Frobenius map under which $\uz \mapsto \uz^p=(z_1^p,\dots,z_{r}^p)$ induces an action $F_p(\underline{z})$ on the middle cohomology that is compatible with the Gauss--Manin connection and whose characteristic polynomial is the numerator of the local zeta function:
\begin{align}
    P_3\left( X_\uz / \mathbb{F}_p, T \right)=\text{det}(1-TF_p(\uz))\vert_{\uz=\text{Teich}(\underline{z})}\,.
\end{align}
Here, $\text{Teich}(\uz)\in \mathbb{Z}_p$ denotes the component-wise Teichmüller lift of $\uz$ with $\text{Teich}(\uz) =  \text{Teich}(\uz^p)$ and the Frobenius action is 
\begin{align}
\label{eq:frob}
   F_p(\uz)=\boldsymbol{\Pi}(\uz^p)^{-1}V_0\boldsymbol{\Pi}(\uz),
\end{align}
with the period matrix $\boldsymbol{\Pi}(\uz)$ as defined in \eqref{eq:generalperiodmatrix} using the Frobenius basis \eqref{eq:generalfrobeniusbasis}. The matrix $V_0$ depends on $p$ only and is (in part conjecturally) given by
\begin{align}
    V_0=\begin{pmatrix}
1 & 0 & 0 & 0 \\
0 & p\mathds{1}_{r\times r} & 0 & 0 \\
0 & 0 & p^2\mathds{1}_{r\times r} & 0 \\
c\,\zeta_p(3) p^3& 0 & 0 & p^3
\end{pmatrix}
\end{align}
where the constant $c$ in the chosen basis is given by the Euler number $\chi$ and $\zeta_p(3)$ is a $p$-adic zeta value. 

When working with Calabi--Yau operators of one-parameter models in the AESZ list where the underlying geometry and topological data is not known, instead of using the Frobenius basis of solutions normalised with the double- and triple-logarithmic periods multiplied by the triple intersection number $\kappa$ we use the basis  
\begin{align}
\label{eq:Frob1param}
    \underline{\omega}(z)=\begin{pmatrix}
\hfill \sigma_0(z) \\
\hfill \sigma_0(z) \log(z) + \sigma_1(z) \\
\hfill \frac{1}{2}\sigma_0(z) \log^2(z)+ \sigma_1(z) \log(z) + \sigma_2(z) \\
\hfill\frac{1}{6} \sigma_0(z) \log^3(z)+\frac{1}{2} \sigma_1(z) \log^2(z) + \sigma_2(z) \log(z) + \sigma_3(z)
\end{pmatrix}.
\end{align}
with $\sigma_0(0)=1$ and $\sigma_1(0)=\sigma_2(0)=\sigma_3(0)=0$ and for the Wronskian $\boldsymbol{\Pi}_{ij}(z)=\theta^j\omega_i(z)$. The value of $c$ is then $\chi/\kappa$ and its ratio can be determined as follows. The logarithms in the expression \eqref{eq:frob} cancel and so the entries of $F_p$ are elements in $\mathbb{Q}_p\!\left[\!\left[ z \right]\!\right]$. Due to bounds on the coefficients of $P_3\left( X_z / \mathbb{F}_p, T \right)$ implied by the Weil conjectures it suffices to calculate to finite $p$-adic order. To finite $p$-adic precision the entries of the Frobenius converge to rational functions, whose denominator  
is known conjecturally as some power of the discriminant, the exponent growing linearly with $p$. Hence it suffices to calculate periods to a finite order, growing as well linearly with $p$, to compute the local zeta function anywhere in the moduli space. The convergence of the Frobenius to a rational function only happens for the correct value of $c$, thus offering a way to calculate its value by demanding the Frobenius to converge to finite $p$-adic precision to a rational function of the expected form. More precisely, for some operators in the AESZ list, such as the operator $\mathcal{L}^{(4)}_{2.45}$ discussed in \cref{sec:AntisymProducts}, the monodromy around some points contain values in field extensions $\mathbb{Q}(\sqrt{D})$. As observed in \cite{Thorne2018} for those primes $p$ for which $D$ is not a quadratic residue modulo $p$ the Frobenius is then not a rational function. Instead, the denominator contains square roots of the corresponding irreducible factors of the discriminant.

As noted in \cref{sec:fivefoldsandgeneral} the definition of the Wronskian used in \cite{Candelas:2024vzf} differs from our period matrix in that only the classical part of the couplings is used. This does not affect the result. While the rational function expression of the Frobenius may depend on the convention that is used, the difference corresponds to a rational basis change $U\in \text{Mat}_{b_3\times b_3}(\mathbb{Q}(\uz))$ of the Frobenius that drops out of its characteristic polynomial $P_3(X_\uz/\mathbb{F}_p,T)$ upon inserting the Teichmüller lift.

In \cite{Jockers:2023zzi} the above method to calculate the local zeta functions was generalised to one-parameter four-folds. The Frobenius takes for operators with horizontal middle cohomology the analogous form as for three-folds with the matrix $V_0$ given conjecturally by 
\begin{align}
    V_0=\begin{pmatrix}
1 & 0 & 0 & 0 & 0 \\
0 & p & 0 & 0 & 0 \\
0 & 0 & p^2 & 0 & 0 \\
c\,\zeta_p(3) p^3& 0 & 0 & p^3 & 0 \\
0 & c\,\zeta_p(3) p^4 & 0 & 0 & p^4
\end{pmatrix},
\end{align}
with the Frobenius basis chosen analogous to \eqref{eq:Frob1param} with Wronskian $\boldsymbol{\Pi}_{ij}(z)=\theta^j\omega_i(z)$ and for all the examples discussed in \cref{sec:AntisymProducts} the constant $c$, determined by demanding the Frobenius to converge to the necessary $p$-adic precision to a rational function, takes the value $c_3 \cdot D/\kappa$.

\newpage

\bibliography{refs.bib}
\end{document}